\documentclass[12pt,twoside]{report}
\usepackage[english]{babel}

\usepackage{geometry}
\usepackage{setspace}
\usepackage{titlesec}
\usepackage[subfigure]{tocloft}

\usepackage[toc,page]{appendix}

\usepackage[hyphens]{url} 

\usepackage{multirow,dsfont,pifont}
\usepackage{color}
\usepackage{array}

\usepackage[
			sort,
			round,
			semicolon]{natbib}

\usepackage{bibentry}
\nobibliography*

\usepackage[linesnumbered,algoruled,boxed,lined]{algorithm2e}

\SetKwInput{KwInput}{Input}
\SetKwInput{KwOutput}{Output}

\usepackage[caption=false]{subfig} 
\usepackage{epsfig}
\usepackage{booktabs}
\usepackage{multicol}
\usepackage{listings}
\usepackage[bf,hang]{caption}

\usepackage{amsthm}
\usepackage{amsmath}
\usepackage{amssymb}

\usepackage{palatino} 
\usepackage{microtype}
\usepackage{textcomp}

\usepackage{fancyhdr}

\usepackage[dvipsnames,table]{xcolor} 
\usepackage[
			colorlinks=true,
			citecolor=MidnightBlue,
			urlcolor=Magenta,
			linkcolor=MidnightBlue,
			breaklinks=True,
			linktocpage=true 
			]{hyperref}

\usepackage{Definitions}

\usepackage{hhline}

\usepackage{rotating}
\usepackage{pdflscape}
\usepackage{afterpage}
\newcommand\blankpage{%
    \null
    \thispagestyle{empty}%
    \addtocounter{page}{-1}%
    \newpage}

\usepackage{chngcntr}
\counterwithout{footnote}{chapter}

\usepackage{enumitem}
\usepackage[T1]{fontenc}

\titleformat{\chapter}[display]
{\normalfont\huge\filleft\bfseries}
{
\vspace{1ex} 
\textcolor{gray}{\Large \textsc \chaptertitlename}
\textcolor{gray}{\fontsize{70}{70} \selectfont \thechapter}
}
{10pt} 
{\Huge}
[
\vspace{0ex}
{\titlerule[1pt]}
]

\titleformat{name=\chapter,numberless}[display]
{\normalfont\huge\filleft\bfseries}
{}
{0pt}
{
\vspace{1ex}%
\Huge}
[\vspace{0ex}
{\titlerule[1pt]}
] 

\titlespacing*{\chapter} {0pt}{20pt}{30pt}   
\titlespacing*{name=\chapter,numberless} {0pt}{20pt}{30pt}   

\titlespacing{\paragraph}{0in}{0.08in}{0.07in}

\usepackage[group-separator={,},per-mode=symbol,binary-units=true]{siunitx}

\xspace{}

\renewcommand{\cite}{\citep}

\newcommand{\dsa}[0]{$\mathcal{A}$\xspace}
\newcommand{\dsb}[0]{$\mathcal{B}$\xspace}
\newcommand{\dsc}[0]{$\mathcal{C}$\xspace}
\newcommand{\dsd}[0]
{$\mathcal{D}$\xspace}
\newcommand{\dse}[0]
{$\mathcal{E}$\xspace}
\newcommand{\DefTxFee}[0]{\SI{1}{sat{\per}B}}

\newcommand{\feeunit}{BTC/kB}
\newcommand{\gbt}{\stress{GetBlockTemplate}\xspace}

\newcommand{\mpool}{Mempool\xspace}
\newcommand{\pow}[0]{proof-of-work\xspace}

\newcommand{\usd}[1]{\SI{#1}{\second}}
\newcommand{\uGB}[1]{\SI{#1}{\giga\byte}}
\newcommand{\uGHz}[1]{\SI{#1}{\giga\hertz}}

\newcommand{\uMB}[1]{\SI{#1}{MB}}
\newcommand{\uTB}[1]{\SI{#1}{\tera\byte}}

\newcommand{\uBTC}[1]{\SI{#1}{BTC}}

\newcommand{\uTxFee}[1]{\SI{#1}{BTC{\per}kB}}

\newlength{\onecolgrid}
\newlength{\twocolgrid}
\newlength{\threecolgrid}
\newlength{\fourcolgrid}
\newcommand{\tabcap}[1]{\caption{\textit{#1}}}
\newcommand{\thead}[1]{\textbf{\textit{#1}}}
\newcommand{\term}[1]{\emph{#1}}
\newcommand{\newterm}[1]{\emph{#1}}
\newcommand{\stress}[1]{\textit{#1}}

\definecolor{maroon}{rgb}{0.5, 0.0, 0.0}
\newcommand{\attention}[1]{\textcolor{maroon}{\textbf{#1}}}%

\newcommand{\red}{\textcolor[HTML]{e41a1c}}
\newcommand{\green}{\textcolor[HTML]{4daf4a}}
\newcommand{\blue}{\textcolor[HTML]{377eb8}}
\newcommand{\grey}{\textcolor[HTML]{b7b7b7}}
\newcommand{\purple}{\textcolor[HTML]{6e18ee}}

\newcommand{\tsup}[1]{\ensuremath{^{\textrm{#1}}}}

\newcommand{\parai}[1]{\vspace{0.03in}\noindent{\textit{#1}}\quad}
\newcommand{\paraib}[1]{\vspace{0.03in}\noindent{\textit{\textbf{#1}}}\quad}

\newcommand{\point}{{\noindent}~$\blacktriangleright$\,}

\usepackage{mdframed}

\begin{document}

\pagenumbering{roman}

\pagestyle{plain} 

%
\begin{titlepage}
\begin{center}


\rule{\linewidth}{2pt}
\begin{singlespace}
\LARGE \bf
On Fairness Concerns in the Blockchain Ecosystem
\end{singlespace}
\rule{\linewidth}{2pt}

\vspace{150pt}
\begin{singlespace}
\noindent \Large
A dissertation submitted towards the degree\\
Doctor of Engineering\\
of the Faculty of Mathematics and Computer Science of\\
Saarland University
\end{singlespace}

\vspace{150pt} 
\large by\\
\large Johnnatan Messias Peixoto Afonso\\
\end{center}

\begin{center}
\begin{singlespace} 
\large Saarbrücken\\
\large  2023
\end{singlespace}
\end{center}

\end{titlepage}
%

\null
\vfill
\begin{flushbottom}
\begin{singlespace}
\begin{tabular}{ p{ 0.5\linewidth} l}
{\bf Date of Colloquium:} &  April 25\tsup{th}, 2024\\

{\bf Dean of Faculty:} & Univ.-Prof. Dr. Roland Speicher\\
\\
{\bf Chair of the Committee:} & Prof. Dr. Anja Feldmann\\
{\bf Reporters} & \\
{\bf First Reviewer:} & Prof. Dr. Krishna P. Gummadi\\
{\bf Second Reviewer:} & Prof Dr. Ingmar Weber\\
{\bf Third Reviewer:} & Prof. Dr. Balakrishnan Chandrasekaran\\
{\bf Fourth Reviewer:} & Prof. Dr. Patrick Loiseau\\
{\bf Academic Assistant:} & Dr. Abhisek Dash
\end{tabular}
\end{singlespace}
\end{flushbottom}

\clearpage

%

\begin{center}
\begin{singlespace}
\copyright 2023\\
Johnnatan Messias Peixoto Afonso \\
ALL RIGHTS RESERVED
\end{singlespace}
\end{center}

\clearpage


\chapter*{Abstract}

Blockchains revolutionized centralized sectors like banking and finance by promoting decentralization and transparency. In a blockchain, information is transmitted through transactions issued by participants or applications. Miners crucially select, order, and validate pending transactions for block inclusion, prioritizing those with higher incentives or fees. The order in which transactions are included can impact the blockchain final state.

Moreover, applications running on top of a blockchain often rely on governance protocols to decentralize the decision-making power to make changes to their core functionality. These changes can affect how participants interact with these applications. Since one token equals one vote, participants holding multiple tokens have a higher voting power to support or reject the proposed changes. The extent to which this voting power is distributed is questionable and if highly concentrated among a few holders can lead to governance attacks.

In this thesis, we audit the Bitcoin and Ethereum blockchains to investigate the norms followed by miners in determining the transaction prioritization. We also audit decentralized governance protocols such as Compound to evaluate whether the voting power is fairly distributed among the participants. Our findings have significant implications for future developments of blockchains and decentralized applications.

\clearpage

%

\chapter*{Zusammenfassung}

Blockchain-Technologien revolutionierten zentralisierte Bereiche wie Bankwesen und Finanzen, indem sie Dezentralisierung und Transparenz förderten. In einer Blockchain wird Informationen durch Transaktionen übertragen, die von Teilnehmern oder Anwendungen ausgestellt werden. Miner wählen Transaktionen aus, ordnen sie an und validieren sie für die Aufnahme in einen Block. Dabei priorisieren sie jene Transaktionen mit höheren Gebühren. Die Reihenfolge, in der Transaktionen aufgenommen werden, kann den endgültigen Zustand der Blockchain beeinflussen.

Anwendungen, die auf einer Blockchain laufen, oft auf Governance-Protokolle angewiesen, um die Entscheidungsbefugnis zur Änderung ihrer Kernfunktionalität zu dezentralisieren. Diese Änderungen können beeinflussen, wie Teilnehmer mit diesen Anwendungen interagieren. Da ein Token einem Stimmrecht entspricht, haben Teilnehmer mit mehreren Tokens eine höhere Abstimmungsbefugnis, um die vorgeschlagenen Änderungen zu unterstützen oder abzulehnen. Fraglich ist, inwieweit diese Abstimmungsbefugnis verteilt ist.

In dieser Arbeit prüfen wir die Bitcoin- und Ethereum-Blockchains, um die Normen zu untersuchen, denen Miner folgen, um die Priorisierung von Transaktionen festzulegen. Wir überprüfen dezentrale Governance-Protokolle wie Compound, um festzustellen, ob die Abstimmungsbefugnis fair unter den Teilnehmern verteilt ist. Unsere Ergebnisse haben wesentliche Auswirkungen auf zukünftige Entwicklungen von Blockchains und dezentralen Anwendungen.

\clearpage

%

\chapter*{Publications}

\noindent {\bf Parts of this thesis have appeared in the following publications and technical reports.}

\begin{itemize}
\item
``Understanding Blockchain Governance: Analyzing Decentralized Voting to Amend DeFi Smart Contracts''. 
{\bf J. Messias}, V. Pahari, B. Chandrasekaran, K. P. Gummadi, and P. Loiseau. 
This work has been submitted and we are currently awaiting a decision.

\item
``Dissecting Bitcoin and Ethereum Transactions: On the Lack of Transaction Contention and Prioritization Transparency in Blockchains''. 
{\bf J. Messias}, V. Pahari, B. Chandrasekaran, K. P. Gummadi, and P. Loiseau. 
In \stress{Proceedings of the 27\tsup{th} Financial Cryptography and Data Security (FC)}, Bol, Brač, Croatia, May 2023.

\item
``Selfish \& Opaque Transaction Ordering in the Bitcoin Blockchain: The Case for Chain Neutrality''. 
{\bf J. Messias}, M. Alzayat, B. Chandrasekaran, K. P. Gummadi, P. Loiseau, and A. Mislove. 
In \stress{Proceedings of the 21\tsup{st} ACM SIGCOMM Internet Measurement Conference (IMC)}, Virtual Event, November 2021.

\item
``On Blockchain Commit Times: An analysis of how miners choose Bitcoin transactions''. 
{\bf J. Messias}, M. Alzayat, B. Chandrasekaran, and K. P. Gummadi. 
In \stress{2\tsup{nd} International KDD Workshop on Smart Data for Blockchain and Distributed Ledger (SDBD)}, Virtual Event, August 2020.
\end{itemize}

\noindent {\bf Additional publications and technical reports while at MPI-SWS.}

\begin{itemize}
\item
``Modeling Coordinated vs. P2P Mining: An Analysis of Inefficiency and Inequality in Proof-of-Work Blockchains''. 
M. Alzayat, {\bf J. Messias}, B. Chandrasekaran, K. P. Gummadi, and P. Loiseau.
June 2021. (\textbf{Technical report})

\item
``(Mis)Information Dissemination in WhatsApp: Gathering, Analyzing and Countermeasures''. 
G. Resende, P. Melo, H. Sousa, {\bf J. Messias}, M. Vasconcelos, J. Almeida, and F. Benevenuto.
In \textit{Proceedings of the 28\tsup{th} Web Conference (WWW)}, San Francisco, USA, May 2019.

\item
``WhatsApp Monitor: A Fact-Checking System for WhatsApp''. 
P. Melo, {\bf J. Messias}, G. Resende, K. Garimella, J. Almeida, and F. Benevenuto.
In \textit{Proceedings of the 13\tsup{th} International AAAI Conference on Web and Social Media (ICWSM)}, Munich, Germany, June 2019.

\item
``Search Bias Quantification: Investigating Political Bias in Social Media and Web Search''. 
J. Kulshrestha, M. Eslami, {\bf J. Messias}, M. B. Zafar, S. Ghosh, K. P. Gummadi, and K. Karahalios.
In \textit{Information Retrieval Journal}, Springer. Volume 22, Issue 1-2, April 2019.

\item
``On Microtargeting Socially Divisive Ads: A Case Study of Russia-Linked Ad Campaigns on Facebook''. 
F. N. Ribeiro, K. Saha, M. Babaei, L. Henrique, {\bf J. Messias}, F. Benevenuto, O. Goga, K. P. Gummadi, and E. M. Redmiles.
In \textit{Proceedings of the Conference on Fairness, Accountability, and Transparency (FAT*)}, Atlanta, Georgia. January 2019.

\end{itemize}

\clearpage


I dedicate my thesis to the following people who have played an important role in my life: my wife Aline, my father J. Missias~\cite{JotaMissias@Wiki}, my mother Varlene, and my brothers Joarlens and Jeanderson.


\chapter*{Acknowledgements}

I would like to express my gratitude to my advisor, Krishna P. Gummadi, for his invaluable feedback and unwavering support throughout my PhD journey. I am also deeply thankful to Balakrishnan Chandrasekaran and Patrick Loiseau for their constructive insights and encouragement in my studies. I must also extend my appreciation to my previous advisor, Fabrício Benevenuto, whose consistent guidance and support carried me through this PhD journey.

Working alongside remarkable individuals from our Networked Systems group has been a privilege. I would like to thank Ayan Majumdar, Abhisek Dash, Camila Kolling, David Miller, Junaid Ali, Sepehr Mousavi, Till Speicher, Vabuk Pahari, Vedant Nanda, Nina Grgić-Hlača, and many others who have made an impact in my PhD journey.

I extend my thanks to the exceptional research assistants at MPI-SWS, who enthusiastically celebrated every milestone in my doctoral studies. A special mention goes to Mohamed Alzayat, whose feedback and collaboration were instrumental as I embarked on my research topics. I am also grateful to Ahana Ghosh, Andi Nika, Andrea Borgarelli, Angelica Goetzen, Chao Wen, Debasmita Lohar, George Tzannetos, Heiko Becker, Jan-Oliver Kaiser, Lennard Gäher, Matheus Stolet, Michael Sammler, Mihir Vahanwala, Nils Müller, Oshrat Ayalon, Pierfrancesco Ingo, Ralf Jung, Rati Devidze, Roberta De Viti, Victor Alexandru Padurean, Vaastav Anand, and many others for their steadfast support.

I am grateful for the vibrant interactions I had with interns who shared and discussed intriguing research ideas, with a special shoutout to Aleksa Sukovic, Ani Saxena, Ana-Andreea Stoica, Baltasar Dinis, Barbara Gomes, Daniel Kansaon, Diogo Antunues, Isadora Salles, Ignacio Tiraboschi, Lucas Costa De Lima, Maria Petrisor, Pedro Las-Casas, Ruchit Rawal, Sapana Chaudhary, and many others.

I am thankful to all MPI-SWS staff, in particular, Annika Meiser, Carina Schmitt, Christian Klein, Claudia Richter, Gretchen Gravelle, Krista Ames, Maria-Louise Albrecht, Rose Hoberman, and Sarah Naujoks for their sincere efforts and help during my PhD studies.

My appreciation extends to my wife Aline for her unwavering love and support throughout this journey. I also want to extend my thanks to my father, Jota Missias, my mother, Varlene, and my brothers, Joarlens and Jeanderson, for their unconditional support. 

\clearpage


%
\phantomsection
\renewcommand{\contentsname}{}
\chapter*{Table of contents}

\vspace*{-35mm}

\renewcommand{\cftchapfont}{\normalfont}
\renewcommand{\cftchappagefont}{\normalfont}
\renewcommand{\cftchapleader}{\cftdotfill{\cftdotsep}}

\setlength{\cftbeforechapskip}{15pt}
\setlength{\cftbeforesecskip}{10pt}
\setlength{\cftbeforesubsecskip}{10pt}
\setlength{\cftbeforesubsubsecskip}{10pt}

\begin{singlespace}
\tableofcontents
\end{singlespace}

\clearpage


%

\chapter*{List of figures}
\renewcommand{\listfigurename}{}

\addcontentsline{toc}{chapter}{List of figures}

\setlength{\cftbeforeloftitleskip}{-11pt}
\setlength{\cftafterloftitleskip}{22pt}
\renewcommand{\cftloftitlefont}{\hfill\Large\bfseries}
\renewcommand{\cftafterloftitle}{\hfill}

\setlength{\cftbeforefigskip}{10pt}
\cftsetrmarg{1.0in}

\begin{singlespace}
\listoffigures
\end{singlespace}
\clearpage


%
\chapter*{List of tables}
\renewcommand{\listtablename}{}

\addcontentsline{toc}{chapter}{List of tables}

\setlength{\cftbeforelottitleskip}{-11pt}
\setlength{\cftafterlottitleskip}{22pt}
\renewcommand{\cftlottitlefont}{\hfill\Large\bfseries}
\renewcommand{\cftafterlottitle}{\hfill}

\setlength{\cftbeforetabskip}{10pt}

\begin{singlespace}
\listoftables
\end{singlespace}

\clearpage
\afterpage{\blankpage}



\pagenumbering{arabic}

\pagestyle{fancy} 

%
\chapter{Introduction}\label{sec:intro}

%

Blockchains have the potential to transform traditional and centralized sectors of great societal importance, such as banking and finance~\cite{Daian@S&P20,adams2021uniswap,Qin@FC21,Perez@FC21}. They provide a secure means of ensuring compliance via contracts (i.e., established agreements) and tamper-proof mechanisms, especially in situations where participants cannot trust each other~\cite{Nakamoto-WhitePaper2008,Wood@Ethereum,van2013cryptonote,sasson2014zerocash}. As a result, there are many blockchains available such as Bitcoin~\cite{Nakamoto-WhitePaper2008}, Ethereum~\cite{Wood@Ethereum}, Polkadot~\cite{wood2016polkadot}, Zcash~\cite{sasson2014zerocash}, Monero~\cite{van2013cryptonote}, among others. Bitcoin and Ethereum stand out as the most widely used blockchains, with market capitalization surpassing \$536.72B and \$228.57B as of May 2023~\cite{CoinMarketCap-URL2023}, respectively.
In addition, blockchains have not only been used to implement cryptocurrencies, but are increasingly being adopted across a wide range of domains including insurance~\cite{Ruubel-WebArticle2018}, education~\cite{Schmidt-WebArticle2015}, healthcare~\cite{Ekblaw-VC2017}, supply-chain management~\cite{Provenance-WebArticle2015,Hackett-WebArticle2017}, decentralized governance~\cite{leshner2019compound, adams2021uniswap, Governance@MakerDAO}, and decentralized finance (DeFi) applications through smart contracts such as exchanges~\cite{Daian@S&P20,UniswapDEX}, lending~\cite{Qin@FC21,Perez@FC21}, and auctions~\cite{NFTs}.

In the blockchain space multiple parties interact with each other. These include: (i) transaction issuers who are responsible for issuing transactions via interactions with the blockchain and its applications through smart contracts; (ii) miners or block validators who ensure the validity of forthcoming information or blocks for inclusion within all its transactions; and (iii) smart contract applications that are software programs running atop a blockchain, capable of executing predetermined actions, creating or transferring tokens, enabling voting for smart contract amendments, etc. In any real-world scenario involving a diverse group of individuals with varying roles, establishing a foundation of trust and fairness becomes paramount to ensure that no one can take advantage of others. 

Unlike the past, where interactions occurred mostly between individuals who knew and trusted each other, the rise of blockchain has enabled interactions within a decentralized system, devoid of inherent trust. However, in such a trustless environment, the potential for unfairness arises. One of the captivating aspects of blockchain systems is the interaction among participants who are strangers to each other, lacking pre-established trust. This raises questions about whether interactions are conducted fairly and what fairness concerns exist. For instance, in this thesis we focus on three primary unfairness issues: (i) Transaction ordering; (ii) Transaction transparency; and (iii) Fair distribution of voting power for smart contract amendments.


\paraib{Fairness related to transaction ordering.} The sequence in which transactions are processed is crucial, as everyone seeks timely processing. Ensuring fairness in this ordering presents challenges. For instance, how do we know that the ordering is fair?

In other words, noticeably absent from Bitcoin, Ethereum, and other decentralized blockchains is the requirement of any a priori trust between the users issuing transactions (i.e., registers persisted in the blockchain), the miners confirming transactions, and the peer-to-peer (P2P) nodes maintaining the blockchain.
Despite their widespread use in ordering critical applications~\cite{Mccorry@FC17,Daian@S&P20,pilkington2016blockchain,kharif2017cryptokitties,UniswapDEX,Perez@FC21}, blockchain protocols formally specify \stress{neither} the manner by which miners should select transactions for inclusion in a new block from the set of all available transactions, \stress{nor} the order in which they should be included in the block.
While informal conventions or \stress{norms} for prioritizing transactions exist, to our knowledge, before us no one has systematically verified if these norms were being followed by miners in practice~\cite{Messias@SDBD2020,Messias@IMC2021,Messias@FC2023}.

Studying this problem has significant implications for both blockchain users and miners.
Specifically, when setting fees for their transactions, transaction issuers (i.e., through their wallet software) assume that the fees offered by all their competing transactions are fully transparent---our findings contradict this assumption.
Similarly, when transactions offer different confirmation fees to different miners, it raises significant unfairness concerns with respect to the order in which these transactions are included.
We also show that mining pools collude when prioritizing self-interested transactions for inclusion which can exacerbates the growing concerns about the concentration of hash rates amongst a few miners in proof-of-work blockchains (PoW)~\cite{Gervais@CCS-16,bahack2013theoretical}.

\paraib{Fairness related to transaction transparency.} Assumptions dictate that all participants can observe public transactions. This transparency impacts transaction prioritization and fee-setting. However, reality often diverges, leading to concerns about fairness in transaction transparency.

For instance, the lack of transparency in blockchains arises from genuine concerns of transactions issuers, which cannot be overlooked.
One significant concern is the risk of transactions being front-run by bots~\cite{Daian@S&P20,Eskandari@FC-2020,Christof@USENIX,Weintraub@IMC2022}, which creates the need for transaction privacy.
Mining pools that address this need also facilitate, unsurprisingly, off-chain payments via which transaction issuers can (privately) incentivize the miners~\cite{BTC@accelerator,Messias@IMC2021,ViaBTC@accelerator}.
We consider these developments as natural and logical steps in the evolution of blockchains and back our assertions with empirical observations.
In contrast to prior research~\cite{Daian@S&P20,strehle2020exclusive}, we argue that the fundamental threat to blockchain stability lies in the opacity of the overall fees issued by transaction issuers.
Most wallet software and crypto-exchanges currently rely on reconstructing the current public \mpool state to suggest an appropriate fee to transaction issuers.
As a result, transaction issuers cannot precisely determine the fee required to ensure the inclusion of their transaction in the next block.
Consequently, miners can overcharge them, as the ``real'' fees are opaque to the rest of the network~\cite{Weintraub@IMC2022}.

\paraib{Fairness related to voting power to amend smart contracts.} Smart contracts, serving as trust-enforcing mechanisms, entail participants' agreement with stipulated rules. However, as these smart contracts can be upgraded (or changed) it raises the question of who possesses the authority to modify them.

Put differently, blockchains face challenges related to decentralized decision-making processes for amending smart contracts. For example, blockchains have been explored by many prior works who studied different types of security vulnerabilities that arise from incorrect implementations or unintended (or undesired) executions of smart contracts over blockchains, particularly in the context of DeFi applications~\cite{BuildFinance@Twitter,Qin@FC21,Weintraub@IMC2022,Christof@USENIX,Daian@S&P20}.
However, few studies, if any, focused, however, on vulnerabilities that may originate in the design of the procedures to \stress{amend}, i.e., change, smart contracts through \stress{governance protocols}, and/or stem from the execution of these procedures in practice.
These governance protocols intend to eliminate (or at least minimize) centralized decision-making in blockchains.
Their effectiveness in achieving that goal can, however, be compromised depending on how the tokens (i.e., voting power---typically one token equals one vote) are distributed which can lead to voting concentration.
Such concentration poses a threat to the overall governance of smart contract applications in blockchains leading, for example, to governance attacks~\cite{BuildFinance@Twitter}.

Therefore, in this thesis, we also provide an in-depth analysis of the voting patterns, delegation practices, and outcomes of proposals in one of the widely used governance protocols: Compound~\cite{leshner2019compound,Governance@Compound}. 
Since Compound records the votes cast transparently on a blockchain (i.e., it uses on-chain voting), we conduct measurements studies to analyze the extent to which this voting is decentralized, i.e., how small or large are the set of voters that determine the outcomes for the amendments.

It is important to acknowledge that our focus on these fairness concerns does not imply exclusivity. Additional concerns exist, such as fairness in compensating miners proportionately to their contributions. Nonetheless, in this thesis, we focus on the three aforementioned fairness concerns: (i) Transaction ordering; (ii) Transaction transparency; and (iii) Fair distribution of voting power for smart contract amendments.

\section{Overview of thesis contributions}\label{sec:contrib}

This thesis aims to address the fairness challenges mentioned above. We outline our research contributions in pursuit of this objective below.

\subsection{Transaction prioritization norms~\cite{Messias@IMC2021,Messias@SDBD2020}} \label{sec:intro_contribution_1}

\begin{figure*}[t]
	\centering
		\includegraphics[width={\onecolgrid}]{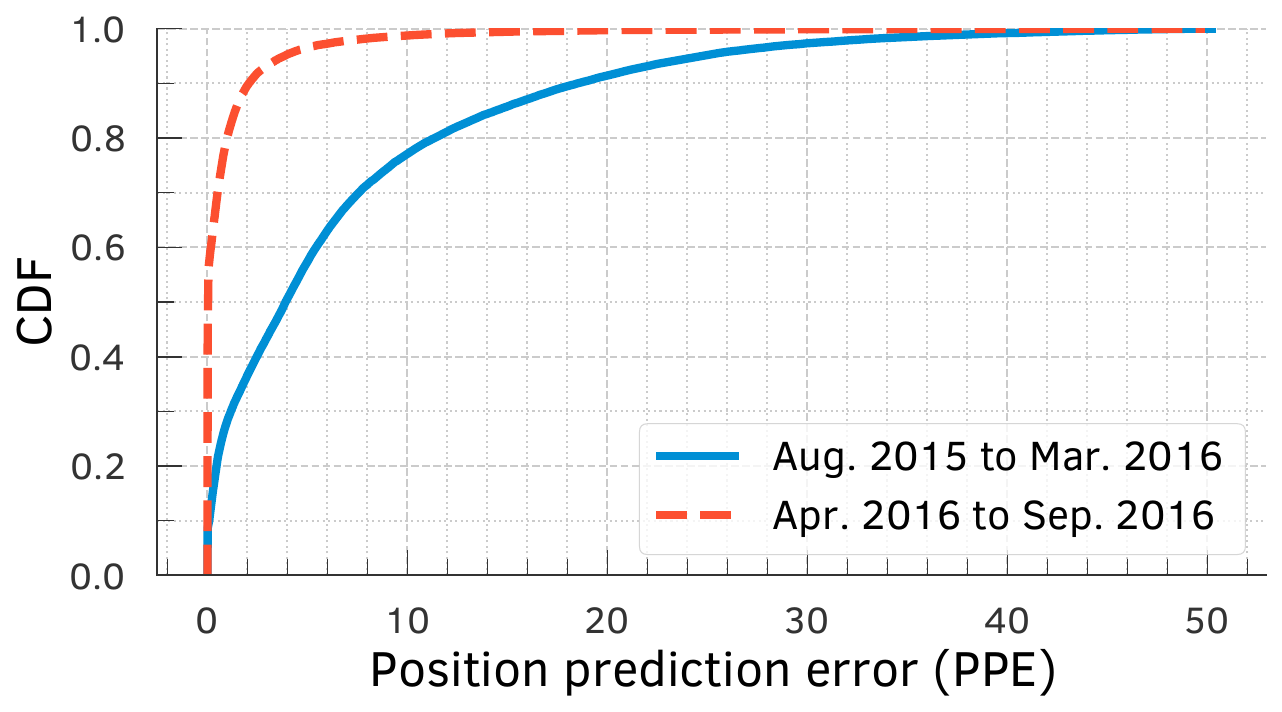}
	\caption{
  CDF of the error in predicting where a transaction would be positioned or ordered within a block according to the greedy fee-rate-based norm. Bitcoin Core code shifted completely to the fee-rate-based norm starting April 2016: Transaction ordering in Bitcoin closely tracks the fee-rate-based norm from April 2016, but differs significantly from it prior to April 2016 when a different norm was in place.
	}
    \label{fig:different-norms-in-btc}
\end{figure*}

The conventional wisdom today is that many miners follow the prioritization norms, implicitly, by using widely shared blockchain software like the Bitcoin Core~\cite{BitcoinCore-2021,CoinDance-2021}.
In Bitcoin, the presumed ``norm'' is that miners prioritize a transaction for inclusion based on its offered \stress{fee rate} or fee-per-byte, which is the transaction's fee divided by the transaction's size in bytes.
We show evidence of this presumed norm in Figure~\ref{fig:different-norms-in-btc}.
The norm is also justified as ``incentive compatible'' because miners wanting to maximize their rewards, i.e., fees collected from all transactions packed into a size-limited block, would be incentivized to include preferentially transactions with higher fee rates.
Assuming that miners follow this norm, Bitcoin users are issued a crucial recommendation: To accelerate the confirmation of a transaction, particularly during periods of congestion, they should increase the transaction fees.
We show that miners are, however, free to deviate from this norm and such norm violations cause irreparable economic harm to users.

We summarize our contributions as follows.

\point{}
To quantify the deviation from the norm, we propose two measures that we call \textit{\textbf{signed position prediction error (SPPE)}} and \stress{\textbf{position prediction error (PPE)}}.
These measures allow us to quantify the transaction deviation (or acceleration) for all transactions in block. It can also be applied to any blockchain that order transactions based on fees-incentive.

\point{}
We perform an extensive empirical audit of the miners' behavior to check whether they conform to the norms.
At a high-level, we find that transactions are indeed primarily prioritized according to the assumed norms. 
We also, nevertheless, offer evidence of a non-trivial fraction of priority-norm violations amongst confirmed transactions.
An in-depth investigation of these norm violations uncovered many highly troubling misbehavior by miners.

\point{}
Multiple large mining pools tend to \stress{selfishly prioritize} transactions in which they have a vested interest; e.g., transactions in which payments are made from or to wallets owned by the mining pool operators. Some even \stress{collude} with other large mining pools to prioritize their transactions.

\point{}
Many large mining pools accept additional \stress{dark (opaque) fees} to accelerate transactions via non-public side-channels (e.g., their websites). Such dark-fee transactions violate an important, but unstated assumption in blockchains that confirmation fees offered by transactions are transparent and equal to all miners.

\point{}
We release the data sets and the scripts used in our analyses to facilitate others to reproduce our results~\cite{Messias-DataSet-Code-2023}.

\subsection{Transaction Prioritization and Contention Transparency~\cite{Messias@FC2023}}\label{sec:intro_contribution_2}

In the context of the \stress{lack of contention transparency},
not all transactions are publicly broadcasted.
Instead, users can submit transactions to a subset of miners or mining pools through \stress{private channels} or \newterm{relays} that are not visible to the public.
In this case, transactions remain private to the relay until they are committed into a block.
Additionally, users may choose to submit their transactions exclusively to a particular mining pool that guarantees a fast commit time.
This thesis aims to shed light on the growing prevalence of private mining practices, where transactions are submitted to only a subset of the miners.
Furthermore, it analyzes the distinct characteristics of these private transactions.

Moreover, with the \stress{lack of prioritization transparency}, the fees offered by a transaction can be significantly higher than what is publicly declared.
For example, a transaction can privately offer additional fees to a miner in order to ``accelerate'' its inclusion in a block.
Many such transaction-accelerator (or \newterm{front-running as a service (FRaaS)}) platforms exist for Bitcoin~\cite{BTC@accelerator,ViaBTC@accelerator} and Ethereum~\cite{Eskandari@FC-2020,Flashbots@Ethereum,Taichi@accelerator,strehle2020exclusive}.
Furthermore, the same transaction can offer different fees to different mining pools through their relays.
The existence of these hidden or dark-fees can undermine the reliability of any fee prediction:
Transaction issuers may end up paying considerably higher fees without receiving proportional or any reduction in commit delays.
This thesis aims to characterize the prevalence of such dark-fee transactions and analyze the most popular private relay network available in Ethereum, Flashbots~\cite{Flashbots@Ethereum}.
We also conduct active experiments in both Bitcoin and Ethereum to validate our assumptions regarding the prioritization transparency.

In addition to demonstrating the non-uniformity of transaction fees across miners, we argue that, given the lack of contention transparency, the lack of prioritization transparency may become even more widespread in the future.

We summarize our contributions as follows.

\point{}
We provide a comprehensive characterization of the lack of contention transparency in both Bitcoin and Ethereum.
Our analysis reveals the widespread use of private channels or relay networks to submit transactions directly to a subset of miners.
This practice has the potential to undermine prioritization transparency, as transaction issuers may not be able to estimate the appropriate fees once none of which is publicly visible.

\point{}
We investigate the prevalence of private transaction fees, with a particular focus on Flashbots bundles in Ethereum.
Our findings indicate that Flashbots bundles represent a significant portion (\num{52.11}\%) of all Ethereum blocks.
This lack of prioritization transparency may enable miners to overcharge users when they send their transactions privately. 

\point{}
We investigate whether public transactions are bundled together with private transactions using the Flashbots private relay to exploit arbitrage opportunities through \stress{Maximal Extractable Value (MEV)}.
Interestingly, we find that the public transactions within these bundles are, for example, associated with oracle\footnote{Decentralized Oracle Networks facilitate off-chain data access for blockchains, including exchanges prices, weather forecast, and more~\cite{breidenbach2021chainlink}.
Typically, blockchain applications rely on these oracles to gather information they need.} updates, specifically involving the adjustment of prices for particular token pairs.
This is made possible by the sequential execution of transactions within the bundle by miners.
Consequently, the transaction that is executed right after the oracle update gain immediate access to the updated price information as soon as it is recorded on the blockchain.

\point{}
We  demonstrate evidence of collusion among Bitcoin miners, collectively possessing more than \num{50}\% of the network's total hashing power, particularly concerning the inclusion of dark-fees transactions.

\point{}
To promote transparency and facilitate the scientific reproducibility of our results, we publicly release our data sets and scripts used in our analysis~\cite{Messias-DataSet-Code-2023}.

\subsection{Decision-making power distribution for blockchain governance~\cite{Messias@FC2024}} \label{sec:intro_contribution_3}

This thesis also provides an in-depth analysis of the voting patterns, delegation practices, and outcomes of proposals in one of the widely used governance protocols: Compound~\cite{leshner2019compound,Governance@Compound}. 
Since Compound records the votes cast transparently on a blockchain (i.e., uses on-chain voting), we conducted measurements studies to analyze the extent to which this voting is decentralized, i.e., how small or large are the set of voters that determine the outcomes for the amendments.
Our goal is to thoroughly examine this protocol to better understand how its governance mechanism operates and identify potential areas for improvement.


\begin{figure*}[t]
	\centering
		\includegraphics[width={\onecolgrid}]{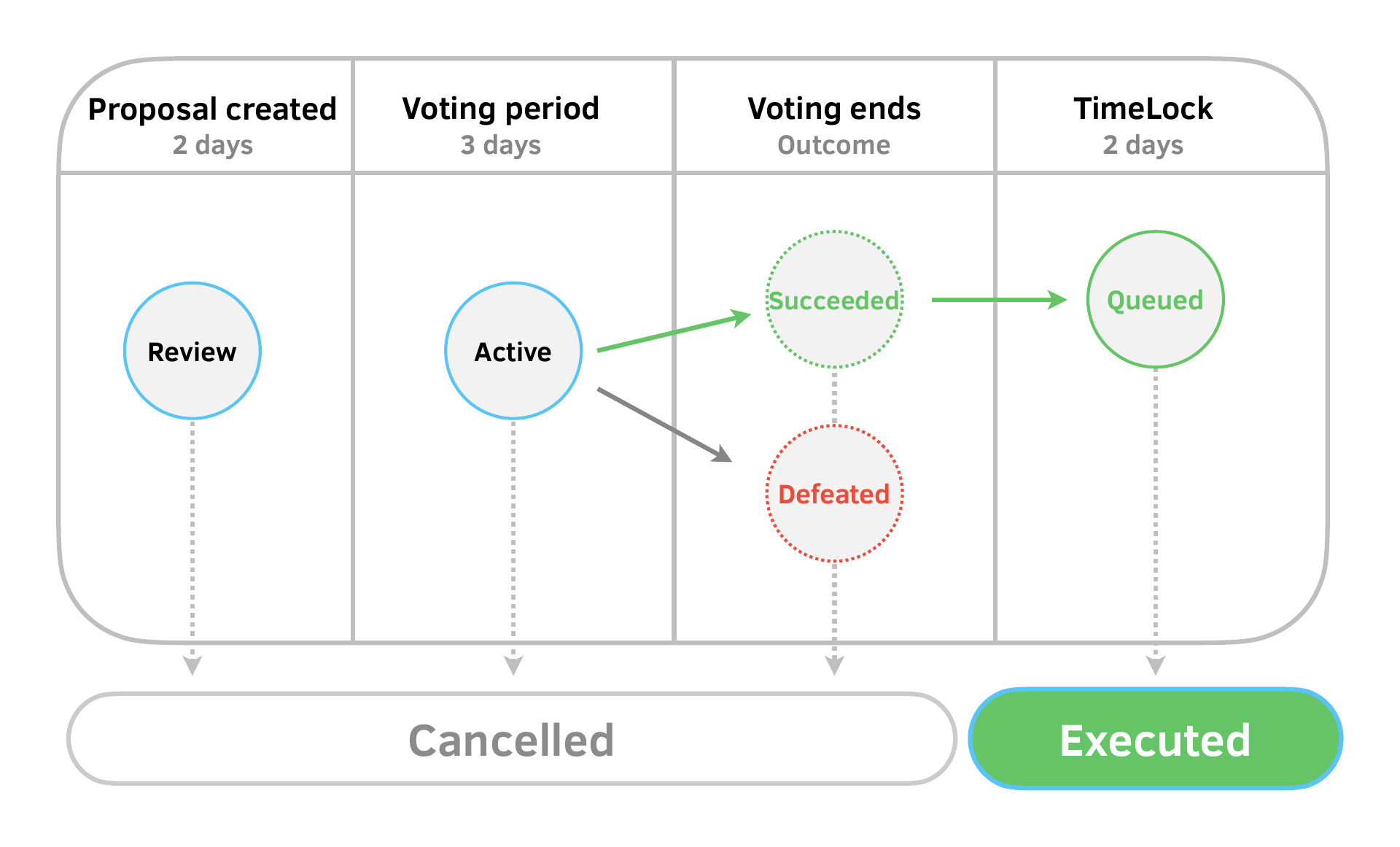}
	\caption{
  The lifecycle of a Compound proposal lasts 7 days. After a proposal is created, it waits for 2 days before the 3-day voting period begins. Once the outcome of the election is decided, it takes 2 more days for the proposal to be executed and become part of the Compound Governance protocol. Proposals can also be cancelled at any time before they are executed.
	}
\label{fig:compound-life-cycle}
\end{figure*}


Compound regulates its voting process via the Compound (COMP) token, an ERC-20 asset, as follows.
First, it allows token holders to participate in governance by proposing and voting on changes to the protocol through an on-chain voting mechanism where voting power of a user is proportional to the amount of delegated tokens held by them---one token equals one vote.
Second, it permits its holders to delegate their tokens to other users, enabling users (who do not wish to exercise their voting rights) to delegate their voting power to others.
The protocol essentially supports a form of \stress{liquid democracy} that combines direct democracy and representative democracy, where voters can delegate their voting power to trusted representatives~\cite{behrens2017origins,carroll1884principles,blum2016liquid}.
Some protocol changes that token holders can propose and vote on include adjustments to the borrowing and lending rates, changes to how new tokens are distributed, and changes to the parameters of the voting process.
They can also change the duration of the proposal's life cycle (per Figure~\ref{fig:compound-life-cycle}, currently taking on average 7 days) and other aspects of the protocol.
To incentivize participation in token lending and borrowing through the protocol, Compound distributes \num{1234} COMP tokens \stress{daily} to users and applications in various markets (e.g., ETH, DAI, and USDC), in proportion to the amount that they lend or borrow~\cite{token-dist@Compound}.

We summarize our contributions as follows.

\point{}
We characterize the Compound protocol's on-chain voting process, showing that it is active and regularly used, with a steady flow of proposals.
The majority of the proposals receive significant support: on average, \num{89.39}\% of votes are in favor.

\point{}
We reveal a substantial variation in voting costs, from \$\num{0.03} to \$\num{294.02}, with an average of \$\num{7.88}.\footnote{All costs are in US dollars, taking into account the exchange rate at the time of casting the vote.}
If we normalize the costs per vote by the count of tokens held by users, we obtain an average cost per vote unit of \$\num{358.54}.
Voting costs on Compound can, hence, be \stress{unfairly} expensive for small token holders, which has fairness implications for the decision-making process.

\point{}
We show that a small group of 10 voters holds a significant amount of voting power (57.86\% of all tokens) and that proposals only required an average of \num{2.84} voters to obtain at least \num{50}\% of the votes. These observations strongly suggest that the voting outcomes in Compound may not reflect the preferences of the broader community.

\point{}
We also discover potential voting coalitions among the top voters, which could further exacerbate concerns of voting concentration.

\point{}
To foster reproducible research and inspire investigations into other aspects of governance protocols, we share our scripts and data sets in a GitHub repository~\cite{Messias-DataSet-Governance-Code-2023}.

\section{Thesis outline}

In summary, this thesis addresses three important fairness concerns. First is the transaction ordering where we examine the prioritization of transactions by auditing the order in which they are included by miners. We investigate whether miners adhere to the existing norms or widely accepted practices in this regard. Second, is the transaction transparency where we explore whether miners ensure transparency in terms of transaction prioritization and transaction contention (i.e., the transaction and its content is accessible by all miners), ensuring equal access for all participants. Finally, is the fair distribution of voting power for smart contract amendments where we delve into the distribution of voting power in decentralized governance protocols, with a particular focus on Compound. Both of these concerns are crucial for establishing a fair blockchain ecosystem.

Specifically, this thesis is organized as follows:

\point{}
In Chapter~\ref{sec:diss_background}, we begin by presenting the background of blockchains and smart contracts. Next, we delve into the background of transaction prioritization norms. Then, we discuss the background of transaction prioritization and contention transparency. Finally, we explore the background of decentralized governance protocols used for amending smart contracts.

\point{}
In Chapter~\ref{chap:tx_norms}, we perform an audit of miners' prioritization norms and evaluate the degree to which they comply with commonly accepted prioritization assumptions. Our findings reveal that miners tend to prioritize transactions based on self-interest, including their own transactions or those from friendly mining pools. Additionally, we highlight the presence of acceleration services provided by miners, enabling off-chain payments to expedite transactions. Unfortunately, the fees associated with these services are dark or opaque to other participants, making it challenging for them to estimate appropriate fees for timely transaction inclusion. Finally, we propose two metrics for verifying transaction ordering misposition within a block. These metrics play a crucial role in determining whether miners adhere to the assumed norms and can be generalized to other blockchains.

\point{}
In Chapter~\ref{chap:tx_prioritization_contention}, we conduct a data-driven analysis focused on the prioritization of dark-fees payments by miners. Our results reveal the existence and prevalence of private relay networks (\eg Flashbots), which allow transaction issuers to privately send their transactions to miners, keeping them hidden from other participants. To assess the level of prioritization provided by miners, we performed two active experiments, where we accelerated the inclusion of transactions by offering dark-fee payments to miners. We discovered evidence of potential collusion among mining pools, where their combined hash rate accounted for over \num{50}\% of the network's hash rate.

\point{}
In Chapter~\ref{chap:governance}, we conduct an in-depth audit of governance protocols, with a specific focus on Compound, to evaluate the extent to which they have succeeded in achieving their primary goal of decentralizing the decision-making process for amending smart contracts. Our analysis reveals that this goal may not have been fully achieved, as we observe a concerning concentration of voting power among a small number of participants, along with the existence of voting coalitions formed by powerful voters. This concentration stands in contrast to the fundamental objective of decentralized governance protocols, which aims to mitigate centralization in the decision-making process.

\point{}
In Chapter~\ref{chap:related}, we review the works in the literature that are relevant to this thesis.

\point{}
In Chapter~\ref{chap:discussion_limit}, we present a detailed discussion of our work and its limitations. Additionally, we explore potential directions for future work.

\clearpage
%
\chapter{Background}\label{sec:diss_background}

In this chapter, we provide the necessary background on blockchains and smart contracts. We then present the context for transaction prioritization norms, followed by transaction prioritization and contention transparency. Next, we present the background information on decentralized governance. Lastly, we introduce the background details concerning Layer 2.0 solutions aimed at enhancing the scalability of blockchains.

%

\section{Blockchains \& smart contracts}\label{sec:background-blockchains}

\begin{figure*}[t]
	\centering
		\includegraphics[width={1.2\onecolgrid}]{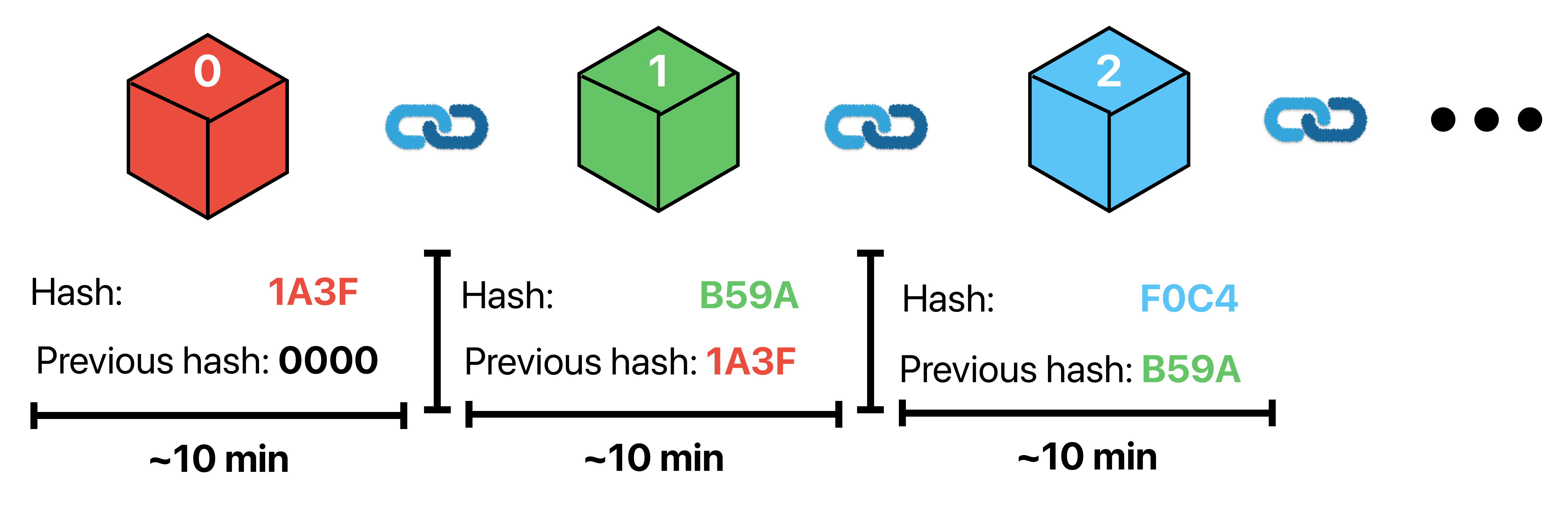}
	\caption{
  Illustration of a blockchain consisting of three blocks, with an extimation generation and inclusion time approximately 10 minutes. It is important to note that every block, except for the Genesis block which is the initial block in the blockchain, also uses the hash of its preceding block to compute its own hash.
	}
\label{fig:blockchain}
\end{figure*}

A blockchain is a decentralized and distributed ledger of cryptographically linked records of transactions stored in blocks. A block is a set of zero\footnote{As miners can also mine a block without including any transaction on it, we refer to those blocks as \stress{empty blocks}.} or more transactions. In the case of Bitcoin, a block also includes a Coinbase transaction, responsible for transferring rewards (or compensation for the miner' efforts) to the miner's wallet. These blocks are interconnected, tracing back to the original (or ``Genesis'') block. Figure~\ref{fig:blockchain} presents an illustration of a blockchain with a block inclusion rate of 10 minutes.
The ledger is maintained and continually extended by the blockchain participants by carrying out various functions.
\stress{Transaction issuers}, for instance, issue transactions and share it with other participants through a peer-to-peer (P2P) network.
Such transactions are deemed \stress{unconfirmed} until they are added permanently to the blockchain.
Some others, called \stress{miners\footnote{Throughout this thesis, we use the terms \stress{miner}, \stress{mining pool}, \stress{mining pool operator (MPO)}, and \stress{block proposer} interchangeably.}} (in proof-of-work blockchains) or \stress{block proposers} (in proof-of-stake blockchains), bundle the transactions into \stress{blocks} for confirmation.
They propose the new block over the P2P network where others can verify it and, if successful, add to their copy of the blockchain---thereby extending the ledger or chain of blocks.
The miner typically collects a reward in the form of newly minted coins as an incentive for their contribution to the network along with the transactions fees provided by the issuers.
We use the term ``fee'' to refer generally to the incentive offered by a user to miners for prioritizing the inclusion of their transaction in a block, albeit its exact form may vary, e.g., \stress{fee rate} (or \stress{fee-per-byte}) in Bitcoin and \stress{gas price} in Ethereum\footnote{Ethereum recently switched its consensus mechanism from proof-of-work to proof-of-stake with \stress{The Merge} hard fork deployed on September 15, 2022, at block number \href{https://etherscan.io/block/15537394}{15537394}~\cite{Eth-PoS,Eth-Merge}.}.
To join a blockchain, miners use a software implementation (along with the hardware) which we refer as a \newterm{node}.
A~node allows a miner to receive broadcasts of transactions and blocks from
their peers, validate the data, and mine a block.
Nodes queue the unconfirmed transactions received via broadcasts in an in-memory
buffer, called the \newterm{\mpool}, from where they are dequeued for inclusion
in a size-limited block.
One can also configure the node to skip mining and simply use it as an observer.

Some blockchains, such as Ethereum, allow for the execution of \stress{smart contracts}---contractual agreements encoded in software programs that run atop the blockchain.
Smart contracts are typically developed using Solidity, a domain-specific programming language~\cite{Solidity@Ethereum}, and they can be executed in the Ethereum Virtual Machine (EVM).
Their implementations abide by standards such as ERC-20~\cite{ERC-20@Ethereum} and ERC-721~\cite{ERC-721@Ethereum} to ensure compatibility and interoperability among them.
These standards define, for instance, the key functions for creating and implementing smart contracts for fungible tokens (e.g., Compound's COMP token~\cite{leshner2019compound}) in the case of ERC-20 or non-fungible tokens (NFTs) in the case of ERC-721.
%

%

\section{Transaction prioritization norms}\label{sec:background-tx-norms}

A crucial detail absent in the design of a blockchain
per~\cite{Nakamoto-WhitePaper2008} is any notion of a formal specification of
transaction prioritization.
Said differently, Nakamoto's design does not formally specify how miners should
select a set of candidate transactions for confirmation from all available unconfirmed
transactions.
Notwithstanding this shortcoming, ``norms'' have originated from miners' use
of a shared software implementation: For example, in Bitcoin, 
miners predominantly use the Bitcoin Core~\cite{BitcoinCore-2021} software for
communicating with their peers (e.g., to advertise blocks and learn about new
unconfirmed transactions) and reaching a consensus regarding the chain.

Of particular note in the popular Bitcoin core’s implementation is the
\texttt{GetBlockTemplate (GBT)} mining protocol, implemented by the Bitcoin
community around February 2012.\footnote{%
Even within mining pools, the widely used Stratum protocol internally uses the \gbt mechanism~\cite{Stratum-v1-2021}.}
\gbt{} rank orders transactions based on the fee-per-byte (i.e., transaction
fees normalized by the transaction's size) metric~\cite{GBT-Bitcoin-2019}.

In Bitcoin, the term \stress{size}, refers to
\newterm{virtual} size, each unit of which corresponds to four \newterm{weight
units} as defined in the Bitcoin improvement proposal
BIP-141~\cite{Lombrozo-BIP141-2015}. The size (or maximum capacity) of a block is also limited to \uMB{1} virtual size.
The predominant use of GBT (through the use of Bitcoin core) by miners coupled
with the fact that GBT is maintained by the Bitcoin community
\stress{implicitly} establishes two norms.
A~third norm stems from a configuration parameter of the Bitcoin core
implementation.
We now elucidate these three norms.

\textbf{I.}
\stress{When mining a new block, miners select transactions for inclusion, from
the \mpool{}, based solely on their fee rates.}

\textbf{II.}
\stress{When constructing a block, miners order (place) higher fee rate transactions before lower fee rate transactions.}

\textbf{III.}
\stress{Transactions with fee rate below a minimum threshold are ignored and never committed to the blockchain.}

The GBT protocol implementation in Bitcoin core is the source of the first two norms.
GBT's rank ordering determines both which set of transactions are selected for inclusion (from the \mpool{}) and in what order they are placed within a block.
GBT dictates that a transaction with higher fee-per-byte \stress{will} be
selected before all other transactions with a lower fee-per-byte.
It also stipulates that within a block a transaction with the highest fee-per-byte appears first, followed by next highest fee-per-byte, and so on.

The third norm stems from the fee-per-byte threshold configuration parameter.
Bitcoin core, by design, will not accept any transaction with fee rate below
this threshold, essentially filtering out low-fee-rate transactions from even
being accepted into the \mpool{}.
The default (and recommended) value for this configurable threshold is set to
$\DefTxFee{}$.\footnote{One Bitcoin (BTC) is equal to $10^{8}$ satoshi (sat).}

%

\section{Transaction prioritization and contention transparency}\label{background-tx-prioritization-contention}

\begin{figure*}[t]
	\centering
		\subfloat[Bitcoin blockchain\label{fig:cdf-tx-blks-btc}]{\includegraphics[width=\twocolgrid]{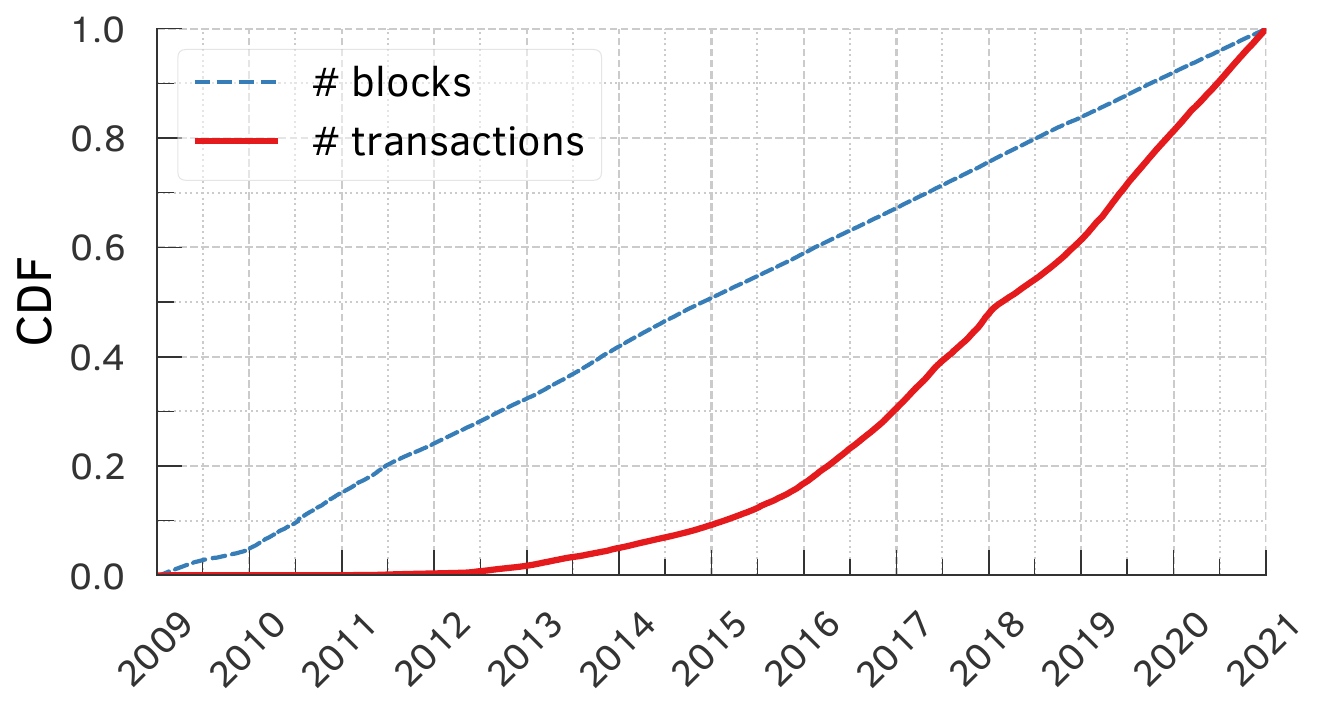}}
		\subfloat[Ethereum blockchain\label{fig:cdf-tx-blks-eth}]{\includegraphics[width=\twocolgrid]{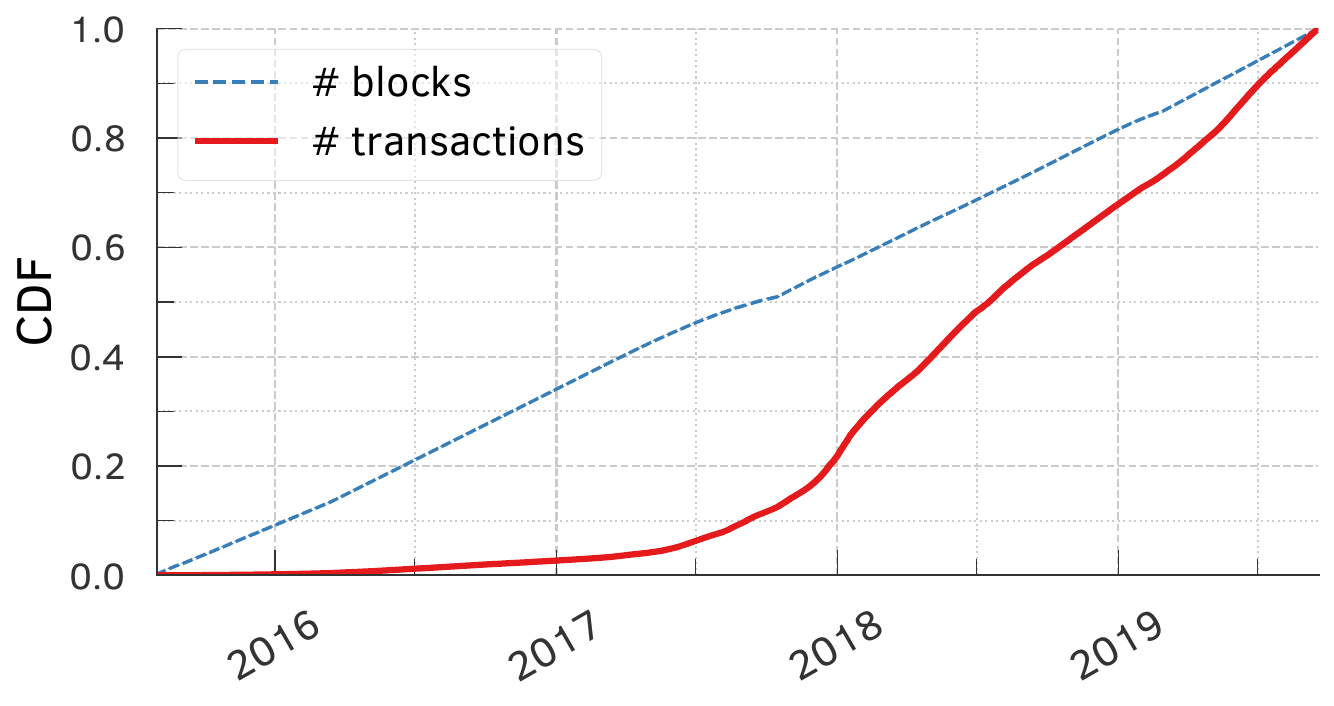}}
	\caption{
  Volume of transactions issued and blocks mined as a function of time, showing that transactions have been issued at high rates for both (a) Bitcoin and (b) Ethereum blockchains.
	}
  \label{fig:cdf-tx-blks}
\end{figure*}

The rate at which users issue transactions in permissionless blockchains, e.g., Bitcoin~\cite{Nakamoto-WhitePaper2008} and Ethereum~\cite{Wood@Ethereum}, is often much higher than the rate at which miners can include them in a block~\cite{Easley19a,Monopolywithoutmonopolist,Lavi-WWW2019,Messias@SDBD2020,Messias@IMC2021}.
Figure~\ref{fig:cdf-tx-blks} shows that \num{50}\% of all Bitcoin transactions were added to the blockchain in just 3 years. Similarly, in Ethereum, \num{80}\% of the transactions were added in just 1.5 years. 
Users typically issue transactions using a wallet software, whose primary functionality is determining an ``appropriate'' fee for a given transaction.
This (prioritization) fee varies, unsurprisingly, as a function of the level of congestion in the blockchain~\cite{Messias@IMC2021} as well as the distribution of fees across available transactions.
Inferring either of these is, however, deceptively complicated.

At first glance, these tasks appear straightforward, since every transaction is broadcast to all miners in the blockchain.
A user could simply gather all transactions broadcast over time and reconstruct the set of uncommitted transactions available to a miner (i.e., contents of the miner's \mpool) at any point of time~\cite{Messias@SDBD2020}.
We refer to this assumption of a public and uniform view (across miners) of all available transactions as \newterm{contention transparency}.
If contention transparency exists, a user could rank order available transactions by their fee (based on which miners should select transactions for inclusion) and estimate the commit delay of any transaction~\cite{Messias@IMC2021}.
Consequently, they could determine the fee that they must pay to guarantee inclusion of their transaction in a given block.
We label this assumption that the (prioritization) fee offered by a transaction is only that publicly declared by that transaction as \newterm{prioritization transparency}.
Neither the contention transparency nor the prioritization transparency, however, holds today in permissionless blockchains.

%

\section{Decentralized governance}\label{sec:background-governance}

Smart contracts underpin many DeFi applications today~\cite{Daian@S&P20,adams2021uniswap,NFTs,Qin@FC21,Perez@FC21}, and it is only natural to have a mechanism for updating these (software) contracts to fix bugs or evolve them over time to cater for new use cases~\cite{Liu@ICSE,DaoAttack@CoinDesk,Zhou@SP23}.
If decisions concerning such updates are made in a centralized manner, e.g., by a regulatory body, or a cabal of developers or miners, it undermines users' trust in the applications that these contracts support.
The updates could, for instance, be tailored to benefit the centralized regulatory body at the expense of others.
Governance protocols address this issue by distributing the decision-making power among all the users of the application or smart contract being updated.

A governance protocol establishes rules and (transparent) mechanisms for changing smart contracts.
It defines the required procedures for creating, voting on, and executing proposals to amend smart contracts.
It facilitates users of a protocol (or, more aptly, \stress{token holders} who hold one or more tokens of the protocol) to propose changes.
The changes are then vetted by and voted by other users, and implemented only if the proposals receive the majority of favorable votes.
The protocols also grant voting power to a user based on the number of tokens held by them---typically one token equals one vote,  essentially capturing the user's stake and/or participation in the protocol.
Some protocols such as Compound~\cite{leshner2019compound} and Uniswap~\cite{adams2021uniswap} allow token holders who do not wish to exercise their voting power to delegate their voting power (i.e., tokens) to others.
This delegation is a form of \stress{liquid democracy}, where voters can participate in decision-making either directly by voting or indirectly by delegating their voting rights to trusted representatives~\cite{behrens2017origins,carroll1884principles,blum2016liquid}.
Governance protocols give every participant the right to propose, support, or oppose any proposal.
They are, hence, crucial for ensuring absolute decentralization of applications running atop blockchains.

%

\subsection{Voting modalities}\label{subsec:background-voting-modalities}

Proposals to change a governance protocol takes birth in the protocol's community forum.
Community members suggest and discuss potential changes to the protocol in the forum and may even conduct an informal poll to gauge the community's support for a proposal.
The proposer then either amends the proposal to incorporate the community's feedback and submit it for a formal vote, or simply abandon it.
The formal voting on the proposal has two modes: \stress{on-chain} and \stress{off-chain} voting.

\paraib{On-chain voting}
In this voting system, participants make all governance decisions via smart contracts on a blockchain.
Under this system, participants cast a vote by issuing a transaction (and paying a fee for committing it) to the blockchain.
The system allows only participants with at least a threshold amount of (governance) tokens to create a proposal, albeit any token holding participant can vote on that proposal.
It executes the proposal on the blockchain only if it receives a significant number of votes in favor and reaches a \stress{quorum}.\footnote{
In Compound, in order for a proposal to be executed, it needs to meet two requirements.
First, it must receive a minimum of \num{400000} votes in favor of the proposal.
This number corresponds to \num{4}\% of the total supply and is known as the \stress{quorum}.
Second, the majority of the votes cast must be in favor of the proposal.
} 
This voting system thus facilitates making transparent and tamper-proof changes to the protocol.
Decentralized governance protocols such as AAVE~\cite{Governance@AAVE}, Compound~\cite{Governance@Compound}, and Uniswap~\cite{Governance@Uniswap} use on-chain voting.

\paraib{Off-Chain Voting}
This system conducts voting on an off-chain third-party platform and, as a consequence, also establishes the rules for voting, aggregating votes, and determining the results off-chain.
Protocols such as Balancer~\cite{Governance@Balancer} and Convex Finance~\cite{Governance@ConvexFinance}, for instance, use Snapshot~\cite{Snapshot} for off-chain voting.
Snapshot stores the voting data on a P2P network called InterPlanetary File System (IPFS)~\cite{IPFS}.
The voting process does not require voters to pay any fees and (unlike on-chain voting) promotes participation across all participants, regardless of their level of participation or investment in the protocol.
After the voting, this system uses a \stress{multi-signature} contract to enact the off-chain voting outcome on the blockchain.
Typically, an \stress{n-of-m multisig} contract requires the transaction to be signed by at least \stress{n} out of the \stress{m} ``admins'' to be executed on-chain.
The system trusts the multisig ``admins'', who are well-known in the community, to implement the voting outcome on the blockchain truthfully.
The admins can also, however, refuse the proposal.
In Convex Finance, for instance, the admins can choose not to execute a proposal if they deem it harmful, even if it had received the majority of votes and reached a quorum~\cite{Docs@ConvexFinance}.
On-chain voting systems, in contrast, prevent such manipulation of voting outcomes by one or more individuals (after the voting process), since all governance decisions (e.g., voting and execution) happen on the chain.

\paraib{Token delegations}
Some governance protocols (e.g., Compound, Uniswap, and AAVE) require a user to own a certain amount of governance tokens for casting a vote.
Users must delegate their tokens either to themselves or, if they do not wish to vote, delegate them to others.
The ability to delegate voting power to others facilitates a form of liquid democracy; the token holder who delegates or sells their tokens to another loses their voting power.
Delegations allow anyone to buy (or sell) tokens and gain (or lose) voting power instantly.
Justin Sun, the founder of (stablecoin) TrueUSD~\cite{TrueUSD}, for instance, allegedly borrowed COMP tokens to create and vote for Compound proposal \#84~\cite{Proposal-84@Compound}, resulting in a \stress{governance attack}~\cite{Thurman@Coindesk}; this proposal was, however, defeated.

\paraib{Token ``locking''}
Protocols such as Balancer and Curve~\cite{Governance@Curve} mandate that a user ``lock'' their tokens into a smart contract for a specified period of time to gain their right to vote.
The user cannot withdraw the locked tokens until the lock-up period expires.
The voting power of a user in this system is proportional to the amount of tokens locked as well as the lock-up period.
In Balancer, for instance, a user receives \num{1} unit of voting power if they lock \num{1} token into the contract for \num{1} year, and only half that voting power if they lock it instead for \num{6} months.

\paraib{Continuous voting}
A few protocols (e.g., MakerDAO~\cite{Governance@MakerDAO}) allow voters to change their votes at any time during the voting period.
Users propose a protocol change by developing a new implementation via a smart contract.
The new implementation is accepted if it receives more votes than the current one, i.e., the winning implementation must always receive the majority of the votes (or tokens).
MakerDAO requires a user to deposit their (MKR) tokens into the (Maker) Governance contract for casting a vote.
The more tokens they deposit, the more voting power they obtain, and they vote for their desired implementation by specifying it as a protocol parameter in the smart contract.
Since the voting process is continuous, if a user withdraws their MKR tokens from the governance contract, their vote will no longer count towards the implementations for which they previously voted.

We refer the reader to \S\ref{sec:types-of-governance} (particularly Tab.~\ref{table:protocol-types}) for a characterization of the voting methods, delegation approaches, and proposal executions in various other governance protocols.
Understanding how voting is conducted (i.e., whether it is on-chain or off-chain) and proposals are executed is fundamental for analyzing how voters, proposers, and others interact with these governance protocols.

%

\section{Blockchain Scalability with Layer 2.0 Solutions}\label{sec:background-layer2}

As observed in Figure~\ref{fig:cdf-tx-blks}, \num{50}\% of all Bitcoin transactions were added to the blockchain in just 3 years. Meanwhile, in Ethereum, \num{80}\% of the transactions were added in just 1.5 years. This highlights a crucial scalability challenge inherent in blockchains: blocks will not be able to accommodate all transaction available. Unfortunately, this issue is projected to aggravate each year. Consequently, the need for Layer 2.0 solutions has arisen to mitigate the escalating blockchain scalability dilemma.

Layer 2.0 consists of an off-chain network, system, or technology that is built on top of a blockchain~\cite{Layer2@Chainlink}. In other words, this approach execute transactions in batches, together, and off-chain. This process helps accelerate the execution of the transactions leaving the main blockchain for persisting the validity (or proofs) that the transactions were correctly executed. As a result, this method accelerates both transaction execution and confirmation processes.

There are two most popular types of Layer 2.0 solutions: Zero Knowledge (ZK)~\cite{goldwasser2019knowledge,goldreich1994definitions} and Optimistic~\cite{Optimistic} rollups.

Within the Zero Knowledge (ZK) rollup, transactions are grouped into batches for execution, leading to a reduction in execution costs. The outcome, which is a proof of transaction validity, gets stored back on the main blockchain (or Layer 1.0). Therefore, rather than persisting the complete transaction data for each individual transaction on the blockchain, this approach conserves space by storing only a proof that verifies the validity of an entire batch of transactions executed on Layer 2.0. Example of ZK rollup is ZKSync~\cite{ZKSync}.

On the other hand, within the Optimistic rollup, it operates under the assumption that all transactions are valid by default, unless certain participants provide evidence to the contrary. Therefore, it is an optimistic approach. Examples of Layer 2.0 using Optimistic rollups are Optimism~\cite{Optimism} and Arbitrum~\cite{Arbitrum}.

In this thesis, we delve into a Layer 2.0 solution designed for Bitcoin, known as Omni Layer~\cite{OmniLayer}. This aims to shed light on off-chain transaction acceleration facilitated by miners. Unlike the batch approach used in both ZK and Optimistic rollups, Omni Layer does not batch transactions. Instead, it creates a Bitcoin transaction for every transaction issued on its layer. This enables us to analysis the extent to which miners accelerate the inclusion of these individual transactions.

\clearpage

%
\chapter{Transaction Prioritization Norms} \label{chap:tx_norms}

%

In this chapter, we delve into the research questions, methodology, and the implications of analyzing and comprehending the norms employed by miners when prioritizing transactions for inclusion.
Understanding these prioritization norms is key to enable transaction issuers to determine the appropriate fees for their transactions to be included within an expected number of blocks.

The order in which miners select transactions for inclusion have significant implications for the ultimate outcome of the transaction execution.
As a result, we aim to audit the Bitcoin blockchain to address the following research questions.

\point{}
\textbf{RQ 1}: \stress{Which transactions are allowed or transmitted over the public P2P network?}
This research question pertains to the \stress{norm III}, as described in \S\ref{sec:background-tx-norms}.
It assumes that transactions with fee rate below a minimum threshold are discarded and never committed to the blockchain.
If miners are considering transactions with incentives below the established threshold, their view of available transactions for inclusion may differ.

\point{}
\textbf{RQ 2}: \stress{How do miners select transactions for inclusion in a block once they enter the miners' \mpool?}
This research question pertains to \stress{norm I}, which states that miners select transactions for inclusion, from the \mpool, based \stress{solely} on their fee rates.
However, if some miners accept additional incentives to accelerate the transaction confirmation such as dark or opaque fees, the miners' view of the offered fees may vary.
As a result, transaction issuers may struggle to estimate the appropriate fees needed for their transaction to be prioritized, as the fees' distribution may differ based on individual miner's view of the system.

\point{}
\textbf{RQ 3}: \stress{In what order do miners include transactions within a block?}
This question aligns with \stress{norm II}, where higher fee rate transactions are prioritized and placed before lower fee rate transactions during the block construction.
If miners misplace transactions within a block, it can result in different outcomes for those transactions when the miner's block is selected as the next block.
Transaction issuers may be susceptible to well-known attacks such as front-running~\cite{Daian@S&P20,Christof@USENIX} or even sandwich attacks~\cite{Qin@FC21,Zhou@SP23}.

Addressing these research questions is crucial for understanding how miners actually prioritize transactions for inclusion. 
Hence, accurate knowledge of the order in which miners include transactions can assist in estimating appropriate fees that issuers need to offer to prioritize their transactions.
Next, we discuss our methodology for gathering the necessary data sets and  detecting accelerated or highly prioritized transactions.

\subsubsection*{Relevant publication}

The results presented in this chapter have been published in~\cite{Messias@IMC2021,Messias@SDBD2020}.

%

\section{Methodology} \label{sec:method_tx_norms}

To understand the importance of transaction ordering to issuers and to investigate when and how miners violate the transaction prioritization ``norms,'' we resort to an empirical, data-driven approach.
Below, we briefly describe three different data sets that we curated from Bitcoin and highlight how we use the data sets in different analyses in the rest of this thesis.
Furthermore, we present our methodology for detecting transaction acceleration, which can be applied to any fee-based incentive blockchains.

\begin{table*}[t]
  \begin{center}
   \small
    \tabcap{Bitcoin data sets (\dsa and \dsb) used for testing miners' adherence to transaction-prioritization norms and (\dsc) for investigating the behaviour of miners with respect to transaction acceleration. \stress{Child-pays-for-parent (CPFP)} transactions are transactions that depend on other transactions to be included into a block.}\label{tab:datasets}
    \begin{tabular}{rrrr}
      \toprule
      \thead{Attributes} & \thead{Data set \dsa{}} & \thead{Data set \dsb{}} & \thead{Data set \dsc{}}\\
      \midrule
      \textit{Time span} & Feb. $20\tsup{th}$ -- Mar. $13\tsup{th}$, 2019 & Jun. $1\tsup{st}$ -- $30\tsup{th}$, 2019 & Jan. $1\tsup{st}$ -- Dec. $31\tsup{st}$, 2020\\
      \textit{Block height} & \num{563833} -- \num{566951} & \num{578717} -- \num{583236} & \num{610691} -- \num{663904} \\
    \textit{Number of blocks} & $\num{3119}$ & $\num{4520}$ & $\num{53214}$ \\
      \textit{\# of txs. issued} & $\num{6816375}$ & $\num{10484201}$ & $\num{112489054}$\\
      \textit{\% of CPFP-txs.} & $26.45\%$ & $23.17\%$ & $19.11\%$ \\
      \textit{\# of empty-blocks} & \num{38} & \num{18} & \num{240}\\
      \bottomrule
    \end{tabular}
  \end{center}
\end{table*}

\begin{figure*}[t]
	\centering
		\subfloat[Data set \dsa{}\label{fig:dist-txs-blks-dataset-a}]{\includegraphics[width=\twocolgrid]{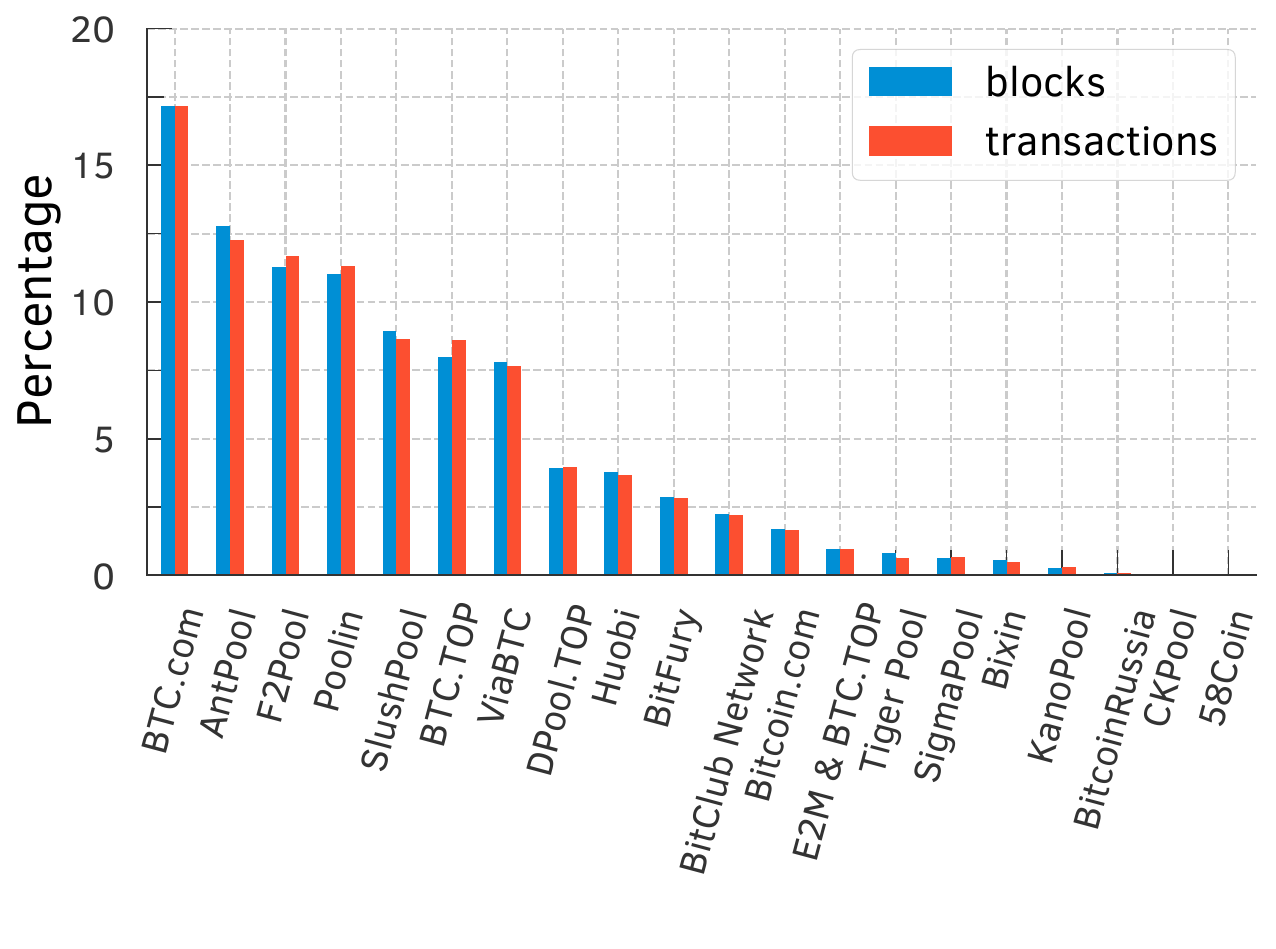}}
		\subfloat[Data set \dsb{}\label{fig:dist-txs-blks-dataset-b}]{\includegraphics[width=\twocolgrid]{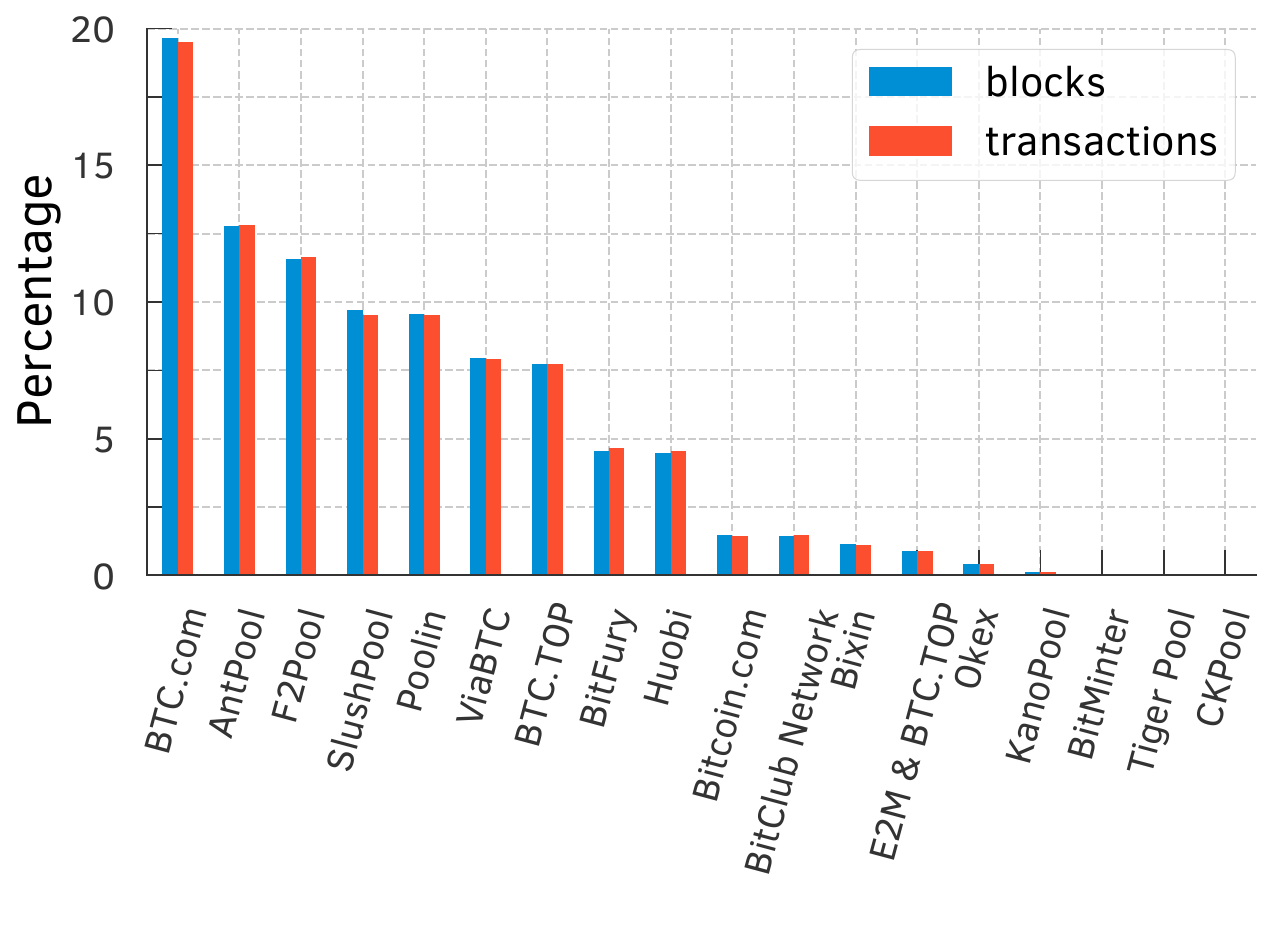}}
        \\
		\subfloat[Data set \dsc{}\label{fig:dist-txs-blks-dataset-c}]{\includegraphics[width=\twocolgrid]{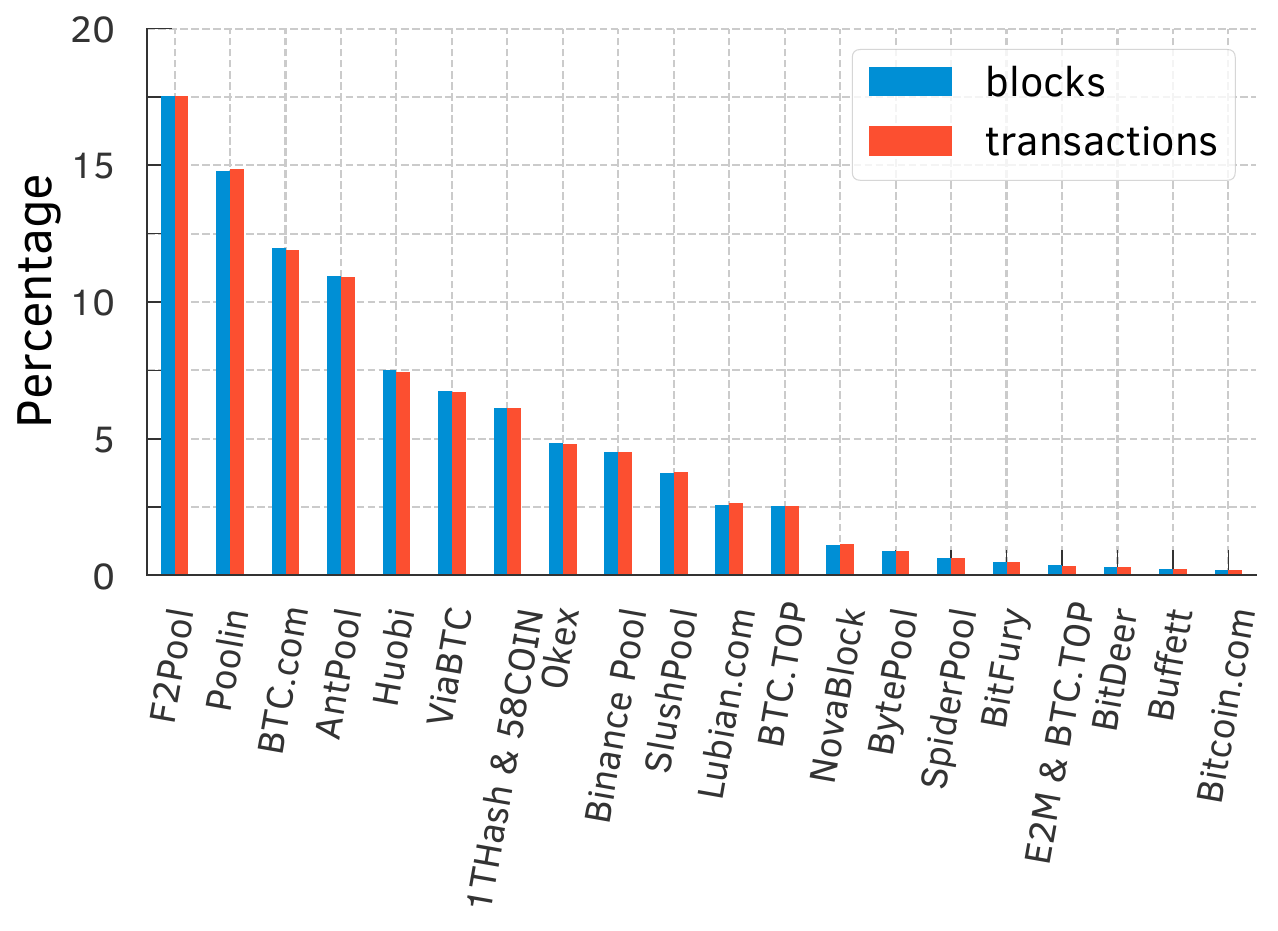}}
	\caption{
        Distribution of blocks mined and transactions confirmed by the top-20 
      MPOs in data sets \dsa{}, \dsb{}, and \dsc{}. Their combined normalized hash-rates account for 94.97\%, 93.52\%, and 98.08\% of all blocks mined in data set \dsa, \dsb, and \dsc, respectively.
      }
  \label{fig:dist-tx-blks-dataset-all}
\end{figure*}

\subsection{Data set collection}\label{subsec:datasets-a-and-b}

\paraib{Data set \dsa{}.}
To check miners' compliance to prioritization norms in Bitcoin, we analyzed all transactions and blocks issued in Bitcoin over a three-week time frame from February 20 through March 13, 2019 (see Table~\ref{tab:datasets}).
We obtained the data by running a \term{full} node, a Bitcoin software that performs nearly all operations of a miner (e.g., receiving broadcasts of transactions and blocks, validating the data, and re-broadcasting them to peers) except for mining.
The data set also contains a set of periodic \term{snapshots}, recorded once per $15$ seconds for the entire three-week period, where each snapshot captures the state of the full node’s \mpool{}.
We plot the distribution of the count of blocks and transactions mined by the top-20 MPOs for data set \dsa in Figure~\ref{fig:dist-txs-blks-dataset-a}. If we rank the MPOs in data set \dsa by the number of blocks ($B$) mined (or, essentially, the approximate hashing capacity $h$), the top five MPOs turn out to be BTC.com ($B$: $\num{536}$; $h$: $17.18\%$), AntPool ($B$: $\num{399}$; $h$: $12.79\%$), F2Pool ($B$: $\num{352}$; $h$: $11.29\%$), Poolin ($B$: $\num{344}$; $h$: $11.03\%$), and SlushPool ($B$: $\num{279}$; $h$: $8.94\%$).
We use this data for checking whether miners adhere to prioritization norms when selecting transactions for confirmation or inclusion in a block.

\paraib{Data set \dsb{}.}
Differences in configuration of the Bitcoin software may subtly affect the inferences drawn from data set \dsa{}.
A full node connects to \(8\) peers, for instance, in the default configuration, and increasing this number may reduce the likelihood of missing a transaction due to a ``slow'' peer.
The default configuration also imposes a minimum fee rate threshold of $\DefTxFee$ (or 1 satoshi-per-byte) for accepting a transaction.
We instantiated, hence, another full node to expand the scope of our data collection.
We configured this second node, for instance, to connect to as many as \(125\) peers. We also removed the fee rate threshold to accept even zero-fee transactions.
\dsb{} contains \mpool{} snapshots of this full node, also recorded once per \usd{15}, for the entire month of June 2019 (refer Table~\ref{tab:datasets}). We notice that $99.7\%$ of the transactions received by our \mpool{} were included by miners.
Figure~\ref{fig:dist-txs-blks-dataset-b} shows the distribution of the count of blocks and transactions mined by the top-20 MPOs for data set \dsb. The top five MPOs are BTC.com ($B$: $\num{889}$; $h$: $19.67\%$), AntPool ($B$: $\num{577}$; $h$: $12.77\%$), F2Pool ($B$: $\num{523}$; $h$: $11.57\%$), SlushPool ($B$: $\num{438}$; $h$: $9.69\%$), and Poolin ($B$: $\num{433}$; $h$: $9.58\%$).

\paraib{Data set \dsc{}.}
The insights derived from the above data motivated us to shed light on the aberrant behavior of mining pool operators (MPOs).
To this end, we gathered all ($\num{53214}$) Bitcoin blocks mined and their \num{112542268} transactions from January 1\tsup{st} to December 31\tsup{st} 2020.
These blocks also contain one Coinbase transaction per block, which the MPO creates to receive the block and the fee rewards.
This data set, labeled \dsc{}, contains $\num{112489054}$ issued transactions (see Table~\ref{tab:datasets}).
MPOs typically include a \stress{signature} or \stress{marker} in the Coinbase transaction, probably to claim their ownership of the block.
Following prior work (e.g., \cite{judmayer2017merged,Romiti2019ADD}), we use such markers for identifying the MPO (owner) of each block.
We failed to identify the owners of $\num{703}$ blocks (or approximately $1.32\%$ of the total), albeit we inferred $30$ MPOs in our data set.
In this thesis, we consider only the top-20 MPOs whose combined normalized hash-rates account for $98.08\%$ of all blocks mined.
Figure~\ref{fig:dist-txs-blks-dataset-c} shows the count of blocks mined by the top-20 MPOs according to \dsc{}.
The top five MPOs in terms of the number of blocks ($B$) mined are F2Pool ($B$: $\num{9326}$; $h$: $17.53\%$), Poolin ($B$: $\num{7876}$; $h$: $14.80\%$), BTC.com ($B$: $\num{6381}$; $h$: $11.99\%$), AntPool ($B$: $\num{5832}$; $h$: $10.96\%$), and Huobi ($B$: $\num{3990}$; $h$: $7.5\%$). 

\subsection{Detecting accelerated transactions.}
Given the high fees demanded by acceleration services, we anticipate that \stress{accelerated transactions would be included in the blockchain with the highest priority}, i.e., in the first few blocks mined by the accelerating miner and amongst the first few positions within the block. 
We would also anticipate that \stress{without the acceleration fee, the transaction would not stand a chance of being included in the block based on its publicly offered transaction fee}.
The above two observations suggest a potential method for detecting accelerated transactions in the Bitcoin blockchain:
An accelerated transaction would have a very high \textit{\textbf{signed position prediction error (SPPE)}} (refer to~\S\ref{subsec:sppe-metric}), as its predicted position based on its public fee would be towards the bottom of the block it is included in, while its actual position would be towards the very top of the block. 

To test the effectiveness of our method, we analyzed all \num{6381} blocks and \num{13395079} transactions mined by BTC.com mining pool in data set \dsc{}. 
We then extracted all transactions with SPPE greater or equal than $100\%$, $99\%$, $90\%$, $50\%$, $1\%$ and checked what fraction of such transactions were accelerated, according to the BTC.com transaction accelerator API~\cite{BTC@accelerator}.
%

%

%

\section{Analyzing norm adherence}
\label{sec:prioritization-norms}

In this section, we analyze whether Bitcoin miners adhere to prioritization norms, when selecting transactions for confirmation. To this end, we first investigate whether transaction ordering matters to Bitcoin users in practice, i.e., \stress{are there times when transactions suffer extreme delays and do users offer high transaction fees in such times to confirm their transactions faster?} We then conduct a progressively deeper investigation of the norm violations, including potential underlying causes, which we investigate in greater detail in the subsequent sections.

\subsection{Does transaction ordering matter?} \label{sec:mempool}

\begin{figure*}[t]
	\centering
		\subfloat[\label{fig:mempool-congestion}]{\includegraphics[width=0.86\twocolgrid]{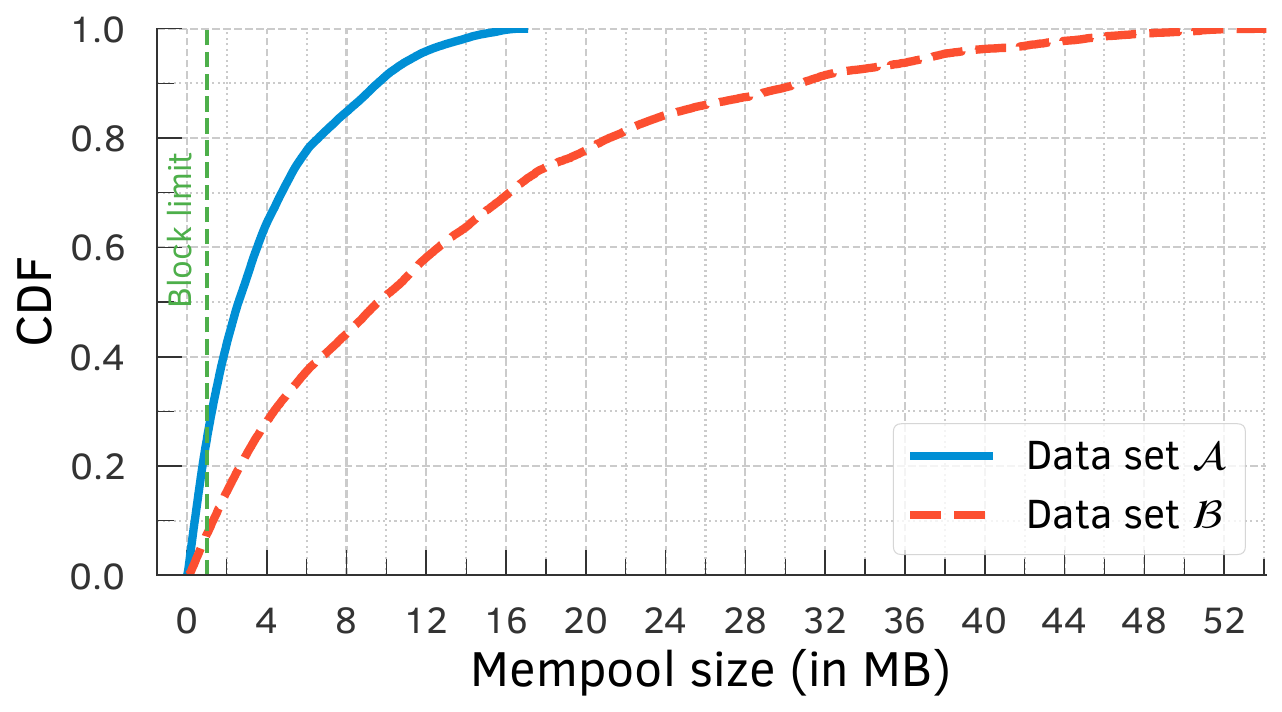}}
		\subfloat[\label{fig:mpool-sz-a}]{\includegraphics[width=\twocolgrid]{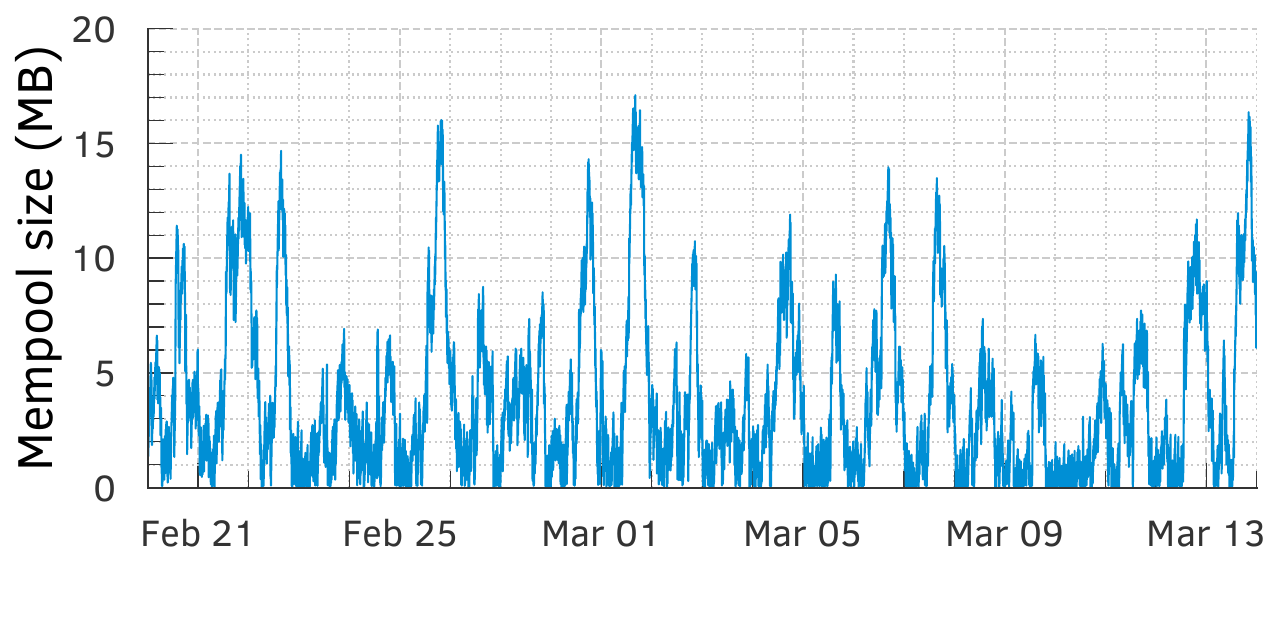}}
	\caption{
        (a) Distributions of \mpool{} size in both data sets \dsa{} and \dsb{}; and (b) the size \mpool in \dsa{} as a function of time, both indicating that congestion is typical in Bitcoin.
      }
\end{figure*}

\begin{figure*}[t]
	\centering
		\subfloat[\label{fig:tx-commit-times}]{\includegraphics[width=\twocolgrid]{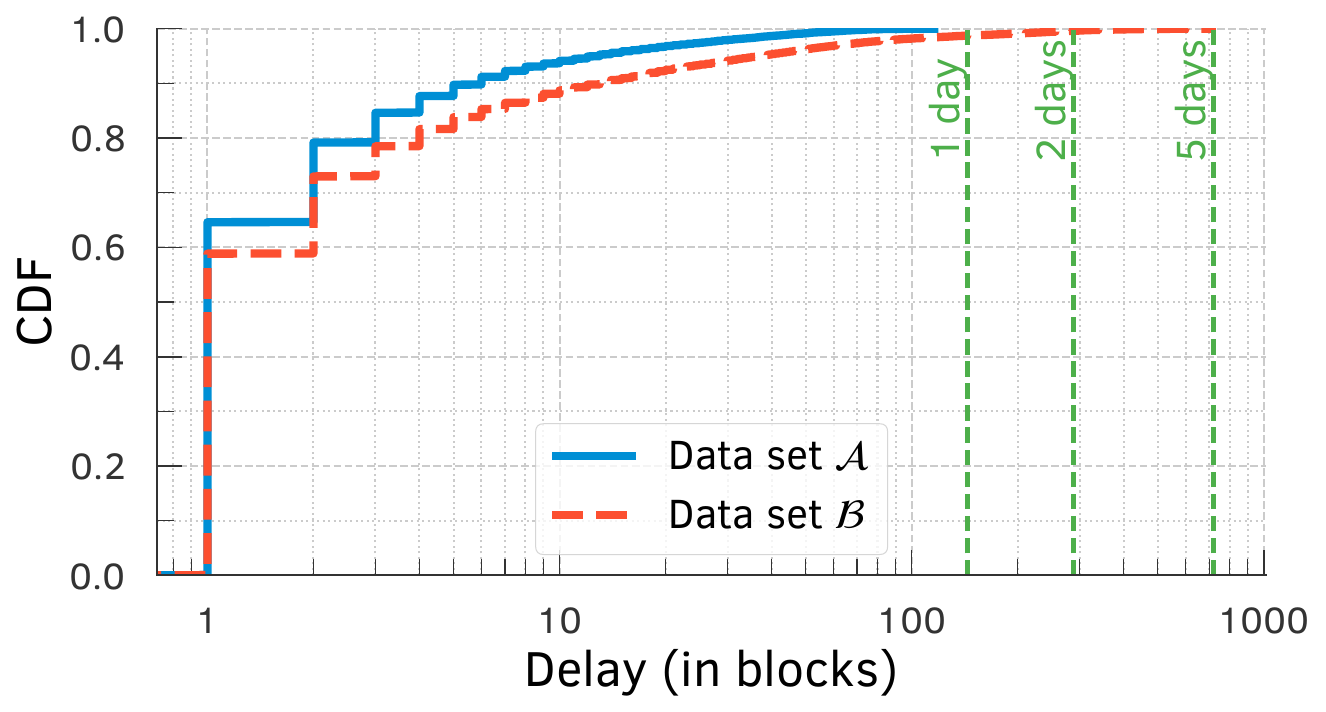}}
		\subfloat[\label{fig:cdf-fee-all}]{\includegraphics[width=\twocolgrid]{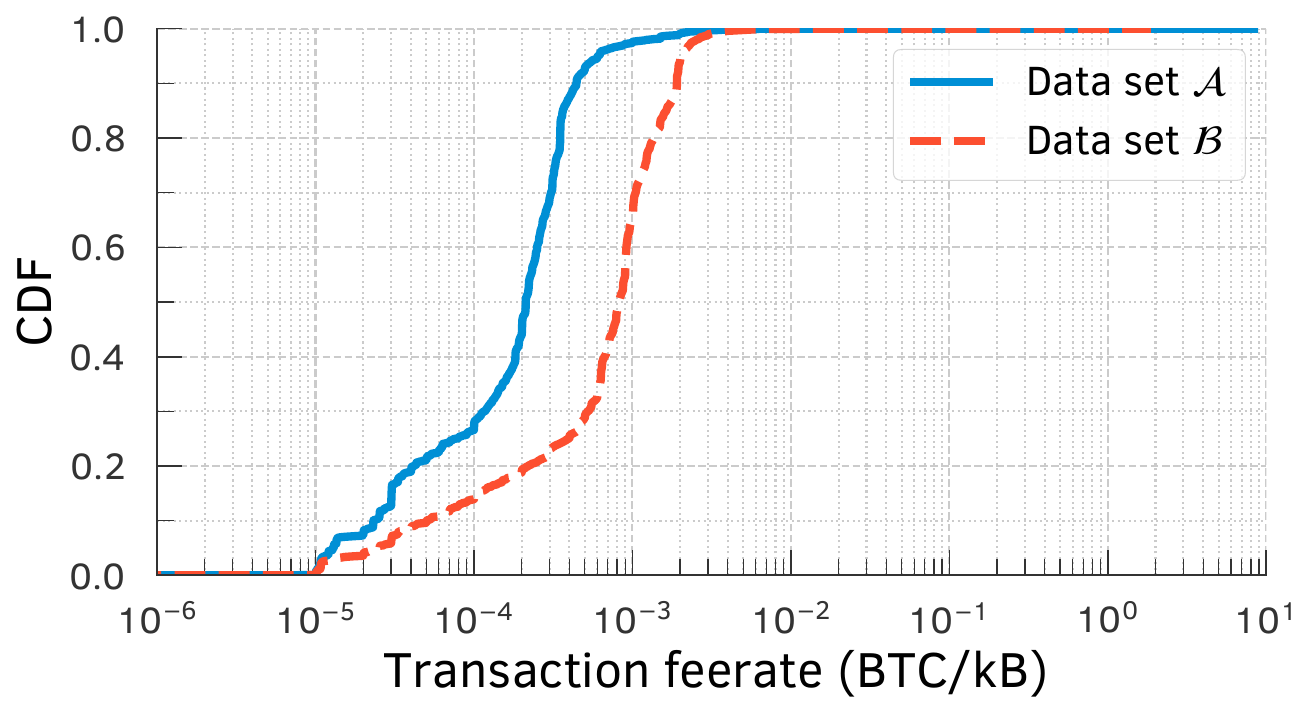}}
        \\
		\subfloat[\label{fig:fee-cong-rel-a}]{\includegraphics[width=\twocolgrid]{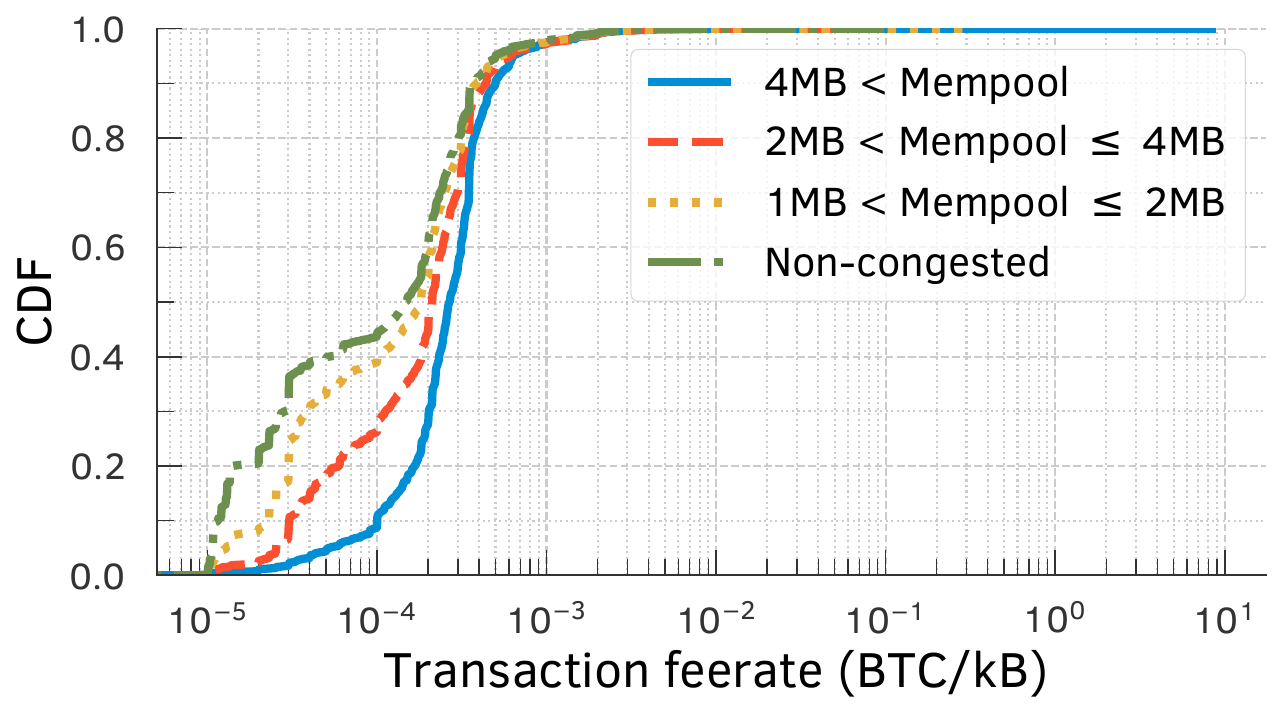}}
	\caption{
        (a) Distributions of delays until transaction inclusion show that a significant fraction of Bitcoin transactions experience at least 3 blocks (or approximately 30 minutes) of delay; Distributions of fee rates for (b) all transactions and (c) transactions (in \dsa{}) issued at different congestion levels clearly indicate that users incentivize miners through transaction fees.
      }
  \label{}
\end{figure*}

A~congestion in the \mpool leads to contention among transactions for
inclusion in a block.
Transactions that fail to contend with others (i.e., win a spot for inclusion)
experience inevitable delays in commit times.
Transaction ordering, hence, has crucial implications for users when the \mpool{} experiences congestion.
For instance, the Bitcoin Core code and most of the wallet software rely on the distribution of transactions' fee rates included in previous blocks to suggest to users the fees that they should include in their transactions~\cite{BitcoinCore-2021,Fees@Coinbase,Lavi-WWW2019}.
Such transaction-fee predictions from any predictor, which assume that miners follow the norm, will be misleading.\footnote{%
Coinbase, one of the top cryptocurrency exchanges, does not allow users to set transaction fees manually. Instead, it charges a fee based on how much they expect to pay for the concerned transaction, which in turn relies on miners following the norm~\cite{Fees@Coinbase}.}
Below, we examine whether \mpool{} in a real-world blockchain deployment
experiences congestion and its impact on transaction-commit
delays.
We then analyze whether, and how, users adjust transaction fees to cope with
congestion, and the effect of these fee adjustments on commit delays.

\subsubsection{Congestion and delays} \label{subsec:cong-delays}

Bitcoin's design---specifically, the adjustment of hashing difficulty to enforce
a constant mining rate---ensures that there is a steady flow of currency
generation in the network.
The aggregate number of size-limited blocks mined in Bitcoin, consequently, increases
linearly over time (Figure~\ref{fig:cdf-tx-blks-btc}).
Transactions, however, are \stress{not} subject to such constraints and have
been issued at much higher rates, particularly, according to Figure~\ref{fig:cdf-tx-blks-btc}, since mid-2017:
$60\%$ of all transactions ever introduced were added in only in the last 3.5 years of the nearly decade-long life of the cryptocurrency.
Should this growth in transaction issued continue to hold, transactions will
increasingly have to contend with one another for inclusion within the limited
space (of \uMB{1}) in a block.
Below, we empirically show that this contention among transactions is
already common in the Bitcoin network.

Using the data sets \dsa{} and \dsb{} (refer~\S\ref{subsec:datasets-a-and-b}), we measured
the number of unconfirmed transactions in the \mpool{}, at the granularity of
\usd{15}.
Per Figure~\ref{fig:mempool-congestion}, congestion in \mpool{} is
\stress{typical} in Bitcoin:
During the three-week period of \dsa, the aggregate size of all unconfirmed
transactions was above the maximum block size (of \uMB{1}) for nearly $75\%$
of the time; per data set \dsb the \mpool was congested for nearly $92\%$ of the
time period.
Figure~\ref{fig:mpool-sz-a} provides a complementary view of the \mpool
congestion in \dsa{}, by plotting the \mpool{} size as a function of time.
The measurements reveal a huge variance in \mpool{} congestion, with size of
unconfirmed transactions at times exceeding 15-times the maximum size of a
block.
Transactions queued up during such periods of high congestion will have to
contend with one another until the \mpool{} size drains below $\uMB{1}$.
These observations also hold in data set \dsb{}, the details of which are in~\S\ref{sec:supp-tx-ord}.

The \mpool congestion, which in turns leads to the contention among transactions
for inclusion in a block, has one serious implication for users: delays in
transaction-commit times.
While $65\%$ ($60\%$) of all transactions in data set \dsa (\dsb) get committed
in the next block (i.e., in the block immediately following their arrival in the
\mpool), Figure~\ref{fig:tx-commit-times} shows that nearly $15\%$ ($20\%$) of
them wait for at least $3$ blocks (i.e., $30$~minutes on average).
Moreover, $5\%$ ($10\%$) of the transactions wait for $10$ or more blocks, or
$100$~minutes on average, in data set \dsa (\dsb).
While no transaction waited for more than a day in data set \dsa, a~small
percentage of transactions waited for up to five days (because of the high
levels of congestion in June 2019) in data set \dsb.

\parai{Takeaways.}
\mpool{} is typically congested in Bitcoin. Transactions, hence, typically
contend with one another for inclusion in a block.
The \mpool{} congestion has non-trivial implications for transaction-commit times.

\subsubsection{Transaction fee rates and delays}\label{subsec:feerates}

To combat the delays and ensure that a transaction is committed ``on time''
(i.e., selected for inclusion in the earliest block), users may include a
transaction fee for incentivizing the miner.
While the block reward from May $11$, 2020 is
$\uBTC{6.25}$, the aggregate fees accrued per block is becoming considerable
(i.e., $6.29\%$ of the total miner revenue in 2020 per Table~\ref{tab:fee-revenue} in~\S\ref{sec:signif-tx-fees}).
Prior work also show that revenue from transaction fees is
clearly increasing~\cite{Easley-SSRN2017}.
With the volume of transactions growing aggressively (Figure~\ref{fig:cdf-tx-blks-btc}) over time and the block rewards, in Bitcoin,
halving every four years, it is inevitable that transaction fees will be an
important, if not the only, criterion for including a transaction, leading possibly to undercutting attacks~\cite{Carlsten@CCS16}.
Below, we analyze whether Bitcoin users incentivize miners via transaction fees and if such incentives are effective today.

Per Figure~\ref{fig:cdf-fee-all} the transaction fee rate of committed
transactions in both data sets \dsa{} and \dsb{} exhibits a wide range, from
$10^{-6}$ to beyond $\uTxFee{1}$.
The fee rate distributions of committed transactions also do not vary much between different mining pool operators (refer Figure~\ref{fig:cdf-fee-top5} in \S\ref{sec:tx-fees-across-mpos}).
A~few transactions ($0.001\%$ in \dsa{} and $0.07\%$ in \dsb{}) were committed,
despite offering fee rates less than the recommended minimum of
$10^{-5}$~\feeunit{}.
A non-trivial percentage of transactions offered fee rates that are two orders
of magnitude higher than the recommended value; particularly, in data set \dsb,
perhaps due to the comparatively high levels of congestion (cf.
Figure~\ref{fig:mpool-sz-a} and Figure~\ref{fig:mpool-sz-b}), $34.7\%$ of transactions offered fee rates higher
than $10^{-3}$~\feeunit{}.
Approximately $70\%$ ($51.3\%$) of the transactions in data set \dsa{} (\dsb{})
offer fee rates between $10^{-4}$ and $10^{-3}$~\feeunit{}, i.e., between one
and two orders of magnitude more than the recommended minimum.
Such high fee rates clearly capture the users’ intents to incentivize the
miners.

Our premise is that the (high) fee rates correlate with the level of
\mpool{} congestion.
Said differently, we hypothesize that users increase the fee rates to curb the
delays induced by congestion.
To test this hypothesis, we separate the \mpool{} snapshots (cf.~\S\ref{subsec:cong-delays}) into $4$ different bins.
Each bin corresponds to a specific level of congestion identified by the \mpool{} size as follows:
lower than \uMB{1} (\stress{no congestion}), in $(1, 2]$ MB (\stress{lowest congestion}), in $(2,
4]$ MB, and higher than \uMB{4} (\stress{highest congestion}).
The fee rates of transactions observed in the different bins or congestion levels, in Figure~\ref{fig:fee-cong-rel-a}, then validates our hypothesis:
Fee rates are strictly higher (in distribution, and hence also on average) for higher congestion levels.

\begin{figure*}[t]
	\centering
		\includegraphics[width={\onecolgrid}]{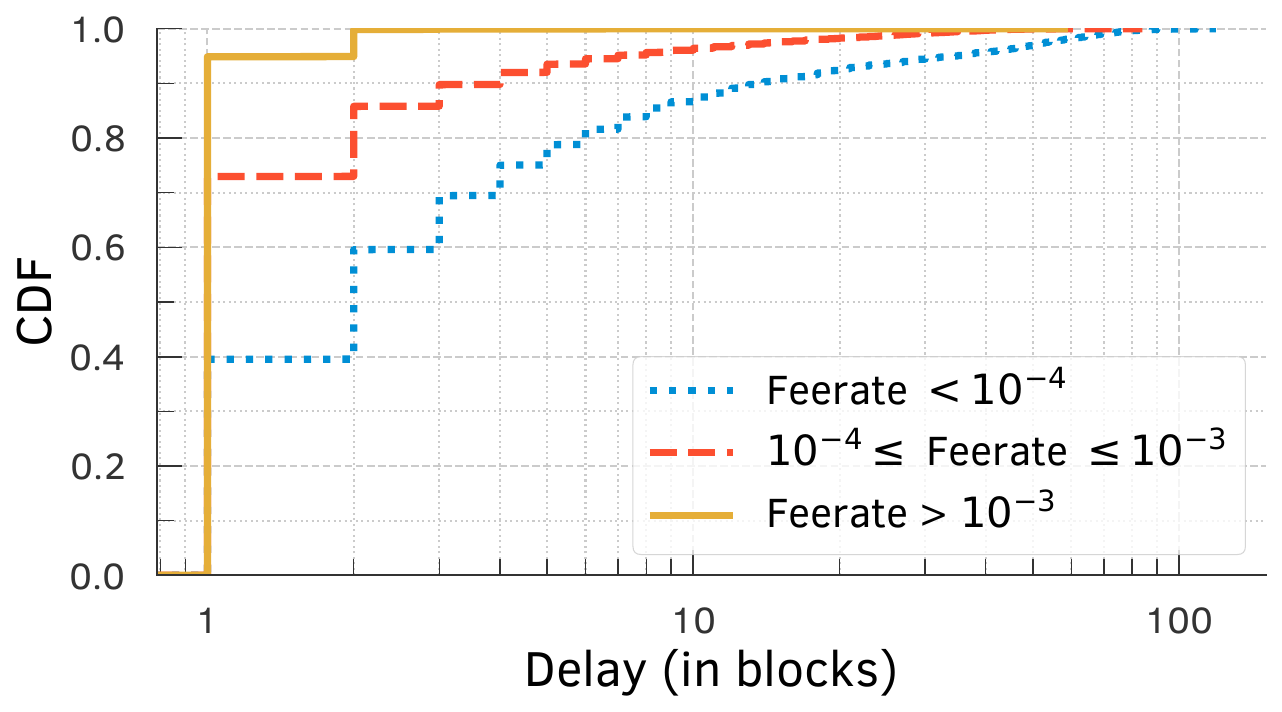}
	\caption{
  Distributions of transaction-commit delays for different fee rates for transactions in \dsa{}; incentivizing miners via fee rates works well in practice.}
\label{fig:fee-delay-rel-a}
\end{figure*}

Figure~\ref{fig:fee-delay-rel-a} shows that users' strategy of increasing
fee rates to combat congestion seems to work well in practice.
Here, we compare the CDF of commit delays of transactions with low (i.e., less
than $10^{-4}$~\feeunit{}), high (i.e., between $10^{-4}$ and
$10^{-3}$~\feeunit{}), and exorbitant (i.e., more than $10^{-3}$) fee rates, in data set \dsa{}.
Similar analysis with data set \dsb{} is provided in~\S\ref{sec:fees-and-cong}.
We observe that an increase in the transaction fee rates is consistently
rewarded (by miners) with a decrease in the commit delays.
This observation suggests that, at least to some extent, miners prioritize
transactions for inclusion based on fee rates or the fee-per-byte metric.

\parai{Takeaways.}
A significant fraction of transactions offers fee rates that are well above the
recommended minimum (i.e., $10^{-5}$~\feeunit{} or simply \DefTxFee).
Fee rates are typically higher at higher congestion levels, and reduce the
commit delays.
These observations suggest that users are indeed willing to spend money to
decrease commit delays for their transactions during periods of congestion.

\subsection{Do miners follow the norms?}\label{subsec:mining-prioritization-based-feerate}

Whether miners follow the transaction prioritization norms (as widely assumed)
has implications for both Bitcoin and its users:
The software used by users, for instance, assumes an adherence to these norms
when suggesting a transaction fee to the user~\cite{BitcoinCore-2021,Fees@Coinbase,Lavi-WWW2019}.
Deviations from these norms, hence, have far-reaching implications for both the
blockchain and crucially for Bitcoin users.

\subsubsection{Fee rate based selection when mining new blocks}

Our finding above show that transactions offering higher fee rates experience lower confirmation
delays suggests that miners tend to account for transaction fee rates when choosing transactions for new blocks. 
We now want to check, however, if transaction fee rate is the primary or the sole determining factor in transaction selection.
To this end, we check our data sets for transaction pairs, where one transaction was issued earlier and has a higher fee rate than the other, but was committed later than the other.
The existence of such transaction pairs would unequivocally show that fee rate alone does not explain the order in which they are selected.

We sampled $30$ \mpool snapshots, uniformly at random, from the set of all available snapshots in data set \dsa{}.
Suppose that, in each snapshot, we denote, for any transaction $i$, the time at
which it was received in the \mpool by $t_i$, its fee rate by $f_i$, and the
block in which it was committed by $b_i$.
We then selected, from each snapshot, all pairs of transactions $(i,j)$ such
that $t_{i} < t_{j}$ and $f_{i} > f_{j}$, but $b_{i} > b_{j}$.
Such pairs clearly constitute a violation of the fee-rate-based transaction-selection norm.

\begin{figure*}[t]
	\centering
		\subfloat[All transactions\label{fig:violation-all}]{\includegraphics[width=\twocolgrid]{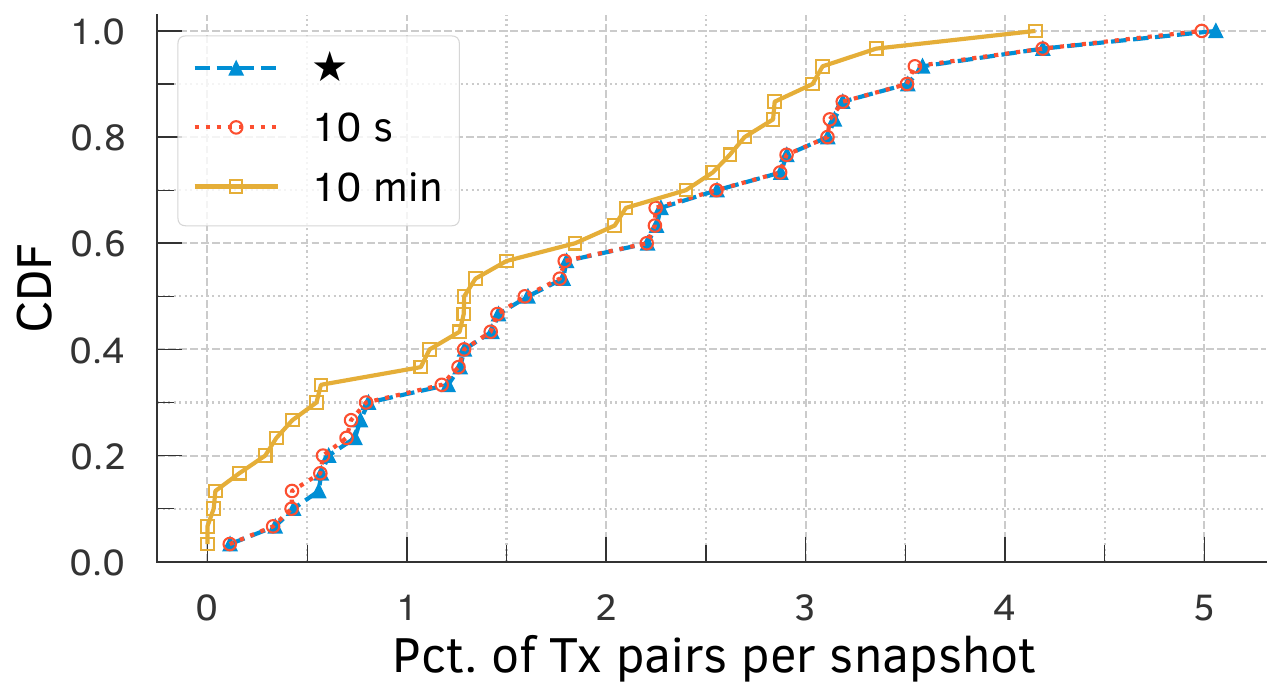}}
		\subfloat[Only Non-CPFP transactions\label{fig:violation-non-cpfp}]{\includegraphics[width=\twocolgrid]{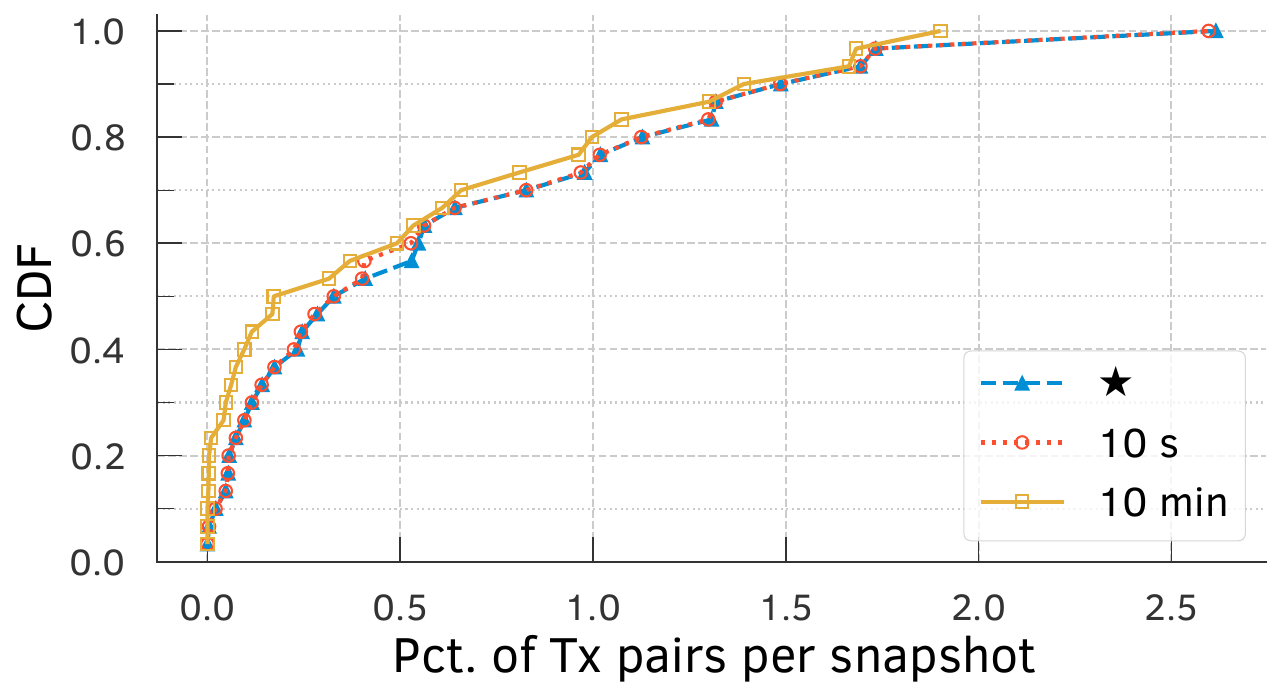}}
	\caption{
        There exists a non-trivial fraction of transaction pairs violating the norm across all		
     snapshots, clearly indicating that miners do \underline{not} adhere to the norm.
      }
  \label{fig:violation}
\end{figure*}


Figure~\ref{fig:violation-all} shows a cumulative distribution of the fraction of all transaction pairs (line labeled ``$\star$'') violating the norm over all sampled
snapshots. 
Across all snapshots, a small but non-trivial fraction of all transaction pairs violate the norm. 
One potential explanation for violations might be that the transactions are received by the mining pools in different order than the one in which our \mpool receives.
To account for such differences, we tighten the time constraint as $t_{i} + \epsilon < t_{j}$ and use an $\epsilon$ of either $10$ seconds or $10$~minutes. 
Even with the tightened time constraints, Figure~\ref{fig:violation-all} shows that a non-trivial fraction of 
all transaction pairs violate the norm.

Another potential source of violations is Bitcoin's dependent (or, parent and child) transactions, where the child pays a high fee to incentivize miners to also confirm the
parent from which it draws its inputs. This mechanism enables users to
``accelerate'' a transaction that has been ``stuck'' because of low
fee~\cite{CoinStaker-2018}. As the existence of such \newterm{child-pays-for-parent (CPFP)} transactions (formally defined in~\S\ref{sec:cpfp-txs})
would introduce false positives in our analysis we decided to discard them.
Figure~\ref{fig:violation-non-cpfp} shows that the violations exist even after discarding all such dependent transaction pairs.

\subsubsection{Fee rate based ordering within blocks}

We now turn our attention to transaction ordering within individual (mined) blocks in Bitcoin.
If a miner followed GBT, transactions would be ordered based on their fee rate.
In this case, given the set of non-CPFP transactions $T = \{T_1, T_2, .... T_n\}$ included in a block $B$, we should be able to predict their position in the block by simply ordering the transactions based on their fee rate (as specified in the GBT implementation in Bitcoin Core).
To quantify the deviation from the norm, we compute a measure that we call \textit{\textbf{position prediction error (PPE)}}: PPE of a block $B$ is the average absolute difference between the predicted and the observed (actual) positions for all transactions in block $B$, normalized by the size of the block ($n$) and expressed as a percentage. More precisely,

\begin{align*}
    PPE(B) = \sum_{i=1}^n \dfrac{(|T^{p}_{i} -  T^{o}_{i})|) \cdot 100}{n}
\end{align*}

where $T^{p}_{i}$ and $T^{o}_{i}$ are the predicted and observed positions of a transaction, respectively.

\begin{figure*}[t]
	\centering
		\subfloat[Overall position prediction error\label{fig:deviation-within-blocks-overall}]{\includegraphics[width=\twocolgrid]{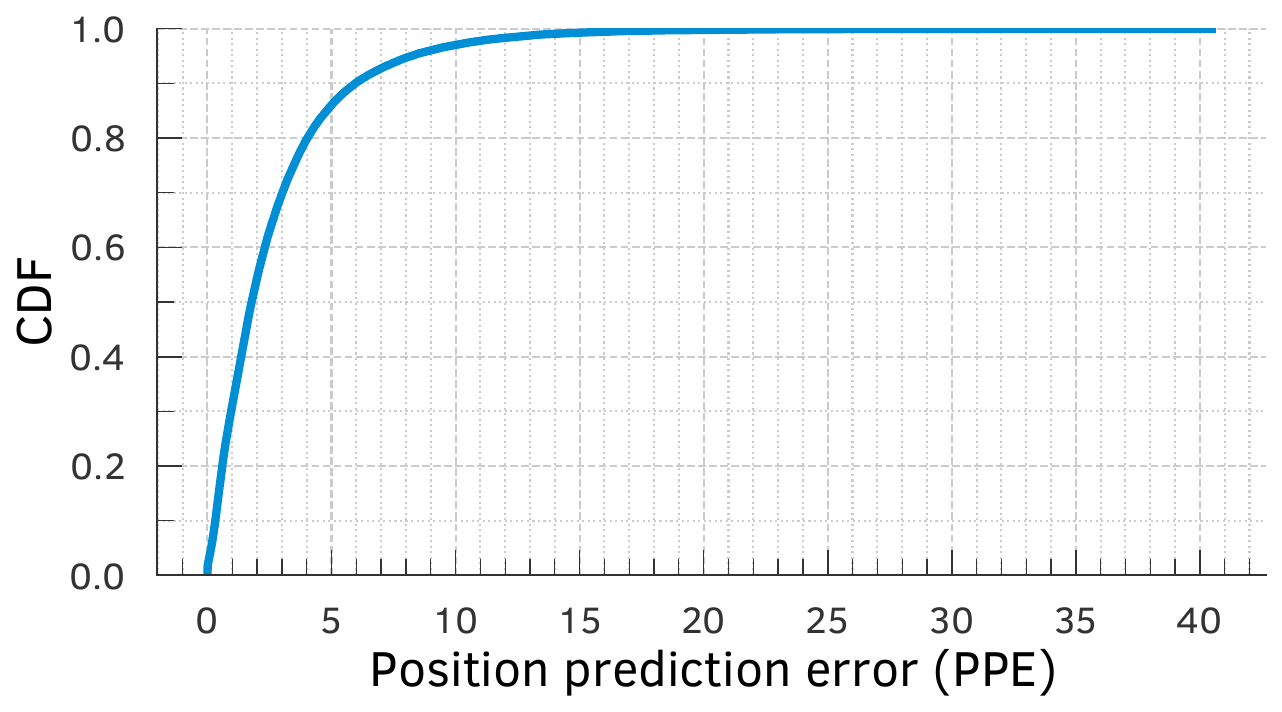}}
		\subfloat[Position prediction error of the top-6 MPOs\label{fig:deviation-within-blocks-top6-mpo}]{\includegraphics[width=\twocolgrid]{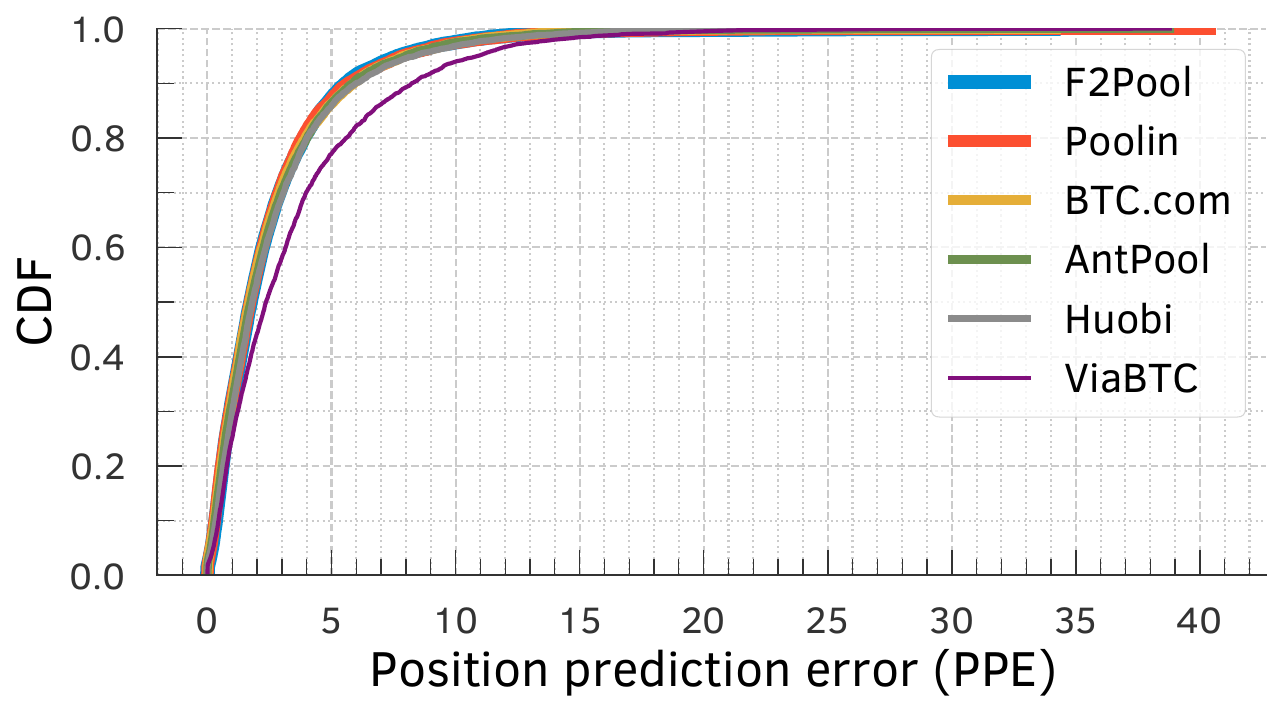}}
	\caption{
        Position prediction error (PPE). (a) There are 52,974 (99.55\%) blocks with at least one non-CPFP txs. The mean PPE is 2.65\%, with an std of 2.89. 80\% of all blocks has PPE less than 4.03\%. (b) The PPEs of blocks mined by the top-6 MPOs according to their normalized hash rate.
      }
  \label{fig:deviation-within-blocks}
\end{figure*}

Figure~\ref{fig:deviation-within-blocks-overall} shows the cumulative distribution of PPE values for each block in our data set \dsc{}, containing \num{53214} blocks. $80\%$ of the blocks have PPE values less than $4.03\%$. The mean PPE across all blocks is $2.65\%$, with a standard deviation of $2.89$.
Per this plot the position of a transaction within a block can be predicted with very high accuracy (within a few percentile position error), suggesting that transactions are by and large ordered within a block based on their fee rate. Figure~\ref{fig:deviation-within-blocks-top6-mpo} shows PPE values separately for each of the $6$ largest mining pools in data set \dsc{}. The plots show that all mining pools by and large follow the norm, though some like ViaBTC seems to deviate slightly more from the norm compared to the other mining pools.

\subsubsection{Fee rate threshold for excluding transactions}

In their default configuration, many nodes in the Bitcoin P2P network drop (i.e., ignore) transactions that offer less than a threshold fee rate (typically, $10^{-5}$ \feeunit). 
As miners select transactions for inclusion from their local Bitcoin P2P node, this (default) norm would result in such low-fee transactions never being included in the blockchain, even during periods of non-congestion (when blocks have spare capacity to accommodate additional transactions).

We collected data set \dsa{} using a default Bitcoin node, and our node, hence, did not accept or record low-fee transactions.
When gathering data set \dsb{}, however, we configured our Bitcoin node to accept all transactions, irrespective of their fee rates.
In data set \dsb{}, our node, consequently, received \num{1084} transactions that offered less than the recommended fee rate and $489$ ($45.11\%$) of them were zero-fee transactions.
From these low fee rate transactions, only \num{53} ($4.89\%$) were confirmed in the Bitcoin blockchain; $9$ ($16.98\%$) were confirmed months after they were observed in our data set.
In contrast, the vast majority ($99.7\%$) of the transactions that offered greater than or equal to the recommended fee rate were all (eventually) confirmed.
Interestingly, the low-fee transactions were confirmed by just three mining pools: F2Pool, ViaBTC, and BTC.com included $38$, $14$, and $1$ low-fee transactions, respectively.
Our findings suggest that while the norm of ignoring transactions offering less than the recommended fee rate is being by and large followed by all miners, a few occasionally deviate from the norm.

%

\section{Investigating norm violations} \label{sec:self-interest-scam-txs}

Our analysis so far showed that while Bitcoin miners by and large follow transaction-prioritization norms, there are many clear instances of norm violations. 
Our next goal is to develop a deeper understanding of the underlying reasons or motivations for miners to deviate from the fee rate based norms, at least for some subset of all transactions. 
To this end, we focus our investigation on the following three types of transactions, where we hypothesize miners might have an incentive to deviate from the current norms, which are well-aligned towards maximizing their rewards for mining.

\begin{enumerate}
  \item \stress{Self-interest Transactions:} Miners have a vested interest in a transaction, where the miners themselves are a party to the transaction, i.e., a sender or a receiver of bitcoins. Miners may have an incentive to selfishly accelerate the commitment of such transactions in the blocks mined by themselves. 
  
  \item \stress{Scam-payment Transactions:} Bitcoins are increasingly being used to launch a variety of ransomware and scam attacks~\cite{Frenkel@nyt17,Mathews@forbest17,Frenkel@nyt20}. A scam attack involved using hijacked Twitter accounts of celebrities to encourage their followers to send bitcoins to a specific Bitcoin wallet address~\cite{Frenkel@nyt20}.
  Given the timely and widespread coverage of this attack in popular press and other similar attacks on crowdsourced websites for reporting scam transactions~\cite{Scam@BitcoinAbuse,Scam@ScamAlert}, and with governments trying to blacklist wallet addresses of entities suspected of illegal activities~\cite{De@Coindesk,Hinkes@Coindesk}, we hypothesize that some miners might decelerate or even absolutely exclude the commitment of scam-payment transactions out of fear or ethical concerns.
  
  \item \stress{Dark-fee Transactions:} Recently, some mining pool operators have started offering transaction acceleration services~\cite{BTC@accelerator,ViaBTC@accelerator,Poolin@accelerator,F2Pool@accelerator,AntPool@accelerator}, where anyone wanting to prioritize their transactions can pay an additional fee to a specific mining pool via a side-channel (often, the MPO's website or via a private-channel~\cite{strehle2020exclusive}). Such transaction fees are ``dark'' or opaque to other mining pools and the public, and we hypothesize that some committed low-fee transactions might have been accelerated by using such services. 
  
\end{enumerate}

To detect whether a mining pool has accelerated or decelerated the above types of transactions, we first design a robust statistical test.
Later, we report our findings from applying the test on the three types of transactions.

\subsection{Statistical test for differential prioritization}\label{subsec:sppe-metric}

Our goal here is to propose a robust statistical test for detecting whether a given mining pool $m$ is prioritizing a given set of committed transactions $c$ \stress{differently} than all other miners.
The basic idea behind the statistical test is as follows. Suppose a mining pool is accelerating (decelerating) transactions in set $c$. In that case, these transactions will have a disproportionately high (low) chance of being included in blocks mined by this mining pool compared to the mining pool's hashing power (or rate).

\subsubsection{Test for differential transaction acceleration}

Consider a miner $m$ with normalized hash rate $h = \theta_0$ (estimated as fraction of blocks mined by $m$). Assume that we are given a set of transactions, denoted as $c$-transactions (for committed transactions), for which we wish to test whether miner $m$ is treating them preferentially.

To test whether $m$ is prioritizing $c$-transactions, we look at all blocks that include at least one $c$-transaction, call them $c$-blocks. Suppose that there are $y$ such blocks.
If $m$ is not prioritizing $c$-transactions, then a fraction $\theta_0$ of all $c$-blocks should be $m$-blocks (i.e., mined by $m$); if $m$ is prioritizing $c$-transactions (compared to other miners) then the fraction will be higher. We want to test whether the true fraction $\theta$ is indeed $\theta_0$ or is higher. We formalize this as follows: We assume that each $c$-block has a probability $\theta$ to be an $m$-block and do the following test.
\begin{align*}
  & H_0: \theta = \theta_0 \\ 
  & H_1: \theta > \theta_0.
\end{align*}
Assuming that the observed number of $c$-blocks that are mined by $m$ is $x$, the $p$-value of the test is 
\begin{align*}
  p = Pr (B \ge x), 
\end{align*}
where $B$ is a binomial distribution of parameter $\theta_0$ and $y$, that is 
\begin{align*}
  p = \sum_{k=x}^y \binom{y}{k} \theta_0^k (1-\theta_0)^{(y-k)}.
\end{align*}
We may fix the size of the test (i.e., the maximal probability of type I error that corresponds to rejecting $H_0$ when $H_0$ is true) to $\alpha = 0.01$. Then $H_0$ should be rejected whenever $p<\alpha$. The smaller $p$, the higher the confidence in rejecting $H_0$, that is declaring that $m$ prioritizes c-transactions. 

The above test is relative in the sense that we can only detect if a miner treats $c$-transactions more preferentially than the rest of the miners. This test cannot conclude on whether it is the miner accelerating the $c$-transactions (relative to their deserved, i.e., fee rate based, priority) or the rest of the miners are decelerating them. 
So, we look at additional empirical evidence from the position of the $c$-transactions within the $c$-blocks that include them.
Specifically, given the set of $c$-transactions $\{c_1, c_2, .... c_n\}$ committed by a miner $m$, 
we compute a measure that we call \textit{\textbf{signed position prediction error (SPPE)}} as the average signed difference between the predicted and observed  positions (measured as percentile rank) for all $c$-transactions within the blocks committed by $m$. More precisely,

\begin{align*}
  SPPE(m) = \dfrac{\sum_{i=1}^n (c^{p}_{i} -  c^{o}_{i}) \cdot 100}{n}
\end{align*}
where $c^{p}_{i}$ and $c^{o}_{i}$ are the predicted and the observed (percentile rank) positions, respectively, of transaction $c_i$ within the blocks committed by $m$.

\subsubsection{Test for differential transaction deceleration}

While the previous test checks for prioritization (or acceleration), one may also want to test for deceleration. To that end, a symmetric test can be used. Specifically, with the previous notation, the test would be
\begin{align*}
  & H_0: \theta = \theta_0 \\ 
  & H_1: \theta < \theta_0;
\end{align*}
and its $p$-value would be 
\begin{align*}
  p = Pr (B \le x), 
\end{align*}
where $B$ is a binomial distribution of parameter $\theta_0$ and $y$, that is 
\begin{align*}
  p = \sum_{k=0}^x \binom{y}{k} \theta_0^k (1-\theta_0)^{(y-k)}.
\end{align*}

\begin{figure*}[t]
	\centering
		\subfloat[Number of MPOs' wallet addresses\label{fig:bar-number-wallet-addresses}]{\includegraphics[width=\twocolgrid]{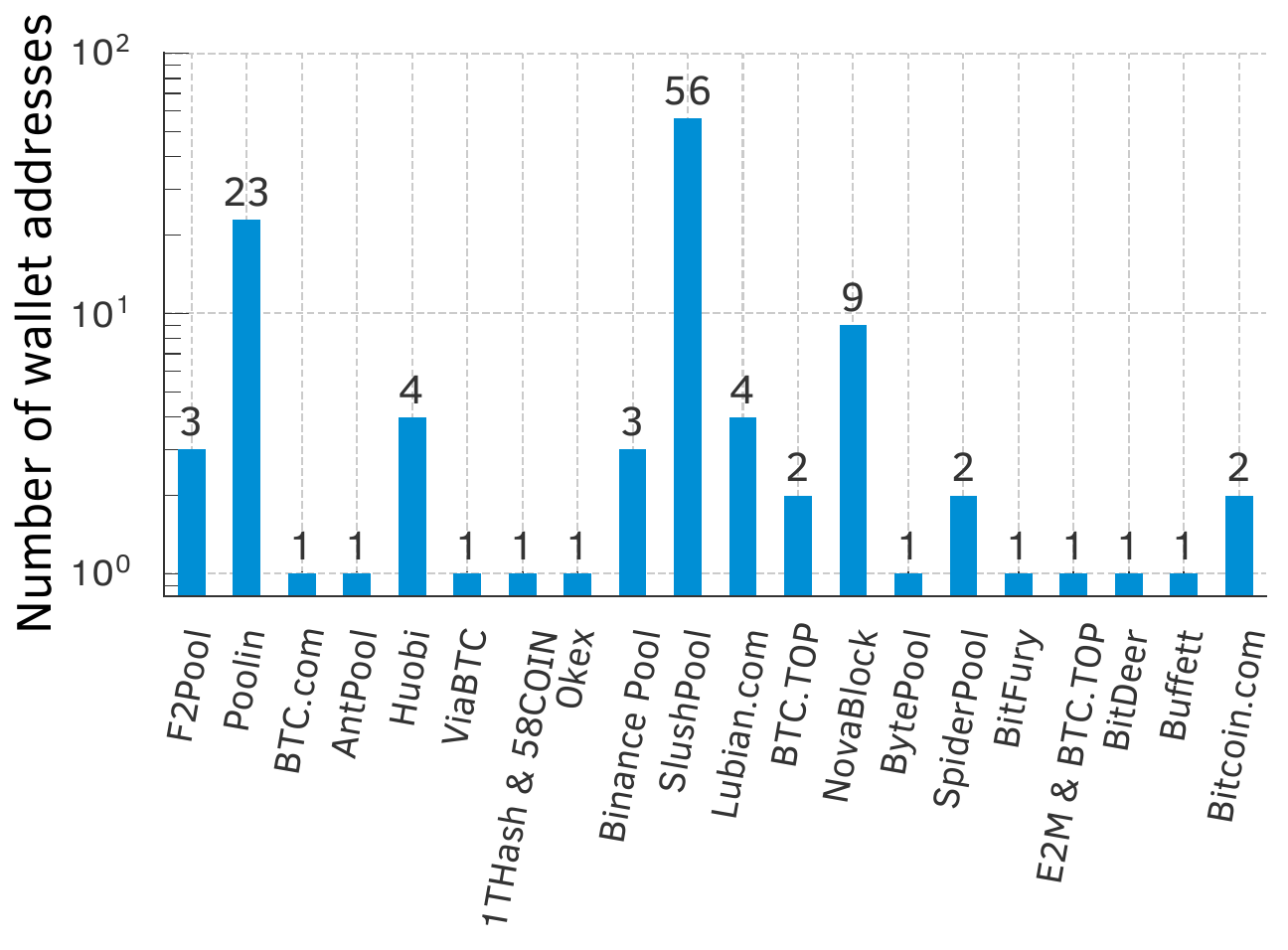}}
		\subfloat[Number of MPOs' transactions\label{fig:bar-num-wallet-addresses-mpo}]{\includegraphics[width=\twocolgrid]{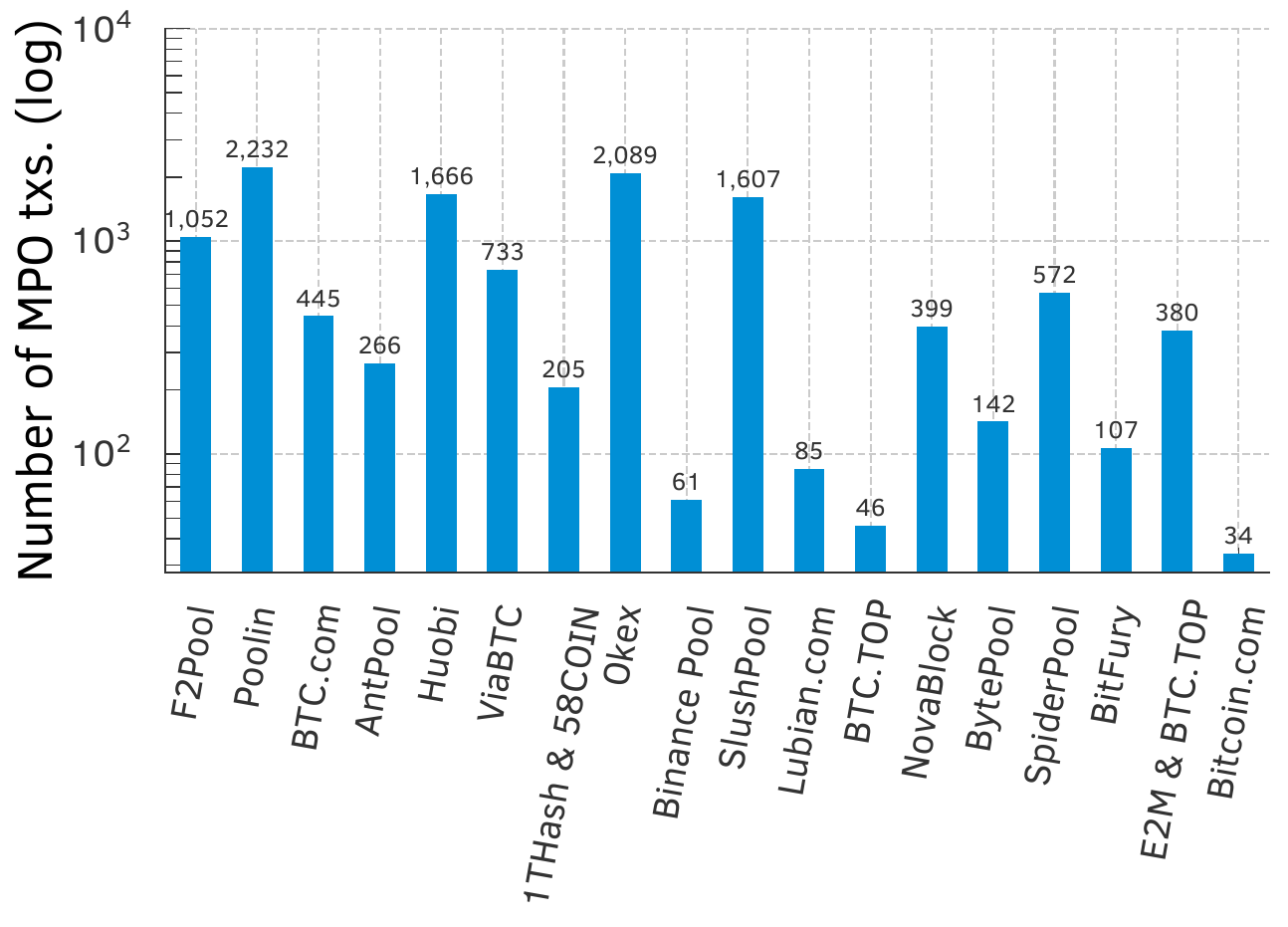}}
	\caption{
        (a) Distribution of the number of wallet addresses in data set \dsc used by each of the top-20 MPOs to receive its block rewards; SlushPool and Poolin, for instance, used 56 and 23 distinct wallet addresses, respectively. (b) The counts of inferred MPO transactions; in total, 12,121 transactions were inferred as MPOs' transactions, which corresponds to 0.011\% of the total issued transactions recorded in the Bitcoin blockchain. Poolin has the majority with 2232 (18.41\%), followed by Okex with 2089 (17.24\%) and Huobi with 1666 (13.74\%) transactions. BitDeer and Buffett have the same wallet address as BTC.com and Lubian.com, respectively. We count the addresses of the former as belonging to the latter.
      }
  \label{fig:bar-wallet-addresses}
\end{figure*}

\subsubsection{Scaling the tests}

While we did not face them in this thesis, our test may have two limitations when scaling to large time windows and/or large numbers of transactions.

First, it may become difficult to compute the $p$-value from the binomial distribution for large values of $y$. In such cases, we can use the following approximation for our analysis: If $y$ is large enough and $\theta_0$ is not close to zero or one (i.e., $x$ and $y-x$ are large enough), the binomial distribution of parameters $\theta_0$ and $y$ is well approximated by the normal distribution with mean $y\theta_0$ and variance $y\theta_0(1-\theta_0)$. Hence, the $p$-value for the acceleration test can be computed as,
\begin{align*}
  p \simeq \Phi \left( \frac{x-y\theta_0}{\sqrt{y\theta_0(1-\theta_0)}} \right),
\end{align*}
where $\Phi$ is the CDF of a standard normal random variable. A similar approximation can be done for the deceleration test.

Second, the hash rates of miners in our $p$-value test are assumed to be more or less constant (i.e., $\theta_0$ is a constant), which is not the case (per Figure~\ref{fig:btc-hashrate} and Figure~\ref{fig:eth-hashrate} in \S\ref{sec:hash-var}). This assumption is a limitation of our test as, in reality, hash rates of miners may vary over time, particularly over large time windows. In such situations, our test results may be affected, particularly when the arrival times of transactions are not regularly spread over the time window of our analysis. We address this issue by confirming the results of the $p$-value test through the SPPE-test, which is not affected by variable hash rates. It is possible, however, to alleviate this limitation of our analysis. One natural way is to divide the total time window into multiple windows such that the hash rate is more or less constant in those shorter time windows; and compute $p$-values in each time window. We can then combine the obtained $p$-values using Fisher's method \cite{Fisher@1992Statistical,Fisher@1948}. We leave the investigation of such extended test procedures to future work, when they might be needed.

\subsection{Self-interest transactions}

To identify transactions where a mining pool is a sender or receiver of transactions, we first need to identify Bitcoin wallets (addresses) that belong to mining pools.
In Bitcoin, whenever a mining pool discovers a new block, it specifies a wallet address to receive the mining rewards. 
This mining pool address is included in the Coinbase transaction (refer~\S\ref{sec:background-blockchains}) that appears at the start of every block.
In our data set \dsc{}, we gathered all the wallet addresses used by the top-$20$ mining pools to receive their rewards.
For each mining pool, we then retrieved all committed transactions, in which coins were sent from the mining pool's wallet.
Figure~\ref{fig:bar-wallet-addresses} shows the statistics for the mining pool wallets and the transactions spending (sending) coins from (to) the wallets, for each of the top-$20$ mining pools in data set \dsc{}.
We found hundreds or thousands of self-interest transactions for most of the mining pools.

\subsubsection{Acceleration of self-interest transactions}

\begin{table*}[t]
    \tiny
    \tabcap{Differential prioritization of self-interest transactions.}\label{tab:self-interest-txs}
    \resizebox{0.95\textwidth}{!}{
        \begin{tabular}{@{}lccccccc@{}}
            \toprule
            \multicolumn{1}{p{0.8cm}}{\thead{Transactions}} & 
            \multicolumn{1}{l}{\thead{mining pool}} &
            \multicolumn{1}{l}{\thead{norm. hash rate}} &
            \multicolumn{1}{c}{\multirow{2}{*}{$\thead{x}$}} &
            \multicolumn{1}{c}{\multirow{2}{*}{$\thead{y}$}} &
            \multicolumn{2}{c}{\thead{p-value}} &
            \multicolumn{1}{c}{\thead{\% SPPE}} \\
            \multicolumn{1}{c}{\thead{of ...}} &
            \multicolumn{1}{c}{\thead{(m)}} &
            \multicolumn{1}{c}{\thead{($\theta_0$)}} &
            \multicolumn{1}{c}{~} & \multicolumn{1}{c}{~} &
            \multicolumn{1}{c}{\thead{(accel.)}} &
            \multicolumn{1}{c}{\thead{(decel.)}} &
            \multicolumn{1}{c}{\thead{(m)}} \\ 
            \midrule
            {\quad\quad \textit{\textbf{F2Pool}}}
                & F2Pool & 0.1753 & 466 & 839 & \attention{0.0000} & 1.0000 & \attention{78.5494} \\
            \arrayrulecolor{gray}\midrule
            {\quad\quad \textit{\textbf{ViaBTC}}}
                & ViaBTC & 0.0676 & 412 & 720 & \attention{0.0000} & 1.0000 & \attention{98.9175} \\
            \arrayrulecolor{gray}\midrule
            \multirow{2}{*}{\quad\quad \textit{\textbf{1THash \& 58Coin}}}
                & ViaBTC & 0.0676 & 34 & 201 & \attention{0.0000} & 1.0000 & \attention{81.4516} \\
                & 1THash \& 58Coin & 0.0611 & 39 & 201 & \attention{0.0000} & 1.0000 & \attention{96.9143} \\
            \arrayrulecolor{gray}\midrule
            \multirow{2}{*}{\quad\quad \textit{\textbf{SlushPool}}}
                & SlushPool & 0.0375 & 214 & 1343 & \attention{0.0000} & 1.0000 & \attention{88.3082} \\
                & ViaBTC & 0.0676 & 140 & 1343 & \attention{0.0000} & 1.0000 & \attention{45.1523} \\
            \arrayrulecolor{black}\bottomrule
        \end{tabular}
    } 
\end{table*}

For self-interest transactions belonging to each of the top-20 mining pools, we separately applied our statistical test to check whether any of the top-10 mining pools (that mined at least $4\%$ of all mined blocks in data set \dsc{}) are preferentially accelerating or decelerating the transactions.
In Table~\ref{tab:self-interest-txs}, we report the statistics from our test for mining pools that were found to preferentially treat transactions belonging to their own or other mining pools.
Strikingly, Table~\ref{tab:self-interest-txs} shows that 4 out of the top-10 mining pools namely, F2Pool, ViaBTC, 1THash \& 58Coin, and SlushPool \stress{selfishly accelerated} their own transactions, i.e., coin transfers from or to their own accounts (p-value for acceleration test is less than $0.001$).
Equally, if not more interestingly, Table~\ref{tab:self-interest-txs} shows collusive behavior among mining pools. 
Specifically, it shows that transactions issued by 1THash \& 58Coin and SlushPool were \stress{collusively accelerated} by ViaBTC (p-value for acceleration test is less than $0.001$).
That these mining pools were accelerating the transactions is further confirmed by the SPPE measure, which clearly shows that in each of the above cases, the self-interest transactions were also being included within the blocks ahead of other higher fee rate transactions.

\subsection{Scam-payment transactions}\label{subsec:scam_payment_txs}

\begin{table*}[t]
    \begin{center}
        \tiny
        \tabcap{Differential prioritization of scam-payment transactions}\label{tab:twitter-scam-txs}
        \resizebox{0.75\textwidth}{!}{%
            \begin{tabular}{@{}ccccccr@{}}
                \toprule
                \multicolumn{1}{c}{\thead{mining pool}} &
                \multicolumn{1}{l}{\thead{norm. hash rate}} &
                \multicolumn{1}{c}{\multirow{2}{*}{$\thead{x}$}} &
                \multicolumn{1}{c}{\multirow{2}{*}{$\thead{y}$}} &
                \multicolumn{2}{c}{\thead{p-value}} &
                \multicolumn{1}{c}{\thead{\% SPPE}} \\
                \multicolumn{1}{c}{\thead{(m)}} &
                \multicolumn{1}{c}{\thead{($\theta_0$)}} &
                \multicolumn{1}{c}{~} & \multicolumn{1}{c}{~} &
                \multicolumn{1}{c}{\thead{(accel.)}} &
                \multicolumn{1}{c}{\thead{(decel.)}} &
                \multicolumn{1}{c}{\thead{(m)}} \\ 
                \midrule
                Poolin & 0.1528 & 10 & 53 & 0.2856 & 0.8227 & 	$-3.9787$ \\
                F2Pool & 0.1450 & 10 & 53 & 0.2323  & 0.8629 & $0.8735$ \\
                BTC.com & 0.1147 & 9  & 53 & 0.1483  & 0.9233 & $-2.8333$ \\
                AntPool & 0.1093 & 4  & 53 & 0.8450 & 0.2989  & $31.5000$  \\
                Huobi  & 0.0955 & 1  & 53 & 0.9951 & 0.0323 &  $-1.6428$ \\
                Okex & 0.0698 & 3  & 53 & 0.7248 & 0.4890  & $-5.0000$  \\
                1THash \& 58COIN & 0.0684 & 8  & 53 & 0.0268  & 0.9907  & $-0.5000$ \\
                Binance Pool   & 0.0590 & 3  & 53 & 0.6120 & 0.6180  & $-2.6000$ \\
                ViaBTC & 0.0552 & 1  & 53 & 0.9507 & 0.2020 &  $-4.0000$ \\ \bottomrule
            \end{tabular}%
        } 
    \end{center}
\end{table*}

Next, we investigate whether any mining pool attempted to decelerate or exclude scam-payment transactions.  

On July 15, 2020, multiple celebrities'
accounts on Twitter fell prey to a scam attack.
The scammers posted the message that anyone who transferred bitcoins to a specific
wallet will receive twice the amount in return~\cite{Frenkel@nyt20}.
In response, several people sent, in total, $12.87051731$ bitcoins---then
worth nearly $\num{142000}$ (USD)---to the attacker's wallet via $386$
transactions, which were confirmed across $53$ blocks by $12$ miners.

To examine the miners' behavior during this scam attack, we selected all blocks mined from July 14 to August 9, 2020 (i.e., \num{3697} blocks in total, containing \num{8318621} issued transactions as described in~\S\ref{sec:supp-scam-txs}) from our data set \dsc.
Once again, we applied our statistical test to check whether any of the top-$9$ mining pools (that mined at least $5\%$ of all mined blocks from this data) are preferentially accelerating or decelerating the transactions.
Table~\ref{tab:twitter-scam-txs} shows the test statistics.
Interestingly, we find no statistically significant evidence (i.e., p-value less than $0.001$) of scam-payment acceleration or deceleration across all top mining pools. 
Looking at SPPE measure across the mining pools, we find no evidence of mining pools (other than AntPool) preferentially ordering the scam-payment transactions within blocks.
In short, our findings show that most mining pool operators today do not distinguish between normal and scam-payment transactions.

\subsubsection{Inferring other scam payment transactions from crowded source data}\label{sec:inferring_scam_payment_txs}

\begin{figure*}[tb]
	\centering
		\subfloat[Monthly reports\label{fig:scam-abuse-reports-monthly}]{\includegraphics[width=\twocolgrid]{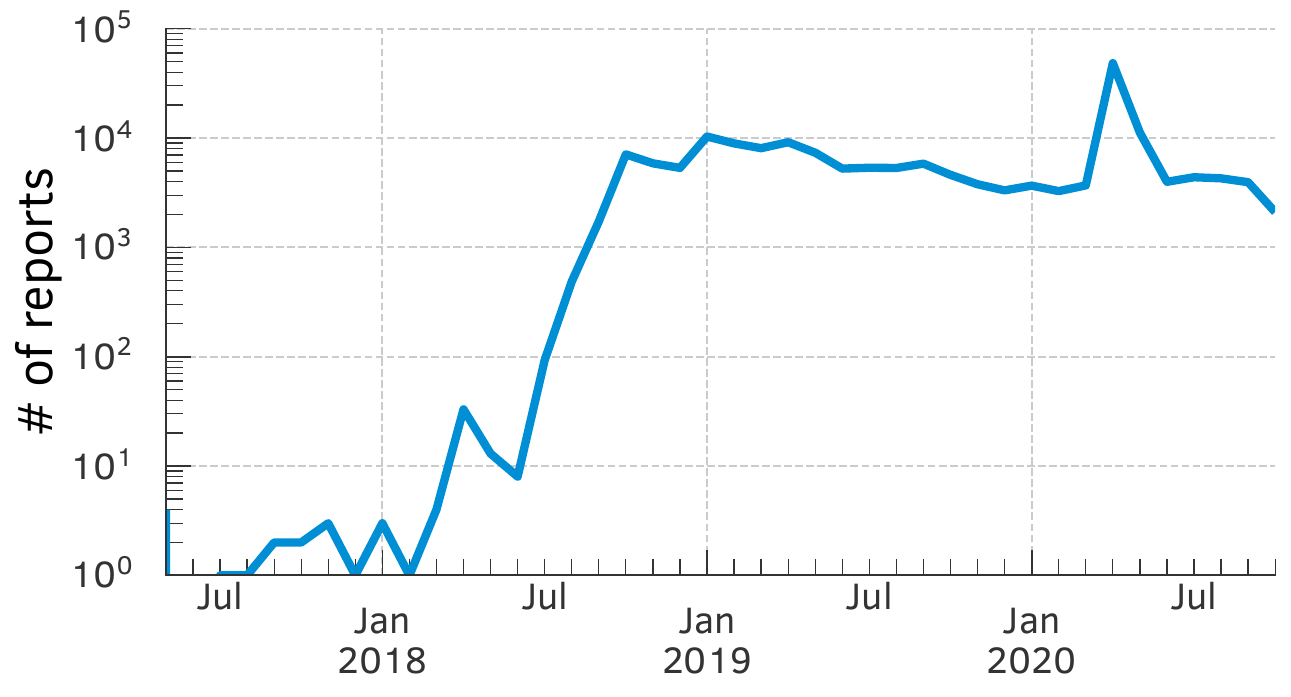}}
		\subfloat[Reports per country\label{fig:scam-abuse-reports-per-country}]{\includegraphics[width=0.86\twocolgrid]{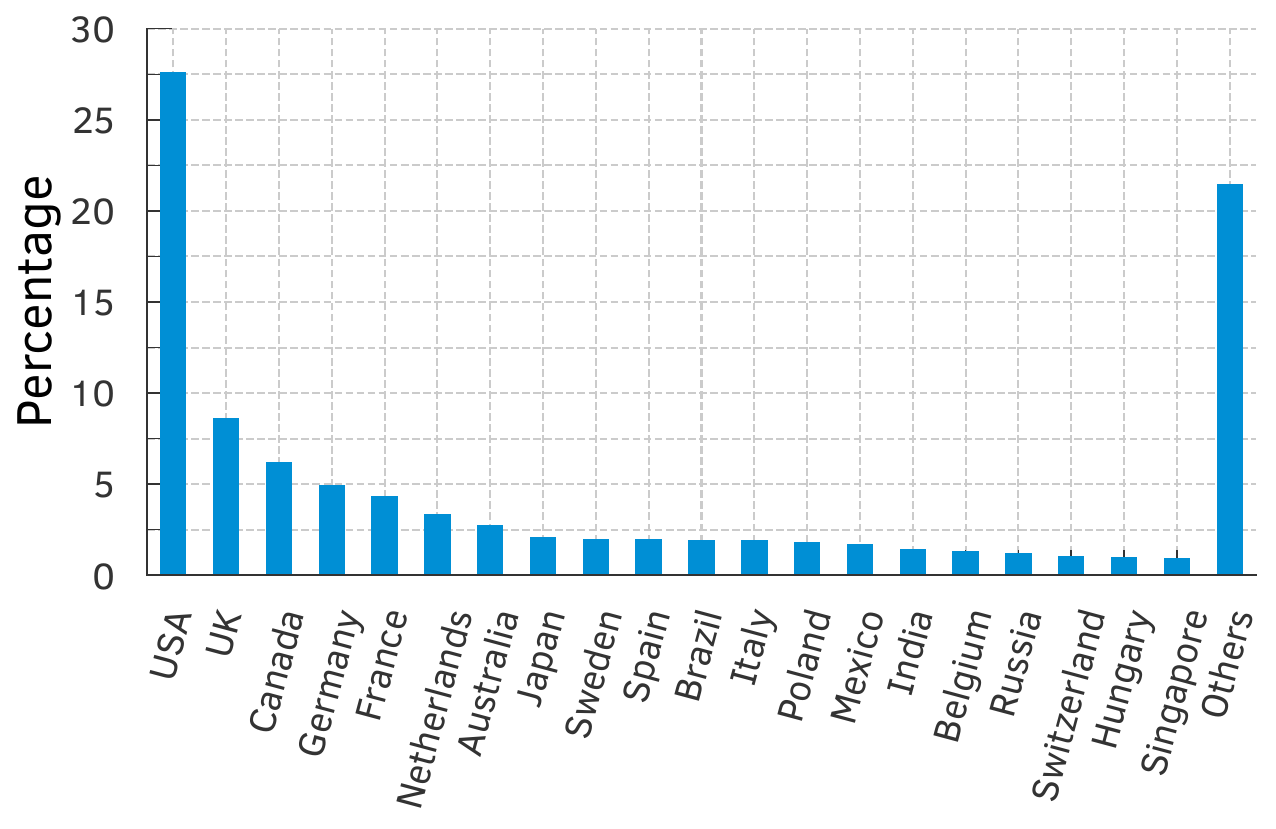}}
	\caption{
        (a) Distribution of the number of reports per month. (b) USA is the country with more reports accounting for 27.6\% of all reports available followed by UK and Canada.
      }
  \label{fig:scam-abuse-reports}
\end{figure*}

We also want to investigate whether other types of scams payments have been recorded on the Bitcoin blockchain.

To conduct this experiment, we gathered data from the Bitcoin Abuse~\cite{BitcoinAbuse} crowdsource platform. This platform allows users to report Bitcoin wallet addresses that they suspect are associated with scams. From May 16, 2017, to October 15, 2020, we gathered a total of \num{186731} reports from users. These reports identified \num{54032} unique wallet addresses that could potentially be linked to scams.
The monthly trend of scam reports is illustrated in Figure~\ref{fig:scam-abuse-reports-monthly}. Notably, there is an increase in the number of reports starting from July 2018. The reports predominantly originated from users in the US, followed by users in the UK (refer Figure~\ref{fig:scam-abuse-reports-per-country}). 

To simplify our analysis, we focused on transactions that occurred during the year 2018.\footnote{Due to limitations in data gathering, we considered transactions issued in 2018. This forms a subset of our data set \dsd introduced in~\S\ref{sec:method_tx_prioritization_contention}.} However, out of the total reported wallets, only a small portion, \num{1169} (2.16\%) wallets, were found in our Bitcoin 2018 data set. On average, each wallet address was reported \num{3.46} times, with a std. of \num{14.41}. Both the minimum and the median number of reports for a wallet address were 1, while the maximum number of reports reached 950.

\begin{table}[t]
    \small
    \begin{center}
        \tabcap{Scam types and their occurrences in wallets and transactions.}\label{tab:scam-transactions-wallets}
        \resizebox{.7\textwidth}{!}{%
            \begin{tabular}{rrrrr}
            \toprule
            \multicolumn{1}{c}{\thead{Scam type}} & \thead{\# of txs.} & \thead{\% of txs.} & \thead{\# of wallets} & \thead{\% of wallets} \\ \midrule
            Sextortion & 2,656 & 40.79 & 478 & 86.59 \\
            Terrorism & 1,093 & 16.79 & 1 & 0.18 \\
            Dark Web Shop & 1,019 & 15.65 & 8 & 1.45 \\
            Fake Giveaway & 739 & 11.35 & 18 & 3.26 \\
            Fake Exchange & 235 & 3.61 & 3 & 0.55 \\
            Other & 218 & 3.35 & 16 & 2.90 \\
            Malware & 195 & 2.99 & 4 & 0.72 \\
            Fake Investment & 164 & 2.52 & 6 & 1.09 \\
            Ransomware & 139 & 2.13 & 17 & 3.08 \\
            Ponzi Scheme & 53 & 0.82 & 1 & 0.18 \\
            \thead{Total} & \thead{6,511} & \thead{100} & \thead{552} & \thead{100} \\
            \bottomrule
            \end{tabular}
        } 
    \end{center}
\end{table}

To ensure accuracy in our analysis, we complemented the data from Bitcoin Abuse with information from another platform called Scam Alert~\cite{ScamAlert}. Scam Alert verifies the wallet addresses reported by users to determine if they are indeed associated with scams. Initially, we had a set of \num{1169} identified wallets reported by Bitcoin Abuse. Upon cross-checking with Scam Alert, we discovered that \num{100} were not involved with scams. Out of the remaining \num{1069} wallet addresses, \num{473} were confirmed to be scams, \num{79} were labeled as probable scams, and \num{517} were pending review at the time of our analysis. Thus, for our analysis, we only considered wallet addresses confirmed or labeled as probable scams by Scam Alert. This resulted in a final scam wallet data set containing \num{552} unique wallet addresses.

Table~\ref{tab:scam-transactions-wallets} provides information about the number of wallet addresses and the number of transactions that sent coins to at least one of the scam wallets during 2018. We identified several types of scams using the Scam Alert data set, with Sextortion comprising \num{40.79}\% of the total, followed by Terrorism at \num{16.79}\%, and Dark Web Shop at \num{15.65}\%. Together, these three types of scams accounted for \num{73.23}\% of the total scams identified in our study.
Additionally, Table~\ref{tab:most-used-wallets-for-scam} shows the top-20 most frequently used wallet addresses for scam payments. Notably, the wallet 1LaN$\cdots$Qctq has been associated with a terrorist organization using it to collect donations~\cite{Terrorism@Cointelegraph20}. This specific wallet address was involved in a total of by \num{1093} transactions.

\begin{figure*}[tb]
	\centering
		\subfloat[Distribution of blocks\label{fig:blocks-and-tx-dist-scam-blocs}]{\includegraphics[width=\twocolgrid]{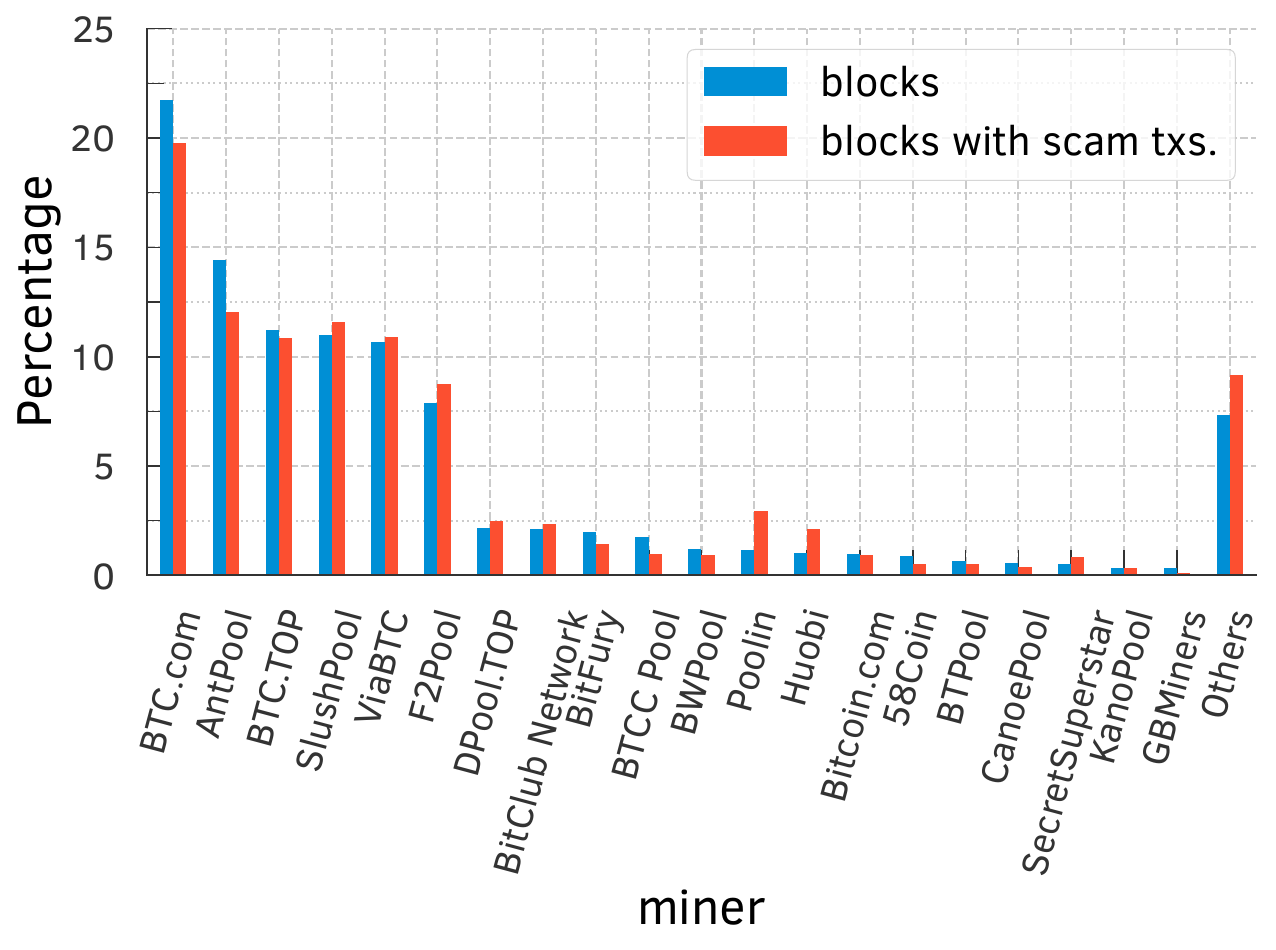}}
        \subfloat[Distribution of transactions\label{fig:blocks-and-tx-dist-scam-txs}]{\includegraphics[width=\twocolgrid]{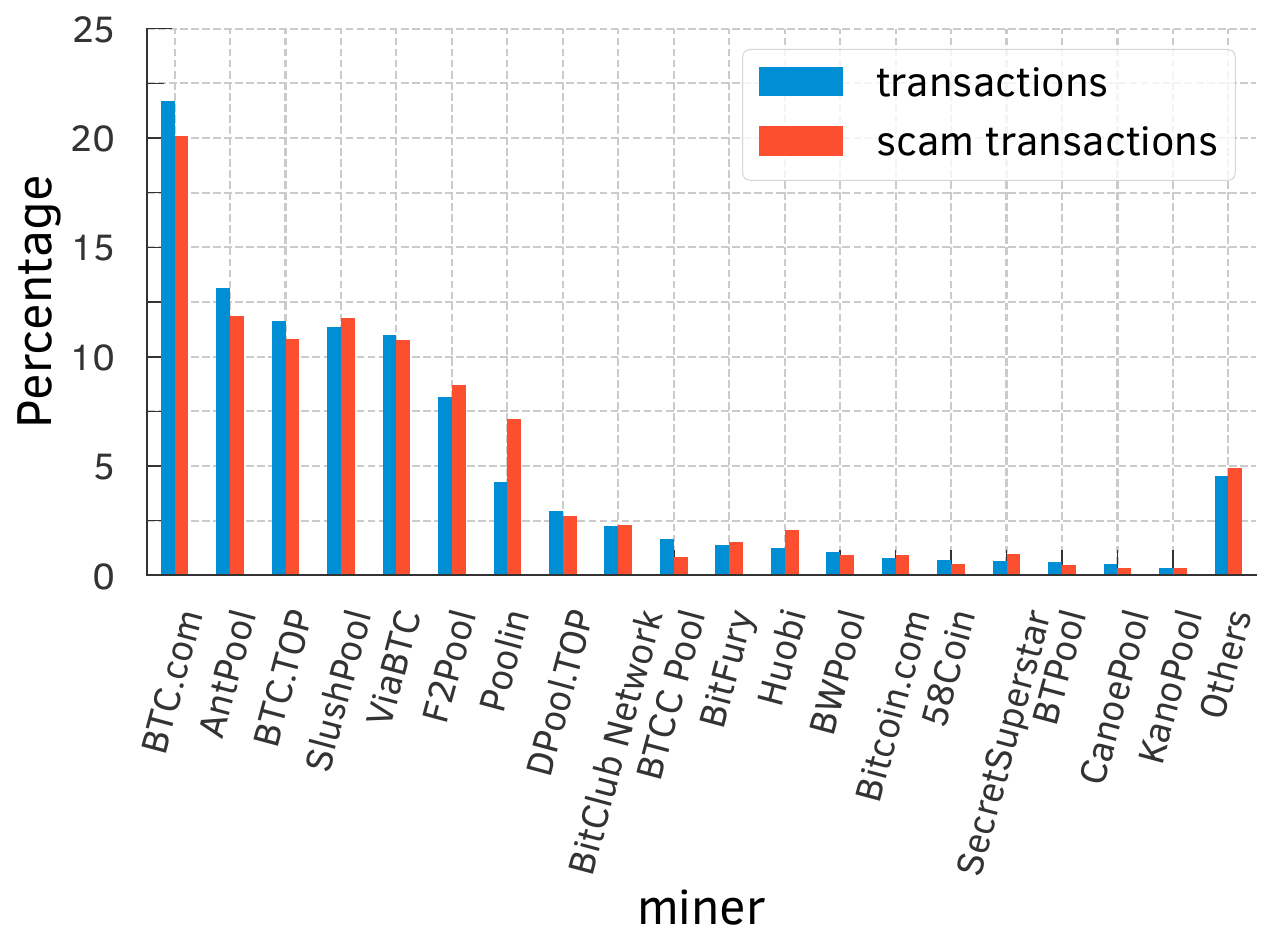}}
	\caption{
  Distribution of (a) blocks mined per each mining pool in comparison to the fraction of blocks that contains at least one scam transaction; and (b) transactions included by each mining pool in comparison to their share of scam transaction inclusion. BTC.com included 20.09\% out of the 6511 scam transactions in 2018.
	}
  \label{fig:blocks-and-tx-dist-scam}
\end{figure*}

Moreover, we observed that despite these scam wallets being publicly available, most of the mining pools still included them in their blocks. Figure~\ref{fig:blocks-and-tx-dist-scam} shows the distribution of blocks and transactions associated with scam payments that were included by each of the mining pools. BTC.com included alone \num{20.09}\% out of the \num{6511} scam transactions in 2018.

\begin{table}[t]
    \small
    \begin{center}
        \tabcap{Top-20 most used wallets for scam payments.}\label{tab:most-used-wallets-for-scam}
        \resizebox{0.7\textwidth}{!}{%
            \begin{tabular}{rrr}
            \toprule
            \multicolumn{1}{c}{\thead{Wallet address}} & \thead{Scam type} & \thead{\# of txs.} \\ \midrule
\href{https://www.blockchain.com/btc/address/1LaNXgq2ctDEa4fTha6PTo8sucqzieQctq}{1LaNXgq2ctDEa4fTha6PTo8sucqzieQctq} & Terrorism & 1,093 \\
\href{https://www.blockchain.com/btc/address/167uU5Q3cCPijsfwmmH6ZAQj8yYxQdmzoN}{167uU5Q3cCPijsfwmmH6ZAQj8yYxQdmzoN} & Dark Web Shop & 359 \\
\href{https://www.blockchain.com/btc/address/17v1cviCPNuGY73wNGvatS3CEZzrcPnXPy}{17v1cviCPNuGY73wNGvatS3CEZzrcPnXPy} & Fake Giveaway & 254 \\
\href{https://www.blockchain.com/btc/address/1Gs7Aztizk2rNNSE6AbpK4K7yAFTCZKV9a}{1Gs7Aztizk2rNNSE6AbpK4K7yAFTCZKV9a} & Dark Web Shop & 251 \\
\href{https://www.blockchain.com/btc/address/1EU1Ly84tYpTCcjWtvF4tYosRNN2xYYSGF}{1EU1Ly84tYpTCcjWtvF4tYosRNN2xYYSGF} & Dark Web Shop & 207 \\
\href{https://www.blockchain.com/btc/address/15ESgUNQ9Hgn2h2FDMJi9NwE4g7ZWRAGJE}{15ESgUNQ9Hgn2h2FDMJi9NwE4g7ZWRAGJE} & Fake Giveaway & 170 \\
\href{https://www.blockchain.com/btc/address/3Lo4nDzH7Bi572T7t8pQGU2Ax9jVymHeC6}{3Lo4nDzH7Bi572T7t8pQGU2Ax9jVymHeC6} & Fake Exchange & 137 \\
\href{https://www.blockchain.com/btc/address/13hjTSbwVJfsDgL3qaQSu3fs2qmHQCHRXT}{13hjTSbwVJfsDgL3qaQSu3fs2qmHQCHRXT} & Sextortion & 131 \\
\href{https://www.blockchain.com/btc/address/1Hy6BcTtNwrCLQK8ViEP742jRgx8Zpfoja}{1Hy6BcTtNwrCLQK8ViEP742jRgx8Zpfoja} & Dark Web Shop & 115\\
\href{https://www.blockchain.com/btc/address/343CXYVBKXT2VgELCdjEeMyPpfiKwkzUNg}{343CXYVBKXT2VgELCdjEeMyPpfiKwkzUNg} & Other & 103\\
\href{https://www.blockchain.com/btc/address/3L5o1AHLTKUeJDF8U2s5dgQCwoGknVyycn}{3L5o1AHLTKUeJDF8U2s5dgQCwoGknVyycn} & Fake Giveaway & 98\\
\href{https://www.blockchain.com/btc/address/16EegrNMdZ9Rxku6Za5neEFjMW57wkQr1S}{16EegrNMdZ9Rxku6Za5neEFjMW57wkQr1S} & Malware & 89\\
\href{https://www.blockchain.com/btc/address/1C4SvJQexhAEZzm3f6E6PMQT2xWtJdKKvp}{1C4SvJQexhAEZzm3f6E6PMQT2xWtJdKKvp} & Fake Exchange & 80\\
\href{https://www.blockchain.com/btc/address/1HYoMM6mfFiDvkRe5z9RsSo3sugnqaDps3}{1HYoMM6mfFiDvkRe5z9RsSo3sugnqaDps3} & Fake Investment & 77 \\
\href{https://www.blockchain.com/btc/address/1B7aczSxaMbRsPJXx22TP1foaHQ6FENwTA}{1B7aczSxaMbRsPJXx22TP1foaHQ6FENwTA} & Fake Giveaway & 62 \\
\href{https://www.blockchain.com/btc/address/3NYKHbX3zRbcZeASxjZmb4bpF8kZytnuvi}{3NYKHbX3zRbcZeASxjZmb4bpF8kZytnuvi} & Malware & 54 \\
\href{https://www.blockchain.com/btc/address/1JTtwbvmM7ymByxPYCByVYCwasjH49J3Vj}{1JTtwbvmM7ymByxPYCByVYCwasjH49J3Vj} & Sextortion & 54 \\
\href{https://www.blockchain.com/btc/address/16JL8g7QQthYorTCkjJNE7Yhm7M3DyVyNZ}{16JL8g7QQthYorTCkjJNE7Yhm7M3DyVyNZ} & Ponzi Scheme & 53 \\
\href{https://www.blockchain.com/btc/address/1GL9JtXPRTPetxgiJ8UcgrEECp12spD4tt}{1GL9JtXPRTPetxgiJ8UcgrEECp12spD4tt} & Sextortion & 52\\
\href{https://www.blockchain.com/btc/address/122wvcbWhBux5jcf2iyzFLmW7Jex7iSpef}{122wvcbWhBux5jcf2iyzFLmW7Jex7iSpef} & Sextortion & 49\\
            \bottomrule
            \end{tabular}
        } 
    \end{center}
\end{table}

%

\section{Dark-fee transactions} \label{sec:dark-fee-txs}

We refer to transactions that offer additional fees to specific mining pools through an opaque and non-public side-channel payment as dark-fee transactions.
Many large mining pool operators allow such side-channel payments on their websites for users wanting to ``accelerate'' the confirmation of their transactions, especially during periods of congestion.
Such private side-channel payments that hide the fees a user pays to miners from others have other benefits for the users~\cite{BTC@accelerator,Taichi@accelerator,F2Pool@accelerator,Poolin@accelerator,AntPool@accelerator}. One well-known advantage is, for instance, avoiding the fee rate competition in transaction inclusion, particularly during periods of high \mpool congestion; private side-channel payments would reduce a user's transaction cost volatility and curb front-running risks~\cite{Daian@S&P20,strehle2020exclusive,Eskandari@FC-2020}.
We use the data set \dsc{} to first investigate how such transaction acceleration services work and later propose a simple test for detecting accelerated transactions in the Bitcoin blockchain.

\subsubsection{Investigating transaction acceleration services}

We examined transaction acceleration services offered by $5$ large Bitcoin mining pools namely, BTC.com~\cite{BTC@accelerator}, AntPool~\cite{AntPool@accelerator}, ViaBTC~\cite{ViaBTC@accelerator}, F2Pool~\cite{F2Pool@accelerator}, and Poolin~\cite{Poolin@accelerator}.
Specifically, we queried BTC.com for the prices of accelerating all transactions in a real-time snapshot of the \mpool in data set \dsc (see~\S\ref{sec:tx-accelerator-comparison}). %
We found that the dark fee requested by BTC.com to accelerate each transaction is so high that if it was added to the publicly offered transaction fee, the resulting total fee rate would be higher than the fee rate offered by any other transaction in the \mpool snapshot.
Put differently, had users included the requested acceleration fees in the publicly offered fee when issuing the transaction, every miner would have included the transaction with the highest priority.

The above observation raises the following question: \stress{why would rational users offer a dark fee to incentivize a subset of miners to prioritize their transaction rather than publicly announce the fee to incentivize all miners to prioritize their transaction?}
One potential explanation could be that as payment senders determine the publicly offered transaction fees, payment receivers might wish to accelerate the transaction confirmation by offering an acceleration fee. 
Another explanation could be that the user issuing the transaction might want to avoid revealing the true fees they are willing to offer publicly, to avoid a fee rate battle with transactions competing for inclusion in the chain during congestion. Opaque transaction fees can reduce transaction cost volatility, but they may also unfairly bias the level playing field amongst user transactions attempting to front-run one another~\cite{strehle2020exclusive,Daian@S&P20}.   

On the other hand, every rational mining pool has clear incentives to offer such acceleration services.
They receive a very high fee by mining the accelerated transaction. 
Better still, they keep the offered fee, even if the accelerated transaction were mined by some other miners. 

\subsubsection{Detecting accelerated transactions}

\begin{table}[t]
    \small
    \begin{center}
        \tabcap{[Data set \dsc{}] For an SPPE $\ge$ 99\%, we observe that 64.98\% of BTC.com transactions were accelerated; the fourth column values are derived by dividing the values in the second with those in the third. The number of accelerated transactions decreases to 18.12\% for an SPPE $\ge$ 90\% and to 1.06\% for an SPPE $\ge$ 50\%.}\label{tab:sppe-tx-violation-acceleration}
        \resizebox{.5\textwidth}{!}{%
            \begin{tabular}{rrrr}
            \toprule
            \multicolumn{1}{c}{\thead{SPPE ($\ge$)}} & \thead{\# txs} & \thead{\# acc. txs} & \thead{\% acc. txs} \\ \midrule
            $100\%$                     & \num{628}     & \num{464}     & $73.89$       \\
            $99\%$                      & \num{1108}     & \num{720}     & $64.98$       \\
            $90\%$                      & \num{5365}   & \num{972}      & $18.12$       \\
            $50\%$                      & \num{95282}  & \num{1007}      & $1.06$          \\ 
            $1\%$                       & \num{657423}  & \num{1029}      & $0.16$          \\ 
            \bottomrule
            \end{tabular}
        } 
    \end{center}
\end{table}

Given the high fees demanded by acceleration services, we anticipate that \stress{accelerated transactions would be included in the blockchain with the highest priority}, i.e., in the first few blocks mined by the accelerating miner and amongst the first few positions within the block. 
We would also anticipate that \stress{without the acceleration fee, the transaction would not stand a chance of being included in the block based on its publicly offered transaction fee}.
The above two observations suggest a potential method for detecting accelerated transactions in the Bitcoin blockchain:
An accelerated transaction would have a very high \textit{\textbf{signed position prediction error (SPPE)}}, as its predicted position based on its public fee would be towards the bottom of the block it is included in, while its actual position would be towards the very top of the block. 

To test the effectiveness of our method, we analyzed all \num{6381} blocks and \num{13395079} transactions mined by BTC.com mining pool in data set \dsc{}. 
We then extracted all transactions with SPPE greater or equal than $100\%$, $99\%$, $90\%$, $50\%$, $1\%$ and checked what fraction of such transactions were accelerated.
Given a transaction identifier, BTC.com's acceleration service~\cite{BTC@accelerator} allows anyone to verify whether the transaction has been accelerated. 
Our results are shown in Table~\ref{tab:sppe-tx-violation-acceleration}. 
We find that more than $64\%$ of the \num{1108} transactions with SPPE greater or equal than $99\%$ were accelerated, while only $1.06\%$ of transactions with SPPE greater or equal than $50\%$ were accelerated.
In comparison, we found no accelerated transactions in a random sample of \num{1000} transactions drawn from the \num{13395079} transactions mined by BTC.com.
Our results show that large values of SPPE for confirmed transactions indicate the potential use of transaction acceleration services.
In particular, a transaction with SPPE $\ge 99\%$ (i.e., a transaction that is included in the top $1\%$ of the block positions, when it should have been included in the bottom $1\%$ of the block positions based on their public fee rate) has a high chance of being accelerated.

\subsection{Layer 2.0 transactions}

\begin{figure*}[tb]
	\centering
		\subfloat[Transaction position percentile\label{fig:omni-tx-percentile}]{\includegraphics[width=\twocolgrid]{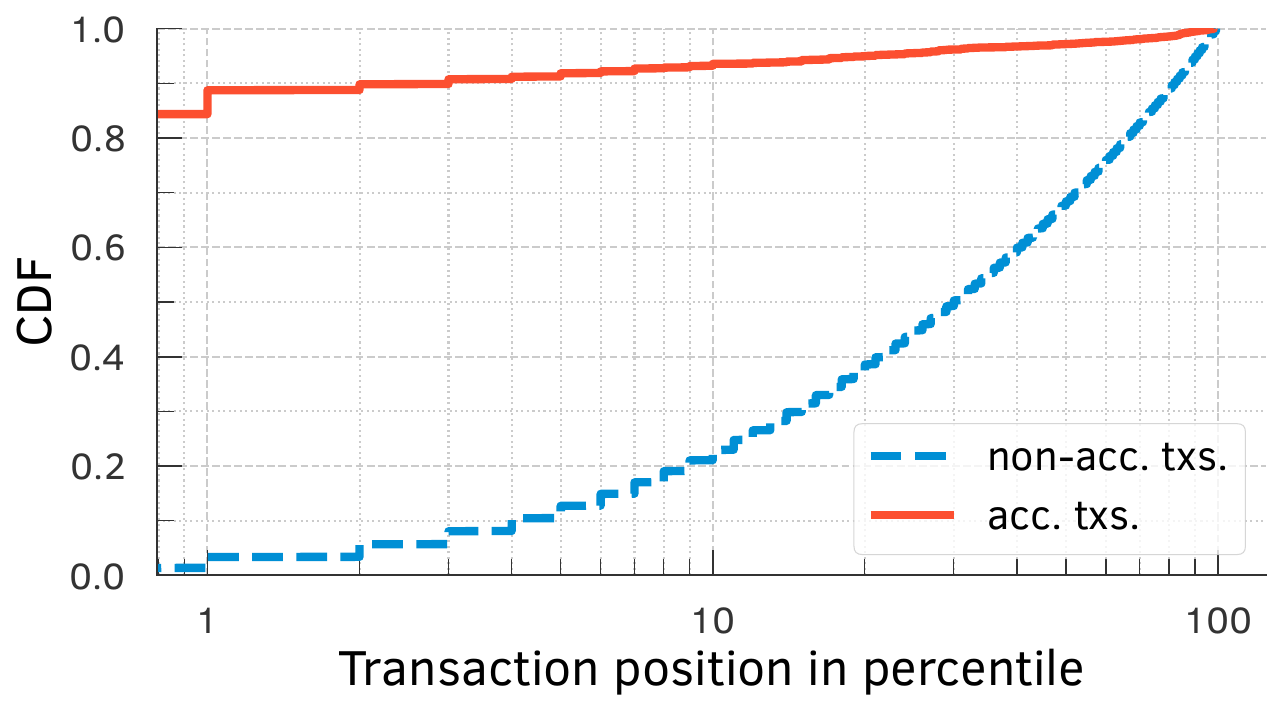}}
		\subfloat[Amount of US dollar transferred\label{fig:omni-vs-btc-transactions}]{\includegraphics[width=\twocolgrid]{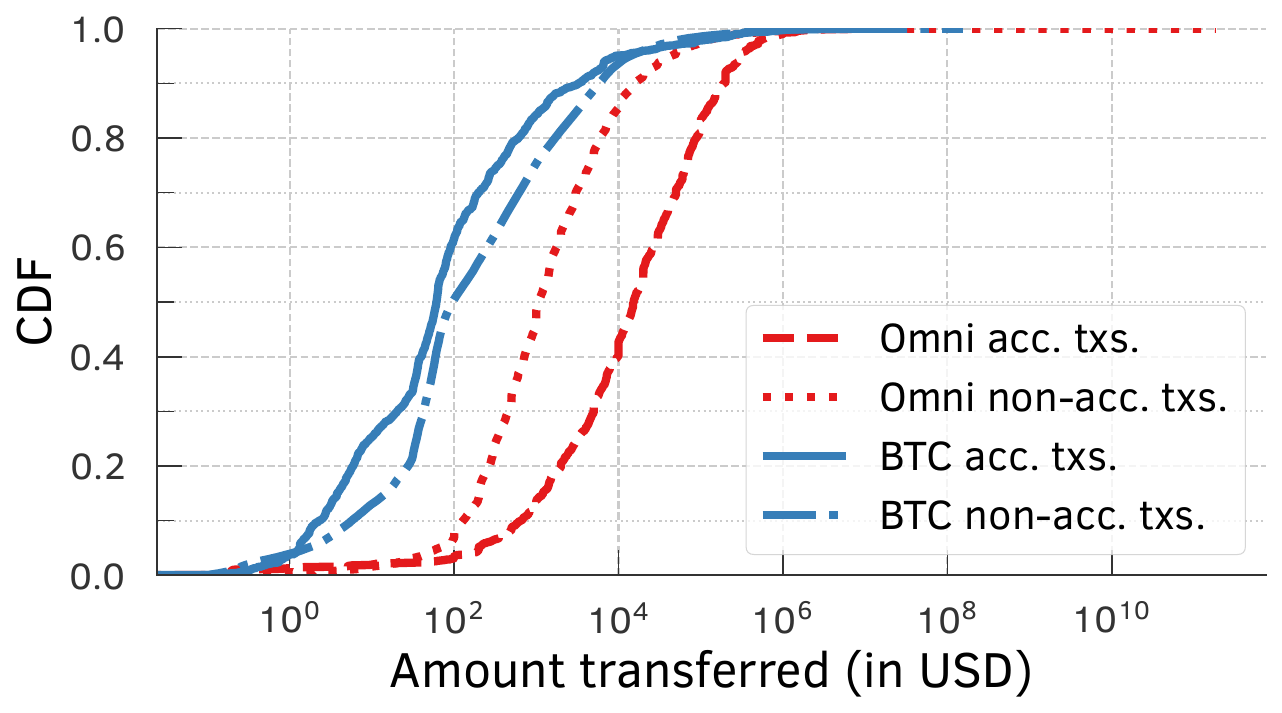}}
	\caption{
        Cumulative distribution function for (a) transaction position percentile: 84.30\% of Omni transactions were positioned right at the top of their respective blocks. This means they were the very first transactions to be included in those blocks; (b) comparison of Omni transfers and Bitcoin value transfers: The amount transferred in Omni to the corresponding value in Bitcoin for accelerated transactions was 259.97 times higher than the value announced in the Bitcoin blockchain.
      }
  \label{fig:omni-layer-txs}
\end{figure*}

Bitcoin offers a unique operation code, or simply opcode~\cite{OPCode@Bitcoin}, known as \stress{OP\_Return}, which allows anyone to write arbitrary data to the Bitcoin blockchain. This opcode was introduced with the release of Bitcoin Core v0.9.0 in 2014. The primary purpose of OP\_Return is to enable participants to mark a transaction output as invalid or to store additional data on the blockchain.
By using OP\_Return, the Bitcoin blockchain can also serve as a Layer 1.0 solution for Layer 2.0 applications like the Omni Layer Protocol~\cite{OmniLayer}. However, this usage can lead to a situation where the true value of a transaction transfer (in the case of Bitcoin) might not be directly visible on the blockchain. Instead, the actual value of a transaction can be determined by interpreting the arbitrary data stored within it.
To better understand these transactions and their purposes, we aim to parse the data written to the blockchain, using the data specification from~\cite{Omni-Specification@GitHub}, and investigate whether transactions with arbitrary data have been accelerated or utilized for specific reasons.

To this end, we considered a 3-year Bitcoin data set named data set  \dsd (refer \S\ref{sec:method_tx_prioritization_contention}), consisting of a total of \num{313737341} transactions (\num{313575387} issued transactions and \num{161954} coinbase transactions), we observed that \num{42994249} transactions (13.70\%) contained at least one OP\_RETURN opcode in their list of transaction output. Focusing solely on the issued transactions, \num{42832713} transactions (13.66\%) included at least one OP\_RETURN opcode.
Regarding the coinbase transactions, from the total of \num{161954} blocks, a majority of \num{161536} coinbase transactions (99.74\%) contained at least one OP\_RETURN opcode. This indicates that miners have also been actively including arbitrary data in the Bitcoin blockchain.

Moreover, we identified \num{17993300} transactions associated with the Omni Layer Protocol~\cite{OmniLayer}, averaging 111 transactions per block. These Omni transactions accounted for 5.74\% of all Bitcoin issued transactions and 42\% of all transactions involving OP\_RETURN opcodes. Notably, a significant portion (97.27\%) of these Omni transactions were related to the Tether USDT token~\cite{TetherUSDT}, a stablecoin pegged to the US dollar.

We found \num{1805} OP\_RETURN transactions among a set of \num{14104} \stress{accelerated} transactions in data set \dsd, using the methodology discussed in the previous section. Out of these accelerated transactions, \num{1740} belong to the Omni Layer Protocol, with \num{1739} of them being related to the Tether token, and the remaining one to the Omni token. These Omni accelerated transactions were consistently placed right at the top of each block, on average within the top \num{3.4}\%, with a std. of 13.89\% and a median of 0\%. This indicates that they were the first issued transactions in their respective blocks. In comparison, non-accelerated transactions were usually placed within the top \num{36.97}\% of the block, with a std. of \num{28.64}\% and a median of \num{30}\%.
Overall, we observed that \num{84.30}\% of all Omni transactions were included at the top of their respective blocks, being the first transactions to be included (refer Figure~\ref{fig:omni-tx-percentile}).

\begin{figure*}[tb]
	\centering
		\includegraphics[width={\textwidth}]{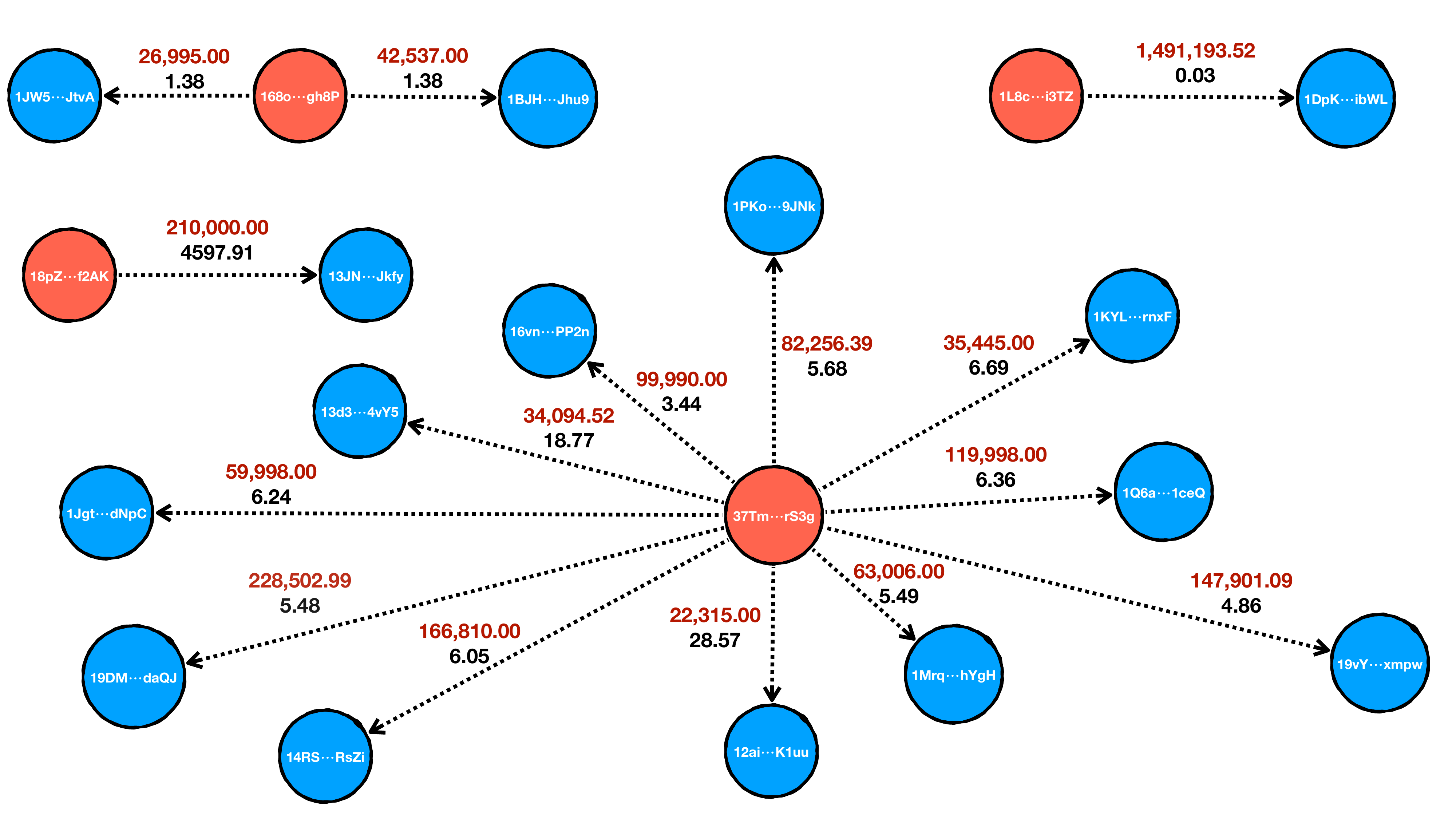}
	\caption{
  Comparison between the value transferred (in USD) in accelerated Omni transactions (shown in the top \red{red} color) and the values from the Bitcoin blockchain (in the bottom black color) for transactions included in block 550,912. Each edge in the graph corresponds to a single BTC transaction. Transaction values available in the Bitcoin blockchain appear to be relatively low. However, in contrast, these transactions in the Omni Layer are notably high-value transactions. Nodes in \blue{blue} indicate receivers, while nodes in \red{red} indicate senders.
	}
    \label{fig:omni-txs}
\end{figure*}

Due to the opacity of the true value of Omni transactions, as it requires interpreting the arbitrary data stored in each transaction, we parsed the data using the protocol transaction specification from~\cite{Omni-Specification@GitHub}. Then, we compared the values transferred in Bitcoin transactions based on the transaction output value with the values stored in the arbitrary data belonging to Omni. The Bitcoin prices were converted to US dollars, considering the exchange rate at the time the transactions were included in the block, obtained from the Yahoo Finance BTC-USD feed~\cite{btc-usd@yahoo}.
As shown in Figure~\ref{fig:omni-vs-btc-transactions}, Omni transactions tend to transfer much higher values among users compared to the values seen in the Bitcoin transaction outputs. For instance, on average, the values transferred in Tether on the Omni layer were \num{501600.69} USD with a median of \num{1053} USD, while the values seen in the Bitcoin blockchain averaged \num{167518.43} USD with a median of \num{84.81} USD. This means that the values transferred in Omni were almost \num{3} times higher on average and \num{12.42} times higher in the median compared to Bitcoin transactions.

We also analyzed the results based on acceleration. For half of the evaluated transactions, the accelerated Omni transaction transferred at least \num{15274.34} USD, while the value reported in Bitcoin was only \num{58.76} USD. This indicates that the value transferred through Omni was at least \num{259.97} times higher for half of the transactions evaluated. In Figure~\ref{fig:omni-txs}, we showcase all accelerated transactions included in block \num{550912}, totaling \num{15} transactions, which belong to Omni. We also highlight the senders and receivers involved in these transactions.
Comparing the value transferred in Bitcoin, we found that the average value was \num{313.22} USD with a std. of \num{1185.35} USD and a median of \num{5.68}, ranging from \num{0.03} to \num{4597.91} USD. In contrast, on the Omni network, the average value transferred was \num{188736.17} USD with a std. of \num{366405.31} USD and a median of \num{82256.39} USD, ranging from \num{22315.00} USD to \num{1491193.52} USD.

%

\section{Concluding remarks}

In this chapter, we conducted an extensive empirical audit of the miners' behavior to check whether they adhere to the established norms. At a high level, our findings reveal that transactions are primarily prioritized based on assumed norms. However, our analysis also uncovers evidence of a significant number of confirmed transactions that violate these priority norms. An in-depth investigation of these norm violations uncovered many highly troubling misbehavior by miners. Our results demonstrate that large values of SPPE for confirmed transactions indicate the potential use of transaction acceleration services. In particular, a transaction with SPPE $\ge 99\%$ has a high chance of being accelerated.

Strikingly, we show that 4 out of the top-10 mining pools namely, F2Pool, ViaBTC, 1THash \& 58Coin, and SlushPool \stress{selfishly accelerated} their own transactions. Furthermore, we uncover instances of collusive behavior between mining pools. Our results are supported by the SPPE metric, which indicates that self-interest transactions were also being included in the blocks ahead of other higher fee rate transactions.

In summary, our findings strongly suggest that several large mining pools tend to give special treatment to transactions that benefit them directly. This included transactions involving payments to or from wallets owned by the mining pool. Some even \stress{collude} with other large mining pools to prioritize their transactions. Additionally, a number of significant large mining pools accept additional \stress{dark (opaque) fees} to accelerate transactions via non-public side-channels (e.g., their websites). This practice of dark-fee transactions contradicts a fundamental, albeit unstated, assumption in blockchain systems: \stress{that the confirmation fees offered by transactions are transparent and equal to all miners}.

In the following chapter, we will explore the implications of the lack of transparency into both the prioritization of transactions and the content of transactions.

\clearpage

%
\chapter{Transaction Prioritization and Contention Transparency} \label{chap:tx_prioritization_contention}

%

In this chapter, we discuss the implications of our findings regarding the lack of transparency in transaction contention and prioritization.
We also argue why our findings and implications would be relevant even in the face of recent changes to blockchain protocols, e.g., Ethereum Improvement Protocol (EIP) 1559~\cite{EIP-1559} and the Ethereum Paris Network Upgrade (a.k.a. the Merge)~\cite{Eth-Merge}.

The lack of transparency in both transaction contention and prioritization has not been thoroughly explored in the literature, resulting in a limited understanding of its implications. This thesis aims to address this gap by investigating the following research questions.

\point{}
\textbf{RQ 1}: \stress{To what extent are private relay networks prevalent in facilitating transactions prioritization?}
Given the rise of transaction attacks like frontrunning and sandwich attacks, it is reasonable to consider that transactions issuers may prefer sending their transactions privately to the miners to avoid such attacks.
This research question aims to explore the current prevalence or widespread adoption of private relay networks as a means for issuers to achieve their goal of protecting their transactions.
However, we also consider the potential downsides of private transaction inclusion, particularly in terms of the lack of transparency in transactions contention and prioritization.
Thus, we investigate this research question to assess the overall benefits and drawbacks of these private relay networks to the broader blockchain ecosystem.

\point{}
\textbf{RQ 2}: \stress{Are private transactions preferentially treated by miners?}
This research question aims to investigate if miners provide preferential treatment to private transactions.
Private transactions offer guaranteed payments (or fees), whereas fees for publicly issued transactions are available to any miners willing to include them.
We hypothesize that miners would likely offer preferential treatment for private transactions due to their guaranteed payment nature.
To address this research question, we conducted an active experiment to assess whether miners exhibit preferential treatment towards private transactions in the context of Ethereum blockchain.

\point{}
\textbf{RQ 3}: \stress{To what extent do transaction bundling practices occur? Do they include public transactions?}
This research question focuses on exploring the frequency and characteristic of the transaction bundling practices, particularly the inclusion of public transactions, in order to exploit MEV opportunities.
Arbitrageurs may have an incentive to create Flashbots~\cite{Flashbots@API} bundles that combine both private and public transactions to capture the financial opportunity derived from the execution of public transactions.
It is worth noting that the fees associated with private transactions remain private to the relay and the miner until the transactions within the bundle are included in a block.
Hence, we investigate the types of public transactions included in these bundles and the specific contracts they call.
Additionally, we also investigate the revenues earned by miners from accepting these bundles.
This analysis is crucial for advancing the transparency goals of blockchain systems.

\point{}
\textbf{RQ 4}: \stress{Has there been collusion among miners to prioritize transaction inclusion?}
This research question focuses on examining whether miners engage in collusion to prioritize the inclusion of transactions.
For instance, if an issuer sends a transaction to a particular miner, we aim to investigate whether other miners share this transaction to accelerate or prioritize its commitment.
If such collaboration exist, it suggests that miners may cooperate not only to accelerate transactions but could also potentially censor specific transactions if they choose to do so.
To address this research question, an active experiment was conducted in the context of Bitcoin.

These research questions are key to investigating the impact of the lack of transaction contention and prioritization in both Bitcoin and Ethereum blockchains.
They allow us to explore the prevalence and extent of private relay networks or acceleration services currently in use within these blockchains.
Furthermore, by examining whether miners collude to prioritize transactions, we gain insight into potential trust issues within blockchains.
For example, this collusion can undermine the trust in the blockchain system, as miners could also censor transactions if they choose to. Next, we discuss our methodology, our findings, and the implications.

\subsubsection*{Relevant publication}

The results presented in this chapter have been published in~\cite{Messias@FC2023}.

%

\section{Methodology} \label{sec:method_tx_prioritization_contention}

In this section, we outline our methodology for evaluating the lack of transparency in transaction contention and prioritization. First, we provide an overview of our Bitcoin and Ethereum data sets utilized in our analysis. Subsequently, we offer detailed information on the active experiments employed to assess preferential treatment of transactions and the aggregated power of colluding miners.

\begin{table*}[t]
  \begin{center}
   \small
    \tabcap{Bitcoin and Ethereum data sets (\dsd and \dse) used to evaluate the lack of contention and prioritization transparency.}\label{tab:datasets-D-E}
    \begin{tabular}{rrr}
      \toprule
      \thead{Attributes} & \thead{Data set \dsd{}} & \thead{Data set \dse{}}\\
      \midrule
      \textit{Time span} & Jan. $1\tsup{st}$, 2018 -- Dec. $31\tsup{st}$, 2020 & Sept. $8\tsup{th}$, 2021 -- Jun. $30\tsup{th}$, 2022\\
      \textit{Block height / number} & \num{501951} -- \num{663904} & \num{13183000} -- \num{15049999} \\
    \textit{Number of blocks} & $\num{161954}$ & $\num{1867000}$ \\
      \textit{Count of transactions issued} & $\num{313575387}$ & $\num{347629393}$\\
      \textit{Percentage of CPFP-transactions} & $21.02\%$ & --- \\
      \textit{Count of empty-blocks} & \num{992} & \num{43069} \\
      \bottomrule
    \end{tabular}
  \end{center}
\end{table*}

\subsection{Data set collection}\label{chapter4:sec-datasets}

\begin{figure*}[t]
	\centering
		\subfloat[Bitcoin distribution\label{fig:dist-tx-blks-bitcoin}]{\includegraphics[width=\twocolgrid]{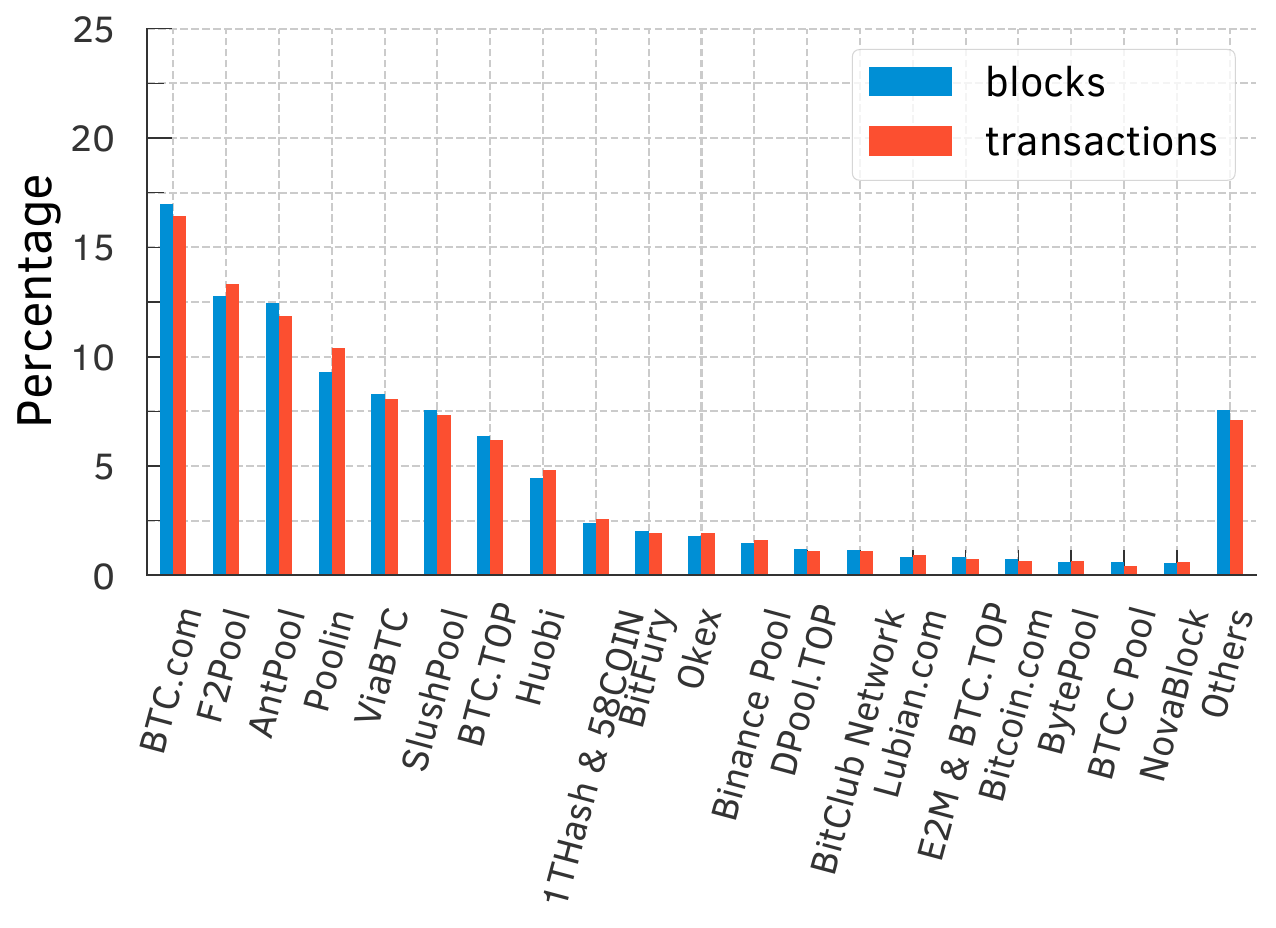}}
        \subfloat[Ethereum distribution\label{fig:dist-tx-blks-ethereum}]{\includegraphics[width=\twocolgrid]{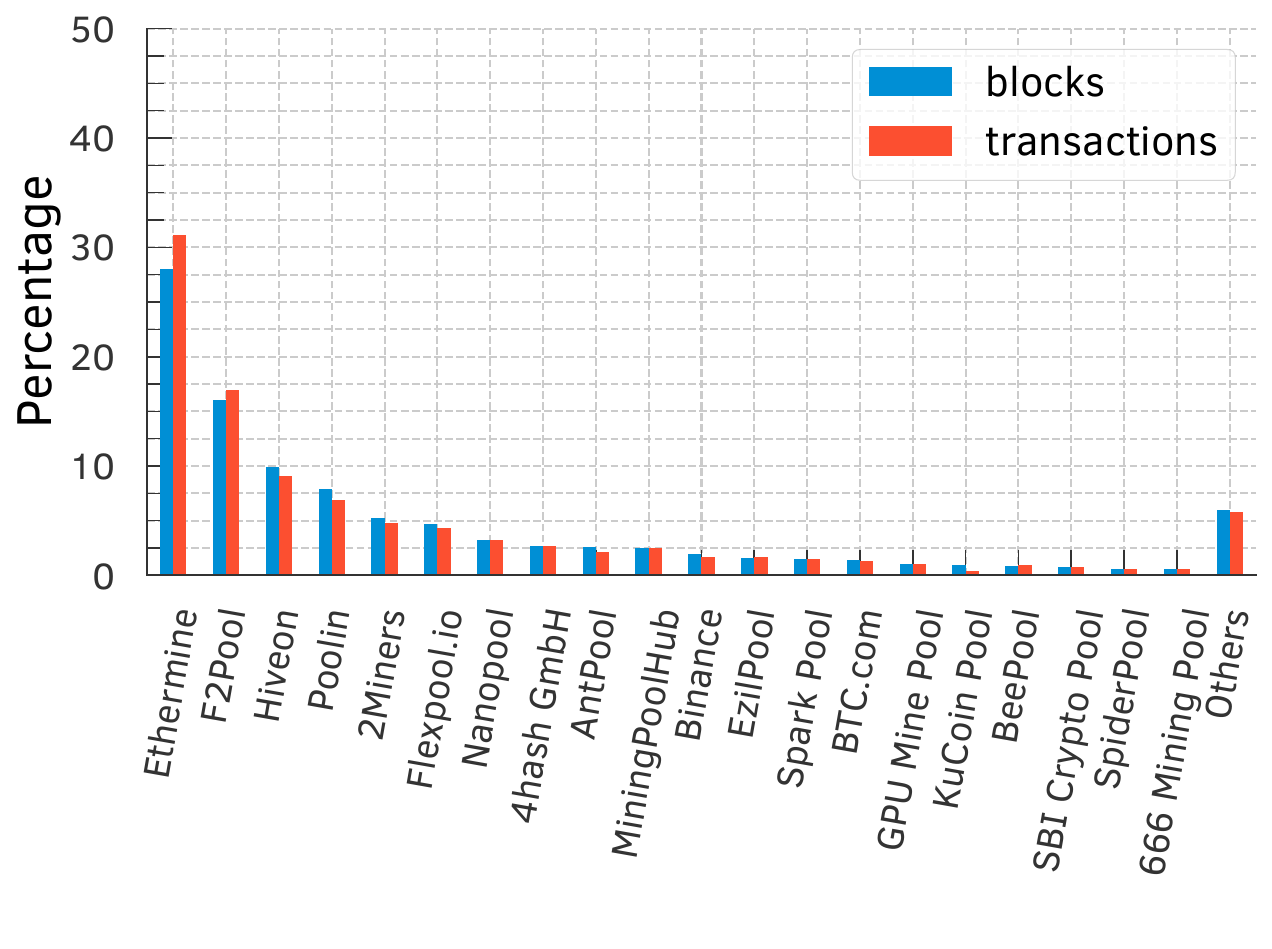}}
	\caption{
  Blocks mined and transactions confirmed in (a) Bitcoin and (b) Ethereum by the top-20 mining pools; ``Others'' consolidates the remaining mining pools.
	}
  \label{fig:eth-btc-blocks-txs-distribution}
\end{figure*}

\paraib{Data set \dsd{}.}
To identify accelerated transactions, we gathered all Bitcoin blocks mined from January $1\tsup{st}$ 2018 to December $31\tsup{st}~2020$. In total, per Table~\ref{tab:datasets-D-E}, there are \num{161954} blocks from block height \num{501951} to \num{663904}, and \num{313575387} transactions. In Bitcoin, mining pools may indicate their ownership of the block by including a \stress{signature} or \stress{marker} in the \stress{Coinbase} transaction (i.e., the first transaction of every block).
We used such markers for identifying the mining pool (owner) of each block following techniques from prior work~\cite{judmayer2017merged,Messias@IMC2021,Romiti2019ADD}.
We failed to identify, however, the owners of $\num{4911}$ blocks (approximately $3\%$ of the blocks) and grouped these blocks under the label ``Unknown.''
Figure~\ref{fig:dist-tx-blks-bitcoin} shows the distribution of the count of blocks mined and transactions confirmed by the top-20 mining pools.
We further removed \num{65902514} (21.02\%) \stress{child-pays-for-parent (CPFP)} transactions from our acceleration analyses.
If we rank the MPOs by the number of blocks ($B$) mined (which essentially approximate the hashing capacity $h$ of the MPOs), the top five MPOs are
BTC.com ($B$: $\num{27534}$; $h$: $17.00\%$), F2Pool ($B$: $\num{20665}$; $h$: $12.76\%$), AntPool ($B$: $\num{20188}$; $h$: $12.47\%$), Poolin ($B$: $\num{15096}$; $h$: $9.32\%$), and ViaBTC ($B$: $\num{13419}$; $h$: $8.29\%$).
But, unsurprisingly, the MPOs' hash rates significantly varied over the years (refer Figure~\ref{fig:btc-hashrate} in \S\ref{sec:hash-var}).
Furthermore, we rely on our SPPE  metric to infer whether a transaction was likely accelerated.

\paraib{Data set \dse{}.}
We gathered all Ethereum blocks mined over a $9$-month time period---from September $8\tsup{th}$, 2021 to June $30\tsup{th}$, 2022---to investigate the behavior of Ethereum mining pools (refer to Table~\ref{tab:datasets-D-E}).
This data set contains \num{347629393} issued transactions and \num{1867000} blocks (from block number \num{13183000} to \num{15049999}).
We used miners' wallet addresses to infer the block owners, but we failed to identify the owners of \num{46895} blocks (or \num{2.51}\% of the total); we grouped the latter into one category, ``Unknown.'' 
Figure~\ref{fig:dist-tx-blks-ethereum} shows the distribution of blocks and transactions mined in Ethereum by the top-20 mining pools.
If we rank the mining pools by the number of blocks ($B$) mined (which approximate the hashing capacity $h$ of the mining pools), the top five mining pools are Ethermine ($B$: \num{523633}; $h$: \num{28.05}\%), F2Pool ($B$: \num{299418}; $h$: \num{16.04}\%), Hiveon ($B$: \num{185495}; $h$: \num{9.94}\%), Poolin ($B$: \num{147983}; $h$: \num{7.93}\%), and 2Miners ($B$: \num{97308}; $h$: \num{5.21}\%) that together account for \num{67.17}\% of all blocks mined. We also report the weekly Ethereum's hash rate in Figure~\ref{fig:eth-hashrate} in \S\ref{sec:hash-var}.
Hash rates of mining pools in Ethereum across the study period did not vary as much as in Bitcoin (refer to Figure~\ref{fig:btc-hashrate} in \S\ref{sec:hash-var}).

Additionally, we also retrieved \num{6937292} transactions ($2\%$ of all issued transactions in Ethereum) contained in \num{3284886} bundles from Flashbots; these are transactions sent privately to miners.

\subsection{Bundling public transactions}

To identify bundles likely sent through the public P2P network, we use a simple heuristic. We concentrate on transaction bundles of sizes 2 and 3, seeking transactions that probably contributed to a publicly transmitted transaction being bundled and signs of sandwich attacks~\cite{Qin@BEV}.

The underlying idea is that miners have no incentive to include transactions offering zero fees, as there is no reward for mining such transactions---unless they receive additional payment via Flashbots coinbase transfer. Consequently, transactions with a non-zero max-priority fee likely underwent public sending.

We discuss the details and our results in \S\ref{subsec:bundling_public_txs}.

\subsection{Aggregated power of colluding miners}

Collusion among mining pools directly challenges the fundamental principle of truly decentralized blockchains. For instance, when powerful mining pools collude to give preference to certain transactions, and they collectively have a hash rate surpassing \num{50}\% of the network, there is no barrier preventing them from also censoring the validation of other transactions. As a result, they could potentially gain substantial control over which transactions are included in the blockchain, creating a significant risk of centralization.

To assess the real-world occurrence of mining pool collusion, we conducted an active experiment within the Bitcoin network. In this experiment, we paid a single mining pool to accelerate a set of public-low-fee-rate transactions. This was done during periods of high congestion in the \mpool. Without this acceleration, these transactions would have faced long delays before being included in the blockchain, due to their public low fees.
However, although we paid just one mining pool to accelerate these transactions, our findings unveiled a concerning revelation: \stress{these accelerated transactions were also included and prioritized within a block by other powerful mining pools}. Collectively, these mining pools possess a hash rate exceeding \num{50}\%. This situation raises significant concerns that reverberate across the entire blockchain ecosystem.

In \S\ref{chapter4:sec_bitcoin_dark_fees}, we delve into a comprehensive discussion of our findings and implications of the mining pool collusion.

%

%

\section{On contention transparency}
\label{sec:contention_transparency}

In this section, we show that contention transparency does not hold in practice as transaction relay networks become popular in Ethereum. This allows miners to include transactions privately and therefore not every miner or even transaction issuers have the full view of all available transactions pending for inclusion.

\subsection{The rise of private relay networks}

With the lucrative market of Decentralized Finance (DeFi) in Ethereum, today, bots engage in predatory front-running behaviors such as sandwich attacks and transaction-replay attacks \cite{Daian@S&P20,kiffer2017stick,Qin@BEV,Qin@FC21,Christof@USENIX,Weintraub@IMC2022,Zhou@S&P2021}. 
Relay networks help users to counter such attacks:
They provide users with a private channel for communicating with miners, who have to prove their identity to participate in the relay.
Relay networks help users completely bypass the P2P network: Users send their transactions to the relay network, which in turn relays them to its participant miners.
The relay network and its participants claim (a) not to front-run these transactions; and (b) to keep them private until they are included in a block~\cite{Flashbots@API}.
These transactions, hence, by construction, experience no front-running issues.
Relay networks are centralized; if miners misbehave, they may lose their network membership and forfeit their future profits.
Multiple relay networks (e.g., bloXroute~\cite{BloXroute@Ethereum}, Taichi Network~\cite{Taichi@accelerator}, and others~\cite{EdenNetwork,EthermineMEVRelay@Ethereum}) exist today, but we focus on Flashbots~\cite{Flashbots2Docs@Ethereum}, the largest relay network for Ethereum.

\subsubsection{Flashbots private relay network}

As discussed previously, at the time of our analysis, Flashbots is the most popular private relay network in Ethereum.
Flashbots's users \stress{bundle} one or more transactions in some specific order~\cite{Flashbots@Ethereum}.
Miners are expected to mine the entire bundle (retaining the ordering of transactions within the bundle) and place it at the top of their blocks.
The miners receive a fee (paid via a direct transfer to their wallets) for including the bundle in addition to the (traditional) fees associated with the transactions in that bundle.
If there are two competing bundles---capturing the same financial opportunity, e.g., liquidations---miners will choose the one with the highest reward (i.e., maximizing financial incentives).
The other bundle is \stress{discarded} (since the financial opportunity no longer exists after having been captured by the included bundle), albeit its transactions do \stress{not} expend \stress{any} gas.
Therefore, except for a network base fee introduced in EIP-1559,\footnote{The EIP-1559 went live in the Ethereum's London hard fork upgrade on August 5\tsup{th}, 2021, at block number \href{https://etherscan.io/block/12965000}{\num{12965000}}.} arbitrageurs and liquidators can participate without having any balance in their wallet:
If they successfully capture a financial opportunity, they pay the miner from the profit secured and pocket the rest~\cite{Flashbots2Docs@Ethereum}.
%
Flashbots is a \stress{free} to use relay network, and they allow anyone to query whether a transaction used their relay network and the private fees paid to the miner (after it has been committed in a block).
We use this publicly available data for analyzing the transactions issued (privately) on Flashbots.
Flashbots, however, does not list the discarded bundles (or its transactions): we have access, hence, only to committed transactions.

\subsection{Characterizing private relay networks}
\label{ss:characterize-pvt-relay}

Flashbots labels its bundles (and constituent transactions) into one of three categories: (i) \stress{flashbots}, which represent those sent through their private relay;  (ii) \stress{rogue}, referring to those delivered to a (Flashbots) miner, but via a different relay network; and (iii) \stress{miner payout}, indicating a bundle containing payouts to users of a mining pool~\cite{Weintraub@IMC2022}.
We found \num{58.82}\%, \num{27.93}\%, and \num{13.25}\% of transactions belonging to the flashbots, miner payout, and rogue categories, respectively.
We also noticed that \num{70260} (\num{1.01}\%) of all Flashbots transactions failed to execute after inclusion in a block.
A small fraction of transactions is, hence, not successfully executed despite using private relays.

Flashbots also claims to have $\approx85\%$ of the total Ethereum hash rate~\cite{Flashbots2Docs@Ethereum}.
Per our analyses, however, the majority of the mining pools ($47$ out of $48$---barring EthPool) use Flashbots, accounting for $99.99\%$ of the total Ethereum hash rate,
A recent work also corroborates our findings~\cite{Weintraub@IMC2022}.

Some of the most powerful mining pools like Spark Pool\footnote{Spark Pool suspended their mining services on September 30\tsup{th}, 2021, due to regulatory requirements introduced by Chinese authorities~\cite{SparkPool@CoinTelegraph}.} (which cooperates with Taichi Network~\cite{Taichi@accelerator}), Ethermine~\cite{EthermineMEVRelay@Ethereum}, and F2Pool (part of Eden Network~\cite{EdenNetwork}) offer their own relay networks. 
As these networks allow transaction issuers to send transactions exclusively to a specific miner, we hypothesize that miners would prefer (or prioritize) these transactions to those sent via the public P2P network. 
Crucially, payments from these private transactions are guaranteed, while those from publicly issued transactions are not---they are available to any miner willing to commit them.
\stress{Miners, hence, would likely offer preferential treatment for private transactions.}

\subsection{On preferential treatment of private transactions} \label{subsec:preferential_treatment}

We substantiate our hypothesis of preferential treatment for private transactions via an active experiment conducted on September 8\tsup{th}, 2021.
We issued $8$ transactions, where $4$ were sent privately via the Taichi Network, powered by Spark Pool, and $4$ through the public Ethereum network (refer Table~\ref{tab:acceleration-experiment-ETH} in~\S\ref{sec:private-txs}).
We spent $100$~Euros for running this experiment.

While running the experiment, we checked if the popular Ethereum blockchain explorers (i.e., Etherscan~\cite{Etherscan@ETH-explorer}, Blockchain.com~\cite{Blockchain@ETH-explorer}, and Blockchair~\cite{Blockchair@ETH-explorer}) observed any of our private transactions;
if they did, it would imply that the Taichi Network leaked the transactions to the public.
While the public transactions appeared in these blockchain explorers, right after we sent them through the public P2P network, the private transactions were not observed by any of them until the transactions were included in a block.
More importantly, our private transactions were \stress{not} flagged by Etherscan (which relies on Flashbots API~\cite{Flashbots@API} and more recently on EigenPhi~\cite{EigenPhi@Ethereum}) as private, \stress{even after inclusion in a block}.
Measuring the prevalence of private transactions is, hence, challenging; it is likely that our estimates of the volume of private transactions based on such tools represent, hence, a lower bound.

Our results show that Babel Pool included \num{2} out of our \num{4} private transactions.
Spark Pool technically supports this mining pool, implying that they ``collaborate'' in committing private transactions sent over the Taichi network~\cite{Babel@Pool}.
Our transactions were included, however, in the appropriate position in the block based on their fees.
We delve into the prioritization of transactions in the next section.

We also characterize the prevalence of private transactions in Ethereum and indicate that mining pools can each have a distinct set of private transactions in their \mpool.
Users, as a result, can no longer rely on the public \mpool alone to estimate their transaction fee.
Given the absence of other data, they are highly likely to end up with a false estimate of the ``appropriate'' transaction fees for their transactions.

%

\section{On prioritization transparency}
\label{sec:prioritization_transparency}

In this section, we delve into our analysis of prioritization transparency within the Ethereum and Bitcoin blockchains. We show that the current assumptions about transparency in blockchains do not hold in practice. Then, we show that transaction relay networks are becoming more popular in Ethereum, with miners creating their own transaction relay networks for private transactions. We also show that miners have different measures for utility of mining a transaction than just the offered fee rate or gas price of a transaction. For instance, transactions issuers that pay miners via a direct transfer to their wallet address or through off-chain fee receive a higher prioritization than their corresponding transaction fee rates would suggest.

\subsection{Prevalence of transaction bundling}\label{subsec:prevalence_of_flashbots}

\begin{figure*}[tb]
	\centering
		\subfloat[Distribution of Flashbots blocks\label{fig:dist-flashbots-blocks-ethereum}]{\includegraphics[width=\twocolgrid]{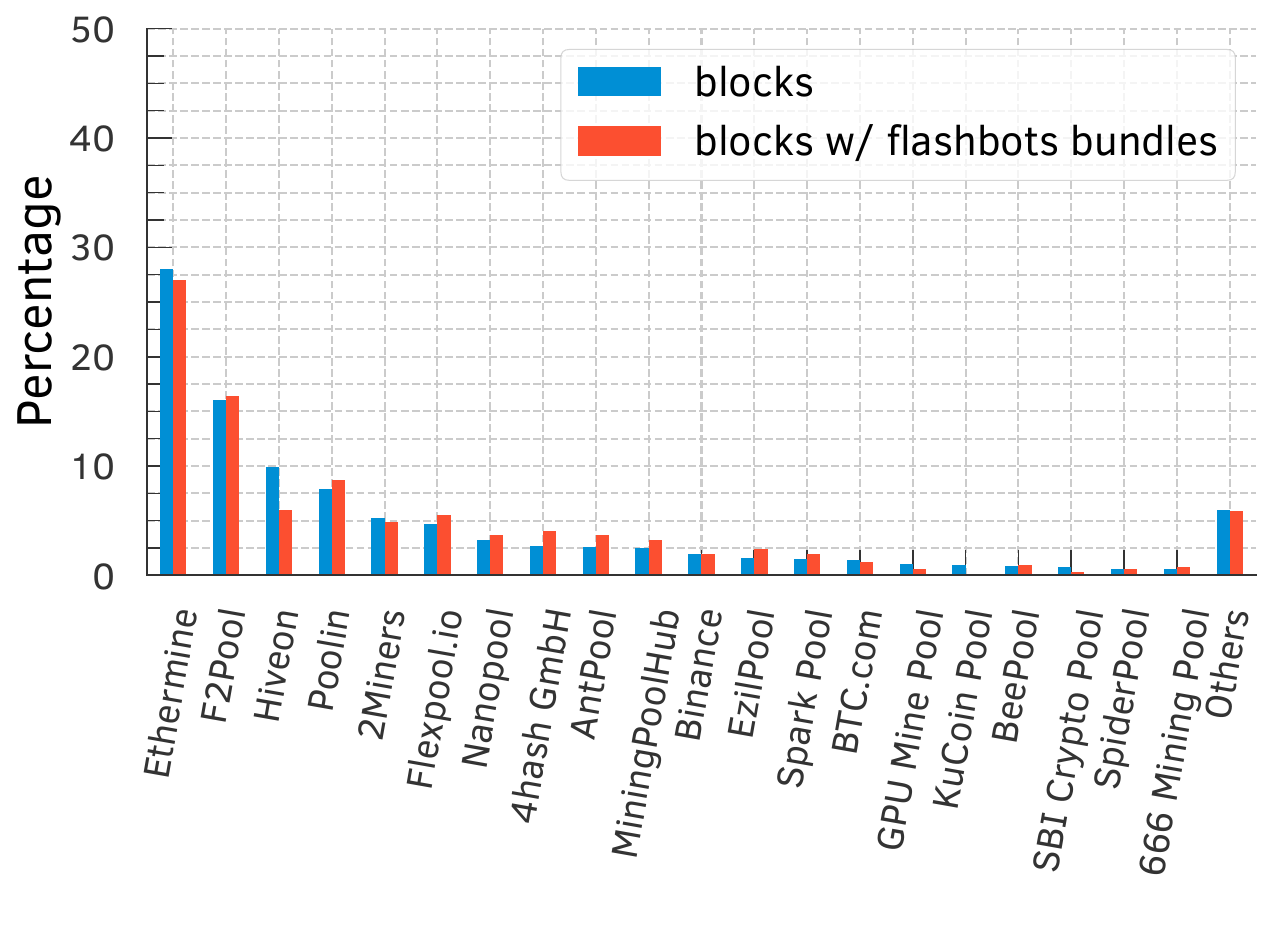}}
		\subfloat[Distribution of Flashbots bundles\label{fig:dist-flashbots-bundle-ethereum}]{\includegraphics[width=\twocolgrid]{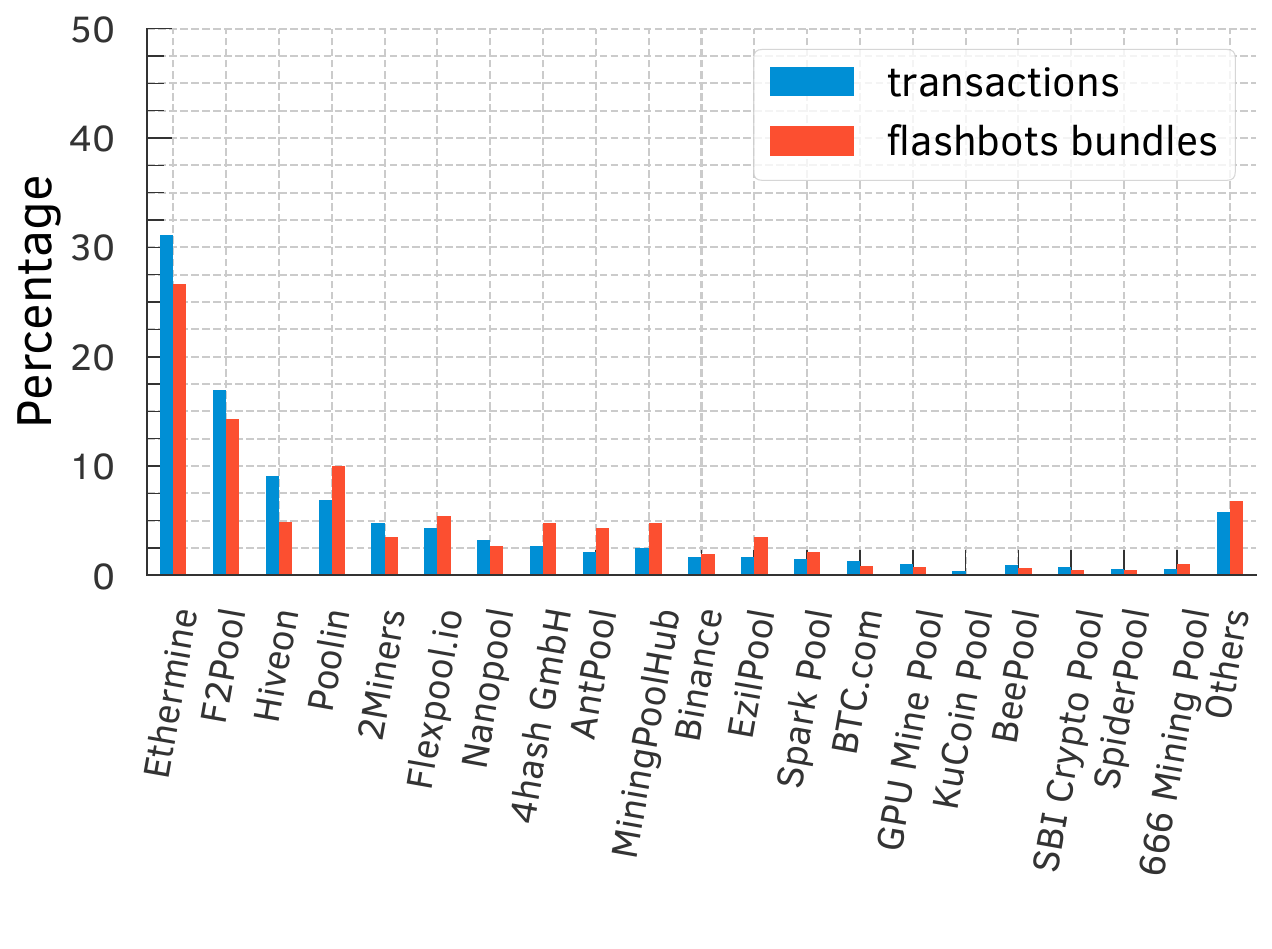}}
	\caption{
  Distribution of (a) blocks with at least one Flashbots bundle; and (b) bundle of transactions per block, per mining pool. Ethermine included 27.05\% of all blocks with a Flashbot bundle and 26.63\% of all Flashbots bundles, while mining around 28.05\% and 31.11\% of all blocks and transactions, respectively.
	}
  \label{fig:dist-flashbots-txs-bundle-ethereum}
\end{figure*}

In this section, we use the Flashbots data set outlined in \S\ref{sec:method_tx_prioritization_contention}, which has \num{6937292} transactions ($2\%$ of all issued transactions in Ethereum) contained in \num{3284886} bundles from Flashbots. These bundles constitute transactions privately sent to miners.
For instance, among all blocks in the data set \dse{}, \num{972911} ($52.11\%$) of blocks have at least one such Flashbots transaction:
\stress{Private transactions are becoming quite common across most of the powerful mining pools in Ethereum.}

Flashbots bundles are quite prevalent in Ethereum, representing \num{99.99}\% of the total Ethereum hash rate (refer~\S\ref{ss:characterize-pvt-relay}).
Our analysis shows that each Flashbots bundle contains at least $1$ transaction and at most \num{631} transactions; on average they contain \num{2.11} transactions, with a median of \num{1} and a std. of \num{6.47}.
We noticed that Ethermine alone included more than a quarter ($26.63\%$) of all \num{3284886} bundles (refer Figure~\ref{fig:dist-flashbots-txs-bundle-ethereum}).
Also, blocks contain at most $40$ bundles, with an average of \num{3.38}, a median of $3$, and a std. of \num{2.64} bundles.

\subsubsection{Miner Incentives in Incorporating Flashbots Bundles}
Flashbots allows users to bundle together a set of transactions, thereby specifying the order in which they are executed.
The bundles can also include public transactions,  propagated over the public P2P network.
A public transaction that buys a coin on a Decentralized Exchange (DEX) can, for example, lead to an arbitrage opportunity~\cite{Qin@FC21}.
A user can include this transaction in a bundle along with one of their own to capture this arbitrage opportunity.
The last transaction in the bundle usually pays the miner (based on the profit made) in \stress{Ether}\footnote{Ether (ETH) is the cryptocurrency used in the Ethereum blockchain to incentivize and reward its participants. Its smallest denomination is called Wei, representing a fraction of $10^{-18}$ ETH. When referencing gas prices or fees, the notation GWei is utilized, equating to a value of $10^{-9}$ ETH.} via a direct transfer (\ie \stress{coinbase transfer})  to their wallet addresses~\cite{Flashbots2Docs@Ethereum}.
This essentially means that miners are being offered different prices for mining the same transaction.
In other words, miners have a financial incentive for including transactions that are in a bundle at the top of a block, even though the public fee offered through gas price in the transaction data is very low (refer Figure~\ref{fig:gas_price_differences}). 
Hence, each transaction in the bundle has a normal gas price and a \stress{bundle gas price}, which is calculated using the total gas used by all transactions in the bundle and the total miner reward for mining the bundle.

\subsubsection{Bundling public transactions}\label{subsec:bundling_public_txs}

\begin{figure*}[t]
	\centering
		\subfloat[Public vs bundled's actual fee\label{fig:public_vs_bundle}]{\includegraphics[width=\twocolgrid]{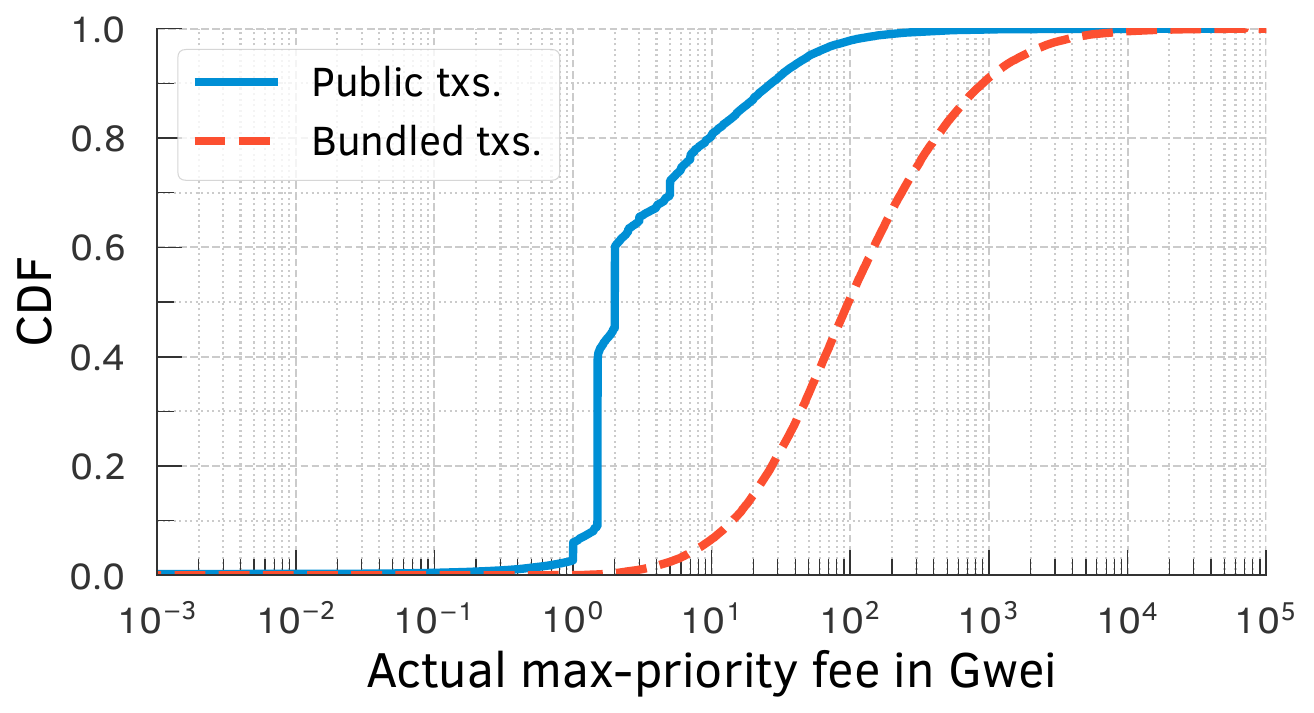}}
		\subfloat[Difference in actual max-priority fee\label{fig:gas_price_diff}]{\includegraphics[width=\twocolgrid]{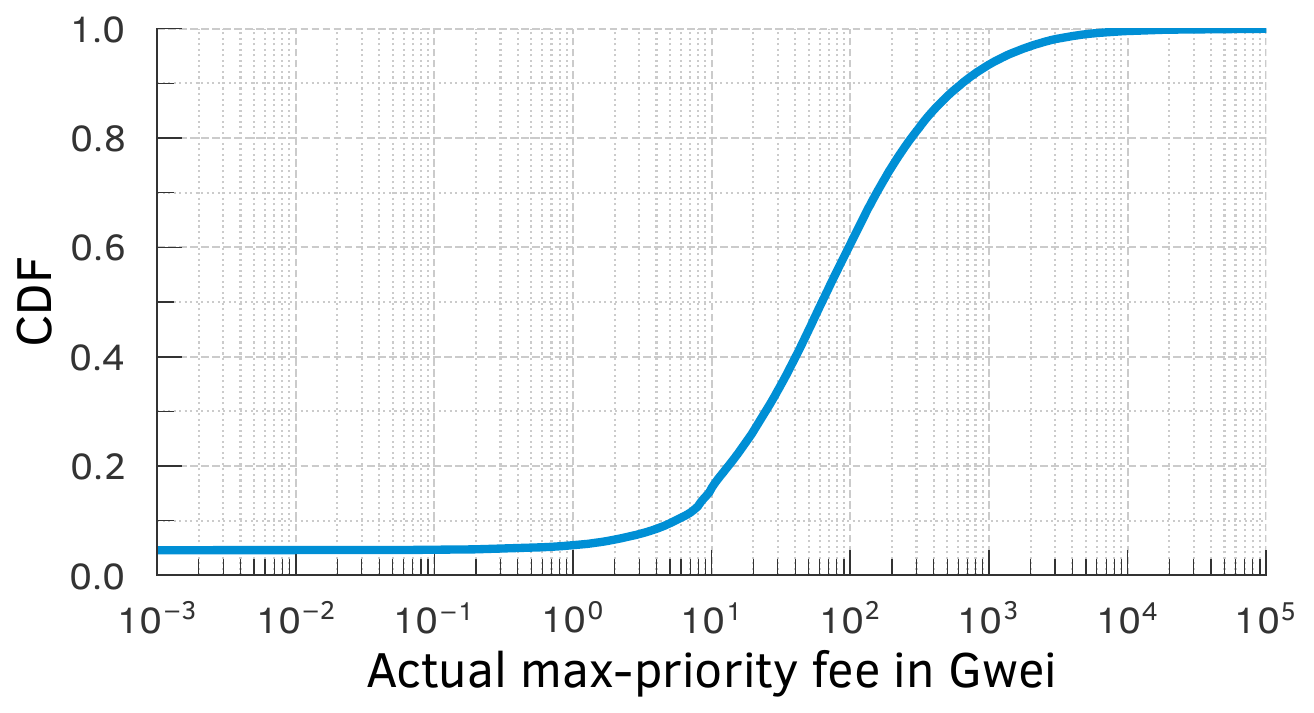}}
	\caption{
  Difference between the actual max-priority fee of public transactions and Flashbots bundles; bundles typically offer a larger \stress{effective} fee to the miners.
	}
  \label{fig:gas_price_differences}
\end{figure*}

To identify bundles with transactions that were probably sent through the public P2P network, we rely on a simple heuristic.
Specifically, we focus on transaction bundles of size \num{2} and \num{3}, and search for transactions that have likely resulted in a publicly sent transaction being bundled.
Then, we find bundles issued from different issuers that include a zero and non-zero \stress{max-priority fee}\footnote{The \stress{max-priority fee} was introduced in EIP-1559 as the unique financial incentive miners get for including publicly announced transactions.
The other fees are burned.} transactions.
The intuition is that miners have no incentive to include transactions that offer a zero max-priority fee, as they receive no rewards for mining these transactions.
Unless they receive extra payment (through Flashbots coinbase transfer). Hence, transactions that have a non-zero max-priority fee were likely sent publicly.

For transaction bundles of size \num{2}, we look for transactions whose issuers are not the same. Furthermore, we look for cases where the first transaction offers a non-zero max-priority fee, with no coinbase transfer to the miner, and the second transaction offers a \num{0} max-priority fee and a non-zero coinbase transfer to the miner.

For transaction bundles of size \num{3}, we look for signs of sandwich attacks~\cite{Qin@BEV}. We look for bundles where the first and last transactions have the same issuer, but the second transaction has a different issuer.
Additionally, we check that the first and third transactions offer a \num{0} max-priority fee, meaning that the miner receives no reward from the gas price for mining these transactions.
Then, we ensure that the second transaction offers miners a non-zero max-priority fee, while the third offers miners a fee through direct coinbase transfer. 
This scenario might be a classic sandwich attack, where public transactions are bundled between two private transactions, sent by the same issuer, and the miner gets paid via a coinbase transfer from the third transaction~\cite{Qin@BEV}.

We found \num{853394} transactions in \num{426697} bundles of length \num{2}, and \num{1231695} transactions in \num{410565} bundles of length \num{3}. 
From those, we found that \num{110401} (\num{25.87}\%) and \num{37447} (\num{9.12}\%) bundles, of lengths \num{2} and \num{3}, respectively, fit our heuristic. 
We then calculate the \stress{actual max-priority fee} for these bundles, as the total gas used by all transactions in the bundle divided by the total miner reward (from gas usage and coinbase transfer). 
Figure~\ref{fig:gas_price_differences} shows the price difference miners get for including publicly and bundled transactions. 
Note that around \num{40}\% of transactions differ in the actual max-priority fee by \num{100} gwei-per-units-of-gas. Flashbots bundles offers much higher gas prices in comparison to the public announced max-priority fee alone.

\subsubsection{Towards liquidations through bundling}

Lending protocols rely on \stress{over-collateralization} of assets:
In order to borrow assets from these protocols, a user has to deposit a collateral of at least \num{150}\% of the borrowed amount.
To borrow \num{1} USDC on AAVE, for example, a user would have to collateralize at least \num{1.5} USDC worth of another asset (\eg in ETH or BTC).
If the ratio of the collateral asset versus the borrowed asset falls below \num{1.5}, the user's position can be liquidated by any other participant until the ratio stabilizes to \num{1.5} again.
The liquidator then pays back a portion of the user's debt to receive the collateral asset at a discount.
In order to assess an asset's on-chain value, lending protocols rely on oracle services, \eg Chainlink Data Feeds~\cite{breidenbach2021chainlink,ChainlinkDataFeeds}.
In the case of the two largest lending platforms, AAVE V2~\cite{AAVE} and Compound~\cite{Compound}, for instance, Chainlink provides the price of each asset in ETH and USD, respectively. 

We found \num{16418} liquidations in AAVE and \num{6387} liquidations in Compound.
Out of these, there were \num{4863} AAVE liquidations and \num{2036} Compound liquidations that were sent privately through Flashbots. 
In AAVE, the three largest collateral assets that were liquidated were WETH (\num{57.58}\%), LINK (\num{11.84}\%), and WBTC (\num{8.99}\%). 
The debt assets paid for, i.e., the assets borrowed by the users, were USDC (\num{33.77}\%), USDT (\num{22.27}\%), DAI (\num{19.39}\%), and GUSD (\num{5.12}\%), all of which are stablecoins and account for over \num{80}\% of the assets repaid by liquidators.
In Compound, the three largest collateral assets that were liquidated were WETH (\num{69.7}\%), WBTC (\num{10.31}\%), and UNI (\num{5.5}\%).
The debt assets were USDC (\num{38.9}\%), DAI (\num{30.45}\%), USDT (\num{23.38}\%), and TUSD (\num{2.7}\%), all of which are stablecoins and account for over \num{90}\% of the assets repaid by liquidators.

\subsubsection{Liquidation with bundled oracle updates}

\begin{figure*}[tb]
	\centering
		\subfloat[Liquidations profit in AAVE\label{fig:oracle-update-AAVE}]{\includegraphics[width=\twocolgrid]{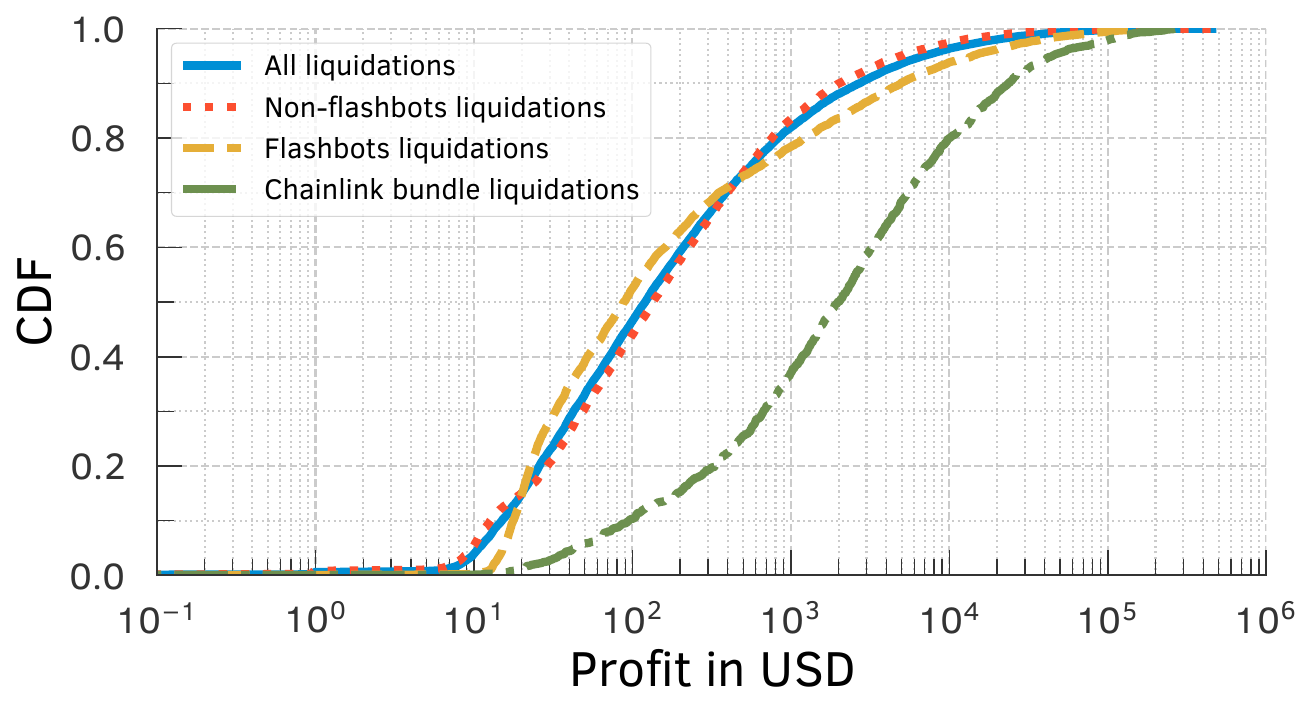}}
		\subfloat[Liquidations profit in Compound\label{fig:oracle-update-Compound}]{\includegraphics[width=\twocolgrid]{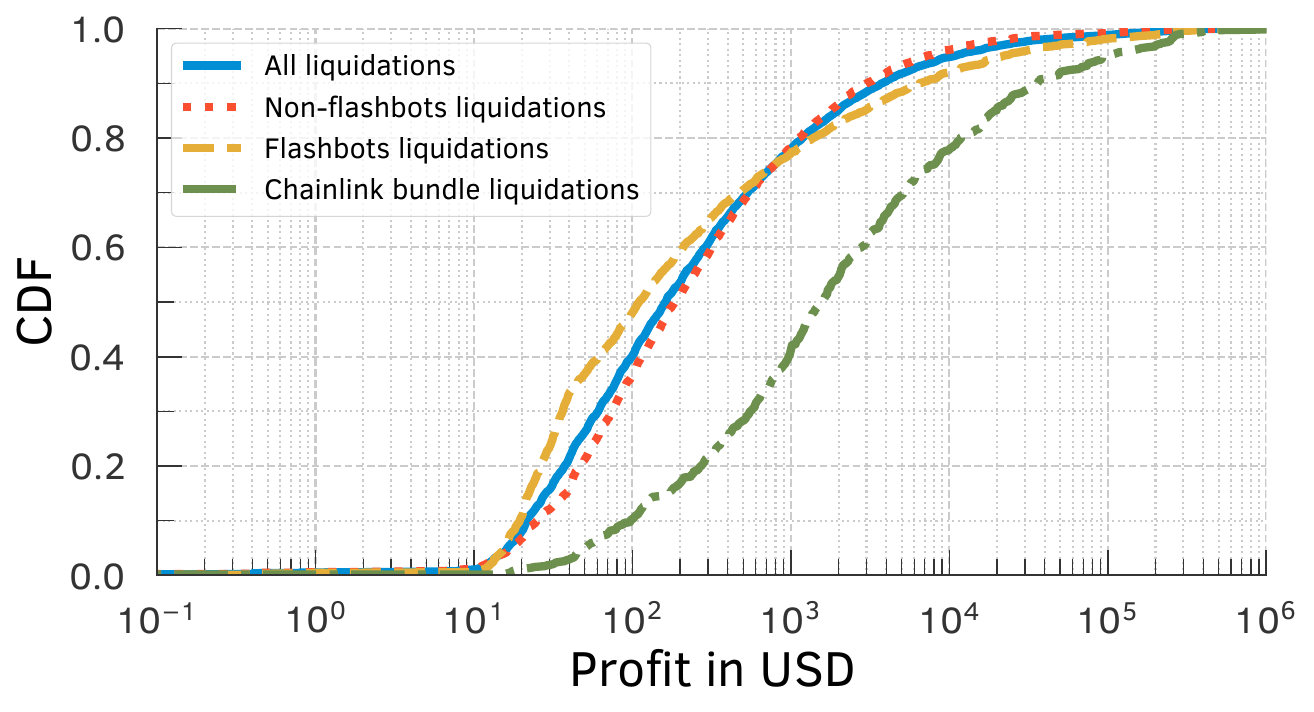}}
	\caption{
  Profits of liquidators in (a) AAVE and in (b) Compound. Liquidations bundled with Chainlink updates generally provide higher profits.
	}
  \label{fig:oracle-update}
\end{figure*}

%
To check the adverse effect of bundling oracle updates, we looked at bundles with Chainlink~\cite{ChainlinkDataFeeds} oracle updates as they are a key part of liquidations.
We identified \num{1165} AAVE liquidations distributed within \num{1154} bundles (\num{2662} transactions including \num{1301} oracle updates) that contained at least one oracle update.
In Compound, we found \num{648} liquidations distributed within \num{641} bundles (\num{1457} transactions including \num{751} oracle updates) that contained oracle updates.
In AAVE, out of \num{1154} bundles, there were \num{994} (86.14\%) bundles that contained an oracle update followed by a liquidation, and \num{52} (4.51\%) with two oracle updates followed by liquidations.
In Compound, out of \num{641} bundles, there were \num{548} (85.49\%) bundles that contained an oracle update followed by a liquidation, and \num{39} (6.08\%) with two oracle updates followed by liquidations.
For details on the specific liquidations for both AAVE and Compound, please refer~\S\ref{sec:liquidations-cll-updates} in the appendix.
Out of the total \num{1813} liquidations in AAVE and Compound we found that only \num{24} were possible in the previous block.
Almost \num{98.68}\% of such liquidations were, hence, only possible because of the Chainlink updates in that block.

In order to calculate the profit made by the liquidators, we get the amount of debt that was repaid and the amount of the underlying collateral that was received by the liquidator. 
We calculate the price of each token at the time of liquidation by looking at the on-chain oracle price from Chainlink at the same block number, where the liquidation took place. 
For AAVE and Compound, we specifically use the Chainlink on-chain price used by AAVE and Compound in their respective protocols. 
AAVE uses the price in ETH as a reference for its tokens, whereas Compound's price oracles are denominated in USD. For AAVE, in order to calculate the profit made by each liquidation, we calculate the profit in ETH, and then multiply the profit by the current Chainlink on-chain price of ETH in USD.
Per Figure~\ref{fig:oracle-update}, liquidations that are bundled with a Chainlink update also have larger profits for liquidators, which implies that the lucrative liquidations are more likely to be bundled together with a Chainlink update.

\subsubsection{Characterizing transaction bundling}

To investigate which DEXes protocols are called within Flashbots bundles, we focus on the following contract calls: 0x Protocol~\cite{0xProtocol}, Balancer~\cite{Balancer}, Bancor~\cite{Bancor}, Curve~\cite{Curve}, SushiSwap~\cite{SushiSwap}, and Uniswap V1 and V3~\cite{Uniswap}.
In our set of \num{3284886} Flashbots bundles, we find that \num{2231051} (\num{67.92}\%) unique Flashbots bundles (and \num{3076760} transactions) called at least one of these contracts. Table~\ref{tab:flashbots-lending} shows the distribution of the number of transactions and the number of bundles for each of these contracts.
We see that Uniswap and SushiSwap are the most bundled DEXes protocols in Flashbots.

\begin{table}[t]
\begin{center}
\tabcap{There are \num{2231051} (\num{67.92}\%) unique Flashbots bundles, and \num{3076760} (\num{44.35}\%) transactions, that called the following decentralized exchange contracts in Ethereum: 0x Protocol, Balancer, Bancor, Curve, SushiSwap, Uniswap V1, or V3. Note that a single transaction or bundle might call one or more contracts.}\label{tab:flashbots-lending}
\resizebox{\textwidth}{!}{%
\begin{tabular}{llllllll}
\toprule
\multicolumn{1}{c}{\thead{}} & \multicolumn{1}{c}{\thead{Balancer}}                          & \thead{Bancor}                                               & \multicolumn{1}{c}{\begin{tabular}[c]{@{}c@{}}\thead{Curve}\\ \thead{v1 \& v2}\end{tabular}} & \multicolumn{1}{c}{\begin{tabular}[c]{@{}c@{}}\thead{Uniswap v2}\\ \thead{\& Sushiswap}\end{tabular}} & \multicolumn{1}{c}{\begin{tabular}[c]{@{}c@{}}\thead{Uniswap}\\ \thead{v3}\end{tabular}} & \multicolumn{1}{c}{\begin{tabular}[c]{@{}c@{}}\thead{0x Protocol}\\ \thead{v1, v2 \& v3}\end{tabular}} & \thead{Total}                                                  \\ \midrule
\# of bundles                 & \begin{tabular}[c]{@{}l@{}}\num{85422}\\ \num{3.83}\%\end{tabular} & \begin{tabular}[c]{@{}l@{}}\num{96122}\\ \num{4.31}\%\end{tabular} & \begin{tabular}[c]{@{}l@{}}\num{53296}\\ \num{2.39}\%\end{tabular}                        & \begin{tabular}[c]{@{}l@{}}\num{1710985}\\ \num{76.69}\%\end{tabular}                                 & \begin{tabular}[c]{@{}l@{}}\num{1337715}\\ \num{59.96}\%\end{tabular}                    & \begin{tabular}[c]{@{}l@{}}\num{28753}\\ \num{1.29}\%\end{tabular}   & \begin{tabular}[c]{@{}l@{}}\num{2231051}\\ \num{67.92}\%\end{tabular} \\
\# of transactions                     & \begin{tabular}[c]{@{}l@{}}\num{87865}\\ \num{2.86}\%\end{tabular} & \begin{tabular}[c]{@{}l@{}}\num{99040}\\ \num{3.22}\%\end{tabular}  & \begin{tabular}[c]{@{}l@{}}\num{58188}\\ \num{1.89}\%\end{tabular}                         & \begin{tabular}[c]{@{}l@{}}\num{2533084}\\ \num{82.33}\%\end{tabular}                                & \begin{tabular}[c]{@{}l@{}}\num{1692485}\\ \num{55.01}\%\end{tabular}                   & \begin{tabular}[c]{@{}l@{}}\num{29100}\\ \num{0.95}\%\end{tabular}                                   & \begin{tabular}[c]{@{}l@{}}\num{3076760}\\ \num{44.35}\%\end{tabular} \\ \bottomrule  
\end{tabular}
}
  \end{center}
\end{table}

\subsection{Side channel (dark-fee) payments and transaction acceleration}\label{chapter4:sec_bitcoin_dark_fees}

In this section, we focus on the Bitcoin blockchain, with a particular emphasis on the data set \dsd{}. Our goal is to build upon our earlier discussion in \S\ref{sec:dark-fee-txs} regarding dark fees transactions.

\subsubsection{Prevalence of transaction acceleration}

As previously discussed in \S\ref{sec:dark-fee-txs}, dark-fee transactions (or accelerated transactions) are transactions that offer additional fees to specific mining pools via an opaque and non-public side-channel payment. In Bitcoin, the top 5 mining pools named BTC.com~\cite{BTC@accelerator}, AntPool~\cite{AntPool@accelerator}, ViaBTC~\cite{ViaBTC@accelerator}, F2Pool~\cite{F2Pool@accelerator}, and Poolin~\cite{Poolin@accelerator}, deploy transaction acceleration services, which enables users to ``accelerate'' the confirmation of their transactions by offering mining pools dark-fees.

These (dark-)fees are paid in fiat currency through a direct bank transfer or via other crypto coins to the mining pool. They are, therefore, opaque or dark to other participants.
Strangely enough, these fees are also non-refundable as the miner receives them regardless of whether they include the transaction in a block or not---a guaranteed payment.
The fees paid by the transaction issuer are, furthermore, not made public:
only the user and the miner knows the actual fee paid by the transaction inclusion.
Since transaction issuers pay the fees off-chain, miners have an incentive for prioritizing these transactions despite the low fee rate offered on-chain.
It also implies that the transaction issuer offers a miner a different fee compared to that offered to other miners for including their transaction in a block.
Miners do not disclose such private fees paid by issuers.
This behavior is different from that of Flashbots in Ethereum: The latter discloses the final dark-fee after the transaction is committed (see \S\ref{subsec:prevalence_of_flashbots}).

\subsubsection{Characterizing transaction acceleration}\label{subsubsec:bitcoin-acceleration}

In order to detect accelerated transactions, we proposed two metrics called \stress{signed position prediction error (SPPE)} and \stress{position prediction error (PPE)} that are described in \S\ref{subsec:mining-prioritization-based-feerate}. 

\begin{figure*}[t]
	\centering
		\includegraphics[width={\onecolgrid}]{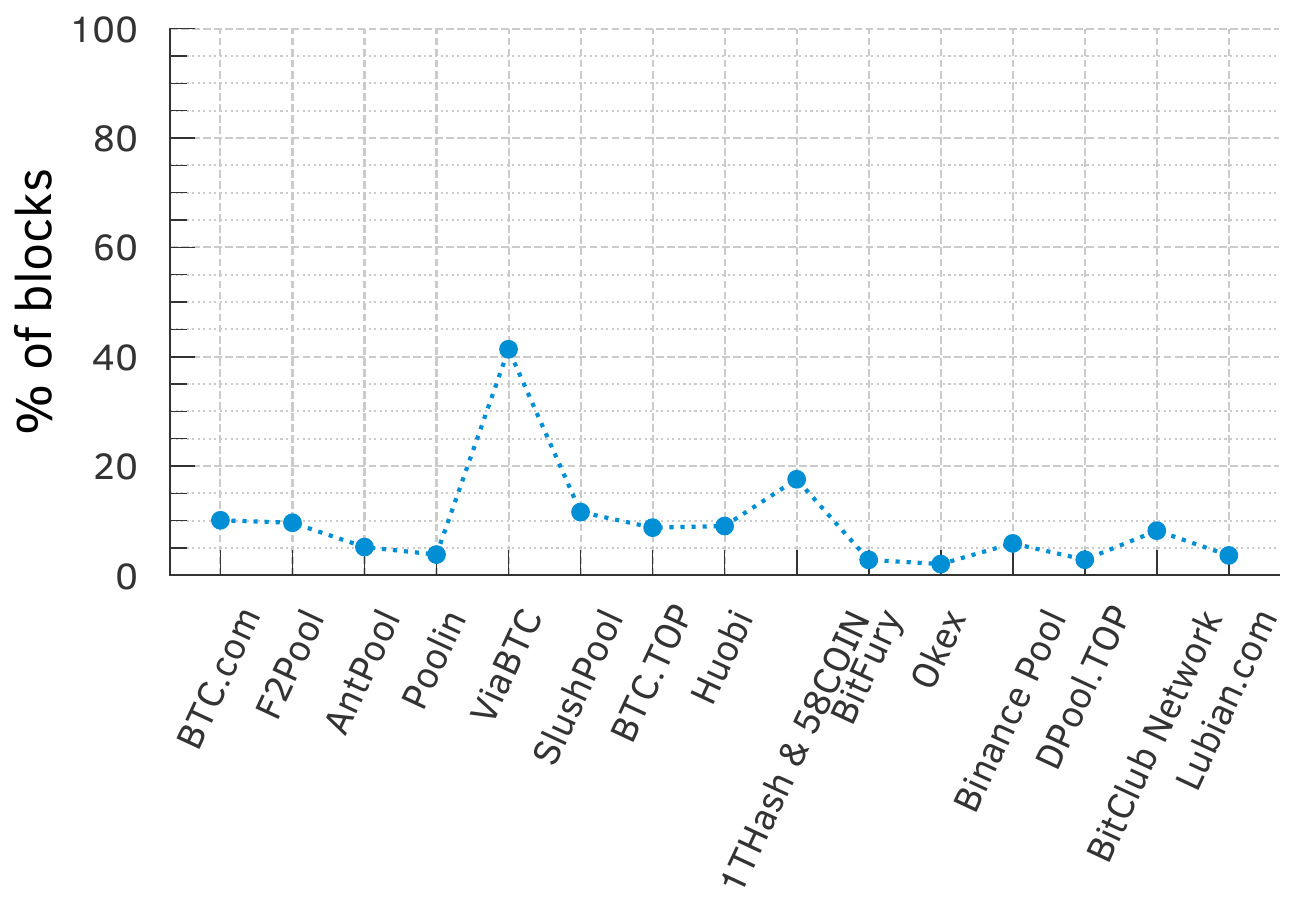}
	\caption{
  Blocks with accelerated transactions (with SPPE $\geq 99\%$) are quite common among the top 15 mining pools. In Bitcoin, the mining pools with a high percentage of such blocks are ViaBTC (41.36\%), 1THash \& 58COIN (17.58\%), SlushPool (11.58\%), BTC.com (10.03\%), and F2Pool (9.63\%).}
\label{fig:block-with-tx-violation}
\end{figure*}

To estimate the prevalence of accelerated transactions in blocks mined by different mining pools, we compute the fraction of blocks mined by the top-15 mining pools, based on their hash rates in our 3-year data set \dsd (refer to \S\ref{sec:hash-var} and Figure~\ref{fig:dist-tx-blks-bitcoin}), that contained transactions with SPPE $\ge 99\%$. 
Per Figure~\ref{fig:block-with-tx-violation},
we find that many large mining pools such as BTC.com, F2Pool, and ViaBTC are likely including accelerated transactions in a sizeable fraction of their mined blocks, with ViaBTC including it in over $40\%$ of their blocks.

If we consider all mining pools' transactions with an SPPE $\geq 50\%$ (\num{1869043} transactions, in total), from $2018$ to $2020$, users transferred in total \num{11631217} BTC (or $\approx 223.55$ billion USD\footnote{Based on the Bitcoin exchange rate on October 19\tsup{th} 2022, 1 BTC = \num{19219.90} USD}). The accelerated transactions accounted for \num{240226} BTC (or $\approx 4.62$ billion USD), corresponding to approximately $2.07\%$. 

\subsubsection{Aggregated power of colluding miners}

In order to check the impact of transactions acceleration services on commit time of transaction, we ran active real-world experiments.
Specifically, we paid ViaBTC~\cite{ViaBTC@accelerator} to accelerate selected transactions (see Table~\ref{tab:acceleration-experiment} in \S\ref{sec:accelerated-txs}) during periods of high congestion between November 26\tsup{th} and December 1\tsup{st}, 2020.
From \num{10} \mpool snapshots during this period, we selected transactions that offered a very low fee rate (i.e., 1--2 sat-per-byte) for acceleration.
To keep our acceleration costs low, we selected transactions with the smallest size (which was \num{110} bytes) within this set.
For each of the \num{10} snapshots, we had multiple transactions with such low fee rates and small size, for a total of \num{212} transactions across all the snapshots.
We randomly selected one transaction from each snapshot (i.e., 10 transactions) and paid ViaBTC $205$ EUR to accelerate them.

\begin{table}[t]
    \begin{center}
    \tabcap{Accelerated transactions have fewer delays and are included at the top of the block, i.e., at higher positions compared to non-accelerated transactions.}\label{tab:active-experiment-delay-position}
    \resizebox{.75\textwidth}{!}{%
        \begin{tabular}{rcccc}
        \toprule
        \multicolumn{1}{c}{\multirow{2}{*}{\thead{metrics}}} & \multicolumn{2}{c}{\thead{delay in \# of blocks}} & \multicolumn{2}{c}{\thead{perc. position in a block}} \\
        \multicolumn{1}{c}{}                         & \thead{acc.}        & \thead{non-acc.}       & \thead{acc.}       & \thead{non-acc.}      \\  \midrule
        minimum                                       & 1                  & 9                    & 0.07                 & 17.47                   \\
        25-perc                                       & 1                  & 148                    & 0.08                 & 75.88                   \\
        median                                        & 2                  & 191                    & 0.09                 & 87.92                   \\
        75-perc                                       & 2                  & 247                    & 0.20              & 95.00                   \\
        maximum                                       & 3                  & 326                    & 4.39               & 99.95                   \\
        average                                       & 1.8                & 198.5                    & 0.79              & 84.46                   \\
        \bottomrule
        \end{tabular}
    } 
    \end{center}
\end{table}

We then compare the priority with which the accelerated transactions and the $202$ ($= 212-10$) non-accelerated transactions with similar fee rates and sizes were included in the Bitcoin blockchain.
The impact of acceleration was strikingly apparent as shown in Table~\ref{tab:active-experiment-delay-position}. 
All $10$ accelerated transactions were included within $1$--$3$ blocks after their acceleration, with an average delay of $1.8$ blocks.
In contrast, the minimum delay for the $202$ non-accelerated transactions of comparable fee rates and sizes was $9$ blocks, with an average delay of 198.5 blocks.
Interestingly, $38$ of the non-accelerated transactions were yet to be included in the blockchain by December 4\tsup{th}, 2020.
Similarly, the accelerated transactions were included in top $0.07$--$4.39$ percentile positions, with an average $0.79$ percentile position, while the non-accelerated transactions were included in the beyond top $17.47$--$99.95$ percentile positions, with an average $84.46$ percentile position.
From the above observations, it is clear that the transactions we accelerated were included with high priority, meaning Bitcoin mining pools take off-chain fees into account when prioritizing transactions.

Although, we accelerated our transactions using ViaBTC mining pool, our $10$ transactions were included by $5$ different mining pools, namely F2Pool, AntPool, Binance, Huobi, and ViaBTC. 
As we accelerated transaction during time of high congestion in Bitcoin, no mining pool would have included a transaction offering $1$--$2$ sat-per-byte, unless they were accelerated. Since we only paid the ViaBTC mining pool, this implies that ViaBTC is colluding with other mining pools to accelerate transactions that offer off-chain fees.
Except for Binance, all these colluding pools rank amongst the top-$8$ mining pools in terms of their hash rates at the time of our experiments.
Table~\ref{tab:active-experiment-hash-rate} shows the individual as well as the combined hash rates of these $5$ colluding mining pools over the last day, last week, and last month before the conclusion of our experiment on December $1\tsup{st}, 2020$. 
The most striking and the most worrisome fact is that \textbf{the combined hash rates of these colluding mining pools exceeds $55\%$ of the total Bitcoin hash rate}. For more details, refer to Figures~\ref{fig:tx-acceleration-active-overtime-month} and \ref{fig:tx-acceleration-passive-active-overtime-month} in \S\ref{sec:accelerated-txs} in the appendix. Additionally, if mining pools are colluding to include accelerated transactions, then they might also potentially collude in malicious ways.  

\begin{table}[t]
    \begin{center}
    \tabcap{If we rank the miners who confirmed the accelerated transactions based on their daily, weekly, and monthly hash rate power, at the time these experiments were conducted, the combined hash power of these mining pools exceeds 55\% of the Bitcoin's total hashing power.}\label{tab:active-experiment-hash-rate}
    \resizebox{.65\textwidth}{!}{%
        \begin{tabular}{rccc}
        \toprule
        \multicolumn{1}{c}{\multirow{2}{*}{\thead{Mining Pool}}} & \multicolumn{3}{c}{\thead{Hash-rate}}                                                                                \\
        \multicolumn{1}{c}{}                     & \multicolumn{1}{c}{\thead{last 24h}} & \multicolumn{1}{c}{\thead{last week}} & \multicolumn{1}{c}{\thead{last month}} \\ \midrule
        F2Pool                                    & \multicolumn{1}{c}{$19.9\%$}   & \multicolumn{1}{c}{$18.7\%$}    & 
        \multicolumn{1}{c}{$19.9\%$}    \\
        AntPool                                   & \multicolumn{1}{c}{$12.5\%$}   & \multicolumn{1}{c}{$10.6\%$}    & \multicolumn{1}{c}{$10.2\%$}    \\
        Binance                                   & \multicolumn{1}{c}{$9.6\%$}    & \multicolumn{1}{c}{$10.3\%$}    & \multicolumn{1}{c}{$10.0\%$}    \\
        Huobi                                     & \multicolumn{1}{c}{$8.1\%$}    & \multicolumn{1}{c}{$9.3\%$}     & \multicolumn{1}{c}{$9.8\%$}     \\
        ViaBTC                                    & \multicolumn{1}{c}{$5.1\%$}    & \multicolumn{1}{c}{$7.1\%$}     & \multicolumn{1}{c}{$7.7\%$}     \\
        \thead{Total}                                     & \multicolumn{1}{c}{\attention{$55.2\%$}}   & \multicolumn{1}{c}{\attention{$56\%$}}      & \multicolumn{1}{c}{\attention{$57.6\%$}}   \\ \bottomrule
        \end{tabular}
    } 
    \end{center}
\end{table}

Furthermore, due to the lack of transparency into their queue, miners can charge higher prices for their acceleration services when colluding. It means that they can overcharge the transaction issuers for including their transactions.

%

\section{Concluding remarks}

In this section, we present the findings derived from our analysis of private relayed transactions, along with the results obtained from our active experiments conducted on Bitcoin and Ethereum blockchains. The main objective of these experiments was to evaluate the lack of transparency in transaction contention and prioritization.

In summary, our findings indicate that private transactions and private relay networks are quite prevalent in both Ethereum and Bitcoin blockchains.
Flashbots, in particular, is extensively used in Ethereum, accounting for a significant portion of $99.99\%$ of the total Ethereum hash rate. It also enables arbitrageurs to exploit MEV opportunities by bundling their private transactions with public transactions like oracle updates or taking advantage of sandwich attacks.
Similarly, in Bitcoin, miners offer transaction acceleration services, allowing users to privately offer a dark-fee to incentivize miners for a faster commit time.
Through active experiments, we show that miners highly prioritize these transactions, on average including them in \num{1.8} blocks, with a range of \num{1} to \num{3} blocks.
Worrisome, we uncover evidence of collusion among miners with a combined hash rate exceeding \num{50}\% to ensure the inclusion of these dark-fee transactions.

In the following chapter, we delve into the voting power distribution for amending smart contracts.

\clearpage

%
\chapter{Decentralized Governance} \label{chap:governance}

%

In this chapter, we present our research questions, methodology, and discuss the implications of our findings regarding the level of decentralization in governance protocols.
To investigate this, we focus on the Compound governance protocol as a case study.
Our analyses reveal that the distribution of voting power in Compound is highly concentrated among a small number of participants, which can significantly hinder the achievement of a fair and decentralized governance system.

The concentration of voting power poses a challenge to achieving truly decentralization in governance protocols.
For example, when a small group of participants holds a majority of the tokens, they can make decisions that benefit themselves at the expense of others.
Therefore, ensuring a fair distribution of tokens becomes crucial to foster decentralization in these protocols.
In this chapter, we aim to analyze transaction data associated with Compound in order to assess the level of decentralization in Compound's voting power.
To guide our analysis, we propose the following research questions.

\point{}
\textbf{RQ 1}: \stress{How frequently are amendments proposed and voted on in the Compound protocol?}
This research question aims to investigate the activity level of the Compound protocol and its community engagement.
For instance, by examining the frequency with which proposals are amended or voted, we can assess the level of participation and the extent to which the community is actively contributing to improving the protocol.

\point{}
\textbf{RQ 2}: \stress{What is the distribution of Compound tokens among its participants?
How small or large is the set of voters who determine the outcomes for the amendments?}
This research question aims to investigate the distribution of Compound tokens among its participants.
Hence, we can assess to which extent the Compound tokens is truly decentralized.
Understanding this distribution is crucial for proposing fairness to the protocol token's distribution.

\point{}
\textbf{RQ 3}: \stress{What is the cost associated with casting a vote in the Compound protocol?}
Voting in on-chain governance protocols, where the entire voting process happens on the blockchain, requires the payment of transaction fees that vary depending on the network congestion.
These voting costs can disproportionately affect small token holders, potentially limiting their participation in the decision-making process.
This research question aims to investigate the impact of voting costs on voter participation in the Compound protocol.
It provides insights into the fairness of the decision-making process and shed light on potential barriers that may discourage certain participants from exercising their voting rights.

\point{}
\textbf{RQ 4}: \stress{What are the voting patterns of delegates, and do voters form coalitions?}
This research question aims to analyze the voting patterns of delegates in the Compound protocol and investigate whether voters form coalitions, where they align their votes as a collective group. 
The formation of coalitions among voters can lead to the marginalization of certain voters, as they consistently find themselves in a minority group.
This undermines the core principle of decentralization and has the potential to compromise the security and effectiveness of the governance protocol.
Specifically, instead of expressing their individual opinions on a proposal, voters may choose to mimic the voting behavior of their peers.
Therefore, exploring the presence and impact of coalition formation can provide valuable insights into the decision-making dynamics among voters and help mitigate the concentration of tokens (or voting power) within the system.

Addressing these research questions is key for improving the protocol's fairness and achieving more decentralization in the distribution of voting power.
In the following section, we discuss our methodology for gathering the necessary data related to Compound protocol from our Ethereum archive node.

\subsubsection*{Relevant publication}

The results presented in this chapter have been submitted, and we are currently awaiting a decision~\cite{Messias@FC2024}.

%

\section{Methodology} \label{sec:method_governance}

\begin{figure*}[t]
	\centering
		\includegraphics[width={\onecolgrid}]{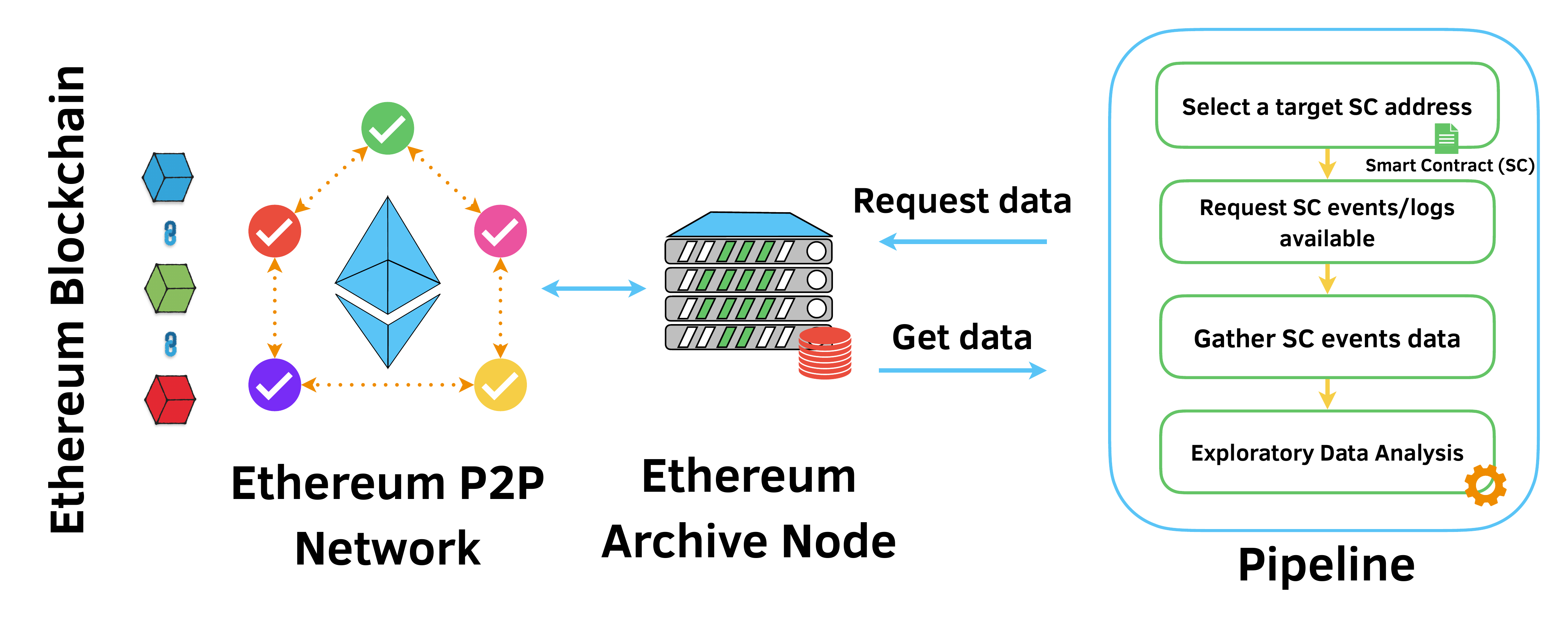}
	\caption{
  Overview of the data collection methodology and analysis.
	}
\label{fig:data-pipeline}
\end{figure*}

To analyze the voting power concentration among Compound token holders (or voters), we adopt a data-driven approach.
Our methodology involves collecting events triggered by transaction executions when voters cast votes, create proposals, cancel proposals, transfer tokens, or delegate their voting rights to another address.
We cover events from the inception of the Compound token and Compound governance protocol.
To address the possibility of a single entity owning multiple addresses, we have developed a methodology to infer address ownership and group them accordingly.
This approach allows us to identify and consolidate addresses that are likely controlled by the same entity.
This process utilizes data from well-known blockchain explorers and publicly disclosed information regarding address ownership. For those addresses that we were not able to infer their ownership we renamed them to their specific wallet addresses.

To gather the data, we deployed an Ethereum \stress{archive} node on a server with \num{64} cores (with a base clock frequency of \uGHz{2.25} that can be boosted to \uGHz{3.4}), \uMB{256} L3 cache, \uGB{252} of RAM,  and \uTB{21} of NVMe-based storage.
The archive node took about \num{4} weeks---a relatively long time, though not unexpected---to fully synchronize with the Ethereum blockchain.
We used Web3.py~\cite{web3py}, a Python library for interacting with Ethereum nodes, to query and retrieve the information that we need from the archive node. Figure~\ref{fig:data-pipeline} summarizes our methodology for the Ethereum data gathering.

\subsection{Smart contract events.}\label{subsec:smart-contract-events}
Smart contracts in Ethereum can generate and dispatch \stress{events} for signaling various types of activities (e.g., ERC-20 token transfers or state changes) within the contract.
We can subscribe to these events, or analyze them later since Ethereum persists the events in the blockchain via the ``logs'' field of the transaction receipt attribute.
In this thesis, we leveraged these logs to filter transactions that triggered specific events, e.g., sending, receiving, or swapping tokens.
We also filtered and analyzed transactions that triggered events related to governance protocols to track the evolution of each proposal, including when it was created, when users started voting, and when it was executed or canceled.

\begin{table*}[t]
\centering
\small
\tabcap{Summary of events related to the Compound (COMP) token that we gathered from the Ethereum blockchain.}
\label{tab:events_tokens}
\begin{tabular}{rrp{10cm}}
\toprule
\thead{Event name} & \multicolumn{1}{c}{\thead{\# of events}} & \thead{Description} \\
\midrule
\stress{Approval} & \num{213220} & \footnotesize{Standard ERC-20 approval event.} \\
\stress{DelegateChanged} & \num{12095} & \scriptsize{Emitted when an account changes its delegate. This means that the delegatee will receive voting power from the sender. Users can only delegate to one address at a time, and the number of votes added to the delegatee's vote count is equal to the user's balance. The delegation of votes will take effect from the current block until the sender either delegates to a different address or transfers their tokens.} \\
\stress{DelegateVotesChanged} & \num{75820} & \footnotesize{Emitted when a delegate account's vote balance changes.} \\
\stress{Transfer} & \num{1886618} & \footnotesize{Emitted when users/holders transfer their tokens to another address.} \\
\bottomrule
\end{tabular}
\end{table*}

\begin{table*}[t]
\centering
\small
\tabcap{Summary of events related to the Compound Governor contracts recorded on the Ethereum blockchain.}
\label{tab:events_governor}
\begin{tabular}{rrp{11cm}}
\toprule
\thead{Event name} & \multicolumn{1}{c}{\thead{\# of events}} & \thead{Description} \\
\midrule
\stress{ProposalCanceled} & \num{17} &  \footnotesize{Emitted when a proposal is canceled.} \\
\stress{ProposalCreated} & \num{133} &  \footnotesize{Emitted when a new proposal is created.} \\
\stress{ProposalExecuted} & \num{101} &   \footnotesize{Emitted when a proposal is executed in the TimeLock.} \\
\stress{ProposalQueued} & \num{105} &   \footnotesize{Emitted when a proposal is added to the queue in the TimeLock.} \\
\stress{VoteCast} & \num{9500} &   \footnotesize{Emitted when a vote is cast on a proposal: \num{0} for against, \num{1} for in-favor, and \num{2} for abstain.} \\
\bottomrule
\end{tabular}
\end{table*}

\subsection{Data set collection}

We gathered various details on Compound tokens and Compound governance contracts between March 3, 2020 (block \#\num{9600000}) and November 7, 2022 (block \#\num{15917000}) from our Ethereum archive node.
This \num{32}-month study period includes Compound's entire lifetime (from its inception).
We illustrate our methodology and data-analysis pipeline in Figure~\ref{fig:data-pipeline}.
We obtained \num{213220} \stress{Approval} events, \num{12095} \stress{DelegateChanged} events, \num{75820} \stress{DelegateVotesChanged} events, and \num{1886618} \stress{Transfer} events for Compound tokens (refer to Table~\ref{tab:events_tokens}).
We also collected various events (refer to Table~\ref{tab:events_governor}) related to the Compound Governance contract for analyzing various aspects of the proposal creation and voting processes.

\subsection{Inferring wallet address ownership.}\label{subsec:dataset-inferred-addresses}

Since an entity can control multiple wallet addresses in the blockchain,  identifying the ownership of these wallets helps in grouping together the accounts that are owned by the same entity.
However, this task of wallet-address ownership determination is challenging due to the inherent anonymity of blockchains~\cite{Antonopoulos-Ethereum,Antonopoulos-Bitcoin}.
This task is further complicated because owners are only identifiable if they choose to voluntarily make their identities public. 
To address this challenge, we combine wallet ownership information from two widely used data sources:
Etherscan~\cite{Etherscan@ETH-explorer} and Sybil-List~\cite{Addresses@Sybil}.
The former is a blockchain explorer that helps in identifying the top holders of various cryptocurrencies, and the latter, a Uniswap governance tool for discovering delegates addresses~\cite{Uniswap@Sybil}. It uses cryptographic proofs for verifying wallet addresses voluntarily disclosed by the wallet owner.
From these two data sources, we gathered the owners of \num{3191} public wallet addresses.
We used these addresses to infer the owners of \num{17} (\num{51.52}\%) of the \num{33} unique addresses associated with proposal creation, \num{114} (\num{3.42}\%) out of \num{3335} proposal voters, and \num{265} (\num{0.13}\%) out of \num{210598} token holders.
By analyzing the top 10 most influential voters for each proposal, determined by the number of delegated tokens they possessed when casting their vote, we were able to infer the ownership of \num{67} (\num{50.37}\%) of these \num{133} unique addresses.
Finally, as an entity can control more than one address, we grouped the addresses we identified belonging to the same entity together to conduct our analysis.

%

%

\section{Attacks on governance}
\label{sec:attacks}

A potential issue in the governance of blockchain networks is the concentration of governance tokens in the hands of a few participants, which can pose a threat to the protocol~\cite{BuildFinance@Twitter}. 
This issue manifested in Balancer~\cite{Governance@Balancer}, a decentralized exchange (DEX) running on top of Ethereum, where a user with large amount of governance tokens voted for decisions that were beneficial for the user but detrimental for the protocol~\cite{Haig@Defiant}. 
When a minority holds a large portion of the tokens, decision-making power can become centralized, which conflicts with the goal of decentralization of governance protocols.

Yet another issue concerns many centralized exchanges that \stress{hold} their users' tokens; they could potentially use these tokens for voting \stress{without} their users' knowledge, compromising the integrity of the voting process~\cite{Binance@Coindesk,Binance@Cointelegraph}.
Alameda Research, a former cryptocurrency trading firm, which was affiliated with FTX, for example, voted on 8 proposals and even initiated three proposals (\#13, \#14, and \#16) on Compound~\cite{Proposal-13@Compound,Proposal-14@Compound,Proposal-16@Compound}.
Eventually, one of the proposals was executed.
Their goal was to raise the collateral of WBTC from 0\% to 40\%, which allowed WBTC to be utilized for borrowing other assets~\cite{Alameda-compound@CoinDesk}.
This change may have been beneficial to Alameda Research as they were one of the biggest WBTC minters and held highly leveraged positions (i.e., borrowed money to invest even more)~\cite{Alameda-wbtc@Forbes,Alameda-wbtc@CoinDesk}.
To alleviate these concerns, centralized exchanges typically promise that they will not use their users' tokens to vote on their behalf~\cite{Binance-denies@coindesk}.
While there is no guarantee that they will keep their promise, we can monitor their public wallet addresses to check if the exchange has delegated these governance tokens to another address, or whether they used the tokens for voting while they were stored on that exchange.

Governance protocols intend to eliminate (or at least minimize) centralized decision-making in blockchains.
Their effectiveness in achieving that goal can, however, be compromised depending on how the tokens (i.e., voting power) are distributed.
This thesis evaluates whether governance protocols uphold their promise of decentralized governance of smart contracts, and, if they do not, investigates exactly how they renege on that promise.

%

\section{Compound's governance}\label{sec:compound}
%

\begin{figure*}[t]
	\centering
		\includegraphics[width={\onecolgrid}]{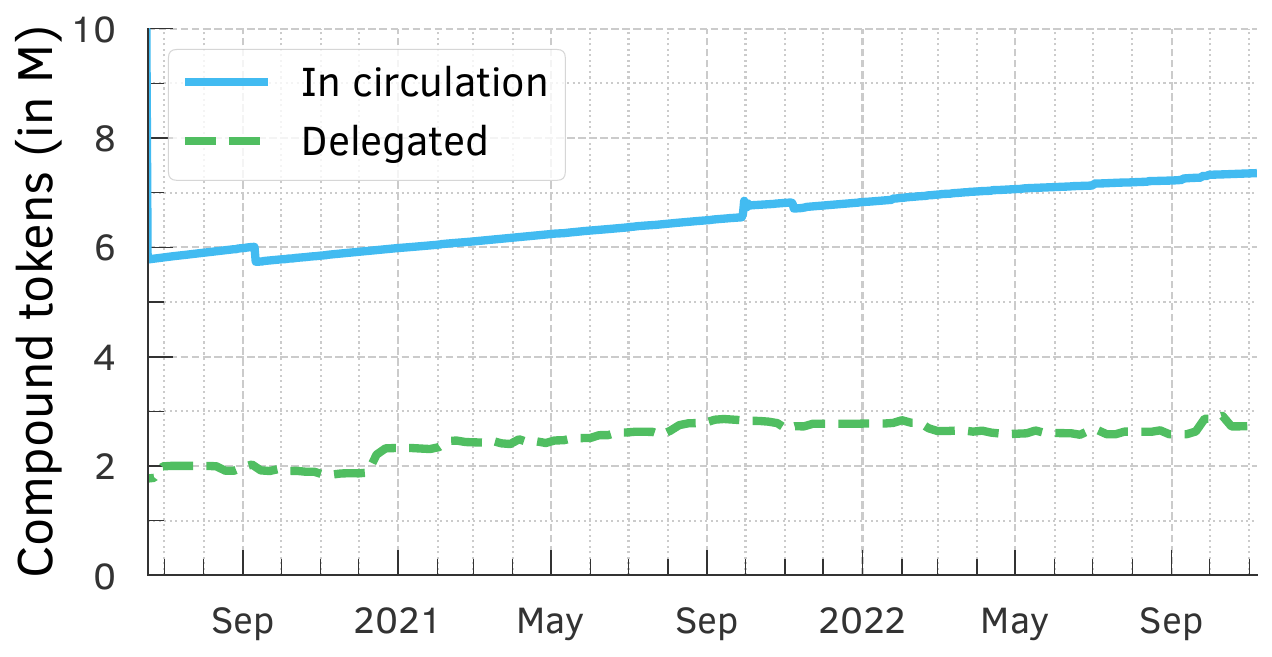}
	\caption{
  Amount of COMP tokens (in millions) in circulation and delegated overtime. Compound tokens have been released to the public since June 15, 2020.
	}
\label{fig:compound-tokens-in-circulation-overtime}
\end{figure*}

Compound~\cite{leshner2019compound} is a decentralized lending protocol that allows users to lend and borrow tokens or assets via smart contracts.
Lenders earn interest (\stress{yield}) by supplying liquidity to the protocol, while borrowers obtain tokens from the protocol and pay interest on the borrowed tokens.

Compound protocol has two versions of its governance contract: \stress{Alpha} and \stress{Bravo}.
\stress{Compound Governor Alpha}, the first version of the governance contract, was deployed on March 4, 2020 (block number \num{9601459}) and was active until March 28, 2021 (block number \num{12126254}).\footnote{The Compound Governor Alpha was deployed at the Ethereum smart contract address \href{https://etherscan.io/address/0xc0dA01a04C3f3E0be433606045bB7017A7323E38\#code}{0xc0dA01a04C3f3E0be433606045bB7017A7323E38}.}
The improved version, \stress{Compound Governor Bravo}, was deployed on March 9, 2021 (block number \num{12006099}) and has been active since April 14, 2021 (block number \num{12235671}).\footnote{The Compound Governor Bravo was deployed at the Ethereum smart contract address \href{https://etherscan.io/address/0xc0Da02939E1441F497fd74F78cE7Decb17B66529\#code}{0xc0Da02939E1441F497fd74F78cE7Decb17B66529}.}
Brave introduced several improvements such as smart-contract upgradability (through proxies), a new option for voters to abstain from voting, and the ability for voters to state the reasons behind their voting choices through text comments attached to on-chain votes.
The Bravo contract was proposed in proposal \#\num{42}, and it received \num{1438679.86} votes from \num{59} voters---all but one vote were in favor of its implementation~\cite{Proposal-42@Compound}.

\begin{figure*}[t]
	\centering
		\includegraphics[width={\onecolgrid}]{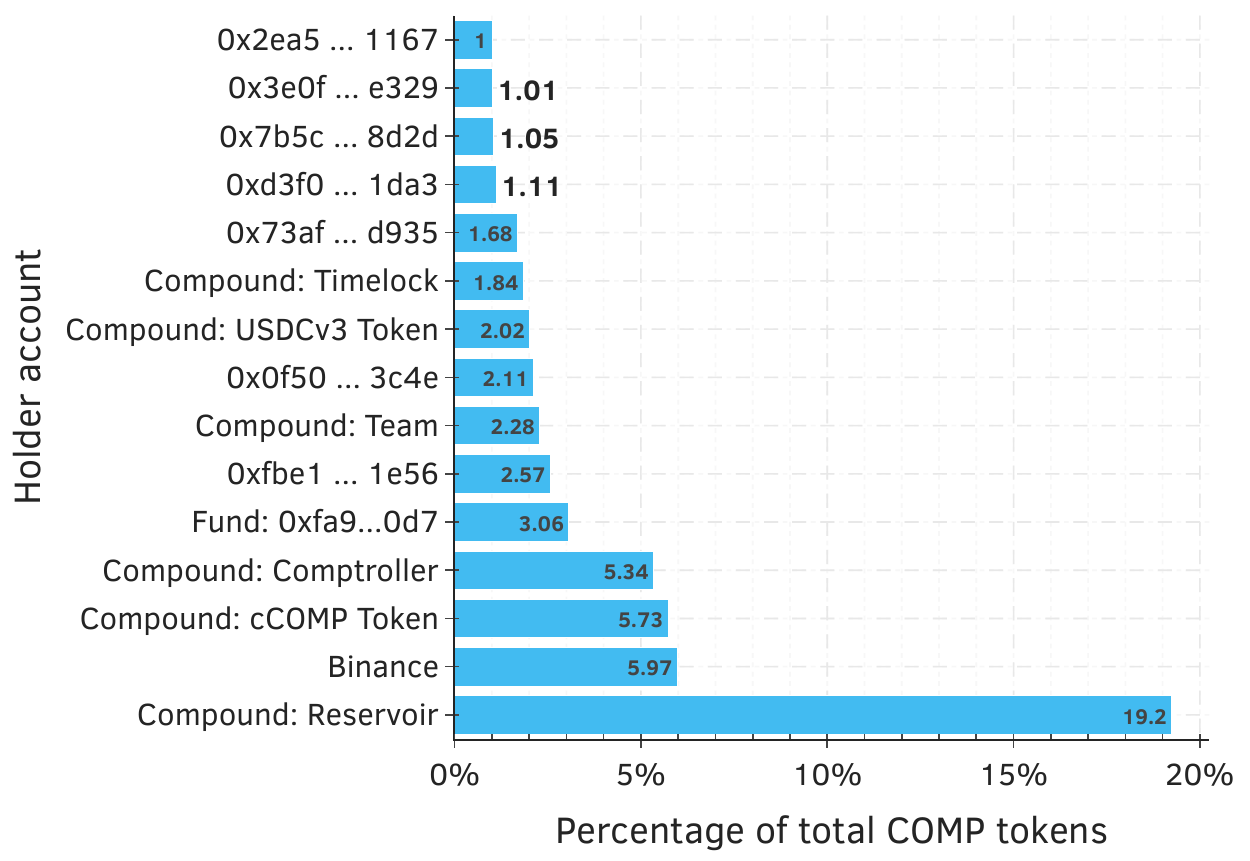}
	\caption{
  Distribution of the top 15 COMP tokens holders. Together, these accounts hold 56.02\% (5.6 million) out of 10 million COMP tokens.
	}
\label{fig:compound-hold-tokens-dist}
\end{figure*}

\subsection{Control of governance tokens}

The voting power of a user in Compound is proportional to the amount of (delegated) tokens held by that user---one token equals one vote.
Below, we examine how these tokens are distributed over time among Compound participants.

\subsubsection{Distribution of token holding}
%

\begin{figure*}[t]
	\centering
		\includegraphics[width={1.1\onecolgrid}]{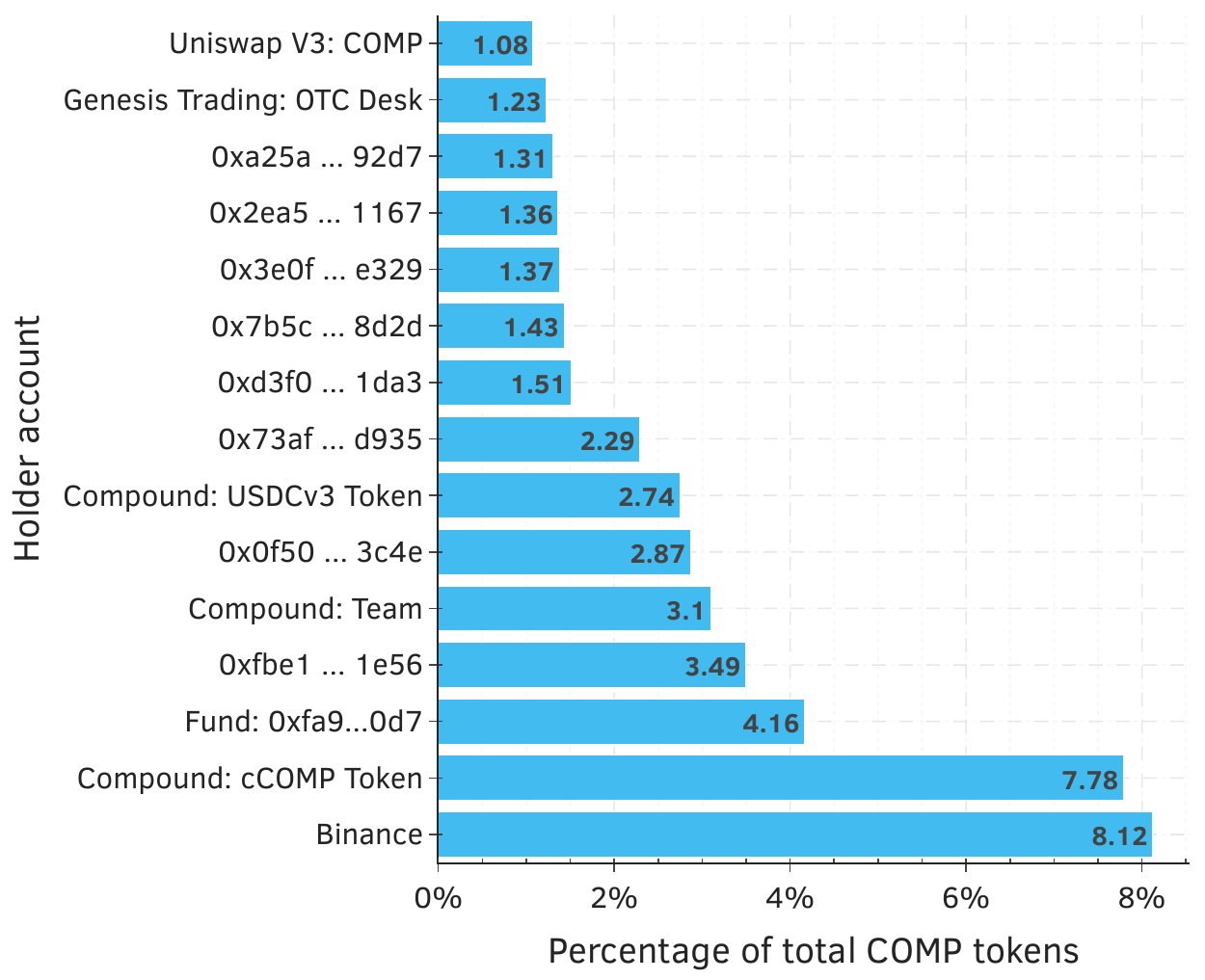}
	\caption{
  Distribution of the top 15 COMP tokens holders (in circulation). These accounts hold 43.83\% (3.2 million) out of 7.3 million COMP tokens in circulation.
	}
\label{fig:compound-hold-tokens-in-circulation-dist}
\end{figure*}

Initially, \num{42.15}\% of the total Compound supply (\num{10} million COMP tokens) was allocated to liquidity mining,\footnote{Liquidity mining is a process where users provide liquidity (i.e., tokens) to a protocol in exchange for rewards or interest.} \num{23.95}\% to shareholders, \num{22.46}\% to the founders and the Compound team, \num{7.73}\% to the community, and \num{3.71}\% to future team members~\cite{Compound@CoinGecko}.
The public release of COMP tokens started only after proposal \#7 was executed on June 15, 2020~\cite{Proposal-7@Compound}.
This proposal enabled the continued distribution of COMP tokens to the protocol users over time (see Figure~\ref{fig:compound-tokens-in-circulation-overtime}).
At the time of our analysis (November 7, 2022), the \num{10} million COMP tokens were distributed among \num{210573} accounts.
The largest holder is \stress{Compound Reservoir} with \num{19.24}\% (\num{1924344.52}) of the tokens followed by Binance (\num{5.97}\% or \num{397289.78} tokens) and cComp (\num{5.73}\% or \num{572723.77} tokens) as shown in Figure~\ref{fig:compound-hold-tokens-dist}.
The Compound Team holds \num{2.28}\% (\num{228061.62}) and Compound Timelock {\num{1.84}}\% (\num{184258.39}) of the tokens.

Of the total supply, only \num{7.3} million COMP tokens are, however, in circulation (Figure~\ref{fig:compound-tokens-in-circulation-overtime}), and we characterize their distribution among a few top token holders in Figure~\ref{fig:compound-hold-tokens-in-circulation-dist}.
In calculating the tokens in circulation, we only included tokens that can be traded or exchanged between users. 
We excluded \stress{locked} tokens from the Compound Reservoir, Comptroller, and Timelock contracts from our analysis~\cite{kybx86@Compound,Governance@Compound}, which are \stress{not} in circulation.
These locked tokens require a governance proposal to be released, although some of them are released daily through the Comptroller as an incentive for users to use the protocol, by lending or borrowing these tokens.

\begin{figure*}[t]
	\centering
		\includegraphics[width={\onecolgrid}]{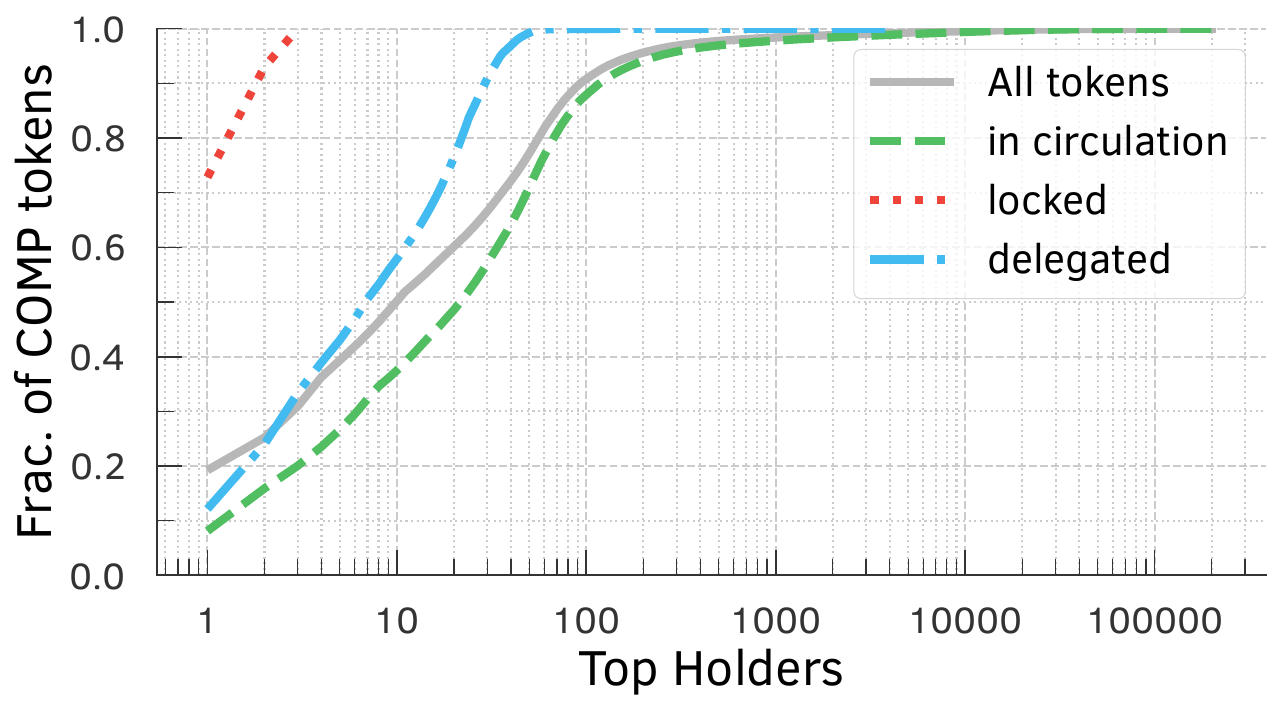}
	\caption{  
  Cumulative distribution of the fraction of COMP tokens held per account. The 10 million tokens available are shared among 210,573 accounts (in \grey{grey}). The dashed \green{green} line shows the distribution of the fraction of 7.3 million (73.57\%) COMP tokens in circulation held by 210,570 accounts. The 2.6 million locked tokens are held by 3 accounts (in dotted \red{red}). Finally, the dash-dotted \blue{blue} line shows the delegated tokens' distribution where 10 out of 4186 accounts have 57.86\% of all delegated COMP tokens available.
	}
\label{fig:compound-hold-tokens}
\end{figure*}

We plot the cumulative distributions of all available COMP tokens along with the locked, delegated, and in-circulation tokens, i.e., the tokens available for users to buy, trade, or sell, in Figure~\ref{fig:compound-hold-tokens}.
The top-15 accounts (in terms of the amount of tokens held) together account for \num{43.83}\% of all tokens in circulation (Figure~\ref{fig:compound-hold-tokens-in-circulation-dist}).
Binance~\cite{Binance@Binance}, a popular centralized cryptocurrency exchange, leads this ranking with \num{8.12}\% of the available tokens.
It is technically feasible for them to delegate these tokens to themselves to vote or propose changes to the protocol (refer~\S\ref{sec:attacks}), but Binance stated that it will not use these tokens to vote on behalf of its users~\cite{Binance-denies@coindesk}.

\stress{Takeaway:
A significant number of tokens were released at the start, and the amount of unlocked tokens continues to increase over time.
A small number of token holders hold the vast majority of all tokens in Compound.}

\subsubsection{Distribution of token delegation}

\begin{figure*}[t]
	\centering
		\includegraphics[width={\onecolgrid}]{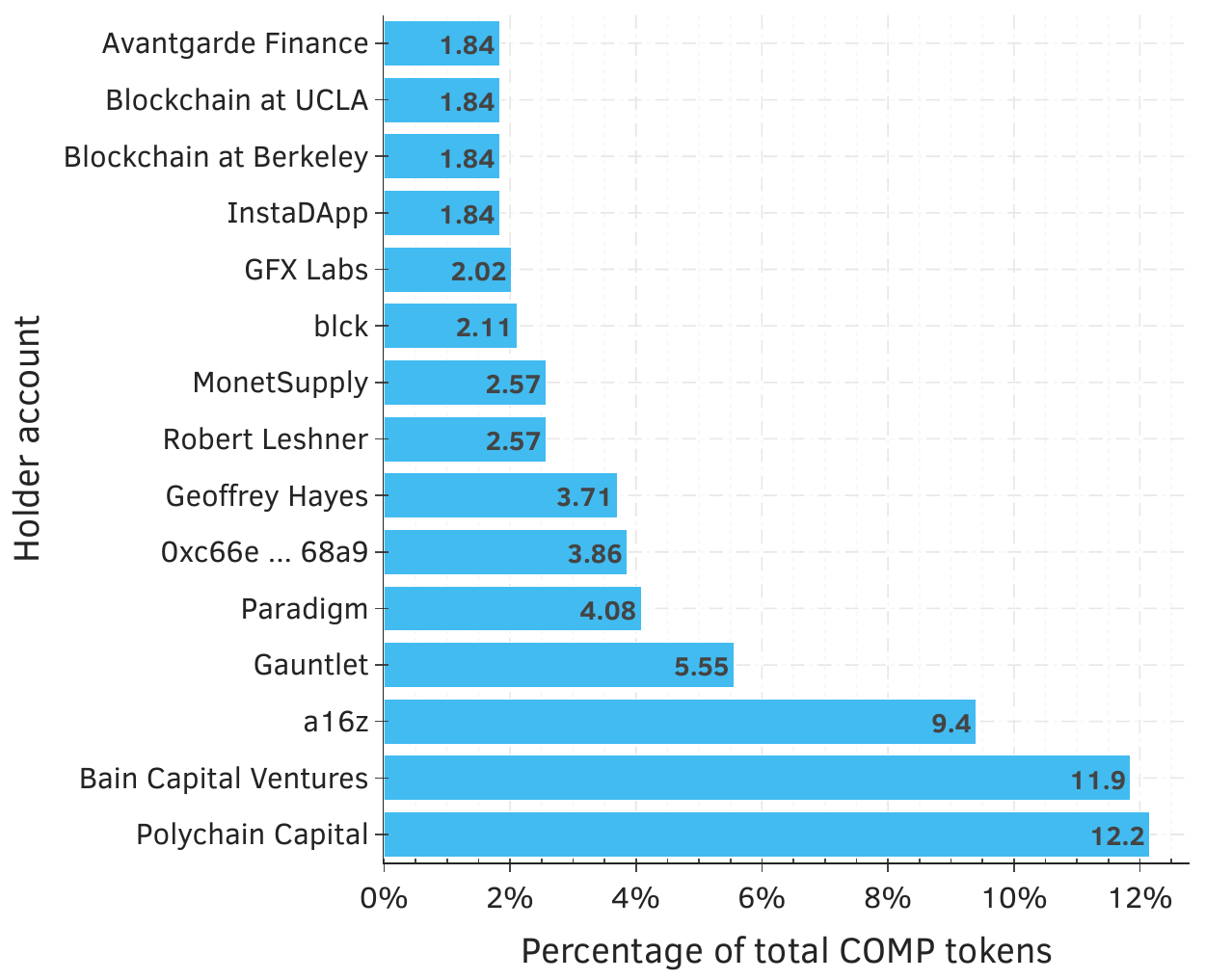}
	\caption{
  Distribution of the top 15 delegated COMP tokens per accounts on November 7, 2022. These addresses have 63.56\% of all 2.7 million delegated tokens.
	}
\label{fig:compound-delegated-tokens-dist}
\end{figure*}

Delegation is a prerequisite for voting (refer~\S\ref{subsec:background-voting-modalities}), and Compound allows its participants to delegate their voting rights to others.
This ability enables users to delegate their voting power to individuals who share their interests, and allows participants with less voting power to pool their votes together and have a significant voting impact.
Users, however, can only delegate \stress{all}, not a fraction, of their tokens.
The protocol, nevertheless, enforces this limitation at the wallet address level.
Users can own multiple wallet addresses and divide their tokens into them, thereby allowing them to delegate a subset of their tokens to others~\cite{fritsch@2022votingpower,Governance@a16z}.
To determine if delegated tokens are held by a few voters, we group together all inferred addresses (as discussed in~\S\ref{subsec:dataset-inferred-addresses}) that belong to the same entity and then count the total number of delegated tokens held by each group.
We observe, per Figure~\ref{fig:compound-hold-tokens}, that delegated tokens are concentrated among few voters, and we show the distribution of delegated tokens across several top token holder accounts in Figure~\ref{fig:compound-delegated-tokens-dist}.
Out of \num{4186} COMP delegatee accounts (or accounts with voting rights), the top 50 (\num{1.19}\%) hold \num{99.23}\% of all delegated tokens, giving them significant decision-making power when voting on proposals.
On November 7, 2022, Polychain Capital held the most delegated tokens, with \num{12.15}\% (\num{330986.09}) followed by Bain Capital Ventures with \num{11.85}\% (\num{322763.87}) and a16z with \num{9.40}\% (\num{256046.13}).
These three addresses together held \num{33.41}\% (\num{909796.10}) of all the \num{2723123.73} delegated tokens in our analysis.

We note that only approximately half of the tokens in circulation are delegated (Figure~\ref{fig:compound-tokens-in-circulation-overtime}).
If we investigate token delegation among the top token holders in Figure~\ref{fig:compound-hold-tokens-in-circulation-dist}, we observe that many of them are crypto exchanges (e.g., Binance and Uniswap V3:COMP) that do \stress{not} delegate their tokens.
This observation assuages concerns that crypto exchanges that hold their users token could abuse their users' trust (\S\ref{sec:attacks}).
Binance publicly stated that they will not abuse their users' voting rights by voting on behalf of them, and our empirical observations, so far, lend credence to that claim.

\subsection{Voting on governance proposals}

To propose changes to the Compound protocol, an address must have at least \num{25000} COMP tokens delegated to it to create a proposal.\footnote{Prior proposal \num{89}, an address should have at least \num{65000} delegated tokens to create proposals~\cite{Proposal-89@Compound}.}
However, as of September 18, 2021, proposal \#60 introduced an exception to this rule, allowing also whitelisted-addresses to create proposals even if they do not have \num{25000} delegated tokens~\cite{Proposal-60@Compound}.

Per Figure~\ref{fig:compound-life-cycle}, when a proposal is created, there is an approximately 2-day voting delay period (or \num{13140} blocks) that is used to allow the community to discuss the proposal before the voting period begins.
During the approximately 3-day voting period (or \num{19710} blocks),\footnote{The duration of the voting period is determined by the number of blocks added to the Ethereum blockchain (specifically, \num{19710} blocks). The actual length of the voting period may be slightly longer than \num{3} days.} voters can cast their votes. 
In order for a proposal to be executed, it needs to meet two requirements.
Firstly, it must receive a minimum of \num{400000} votes in favor of the proposal.
This number corresponds to \num{4}\% of the total supply and is known as the \stress{quorum}.
Secondly, the majority of the votes cast must be in favor of the proposal.
The number of votes each voter has is determined by the number of delegated tokens they held in the block before the voting period began.
This prevents voters from changing their delegated tokens after the voting has begun, which could potentially lead to sudden changes in the outcome of the election.
After a proposal is approved, it is placed in the \stress{TimeLock} for a minimum period of \num{2} days before it can be implemented (or executed)~\cite{Governance@Compound}.
A proposal can be cancelled at any time by the proposer prior to its execution, or by anyone if the proposer fails to maintain at least \num{25000} delegated tokens.

In total, \num{3335} voters cast their votes through \num{9500} transactions with \num{8769} (\num{92.31}\%) for in-favor votes, \num{644} (\num{6.78}\%) for against votes, and finally \num{87} (\num{0.91}\%) for abstained votes.
The majority of voters (\num{51.36}\%) only voted for \num{1} proposal. \num{1}\% of participants voted for at least \num{26.66} proposals.
On average, participants voted on \num{2.85} proposals with a \stress{standard deviation (std.)} of \num{5.23}.
The address \stress{0x84e3$\cdots$5a95} voted on the maximum number of proposals (\num{100}), followed by \stress{MonetSupply} and \stress{blck} who voted on \num{96} and \num{88} proposals, respectively.

\subsubsection{Creation of proposals}\label{subsec:creating_proposals}

In total, \num{33} proposers created the 133 proposals.
Of these proposers, \num{16} (\num{48.48}\%) created one proposal, while \num{10}\% of the proposers created at least \num{8} proposals.
The average number of proposals created per proposer is \num{4.03} proposals, with a std. of \num{5.27} and a median of \num{2}.
The highest number of proposals was created by \stress{Gauntlet}, who created \num{24} proposals, followed by \stress{blck}, who created \num{20} proposals.

\begin{figure*}[t]
	\centering
		\includegraphics[width={\textwidth}]{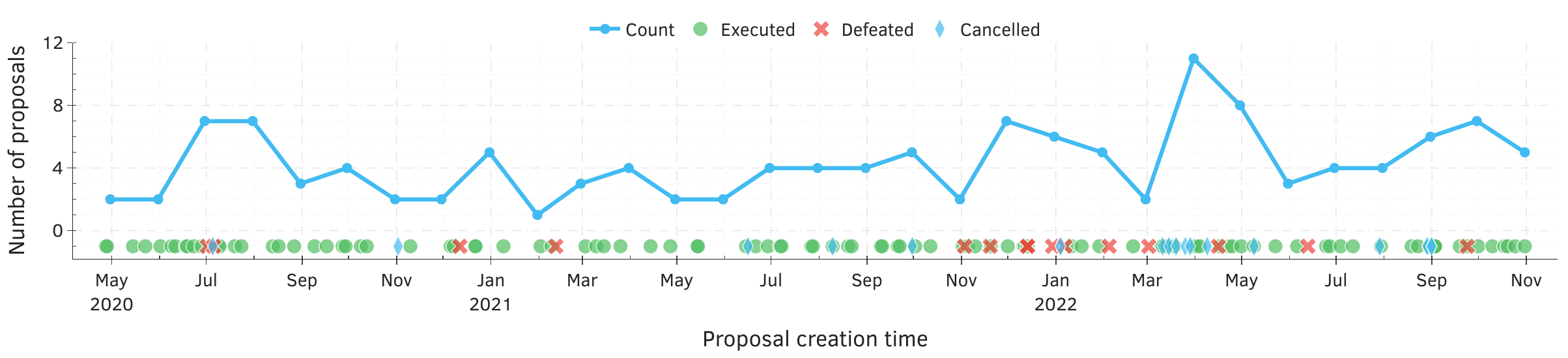}
	\caption{
  Monthly number of Compound proposals created overtime and their respective outcomes (\green{executed}, \red{defeated}, or \blue{cancelled}). Proposals are created, on average, every 6.95 days.
	}
\label{fig:compound-proposals-month}
\end{figure*}

The maximum number of proposals were created in March 2022, \num{11} proposals created (from \#86 to \#96). However, of those, only \num{5} were executed, as \num{1} was defeated and \num{5} were cancelled (see Figure~\ref{fig:compound-proposals-month}).
Proposals were submitted, on average, every \num{6.95} days (std. of \num{6.41}), with a median of \num{5.08} days. 
This may be because the proposal lifecycle lasts \num{7} days, and the voters might not want to consider multiple active proposals at once.
The shortest and longest interval between proposals was \num{0} and \num{31.14} days, respectively.
Additionally, proposals typically take \num{1.64} days (std. of \num{0.72} days) to reach the quorum, as depicted in Figure~\ref{fig:compound-time-to-quorum} in \S\ref{sec:compound-quorum}.

\stress{Takeaway: Compound is actively and regularly used: It received a constant stream of proposals over the course of our study period.}

\subsubsection{Participation in voting}\label{subsec:voting-participation}

\begin{figure*}[t]
	\centering
		\includegraphics[width={\textwidth}]{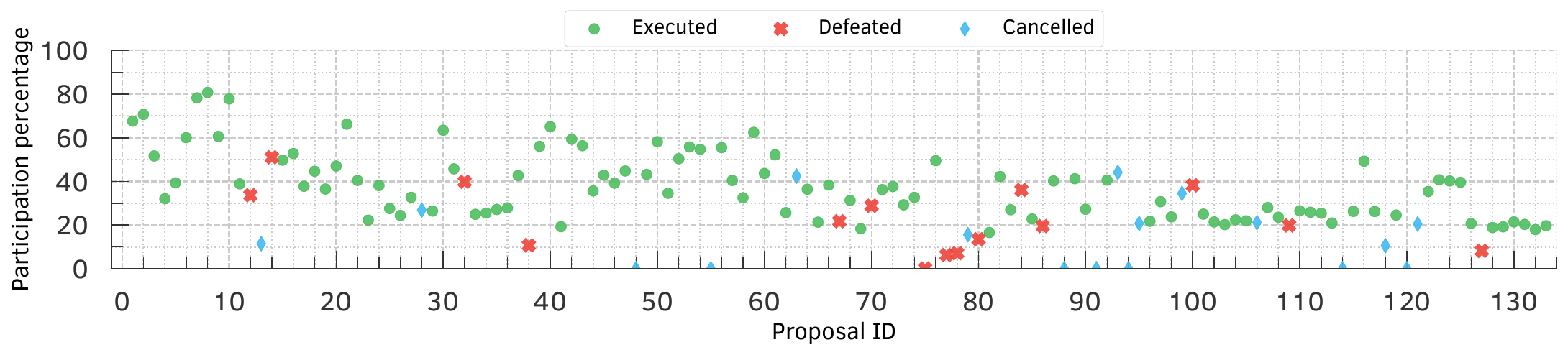}
	\caption{
  Compound's voting participation per proposal in terms of delegated tokens used from all delegated tokens available. Proposals are indicated either as executed (in \green{green}), defeated (in \red{red}), or cancelled (in \blue{blue}).
	}
\label{fig:comp-voting-participation}
\end{figure*}

Next, we computed the voting participation per proposal (see Figure~\ref{fig:comp-voting-participation}).
This metric is calculated by dividing the number of votes (or delegated tokens) cast on a proposal by the total number of delegated tokens eligible to vote on that proposal at the start of the voting period.
This is a crucial measurement as it shows the proportion of all delegated tokens that are used in the governance election process by the voters on proposals.
Also, protocols with low voter turnout are more susceptible to vote-buying, as non-voting users may sell their voting rights to others~\cite{Daian-dark-dao@HackingDistributed}.
Our results show that the average Compound voter turnout is \num{33.25}\% (with a std. of \num{17.61}\%), the median is \num{32.10}\%, and the maximum \num{80.80}\%.
Based on Figure~\ref{fig:comp-voting-participation}, we observe higher voting participation for early proposals compared to recent ones, likely due to the limited availability of tokens to a select few in the beginning.

On average, the 133 proposals had \num{71.43} voters participating in their election, with a standard deviation of \num{98.97} voters.
\num{50}\% of the proposals received votes from \num{38} voters, while the numbers of voters varied between \num{0} (when proposals are cancelled before the voting period begins) and a maximum of \num{619}, as seen in proposal \#111.
This particular proposal received a total of \num{686289.04} votes from \num{615} voters in favor, \num{3} against, and \num{1} abstention.
The next proposals with higher number of voters are proposals \#115 and \#105 that received votes from 579 and 404 voters, respectively.
%

\begin{figure*}[t]
	\centering
		\includegraphics[width={\textwidth}]{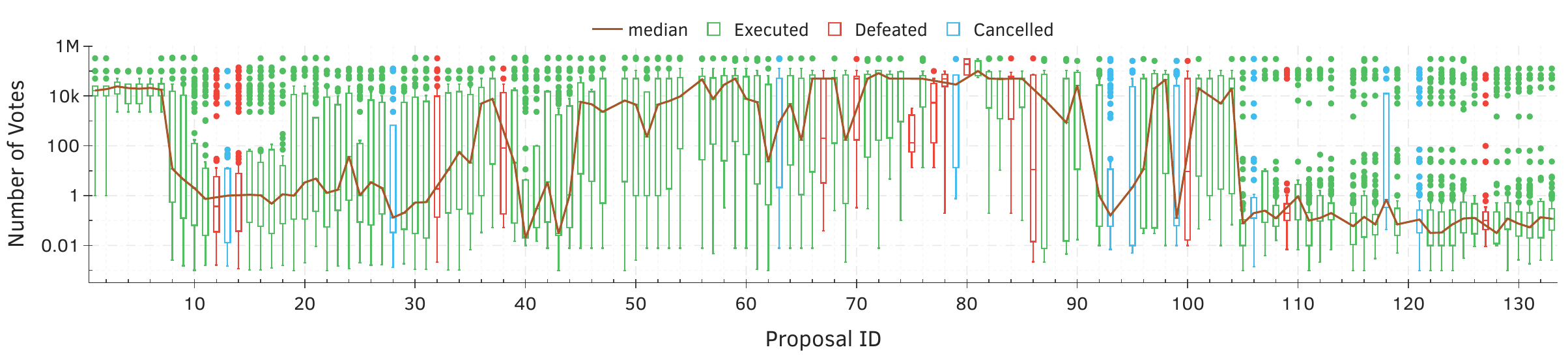}
	\caption{
  Compound's distribution of voting power by voter per proposal. For better illustration, we consider a cutoff of 0.001 votes.
	}
\label{fig:comp-voting-distribution-per-proposal}
\end{figure*}

Each time a voter casts a vote in the Compound governance protocol by issuing a transaction, an event is triggered, as described in \S\ref{subsec:smart-contract-events}.  
We analyzed \num{9500} transactions with events triggered by voters during the voting process.
Of these events, \num{1732} (\num{18.23}\%) were votes cast by voters who did not have any delegated tokens available, resulting in zero voting power or \stress{useless vote}.
Although this is allowed by the protocol, it does not count for or against a proposal.
However, it shows support for the proposal, as these voters still participate in the election despite not having any delegated tokens available.
Therefore, the average number of votes cast (or tokens used to vote) was \num{10961.73}, with a std. of \num{39212.17} and a median of \num{0.1}. 
The range of votes cast was from \num{0} to \num{345067.49} as shown in Figure~\ref{fig:comp-voting-distribution-per-proposal}.
This indicates that most of the voters are small players (or accounts with a low amount of delegated tokens).

\begin{figure*}[t]
	\centering
		\includegraphics[width={\textwidth}]{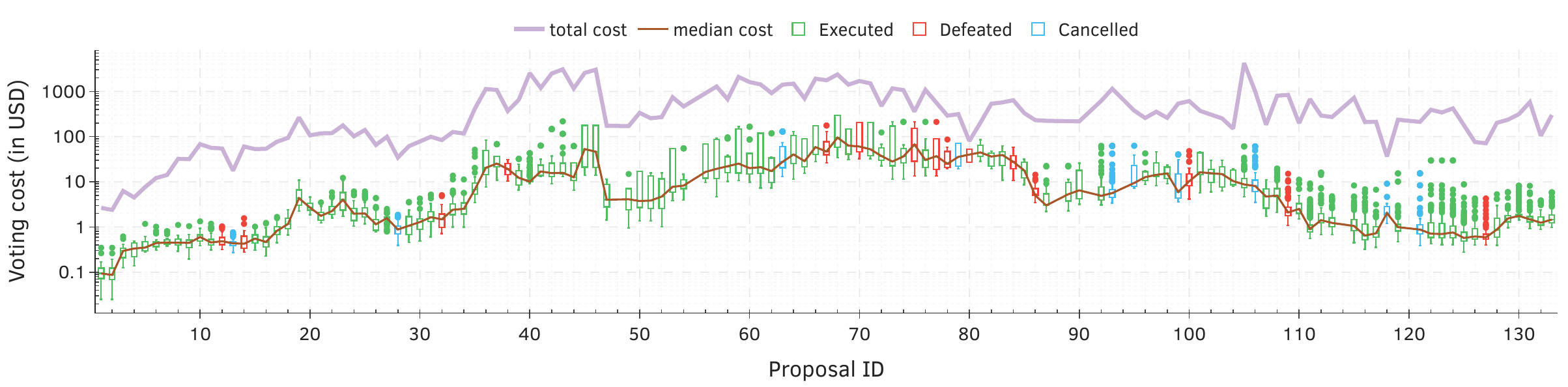}
	\caption{
  Voting cost distribution per proposal. On average, casting a vote costs \$7.88 with a std. of \$22.29.
	}
\label{fig:compound-voting-cost-per-proposal}
\end{figure*}

In addition, when voting in Compound, there is a financial cost involved due to the on-chain transactions required to cast votes.
To determine these costs, we collected the relevant transactions from the Ethereum blockchain and analyzed the fees paid by voters to issue the transactions and cast their votes.
We report the voting cost in US dollars, using the ETH-USD Yahoo Finance data feed~\cite{eth-usd@yahoo} to compute the exchange rate at the time the transaction was included in a block.
In total, voting for the \num{133} Compound proposals, voters paid \$\num{74865.74}.
The average voting cost per proposal is \$\num{7.88} with a std. of \$\num{22.29}.
The median voting cost is \$\num{1.48} with a range from \$\num{0.03} to \$\num{294.02}.
Figure~\ref{fig:compound-voting-cost-per-proposal} shows the voting cost distribution per proposal.
We also computed these metrics at proposal level, on average, each proposal costed \$\num{594.17} with a std. of \$\num{745.62} and a median of \$\num{291.92}.
The cost ranges from \$\num{2.39} to \$\num{4247.25}.

\begin{figure*}[tb]
	\centering
		\includegraphics[width={\textwidth}]{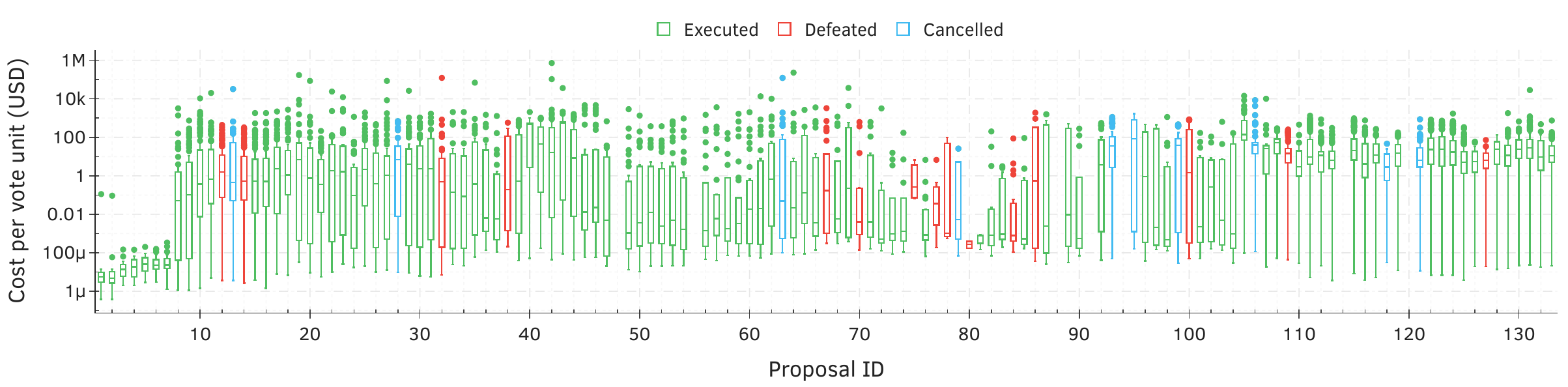}
	\caption{
  Voting cost distribution \stress{normalized} per the voting power. We consider a cutoff of $10^{-6}$ votes for better illustration.
	}
\label{fig:compound-voting-cost-per-proposal-normalized}
\end{figure*}

Voting on proposals can, hence, present a significant cost barrier, especially for voters with relatively few tokens.
In such cases, the cost per token vote (or vote unit) may be too high compared to those with a higher number of delegated tokens.
To better understand this, we normalized the cost of casting a vote by the number of votes cast (measured by the total number of delegated tokens available to voters' addresses).
For this analysis, we focused on voters who cast at least $10^{-6}$ votes in any proposal.
As shown in Figure~\ref{fig:compound-voting-cost-per-proposal-normalized}, some voters faced prohibitively high costs per vote unit.
For example, the cost per vote unit has a mean of \$\num{358.54} and a std. of \$\num{9334.73}, indicating a highly skewed distribution.
However, half of the voters faced a cost per vote unit of only \$\num{6.69}.
The cost per vote unit ranged from \$$3.79\times10^{-7}$ to \$\num{725248.10}.

Additionally, we analyzed the number of voters required for all \num{101} (\num{75.94}\%) executed proposals in our data set to reach the quorum and pass.
Our results show that, for \num{99} proposals, the average number of voters required for a proposal to reach the quorum and pass was \num{3.25}, with a std. of \num{1.65}.
The median number of voters required was \num{2}, and the range of voters required varied from \num{2} to \num{8}.
This sheds light on how centralized these delegated tokens are distributed among a few participants, where for half of the proposals only \num{2} voter casting their votes would be enough to pass (or execute) a proposal.

Furthermore, we analyzed the number of voters needed for proposals to reach \num{50}\% of the total votes cast.
Out of \num{133} proposals, we excluded \num{7} proposals that were cancelled before the voting period, leaving us with \num{126} (\num{94.74}\%) proposals for analysis.
On average, those proposals required \num{2.84} voters with a std. of \num{0.97} and a median of \num{3} voters.
The minimum and maximum number of voters were \num{1} and \num{5}, respectively.
This again suggests that the token distribution is concentrated among few voters who hold a high voting power.
We present the cumulative voting power for the top 10 most powerful voters for each of these \num{126} proposals in our data set in \S\ref{sec:top-voters}.

\subsubsection{Margin of victory/defeat}\label{subsec:margin_victory_defeat}

\begin{figure*}[tb]
	\centering
		\includegraphics[width={\onecolgrid}]{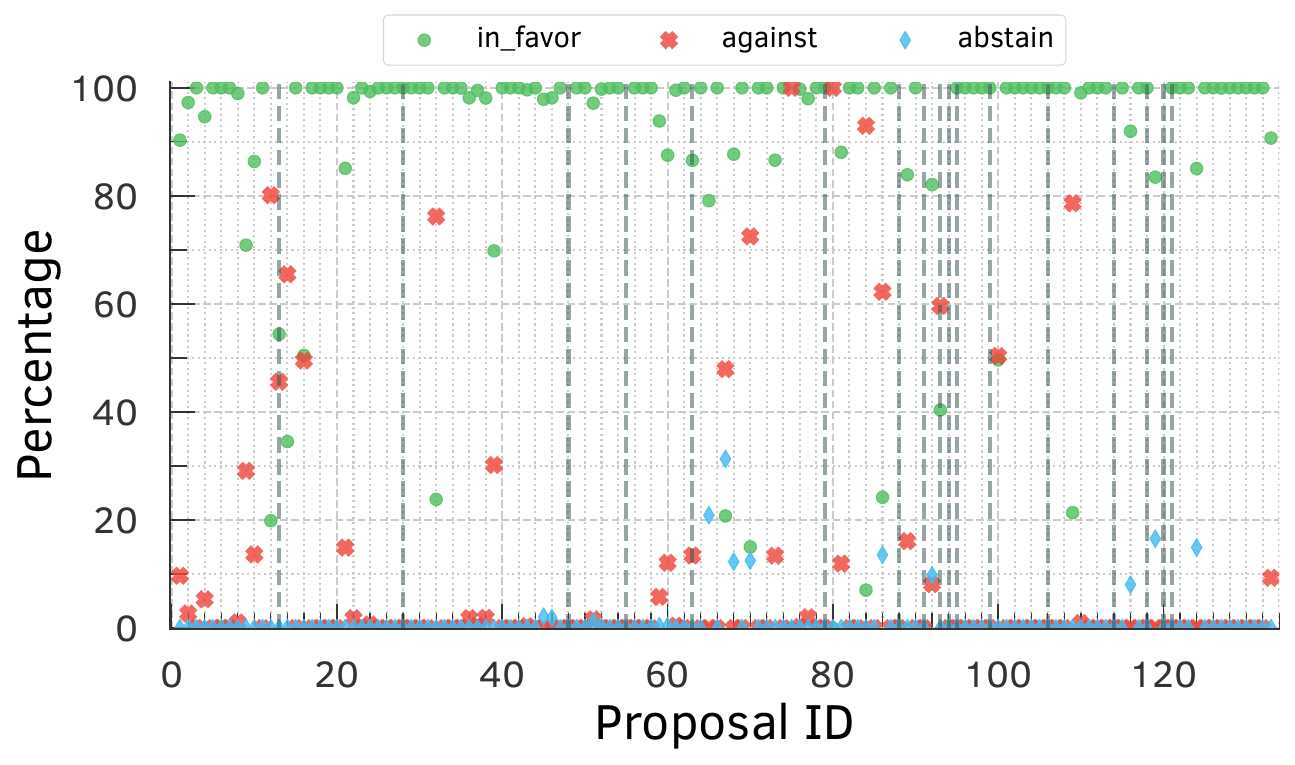}
	\caption{
  Percentage of in-favor (in \green{green}), against (in \red{red}), and abstain (in \blue{blue}) votes for each proposal. A total of 15 (11.28\%) proposals were defeated, and vertical lines represent  17 (12.78\%) cancelled proposals.
	}
\label{fig:compound-votes-proposal-percentage}
\end{figure*}

\begin{figure*}[tb]
	\centering
		\includegraphics[width={\onecolgrid}]{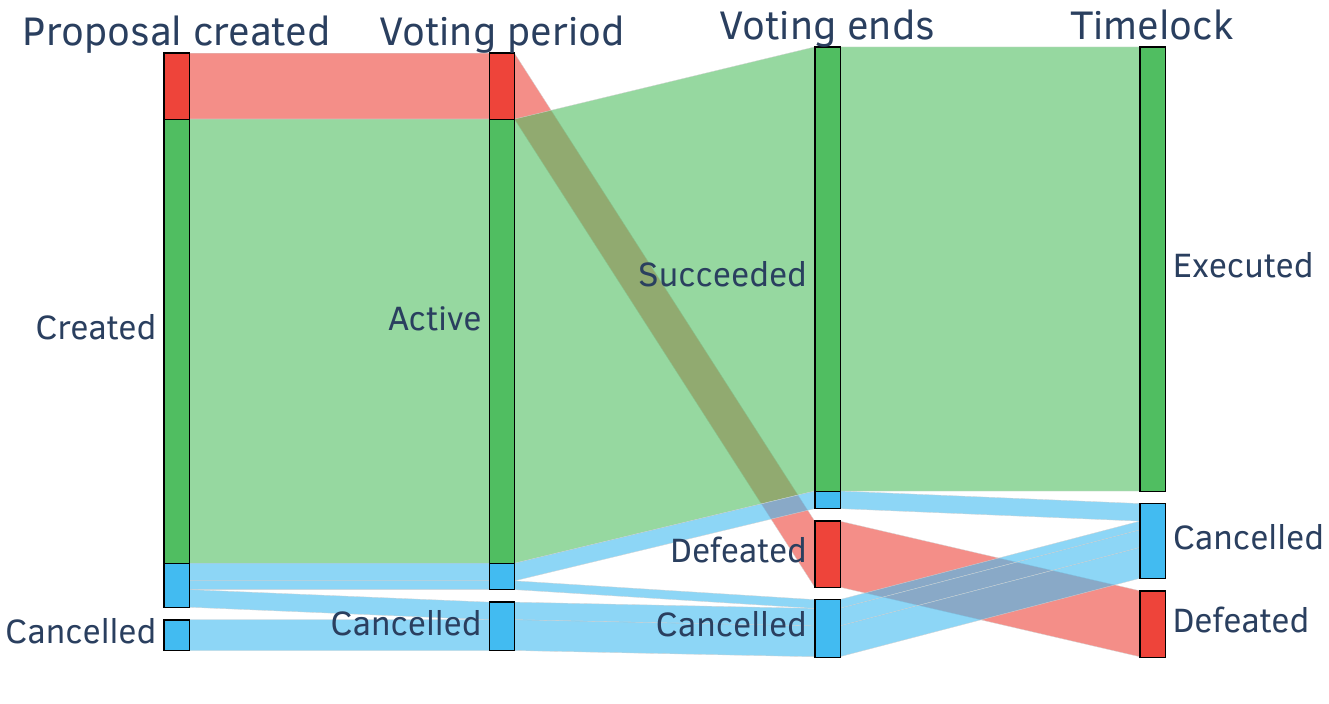}
	\caption{
  Summary of the outcome of 133 Compound proposals at each stage of their lifecycle. There are 101 proposals executed (in \green{green}), 15 defeated (in \red{red}), and 17 cancelled (in \blue{blue}).
	}
\label{fig:compound-proposal-life-cycle}
\end{figure*}

During the analyzed period from March 3, 2020 (block number \num{9600000}) to November 7, 2022 (block number \num{15917000}), \num{133} proposals have been created.
Of these, \num{17} (\num{12.78}\%) were cancelled and \num{15} (\num{11.28}\%) defeated, leaving \num{101} (\num{75.94}\%) executed proposals.
Figure~\ref{fig:compound-votes-proposal-percentage} shows the percentage of in-favor, against, and abstain votes for each proposal. The majority of the proposals received significant support from the voters.
On average, proposals received \num{89.39}\% of the votes in favor, with a std. of \num{23.98}\% votes and a median of \num{99.99}\%.
We highlight the proposals' outcome at each stage of their lifecycle in Figure~\ref{fig:compound-proposal-life-cycle}.
Our analyses show that \num{7} (\num{5.26}\%) out of \num{133} proposals were cancelled right after they were created and, therefore, they had not reached the \stress{Voting Period} meaning they were not available for voting.
Next, 4 proposals were cancelled before the \stress{Voting Ends} stage, meaning they were pulled out before the election finished. 
\num{2} were also cancelled after they succeeded in the election (after the \stress{Voting Ends} stage) but before they were queued in the \stress{Timelock}.
Further, \num{4} proposals were cancelled when in the \stress{Timelock}. These proposals account for \num{6} cancelled proposals after they successfully passed, which could indicate a lack of community consensus~\cite{sharma2023unpacking}. Finally, \num{101} (\num{75.94})\% proposals were successfully executed.
We gathered data from Messari~\cite{Compound@Messari} to categorize these executed proposals and report their importance level in ~\S\ref{sec:proposal_category}.

\subsubsection{Temporal dynamics of voting}

Compound Governor does not allow voters to change their votes once they have been cast.
This means that voters can only vote once on each proposal. Nevertheless, voters can view all votes that have been cast on-chain in real-time.
Thus, understanding how long it takes voters to cast their votes is interesting because it can shed light on whether they want to wait until the last minute to cast their votes.

\begin{figure*}[t]
	\centering
		\includegraphics[width={\textwidth}]{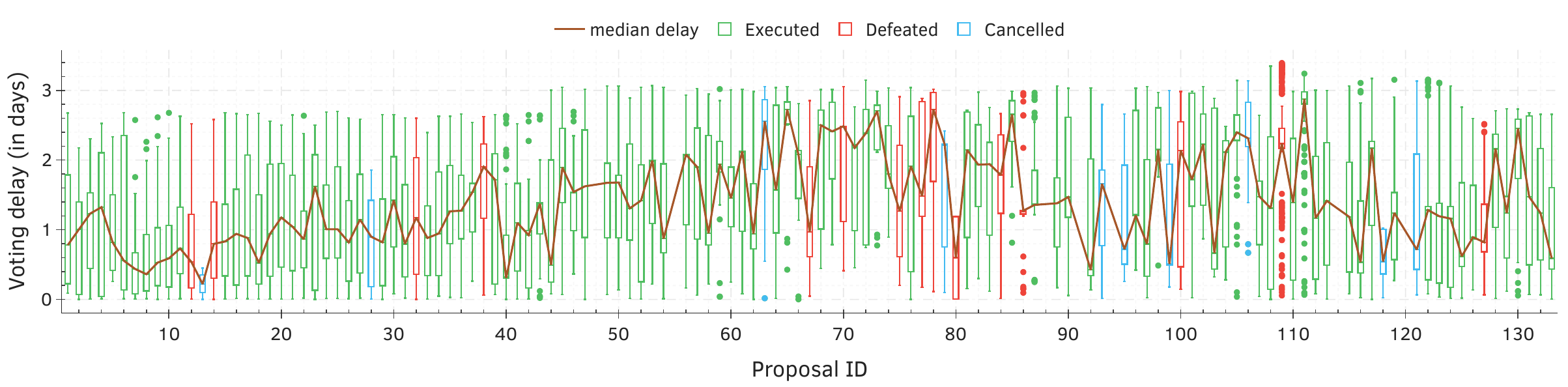}
	\caption{
  Distribution of the number of days it takes voters to cast their votes.
	}
\label{fig:compound-voting-delay}
\end{figure*}

According to our analysis, voters take an average of \num{1.4} days (with a std. of \num{0.95} and a median of \num{1.34} days) to cast their votes after the voting period began.
The shortest and longest recorded delays in our data set are \num{0} and \num{3.39} days, respectively.
Figure~\ref{fig:compound-voting-delay} shows the distribution of the time it takes voters to cast their votes for each proposal.
We also highlight the voting delays for all votes cast per proposal in \S\ref{sec:top-voters}.

\begin{figure*}[t]
	\centering
		\subfloat[All proposals\label{fig:compound-voting-delay-all-proposals-cdf}]{\includegraphics[width=\twocolgrid]{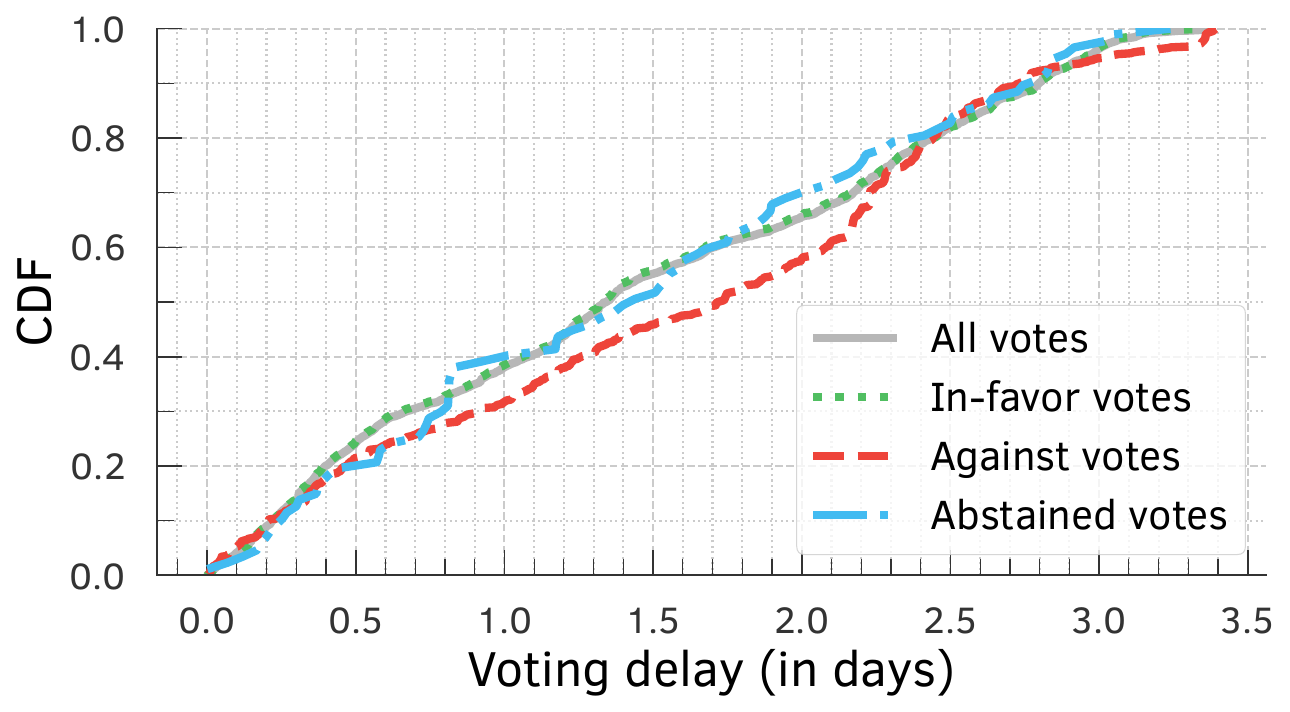}}
		\subfloat[\green{Executed} proposals\label{fig:compound-voting-delay-executed-cdf}]{\includegraphics[width=\twocolgrid]{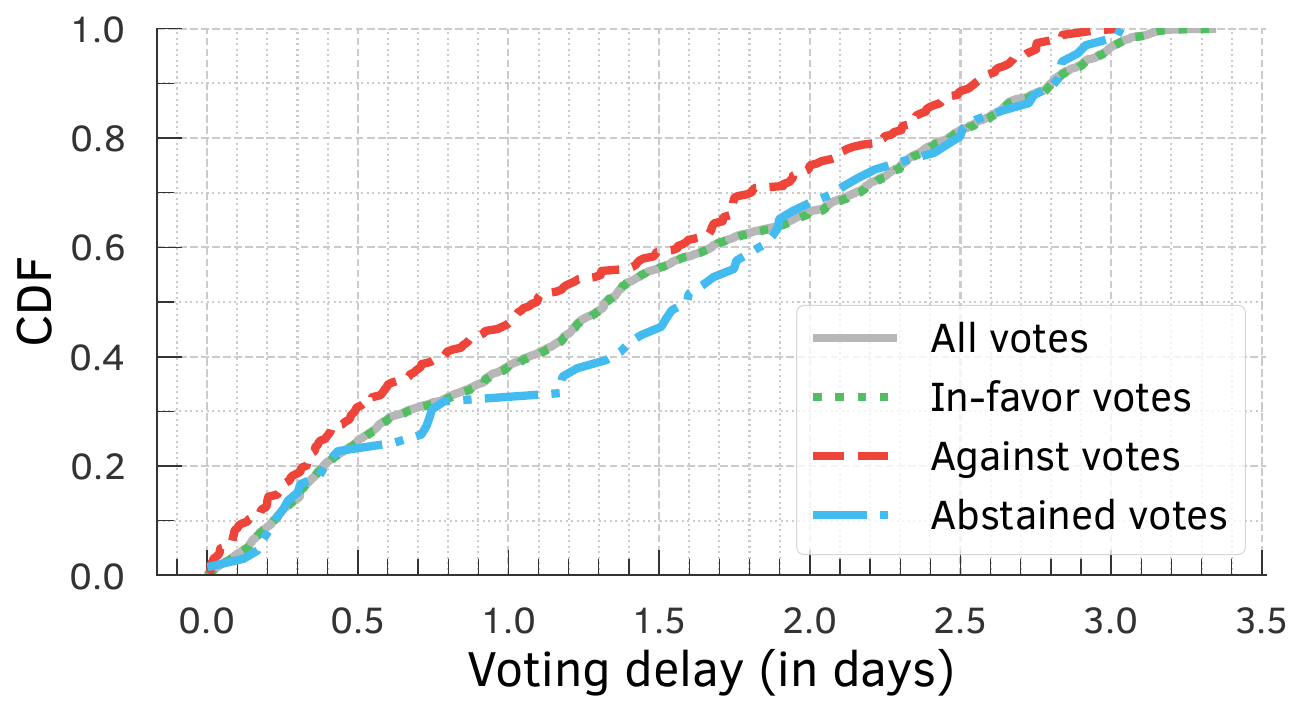}}
        \\
        \subfloat[\red{Defeated} proposals\label{fig:compound-voting-delay-defeated-cdf}]{\includegraphics[width=\twocolgrid]{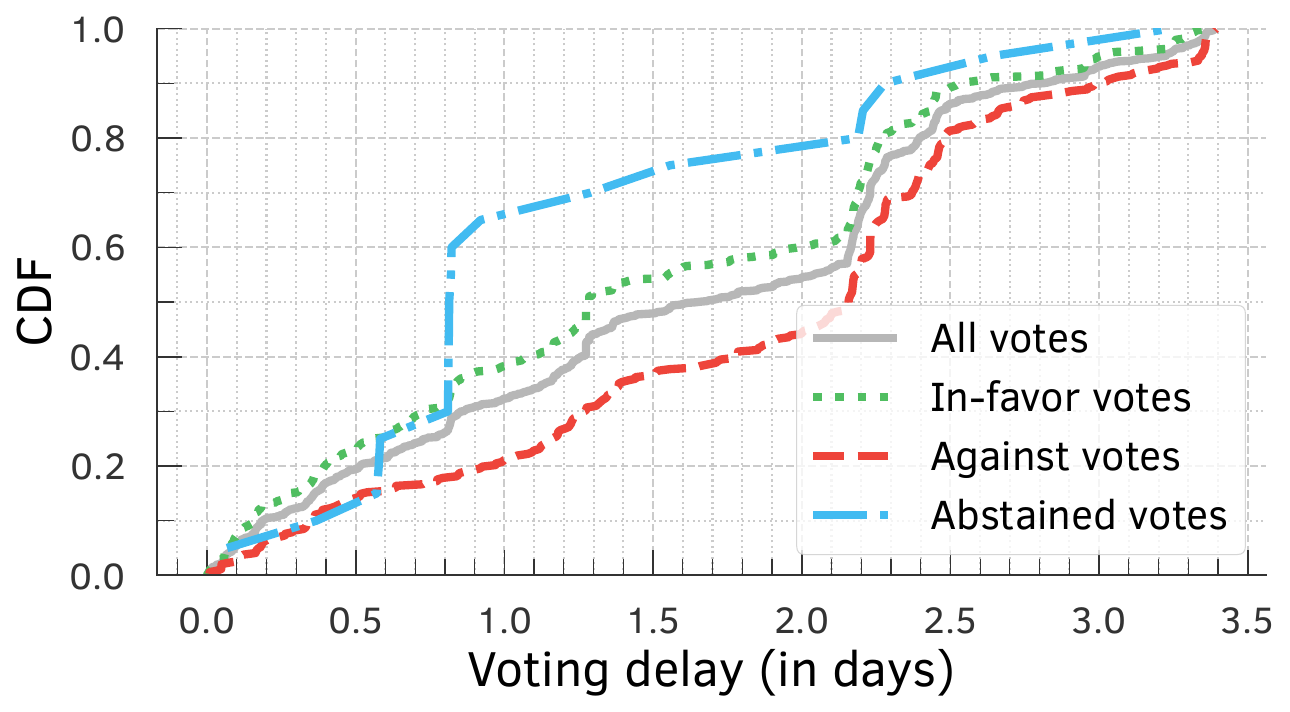}}
        \subfloat[\blue{Cancelled} proposals\label{fig:compound-voting-delay-cancelled-cdf}]{\includegraphics[width=\twocolgrid]{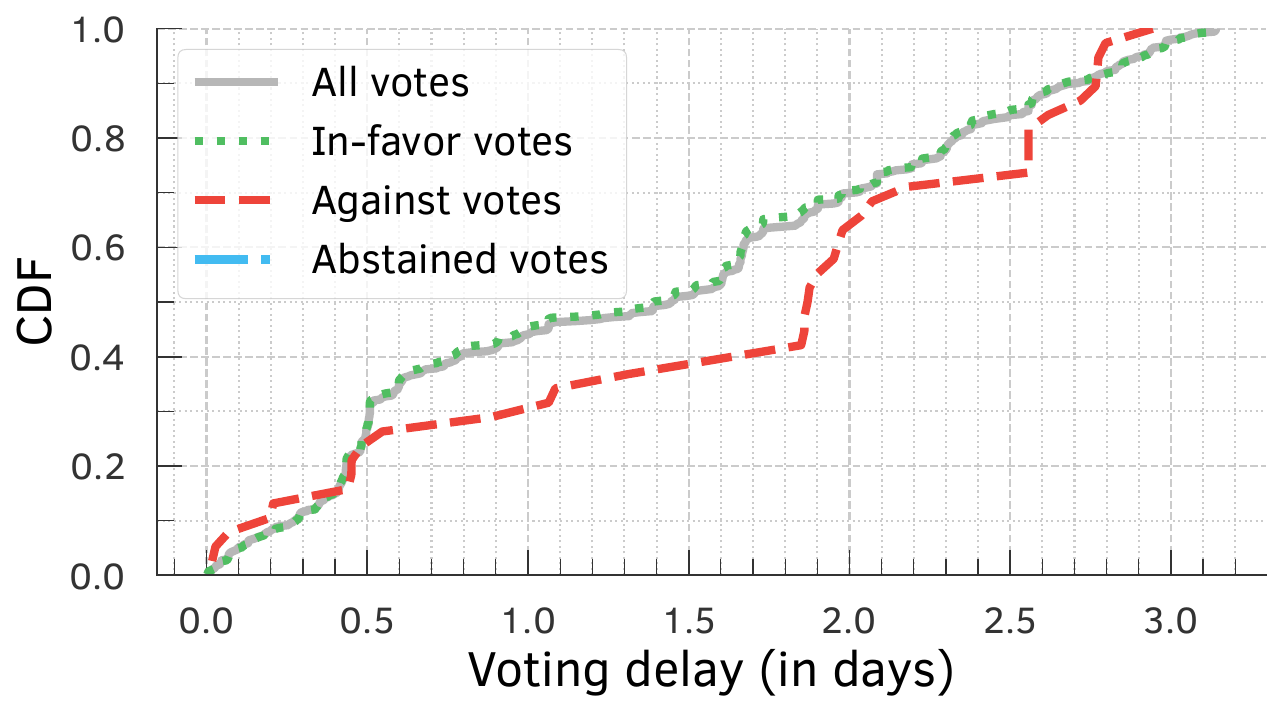}}
	\caption{
  Cumulative distribution function of the time it takes voters to cast their votes since the voting period began considering: (a) All proposals; (b) \green{Executed} proposals; (c) \red{Defeated} proposals; and (d) \blue{Cancelled} proposals.
	}
  \label{fig:compound-voting-delay-cdf}
\end{figure*}

When examining voting delay behavior, voters typically take longer to cast votes against proposals (\num{1.58} days on average) in comparison to all other votes (see Figure~\ref{fig:compound-voting-delay-all-proposals-cdf}).
Considering only executed proposals, voters take longer to abstain but are faster to vote against executed proposal (Figure~\ref{fig:compound-voting-delay-executed-cdf}).
For defeated proposals, on the contrary, they abstain faster and take longer to vote against defeated proposal (Figure~\ref{fig:compound-voting-delay-defeated-cdf}).
Even for cancelled proposals, Figure~\ref{fig:compound-voting-delay-cancelled-cdf} shows that voters take longer to vote against these proposals.
We believe that the executed proposals must have been better discussed prior to the voting period, and therefore voters were more likely to vote for the proposal with high approval rates (Figure~\ref{fig:compound-votes-proposal-percentage}).
Similarly, voters were more likely to vote against proposals that were defeated.

\subsection{Real-world decision-making using Compound governance}\label{subsec:autiting_compound}

Interestingly, Compound has also been utilized for real-world decision-making purposes, such as allocating grants to contributors~\cite{Proposal-40@Compound} or hiring an audit company to review the governance protocol through the Compound code~\cite{Auditing@Compound}.
For instance, on September 29, a bug was introduced in the Comptroller of the Compound Protocol through proposal \#62 that allowed users to claim more COMP tokens than they were entitled to, resulting in a loss of \$50 million worth of COMP tokens~\cite{Bug@Compound,Proposal-62@Compound}.
The Compound community sought to hire, through the Compound governance protocol, a smart contract auditor to audit the protocol~\cite{Auditing@Compound}.
Three companies, ChainSecurity, OpenZeppelin, and Trail of Bits, posted their business plans for discussion and then created proposals via the Compound Governor.
Voters were able to vote for their preferred proposal, and the winning proposal was eventually implemented. The losing proposals would have been cancelled by the community's multi-signature mechanism after the voting period ended, ensuring only one could pass.

OpenZeppelin was the only proposal to reach quorum and get the majority of votes to be implemented.
They audited the Compound code, assisted proposers, participated in community discussions, and reviewed any new proposals formally created by the Compound community~\cite{Compound@OpenZeppelin}. 

\stress{Takeaway: We believe that these governance protocols will be used even more in the future for transparent decision-making in real-world applications like the ones mentioned above.
This will have a positive impact on the use of governance protocols in the everyday life of society.}

\subsection{Voting patterns of delegates}

\begin{figure*}[t]
	\centering
		\includegraphics[width={\textwidth}]{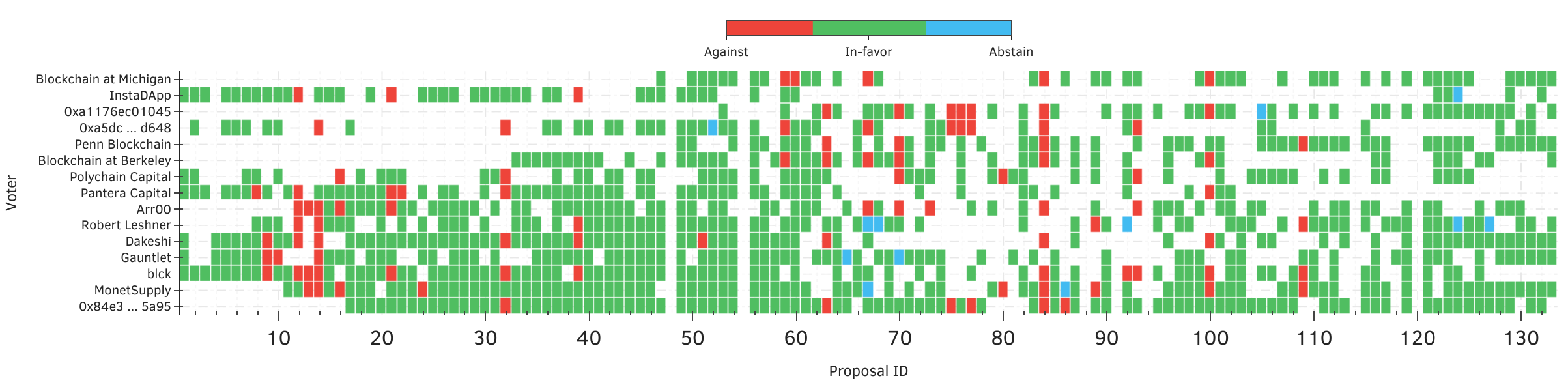}
	\caption{
  Votes cast by the top-15 voters. In-favor votes are in \green{green}, against in \red{red}, and abstain in \blue{blue} color.
	}
\label{fig:comp-votes-top-15-voters}
\end{figure*}

In this section, we analyze the formation of coalitions among voters, where they cast their votes as a group.
This analysis is crucial because such behavior may compromise the security of the governance protocol.
Specifically, instead of expressing their individual opinions on a proposal, voters may choose to mimic the votes of their peers.
The transparency of the Ethereum blockchain used for voting in Compound allows anyone to view the addresses of voters and their corresponding votes (e.g., their voting power and voting preference) during the election process, potentially facilitating this behavior.
As a result, exploring the possibility of coalition formation could provide valuable insights into the decision-making patterns of voters.
Figure~\ref{fig:comp-votes-top-15-voters} shows a heatmap of how each of the top \num{15} voters cast their votes across all \num{133} proposals in our data set.

Further, we use cosine similarity to quantify how similar the voting patterns of different voters are.
Cosine similarity calculates the similarity between two vectors by determining the cosine of the angle between them~\cite{Cosine@ScikitLearn,xia2015learning}.
It is useful in the context of voting because it allows us to compare voting patterns and determine whether and which voters vote for the same proposals.
The cosine similarity value ranges from \num{-1} to \num{1}, with a value of \num{1} indicating a high degree of similarity.

\begin{figure*}[t]
	\centering
		\includegraphics[width={\textwidth}]{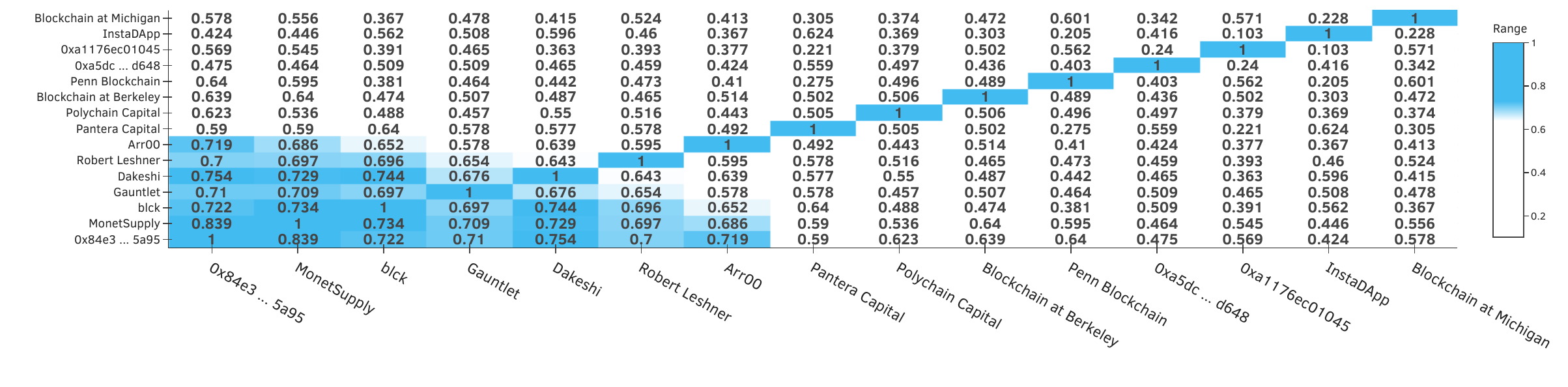}
	\caption{
  Cosine similarity of the top-15 voters voting \green{in-favor} a proposal.
	}
\label{fig:comp-cosine-similarity-top-15-voters-in-favor}
\end{figure*}

Our analysis shows that the top 3 voters (i.e., 0x84e3$\cdots$5a95, MonetSupply, and blck) have a strong cosine similarity in their voting behavior when casting a vote in favor of a proposal, meaning that they cast their votes similarly (see Figure~\ref{fig:comp-cosine-similarity-top-15-voters-in-favor}).
Moreover, Gauntlet, Dakeshi, Robert Leshner, and Arr00 also show a strong similarity with 0x84e3$\cdots$5a95.
We also analyzed the voting similarity when voters cast a vote against a proposal.
However, we cannot make definitive conclusions regarding abstained votes as they are infrequent: only \num{87} (\num{0.91}\%) out of \num{9500} votes.
Regarding votes against proposals, Blockchain at Michigan and Blockchain at Berkeley have the highest cosine similarity with \num{0.73} followed by blck and Dakeshi with \num{0.67}.
These results suggest that these voters have similar voting patterns when indicating their opposition to a proposal.

%

\section{Concluding remarks}

In this chapter, we analyzed data from the Ethereum blockchain related to Compound, a widely used smart contract.
Our analysis is centered on the decentralized governance of Compound, with a particular focus on amendments to the smart contract.
We found that the Compound contract is being actively amended---token holders continuously propose amendments that are then voted on by other token holders.
We observed a striking concentration of tokens (be it in terms of their ownership, their delegation, or their voting participation) in the hands of a few participants, which raises serious concerns about the extent to which governance is decentralized in practice.
For instance, our analysis shows that, on average, only \num{3.25} voters were needed for the proposals to reach \stress{quorum} and pass, and only \num{2.84} voters were needed to reach \num{50}\% of the total votes.
Our analysis also highlights issues with the Compound use of on-chain voting---in particular, the transaction fees voters must pay to cast an on-chain vote can make it prohibitively expensive for voters with fewer tokens.
These costs have implications for voting participation and can affect how voters, proposers, and other stakeholders interact with these protocols.

\clearpage

%
\chapter{Related Work} \label{chap:related}

In this chapter, we examine the literature relevant to this thesis. We explore three main topics: (i) transaction prioritization norms; (ii) transaction prioritization and contention transparency; and (iii) decentralized governance. The latter encompasses works that explore the distribution of decision-making power for blockchain governance.

%
\section{Transaction prioritization norms}

A few recent papers proposed solutions to enforce that transaction ordering follows a certain norm, mostly based on statistical tests of potential deviations \cite{Orda2019,Asayag18a,lev2020fairledger}. These works were, however, mostly of theoretical nature in that they did not contain empirical evidence of deviation by miners, but rather assumed that miners might deviate. Prior efforts also proposed consensus algorithms to guarantee fair-transaction selection~\cite{baird2016swirlds,Kursawe@AFT20,Kelkar@CRIPTO20}. Kelkar \textit{et al.}~\cite{Kelkar@CRIPTO20} proposed a consensus property called \textit{transaction order-fairness} and a new class of consensus protocols called \textit{Aequitas} to establish fair-transaction ordering in addition to also providing consistency and liveness. A number of prior work focused on enabling miners to select transactions. For instance,  SmartPool~\cite{Luu2017} gave transaction selection from mining pools back to the miners. Similarly, an improvement of Stratum, a well-used mining protocol, allows miners to select their desired transaction set through negotiation with a mining pool~\cite{Stratum-2021}. All these prior work are, again, mostly of theoretical nature. In contrast, this thesis provides empirical evidence of deviation from the norm by miners in the current Bitcoin system.

Additionally, fairness issues have been studied in blockchain from the point of view of miners. Pass \textit{et al.}~\cite{Pass@PODC17} proposed a fair blockchain where transaction fees and block rewards are distributed fairly among miners, decreasing the variance of mining rewards. Other studies focused on the security issues showing that miners should not mine more blocks than their ``fair share''~\cite{Eyal-CACM2018} and that mining rewards payout is centralized in mining pools and therefore unfairly distributed among their miners~\cite{Romiti2019ADD}. Chen \textit{et al.}~\cite{Chen@AFT19} studied the allocation of block rewards on blockchains showing that Bitcoin's allocation rule satisfies some properties. It does not, however, hold when miners are not risk-neutral, which is the case for Bitcoin.
In contrast to these prior works, this thesis touches upon fairness issues from the viewpoint of transaction issuers and not miners. 

There is a vast literature on incentives in mining. Most of it, however, considers only block rewards~\cite{Romiti2019ADD,Chen@AFT19,Eyal-CACM2018,Pass_Seeman_Shelat_2017,Zhang_Preneel_2019,sompolinsky2015secure,Kiayias@EC16,Fiat@EC19,Goren@EC19,Noda@EC20}. 
As the block reward halves every four years in the Bitcoin blockchain, some recent work focused on analyzing how the incentives will change when transaction fees dominate the rewards. Carlsen \textit{et al.}~\cite{Carlsten@CCS16} showed that having only transaction fees as incentives will create instability. Tsabary and Eyal~\cite{Tsabary@CCS18} extended this result to more general cases including both block rewards and transaction fees. Easley \textit{et al.}~\cite{Easley19a} proposed a general economic analysis of the system and its welfare with various types of rewards. Those prior works, however, assume that miners follow a certain norm for transaction selection and ordering (mostly the fee rate norm) and look at miners' incentives in terms of how much compute power to exert and when (or some equivalent metric). 
There are also prior studies on the security issues of having transaction fees as the prime miners' incentive~\cite{Carlsten@CCS16,Li@IV18}; and a vast literature on the security of blockchains more generally (e.g., \cite{Gencer-FC2018,Karame-CCS2016,Vasek-FC2014}). Again, however, these studies focus on miners' incentives to mine and not on transaction ordering; for the latter, they assume that miners follow a norm. These prior studies are, hence, somewhat orthogonal to this thesis.

Only a few recent works touched upon the issue of how miners select and order transactions, and how this is interlaced with how the fees are set.
Lavi~\textit{et al.}~\cite{Lavi-WWW2019} and Basu~\textit{et al.}~\cite{Basu-CoRR2019} highlighted the inefficiencies in the existing transaction fee-setting mechanisms and proposed alternatives. They showed that miners might not be trustworthy, but without providing empirical evidence. 
Siddiqui \textit{et al.}~\cite{Siddiqui@AAMAS20} showed through simulations that, with transaction fees only as incentives, miners would have to select transactions greedily, increasing the latency for most of the transactions. They proposed an alternative selection mechanism and performed numerical simulations on it.
This thesis takes a complementary approach: We analyze empirical evidence of miners deviations from the transaction ordering norm in the current ecosystem. 
We also empirically analyze existing collusion at the level of transaction inclusion.

To the best of our knowledge, our study is the first of its kind---showing empirical evidence of norm violations in Bitcoin---and our results help motivate the theoretical studies mentioned above.

%
\section{Transaction prioritization and contention transparency}

As previously mentioned, recent work analyzed the implications of relying on transaction fees separately~\cite{Carlsten@CCS16} and in conjunction with block rewards~\cite{Tsabary@CCS18}, as well as the relationship between such incentives and transaction waiting times~\cite{Easley19a}.
These prior works assume that transactions are broadcast to all miners and the fees offered is uniform across miners.
None of them acknowledge the issue of transparency.
Prior work also analyzed the Ethereum fee (i.e., gas price) mechanism to determine the gas price for a given transaction~\cite{Pierro@IWBOSE,Liu@DSA,Mars@COMPSAC,Turksonmez@COINS}.
However, the fee estimation and fee-based prioritization schemes in these studies do not take into account dark-fees or private mining.

Many transaction-accelerator, or front-running as a service (FRaaS), platforms exist for both Bitcoin~\cite{BTC@accelerator,ViaBTC@accelerator} and Ethereum~\cite{Eskandari@FC-2020,Flashbots@Ethereum,Taichi@accelerator}.
Transaction issuers might resort to such acceleration or off-chain payment channels to hide their true fee from competitors and avoid being front-run~\cite{Daian@S&P20,strehle2020exclusive}.
Tim Roughgarden~\cite{Roughgarden@EC21} discussed the incentives for off-chain agreements (such as dark-fees) between miners and users for first-price auctions and different deviations of the new Ethereum fee mechanism \stress{EIP-1559 protocol}~\cite{EIP-1559}.
Roughgarden showed that miners and users cannot strictly increase their joint utility through off-chain payments under EIP-1559 because on-chain bids can be easily replaced by the off-chain bids. 
However, utility here is only based on the revenue of bidding for block space. The author did not take into account that utility might depend on other factors, such as transaction issuers wanting to keep their actual bids for block space hidden through off-chain payments, which strictly increases their chances of prioritization, as other bidders cannot counter bid, as they are unaware of the bid itself.

There are two work that analyze private mining.
Strehe and Ante~\cite{strehle2020exclusive} investigated \stress{exclusive mining} (or private mining), where transactions issuers and miners collude to include transactions that have been sent through a private network.
In this case, the transactions are not publicly disclosed until they have been included in a block; besides, the fees can remain opaque to everyone forever, as such off-chain agreements may use fiat currencies.
Weintraub~\ea~\cite{Weintraub@IMC2022} measured the popularity of \stress{Flashbots}, the most used private relay network for Ethereum.
This thesis, in contrast, extensively investigates private transactions and dark-fees in the context of Bitcoin and Ethereum blockchains.
Through active measurements, we empirically show that Bitcoin miners collude and highlight the colluding mining pools.
We show that Flashbots bundles are quite prevalent in Ethereum and are mainly used for calling Decentralized Exchanges (DEX) contracts to take advantage of \stress{Maximal Extractable Value (MEV)} opportunities.
Finally, we discuss why our findings are still valid after ``The Merge''---an Ethereum hard fork deployed on September~15\tsup{th},~2022~\cite{Eth-PoS,Eth-Merge}.
%

%
\section{Decentralized governance}

There is rich literature on decentralized governance and social contracts, decentralized autonomous organizations (DAOs), and on-chain governance protocols.
Below, we review prior efforts that is most relevant to this thesis.

\subsection{Decentralized governance and social contracts}
Prior work have studied the potential of blockchain-based (decentralized) governance for replacing centralization in traditional applications and services.
Atzori~\ea{} discussed, for instance, the extent to which blockchain-based governance can mitigate or replace the centralized and hierarchical societal structures and authorities~\cite{atzori2017blockchain}.
Reijers~\ea{} examined the relationship between blockchain governance and social contract theory~\cite{reijers2016governance}.
They analyzed the political implications of the blockchain technology and how it follows or deviates from the governance principles established by philosophers such as Thomas Hobbes~\cite{hobbesleviathan}, Jean-Jacques Rousseau~\cite{rousseau1920social}, and John Rawls~\cite{john1971rawls}.
Chen~\ea presented the trade-offs between decentralization and performance~\cite{chen2021decentralized}.
Arruñada and Garicano suggested new forms of ``soft'' decentralized governance to surpass traditional centralized governance structures~\cite{arrunada2018blockchain}.
Zwitter and Hazenberg conducted a comprehensive review of governance theory and proposed a re-conceptualization of the term governance that is tailored to DAOs~\cite{zwitter2020decentralized}.
These prior work provide valuable insights into decentralized governance structures, albeit they neither confirm the extent to which their (theoretical) observations hold in real-world implementations nor characterize the behavior of governance protocols deployed today.

\subsection{Decentralized Autonomous Organizations (DAOs)}
Several prior studies analyzed the governance structures of DAOs~\cite{beck2018governance,rikken2019governance,hassan2021decentralized}.
Hassan and De Filippi analyzed what DAOs constitute and discuss their key traits~\cite{hassan2021decentralized}.
Rikken~\ea{} identified various political challenges in governance of blockchains~\cite{rikken2019governance}.
Beck~\ea\cite{beck2018governance} presented a case study of a DAO in Swarm City~\cite{swarm-city}, a decentralized commerce platform.
A recent work categorized the governance of several blockchains such as Bitcoin, Ethereum, Tezos, Polkadot, and some governance protocols like Uniswap~\cite{adams2021uniswap}, MakerDAO~\cite{Governance@MakerDAO} and Compound~\cite{leshner2019compound} into different types~\cite{kiayias@2022governance}.
These invaluable prior work do not, nevertheless, empirically examine the data on existing DAOs to characterize how users interact with on-chain governance smart contracts.

There are three works closely related to ours~\cite{feichtinger2023hidden,fritsch@2022votingpower,sharma2023unpacking}.
Their findings agree with our own, e.g., they too found a high concentration of token delegation among a small number of users.
Similarly, they also showed that the largest token holders are more active in voting, further exacerbating the centralization problem.
However, while they analyzed voting participation and the cost of voting on the blockchain for more than 10 DAOs, our study presents a comprehensive and in-depth analysis focused on Compound. Specifically, our analysis reveals the complete life cycles of proposals, highlighting how voting behavior evolves over time for different proposals. We also examine token ownership in detail revealing among which entities the tokens held are concentrated as well as how delegations (by individual entities) affect the concentration of tokens. Finally, we discover a vast inequality in voting costs among the token holders and present its implications for decentralized governance.

\clearpage

%

\chapter{Discussion, Limitations \& Future Work} \label{chap:discussion_limit}

In this chapter, we discuss some consequential points that follow from the prior chapters, mention the limitations of our work, and explore avenues for future work.

%

\section{Transaction ordering}

Our findings have significant implications for both bitcoin users and miners.
Bitcoin users (using their wallet software) typically assume complete transparency regarding the fees associated with competing transactions when setting fees for their own transactions. However, our results challenge this common assumption.
Similarly, the practice of transactions having different confirmation fees for different miners raises notable fairness concerns.

Furthermore, our findings also call for a community-wide debate on defining transaction prioritization norms and enforcing them transparently. Specifically, we highlight three challenging questions that need to be addressed for the future.

\paraib{What are the desired transaction prioritization norms in public \pow{}
blockchains?}
What aspects of transactions besides fee rate should miners be allowed to consider when ordering them? For instance, should the waiting time of transactions also be considered to avoid indefinitely delaying some transactions? Should the transaction value (i.e., amount of bitcoins transferred between different accounts) be a factor in ordering, as fee rate based ordering favors larger value over smaller value transactions? Similarly, while we did not find evidence of miners decelerating or censoring (i.e., refusing to mine) transactions, the current protocols do not disallow such discriminatory behaviors by miners. Should prioritization norms also explicitly disallow discriminating transactions based on certain transaction features like sending or receiving wallet addresses? Such norms would be analogous to \stress{network neutrality} norms for Internet Service Providers (ISPs) that disallow flows from being treated differently based on their source/destination addresses or payload.

\paraib{How can we ensure that the distributed miners are adhering to desired and defined norms?} 
Miners in public proof-of-work blockchains, such as Bitcoin and Ethereum, operate in a distributed manner, over
a P2P network.
This model of operation results in different miners potentially having distinct, 
typically different, views of the state of the system (e.g., set of outstanding transactions).
Given these differences, are there mechanisms (say, based on
statistical tests~\cite{lev2020fairledger,Orda2019,Asayag18a}) that
any third-party observer could use to verify that a miner adheres to the
established norm(s)?

\paraib{How can we model and analyze the impact of selfish, non-transparent, collusive behaviors of miners?}
While the above themes align well with a long-term vision of defining
and enforcing well-defined ordering norms in blockchains, in the short term one could focus on examining the implications of the norm violations in today's blockchains.
Specifically, how can we characterize the ordering that would
result from different miners following different prioritization norms,
especially given an estimate of miners' hashing or mining powers (i.e., their
likelihood of mining a block).
Such a characterization has crucial implications, for example, for Bitcoin users.

%

\section{Transaction transparency}

In this section, we discuss the implications of transactions prioritization and contention transparency in blockchains.
Initially, we highlight the importance of incorporating these aspects into blockchain design to fulfill the overarching goal of transparency.
Subsequently, we explore the implications for publicly mined transactions. Then, we delve into the implications for privately mined transactions.
Lastly, we emphasize that our implications hold both for before or after the introduction of two major improvements to blockchains: EIP-1559 and the Merge.

Our results show that with private mining and accelerated transactions, the promise of the public decentralized blockchain does not hold. 
First, through the Bitcoin active experiment, we show that mining pools with combined hash rates of over $50\%$ are colluding with each other, showing a centralization in the system. Further, these accelerated transactions are highly prioritized by the miners and included mostly on top of their blocks.
This enables miners to also censor certain transactions, breaking the ethos of decentralized public blockchains with no central authority. 
Second, it breaks the assumption that all activities in the blockchain are transparent.
Although this is true for transactions included in the blockchain, prioritization of transactions is becoming more opaque with the rise of private mining and off-chain fees.
Hence, we make the case that to fulfill the transparency promise of public blockchains, prioritization of transactions should be transparent as well.
Third, with private mining in Ethereum, Flashbots is increasingly being used for malicious and predatory activities such as sandwich attacks, which essentially levies a tax on users interacting with financial institutions on the blockchain (e.g., in DEX).
These concerns need to be addressed if public blockchains are going to live up to their promises.

\paraib{Implications for publicly mined transactions.}
Most wallet software and crypto-exchanges today rely on reconstructing the current public \mpool state in order to suggest a suitable fee to transaction issuers. 
With the lack of contention and prioritization transparency, transaction issuers can no longer accurately recreate the current \mpool state for different miners. 
Consequently, they cannot reliably estimate the fees transactions need to pay for their desired prioritization. 
Worse, as the fraction of privately mined and accelerated transactions keeps rising, the transaction fees will become less (reliably) predictable in the future. 

\paraib{Implications for privately mined transactions.}
The problem of reliable fee estimation for a desired level of prioritization is even worse for privately mined transactions that are announced on private relay networks.
When transaction issuers announce on a private relay network today, they are often unsure what fraction of total network power is controlled by the miners listening to the private relay network.
Hence, it is important to estimate the network power controlled by private mining pools to estimate the commit (waiting) times for transactions.
Furthermore, transaction issuers on private relay networks are completely blind to other competing transactions.
This opacity allows miners offering private mining and transaction acceleration services to overcharge and demand exorbitant fees to commit transactions.
For example, in the Ethereum blockchain, users are observed to be overcharged by miners for having their transactions confirmed with high priority through Flashbots bundles~\cite{Weintraub@IMC2022}.

\paraib{Relevance of findings in light of EIP-1559 and the Ethereum Merge.} 
Our observations about the lack of transparency and their implications are fundamental to the current blockchain architectures and hold both before and after the recent major improvements to blockchains, e.g., EIP-1559 and the Ethereum Merge.
While EIP-1559 attempts to improve the estimation of transaction fees that need to be offered, it does not address the problems associated with the lack of transaction contention and prioritization transparency. 
Similarly, after the Ethereum Merge, \stress{validators} that stake a certain amount of Ether (ETH) rather than \stress{miners} would be responsible for selecting and validating transactions to include in the next block~\cite{Eth-PoS}.
Our observations about private mining would still hold for private validation and the implications would still be valid after the Merge.

%

\section{Voting power distribution to amend smart contracts}

An inherent concern in the governance of blockchain networks revolves around the concentration of governance tokens among a select group of participants. This situation can potentially pose a threat to the protocol and compromise its integrity, especially if the voting power or authority to make important changes is proportional to the amount of tokens held by each participant. 
This issue was highlighted in the case of Balancer, a decentralized exchange (DEX) built on top of Ethereum. In this example, a user with a significant amount of governance tokens voted for decisions that were beneficial to the user but detrimental to the protocol~\cite{Haig@Defiant}. 
Therefore, this scenario of a minority holding a significant amount of tokens can lead to a centralization of decision-making power, which is contrary to the goal of decentralizing governance protocols.

While governance protocols in blockchains aim to eliminate (or at least minimize) centralized decision-making, our work reveals that Compound is not effectively achieving its intended goal.
The distribution of tokens, which corresponds to voting power, plays a crucial role in determining the level of decentralization in a protocol.
Our work highlights the importance of measuring and analyzing governance protocols to ensure that they are working as intended.
In addition, this work motivates further research in this area.
For example, our empirical evidence supports recent proposals to redefine voting power based on social rewards, such as a voter's reputation or contributions to the protocol~\cite{sharma2023unpacking,LIU2022103596,Guidi@TCSS}, or the use of a quadratic voting scheme, where voting power is calculated as the square root of the number of tokens held by voters~\cite{Buterin_2019,Lalley@AEA}.

In light of our findings, we argue for integrating these insights into the design of future governance protocols. There, we can effectively increase fairness and decentralization within these protocols.
In addition, it would also be interesting to analyze other widely used governance protocols, such as Uniswap, to ensure that these governance protocols are truly decentralized.

\clearpage

%

\chapter*{Conclusion} \label{chap:conc}

In this thesis, we adopted a data-driven approach to examine fairness within blockchain contexts, focusing on three key aspects: (i) Fairness in ordering; (ii) Fairness in transparency; and (iii) Fairness in voting power to amend smart contract applications.

Our findings reveal a discrepancy between assumed prioritization norms and actual practices within the blockchain community. In particular, miners often deviate from these norms by prioritizing transactions that serve their own interests or friendly miners. This contradicts the principle of exclusively fee-based prioritization.

Through active experiments, we have uncovered instances of miner collusion involving dark-fee transactions. These transactions provide miners with off-chain incentives in a non-transparent manner, contributing to a lack of transparency in the ecosystem. These fees are kept private between the miner and the issuer of a particular transaction, even after the transaction is confirmed on the blockchain. This exacerbates the challenge of accurately estimating fees. As a result, transaction issuers struggle to determine appropriate fees because they do not have a complete view of all transaction fees being offered.

In addition, blockchain applications, or smart contracts, are often amended by governance protocols. These protocols aim to distribute decision-making power among participants. However, we show that the concentration of voting power based on token ownership skews the dynamics of decision-making. A small subset of participants with a significant token stake wields disproportionate influence, allowing them to shape proposals and votes in line with their self-interest. This practice undermines the true decentralization of decision-making power in the blockchain ecosystem.

We believe that our findings provide valuable insights for designing new and more fair blockchains. Additionally, to ensure the reproducibility of our results, we have made the code and data sets used in this thesis publicly available~\cite{Messias-DataSet-Governance-Code-2023, Messias-DataSet-Code-2023}.

\clearpage


\begin{appendices}
%

\chapter{Additional Analysis of Transactions Prioritization Norms}
\label{appendix:tx_norms}

\section{Congestion in \mpool of data set \dsb}\label{sec:supp-tx-ord}

\begin{figure}[t]
	\centering
		\includegraphics[width={\onecolgrid}]{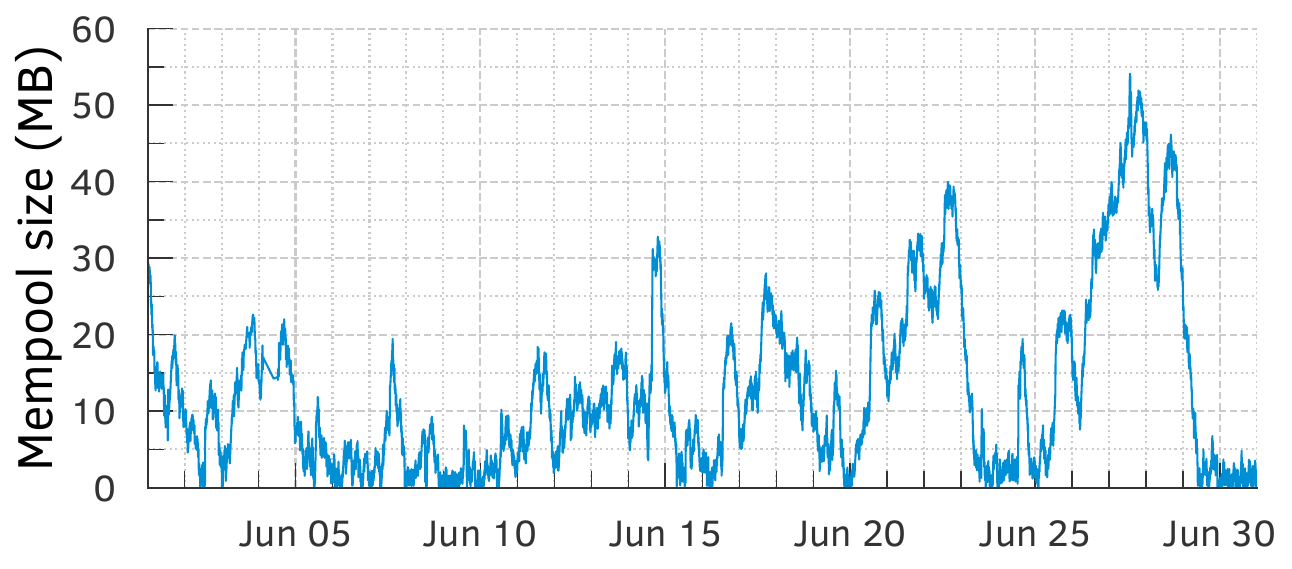}
	\caption{
  \mpool size from \dsb{} as a function of time.}
\label{fig:mpool-sz-b}
\end{figure}

Congestion in \mpool{} is typical not only in \dsa{} (as discussed in~\S\ref{subsec:cong-delays}), but also in \dsb{}.
Indeed, Figure~\ref{fig:mpool-sz-b} reveals a huge variance in \mpool{} congestion,
much higher than that observed in \dsa{}.
\mpool{} size fluctuations in \dsb{} are, for instance, approximately three times
higher than that in \dsa{} (with the size of unconfirmed transactions at one point in time exceeding almost 50 times the maximum block size).
Around June $22\tsup{nd}$, there was a surge in Bitcoin price
following the announcements of Facebook’s Libra\footnote{On June 18\tsup{th}, Facebook
announced its cryptocurrency, Libra, which was later renamed to Diem.~\url{https://www.diem.com}} and another surge
around June $25\tsup{th}$ after the news of US dollar
depreciation~\cite{CNN-BITCOIN-2019}.
These price surges significantly increased the number of transaction issued, which in turn introduced delays.
As a consequence, at times, \mpool{} in \dsb{} takes much longer duration than
in \dsa{} to be drained of all transactions.

\section{Significance of transaction fees} \label{sec:signif-tx-fees}

Table~\ref{tab:fee-revenue} shows the contribution of transaction fees towards miners' revenue across all blocks mined from 2016 to 2020. In 2018, fees accounted for an average of $3.19\%$ of miners' total revenue per block; in 2019 and 2020 were $2.75\%$ and $6.29\%$, respectively. However, if we consider only blocks mined from May 2020 (i.e., blocks with a mining reward of $6.25$ BTC), the fees account for, on average, $8.90\%$ with an std. of $6.54\%$ in total. Therefore, revenue from transaction fees is increasing~\cite{Easley-SSRN2017}, and it tends to continue.

\begin{table}[t]
    \begin{center}
    \tabcap{Miners' relative revenue from transaction fees (expressed as a percentage of the total revenue) across all blocks mined from 2016 until the end of 2020.}\label{tab:fee-revenue}
    \resizebox{.7\textwidth}{!}{%
        \begin{tabular}{rcrccccrc}
        \toprule
        \multicolumn{1}{c}{\multirow{1}{*}{\thead{Year}}} & \multicolumn{1}{c}{\thead{\# of blocks}} & \multicolumn{1}{c}{\thead{mean}} & \multicolumn{1}{c}{\thead{std}} & \multicolumn{1}{c}{\thead{min}} & \multicolumn{1}{c}{\thead{25-perc}} & \multicolumn{1}{c}{\thead{median}} & \multicolumn{1}{c}{\thead{75-perc}} & \multicolumn{1}{c}{\thead{max}}\\  \midrule
            2016 & \num{54851}        & 2.48  & 2.12 & 0    & 0.87    & 1.78   & 3.84    & 92.10 \\
            2017 & \num{55928}        & 11.77 & 7.73 & 0    & 6.33    & 10.49  & 15.58   & 86.44 \\
            2018 & \num{54498}        & 3.19  & 5.85 & 0    & 0.52    & 1.22   & 2.60    & 44.19 \\
            2019 & \num{54232}        & 2.75  & 2.77 & 0    & 0.80    & 1.81   & 3.70    & 24.32 \\
            2020 & \num{53211}        & 6.29  & 6.34 & 0    & 1.37    & 4.00   & 9.71    & 39.46 \\
        \bottomrule
        \end{tabular}
    } 
    \end{center}
\end{table}

\begin{figure}[t]
	\centering
		\includegraphics[width={\onecolgrid}]{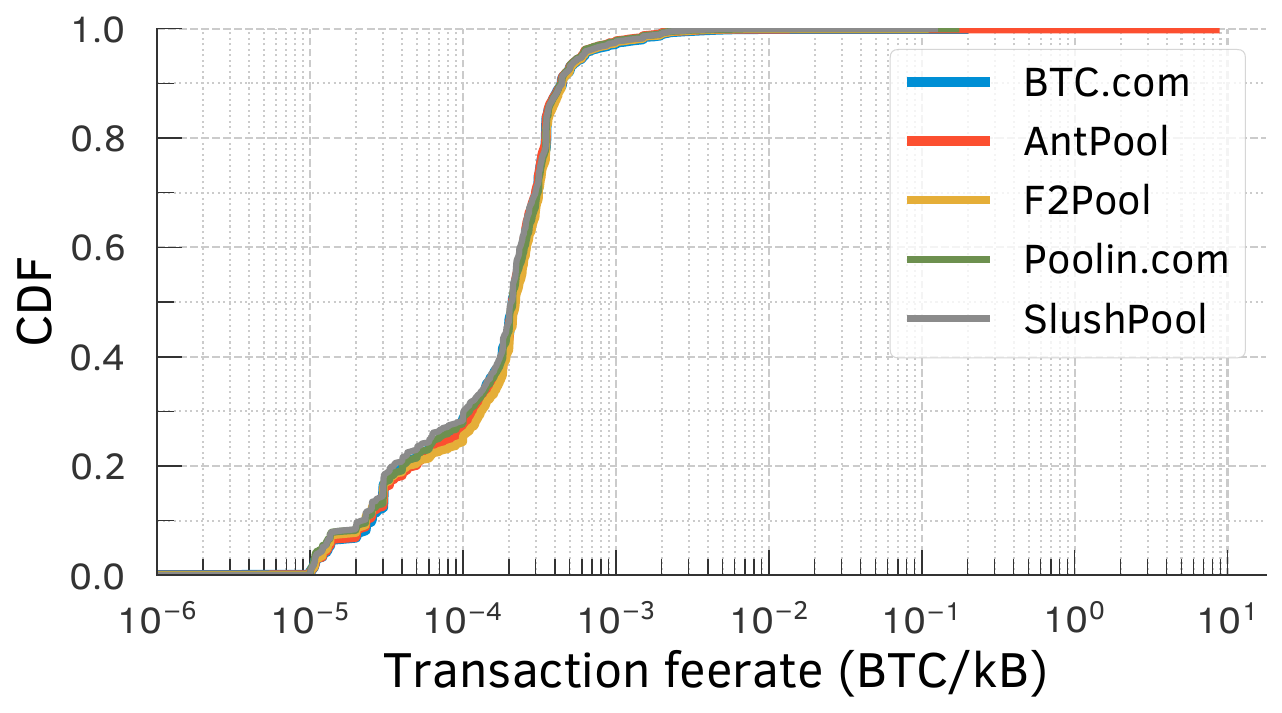}
	\caption{
  Distributions of fee rates for transactions committed by the top-5 mining pools in data set \dsa.}
\label{fig:cdf-fee-top5}
\end{figure}

\section{Transaction fee rates across mining pools} \label{sec:tx-fees-across-mpos}

Transaction fee rate of committed transactions in both data sets \dsa{} and \dsb{} exhibits a wide range, from
$10^{-6}$ to beyond $\uTxFee{1}$.
A comparison of the fee rates of transactions in \dsa{} committed by the top
five mining pool operators (in a rank ordering of mining pool operators based on
the number of blocks mined), in Figure~\ref{fig:cdf-fee-top5}, shows no major
differences in fee rate distributions across the different MPOs. Around $70\%$ of the transactions offer from $10^{-4}$ to $10^{-3}$ \uTxFee{} that is one to two orders of magnitude more than the recommended minimum of $10^{-5}$ \uTxFee{}. We hypothesize that users increase the fee rates offered during high \mpool congestion---they assume that higher the fee rate implies lower the transaction delay or commit time.

\section{On fee rates and congestion} \label{sec:fees-and-cong}

\begin{figure}[t]
	\centering
		\includegraphics[width={\onecolgrid}]{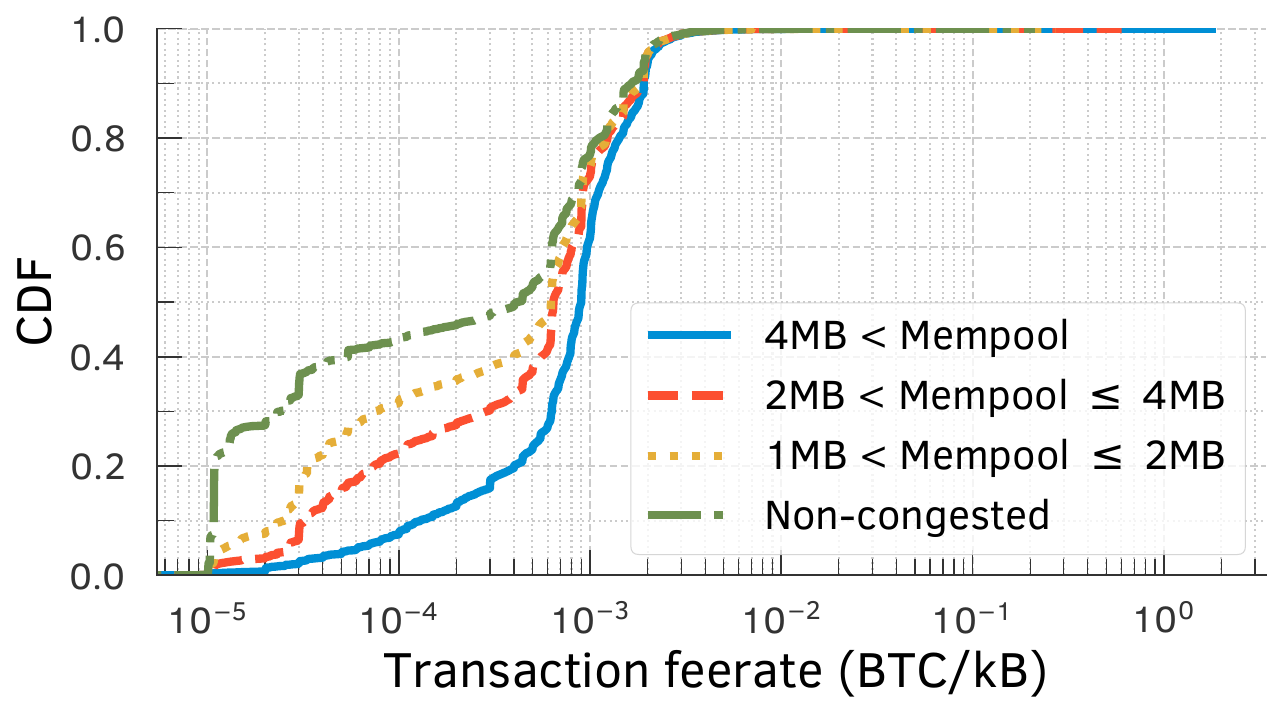}
	\caption{
  Distribution of fee rates for transactions in data set \dsb{} issued at different congestion levels clearly indicate that users incentivize miners through transaction fees.}
\label{fig:fee-cong-rel-b}
\end{figure}

In Figure~\ref{fig:fee-cong-rel-b}, we show the fee rates of transactions observed in $4$ different bins or congestion levels in data set \dsb.
Each bin in the plot corresponds to a specific level of congestion identified by the \mpool size:
lower than \uMB{1} (\stress{no congestion}), in $(1, 2]$ MB (\stress{lowest congestion}), in $(2, 4]$ MB, and higher than \uMB{4} (\stress{highest congestion}).
Fee rates at high congestion levels are strictly higher (in distribution, and hence also on average) than those at low congestion levels.
Users, therefore, increase transaction fees to mitigate the delays incurred during congestion.

\begin{figure}[t]
	\centering
		\includegraphics[width={\onecolgrid}]{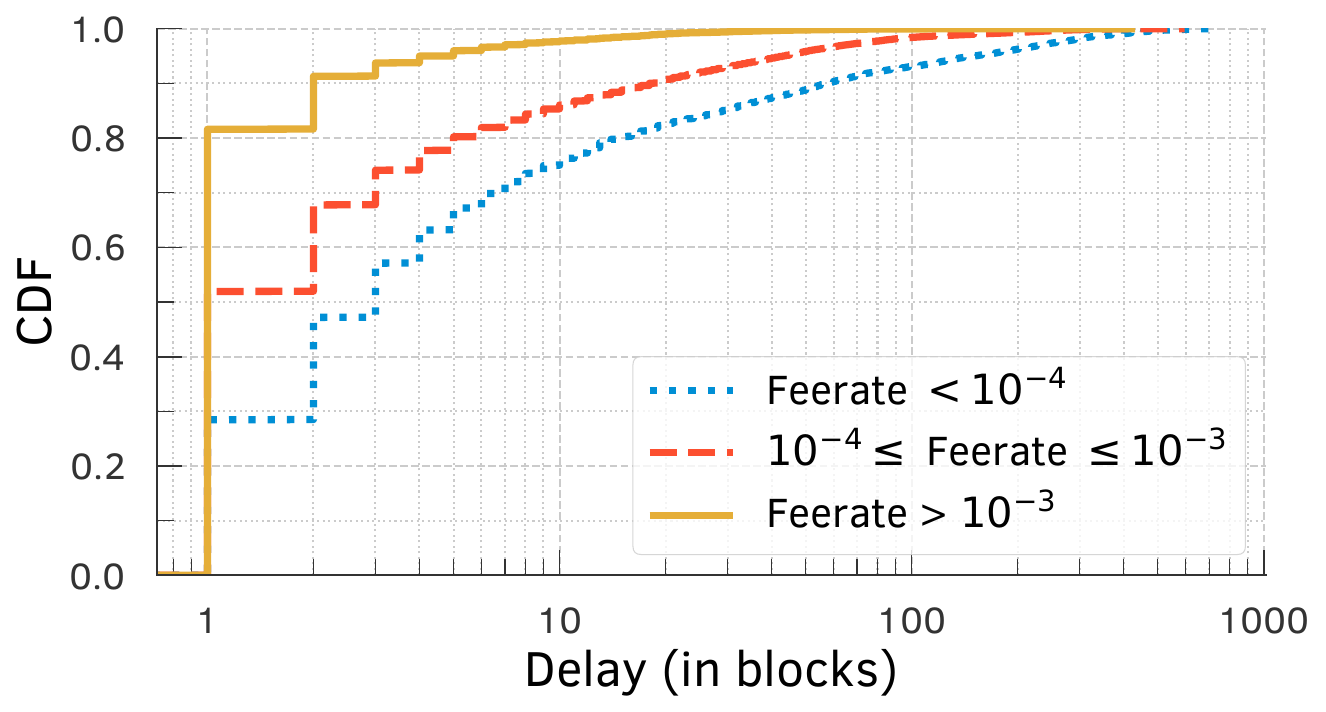}
	\caption{
  Distributions of transaction-commit delays in data set \dsb{} for different transaction fee rates.}
\label{fig:fee-delay-rel-b}
\end{figure}

Figure~\ref{fig:fee-delay-rel-b} shows that users' strategy of increasing
fee rates to combat congestion seems to work well in practice---higher the fee rate, lower the transaction commit delay.
Here, we compare the CDF of commit delays of transactions with low (i.e., less
than $10^{-4}$~\feeunit{}), high (i.e., between $10^{-4}$ and
$10^{-3}$~\feeunit{}), and exorbitant (i.e., more than $10^{-3}$) fee rates, in data set \dsb.
The commit delays for transactions with high fee rates (i.e., greater than $10^{-3}$~\feeunit{}) are significantly smaller than those with low fee rates (i.e., lesser than $10^{-4}$~\feeunit{}).

\section{Child-Pays-For-Parent (CPFP) Transactions} \label{sec:cpfp-txs}

Given any block $B_i$ that contains a set of issued transactions $T = \{t_0, t_1, \cdots, t_n\}$, where each transaction has at least one transaction input identifier $V = \{v_0, v_1, \cdots, v_m\}$, the transaction $t_j \in T$ is said to be a \newterm{child-pays-for-parent transaction (CPFP-tx)} if and only if there exists at least one input $v_k \in V$ that belongs to $T$. In other words, a transaction is a CPFP transaction if and only if it spends from any previous transaction that was also included in the same block $B_i$.

\begin{figure*}[tb]
	\centering
		\includegraphics[width={\onecolgrid}]{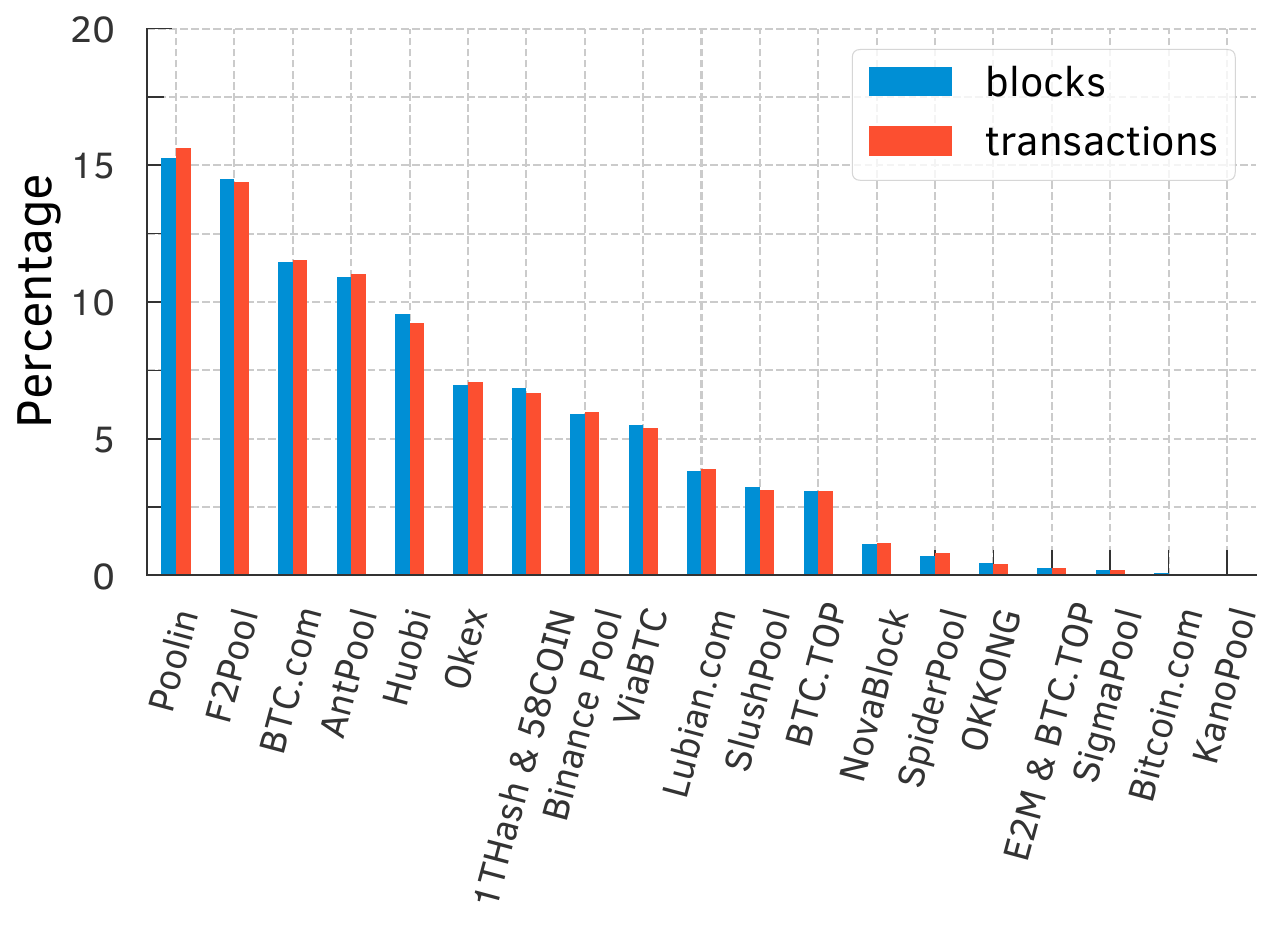}
	\caption{
  Distribution of blocks mined and transactions confirmed by different MPOs during the Twitter Scam attack from July 14\tsup{th} to August 9\tsup{th}, 2020.}
\label{fig:dist-txs-blks-twitter}
\end{figure*}

\section{Miners' behavior during the scam} \label{sec:supp-scam-txs}

To examine the miners' behavior during the Twitter scam attack from July 14\tsup{th} to August 9\tsup{th}, 2020, we selected  all blocks
mined (\num{3697} in total, containing \num{8318621} issued transactions) during this time period from our data set \dsc.
If we rank the MPOs responsible for these blocks by the number of blocks ($B$) mined (or, essentially, the
approximate hashing capacity $h$), the top five MPOs (refer Figure~\ref{fig:dist-txs-blks-twitter}) turn out to be Poolin ($B$: $\num{565}$; $h$: $15.28\%$), F2Pool ($B$: $\num{536}$; $h$: $14.5\%$), 
BTC.com ($B$: $\num{424}$; $h$: $11.47\%$), AntPool ($B$: $\num{404}$; $h$: $10.93\%$), and Huobi ($B$: $\num{353}$; $h$: $9.55\%$).

\begin{figure*}[tb]
	\centering
		\includegraphics[width={\onecolgrid}]{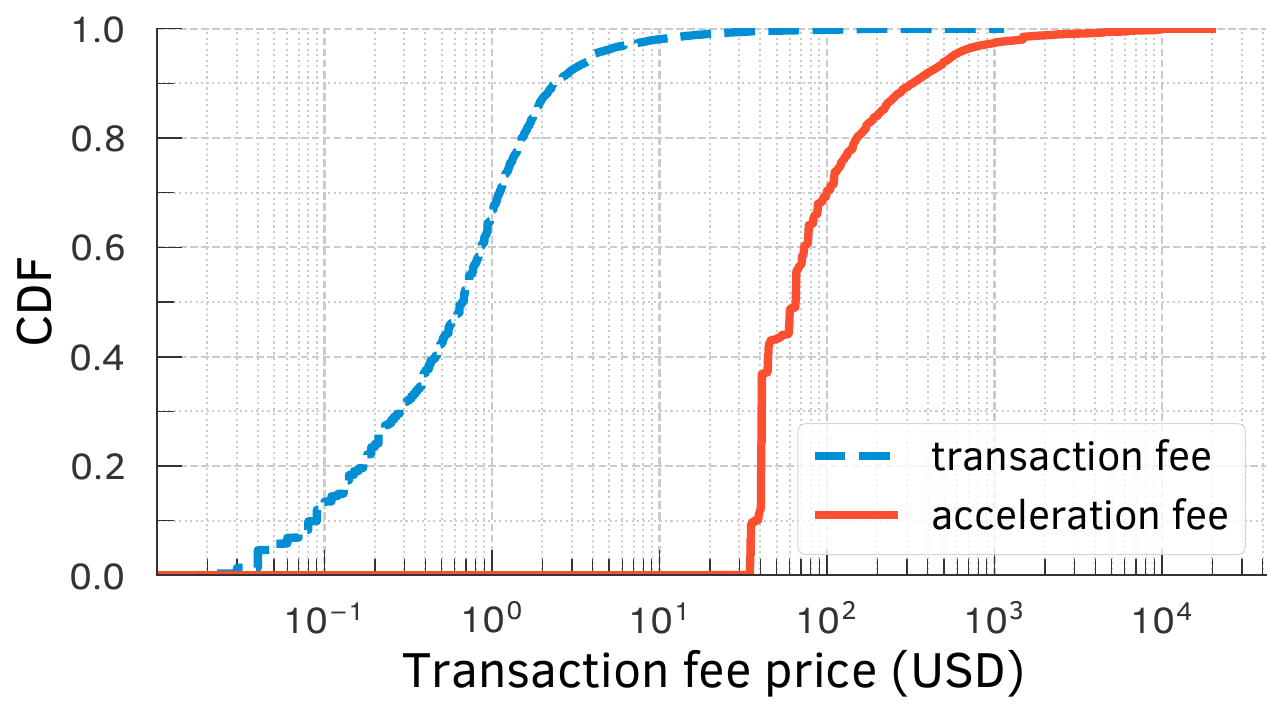}
	\caption{
  Fee price comparison between the transaction fee and the acceleration services from an snapshot of our \mpool on November 24\tsup{th}, 2020. Acceleration service provided by BTC.com is on average 566.3 times higher (4734.67 of std.) and on median 116.64 times higher than the Bitcoin transaction fees. The minimum is 0.54, the 25-perc is 51.64, and the 75-perc and the maximum are 351.8 and 428,800, respectively.}
\label{fig:acceleration-fee-price-comparison}
\end{figure*}

\section{Transaction-acceleration fees}
\label{sec:tx-accelerator-comparison}

In this experiment, we compare the transaction-acceleration fee with the typical transaction fees in Bitcoin.
To this end, we retrieved a snapshot containing \num{26332} unconfirmed transactions from our node's \mpool on November 24\tsup{th} 2020 at 10:08:41 UTC.
Then, for each transaction, we searched its respective transaction accelerator price (or acceleration fee) via the acceleration service provided by BTC.com~\cite{BTC@accelerator}.
We inferred the acceleration fees for \num{23341} ($88.64\%$) out of the \num{26332} unconfirmed transactions.
Figure~\ref{fig:acceleration-fee-price-comparison} shows the CDF of both the Bitcoin transaction fees and the acceleration fees provided by BTC.com.
Acceleration fee is on average $566.3$ times higher (\num{4734.67} of std.) and on median 116.64 times higher than the Bitcoin transaction fees.
At the time of this experiment, 1~BTC was worth \num{18875.10}~USD.

\clearpage

%

\chapter{Additional analysis of transactions prioritization and contention transparency}
\label{appendix:tx_prioritization_contention}

\section{Ethereum private transaction experiment} \label{sec:private-txs}

We conducted $4$ active experiments where we issued $8$ Ethereum transactions; half issued publicly and the other half privately through a private-channel network known as Taichi Network~\cite{Taichi@accelerator}. Table~\ref{tab:acceleration-experiment-ETH} summarizes the transactions in our experiment. Spark Pool and Babel Pool included all private transactions ($2$ transactions each) sent directly to these miners through Taichi Network.

\begin{table}[t]
\caption{We conducted 4 active experiments in Ethereum by simultaneously accelerating transactions privately and publicly via Taichi Network. Private transactions were included only by Spark Pool and Babel Pool. If we rank these mining pools according to their hash-rate, they account for 27.72\% of the total Ethereum hash-rate.}
\label{tab:acceleration-experiment-ETH}
\resizebox{\textwidth}{!}{%
\begin{tabular}{rrrcrrccrccrc}
\hline
\multirow{2}{*}{\thead{\#}} & \multirow{2}{*}{\thead{type}} & \multirow{2}{*}{\thead{tx hash}}                                           & \multirow{2}{*}{\thead{block number}} & \multirow{2}{*}{\thead{miner}} & \thead{tx. position}   & \thead{block delay} & \thead{fee paid}                   & \thead{base fee} & \thead{max fee} & \thead{max priority fee} & \thead{gas price}     & \thead{block timestamp}     \\  
                    &                       &                                                                    &                               &                        & \thead{per \# of txs.} & \thead{(in blocks)} & \thead{(in Ether)}                 & \thead{(Gwei)}           & \thead{(Gwei)}          & \thead{(Gwei)}                   & \thead{(Gwei)}        & \thead{in UTC}              \\ \hline
\multirow{2}{*}{1}  & public                & \href{https://etherscan.io/tx/0xbbe88eae757acf6697d498575dd1d50b3ad9915318cd1ff8d409210d20a4f000}{bbe88e$\cdots$a4f000} & 13,183,516                    & Nanopool               & 305 / 336      & 1           & 0.00190489 & 88.98082939     & 116.52835749   & 1.72836605              & 90.70919543  & 2021-09-08 06:39:18 \\
                    & \thead{private}               & \href{https://etherscan.io/tx/0xc46b7556a20865c9f50166373baf7094104f300ab26ad8e1de894e1318ead538}{c46b75$\cdots$ead538} & 13,183,520                    & Babel Pool             & 29 / 39        & 5           & 0.00225209  & 105.51391459    & 120.56586232   & 1.72836605              & 107.24228063  & 2021-09-08 06:40:29 \\
\multirow{2}{*}{2}  & public                & \href{https://etherscan.io/tx/0x6d994f516f43b8ed3763fe4f81c7cb86146203fda1047cc85e697eefa7c1aadd}{6d994f$\cdots$c1aadd} & 13,183,561                    & Binance                & 209 / 213      & 2           & 0.00244137 & 114.95482846    & 137.64014705   & 1.30100683              & 116.25583529 & 2021-09-08 06:49:26 \\
                    & \thead{private}               & \href{https://etherscan.io/tx/0xa4d4ae2f6f3a798dc6cf5d5f4e15222320d3ee90b023763efe0017e51142ebf5}{a4d4ae$\cdots$42ebf5} & 13,183,565                    & Spark Pool             & 294 / 296      & 6           & 0.00240978 & 113.45059961    & 137.64014705   & 1.30100683              & 114.75160643 & 2021-09-08 06:50:12 \\
\multirow{2}{*}{3}  & public                & \href{https://etherscan.io/tx/0x725743c1700241a6e89b957faf963018f2d169f7f1ec6b9256a92811510a6c45}{725743$\cdots$0a6c45} & 13,183,634                    & Unknown                & 124 / 126      & 2           & 0.00263298 & 123.27216185    & 135.21393222   & 2.10805685              & 125.38021870 & 2021-09-08 07:06:31 \\
                    & \thead{private}               & \href{https://etherscan.io/tx/0xf2beec913ed6c0667fdde4829a004fe9418916af22218d77adf5f38a7c15cdf1}{f2beec$\cdots$15cdf1} & 13,183,635                    & Spark Pool             & 321 / 340      & 3           & 0.00257468 & 120.49562077     & 135.21393222   & 2.10805685              & 122.60367762 & 2021-09-08 07:06:44 \\
\multirow{2}{*}{4}  & public                & \href{https://etherscan.io/tx/0xe21695cc9e1f29f45f38b0fd8323a6e928bd7b55dc84974f217c7042322c1574}{e21695$\cdots$2c1574} & 13,183,679                    & Ethermine              & 280 / 302      & 13          & 0.00223433 & 104.69510748    & 108.95262574   & 1.70164453              & 106.39675202 & 2021-09-08 07:18:37 \\
                    & \thead{private}               & \href{https://etherscan.io/tx/0x4c482b0416b38de9b2995b986d8c0f974018c0aeda02ce6fdc8b196bce87c76f}{4c482b$\cdots$87c76f} & 13,183,690                      & Babel Pool             & 150 / 212      & 24          & 0.00179917  & 83.97323655     & 108.95262574   & 1.70164453              & 85.67488108   & 2021-09-08 07:20:12 \\ \hline
\end{tabular}%
}
\end{table}

\section{Liquidation with Chainlink oracle updates}\label{sec:liquidations-cll-updates}

In AAVE, of \num{1154} bundles, \num{994} (86.14\%) include one Chainlink oracle update followed by a liquidation. There are \num{52} (4.51\%) with two oracle updates followed by liquidations.
Out of \num{1301} oracle updates bundled with liquidations, \num{282} ($21.68\%$) are USDC-ETH, \num{203} ($15.60\%$) are USDT-ETH, \num{169} ($12.99\%$) are DAI-ETH, \num{70} ($5.38\%$) are SUSD-ETH, and \num{60} ($4.61\%$) are LINK-ETH.
In Compound, of \num{641} bundles, \num{548} (85.49\%) contain one Chainlink oracle update followed by one liquidation, while \num{39} (6.08\%) include two oracle updates followed by liquidations.
Out of \num{751} oracle updates bundled with liquidations, \num{311} ($41.41\%$) are ETH-USD, \num{128} ($17.04\%$) are BTC-USD, and \num{53} ($7.06\%$) are UNI-USD.

\section{Hashing rates of mining pools}\label{sec:hash-var}

\begin{figure*}[t]
	\centering
		\includegraphics[width={\textwidth}]{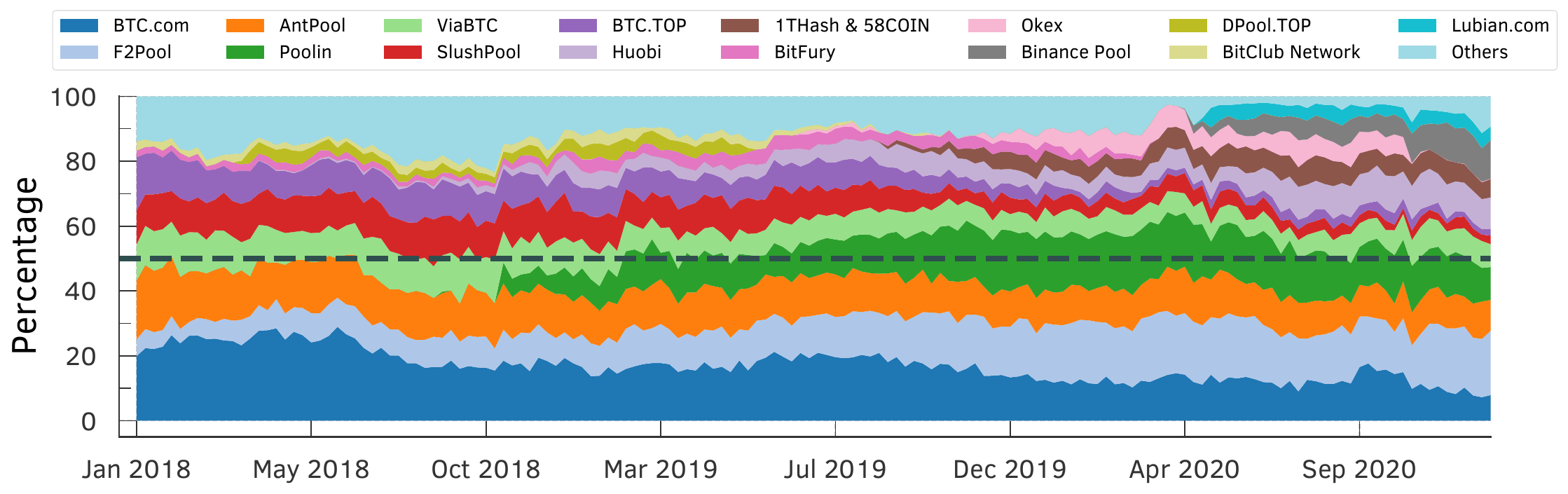}
	\caption{
  Monthly Bitcoin hash rate over the 3-year period.}
\label{fig:btc-hashrate}
\end{figure*}

Per Figure~\ref{fig:btc-hashrate}, the hash rates of Bitcoin mining pools such as BTC.com, F2Pool, and AntPool alone accounted for almost half the total hash rate of the network around May 2018, and roughly a year later, i.e., from March 2019, together with Poolin the four mining pools alone represent more than $50\%$ of the total network hash rate.
At the end of 2020, new MPOs, e.g., Lubian.com and Binance Pool, started mining Bitcoin, which help improve the decentralization of Bitcoin.
However, BTC.com, F2Pool, AntPool, and Poolin still account for almost half of the hash rates showing that a few mining pools control a considerable portion of the Bitcoin hash rate.

\begin{figure*}[t]
	\centering
		\includegraphics[width={\textwidth}]{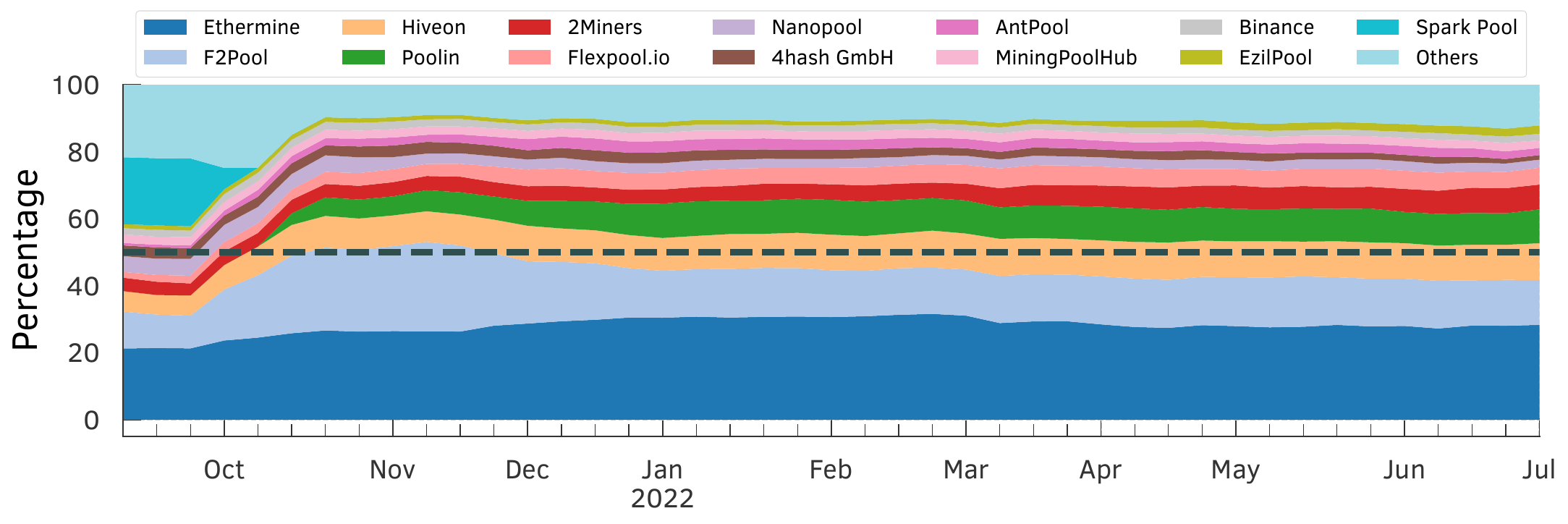}
	\caption{
  Weekly Ethereum hash rate from Sept $8\tsup{th}$, 2021, to Jun $30\tsup{th}$, 2022.}
\label{fig:eth-hashrate}
\end{figure*}

Hash rates of Ethereum mining pools, in contrast to Bitcoin, do \stress{not} show a high variance (see Figure~\ref{fig:eth-hashrate}).
We also observed that Spark Pool, the second-largest Ethereum mining pool, suspended their mining services on September 30, 2021, due to regulatory requirements in response to Chinese authorities~\cite{SparkPool@CoinTelegraph}.

\section{Bitcoin transaction acceleration experiment} \label{sec:accelerated-txs}

\begin{table}[t]
\caption{We conduct 10 transaction acceleration experiments in Bitcoin. If we rank the miners whose included these transactions based on their daily hash-rate power as (D) and weekly hash-rate power as (W), together these mining pools corresponds to a hash-rate power of (D: 55.2\%; W: 56\%).}
\label{tab:acceleration-experiment}
\resizebox{\textwidth}{!}{%
\begin{tabular}{rcccccccccc}
\toprule
\multicolumn{1}{c}{\multirow{2}{*}{\thead{txid}}}                       & \multirow{2}{*}{\thead{block height}} & \multirow{2}{*}{\thead{miner}} & \multirow{2}{*}{\thead{tx. position}} & \thead{delay}       & \thead{acc. cost} & \thead{vsize}  & \thead{fee rate}       & \multicolumn{2}{c}{\thead{Mempool}} & \thead{timestamp}        \\ 
\multicolumn{1}{c}{}                                            &                               &                        &                               & \thead{(in blocks)} & \thead{(BTC)}     & \thead{(byte)} & \thead{sat-per-vsize} & \thead{\# of txs.}    & \thead{vsize (MB)}    & \thead{in UTC}           \\ \midrule
\href{https://explorer.btc.com/btc/transaction/35b18e7a119173c8136c460e45d5d2a87d69304f69546f22ebed2c5f3852dbc1}{35b18e$\cdots$52dbc1} & \num{658805}                        & Huobi                  & 2\tsup{nd}                           & 2           & 0.001254  & 110    & 2             & \num{36644}        & 44.63   & 2020-11-26 19:10 \\
\href{https://explorer.btc.com/btc/transaction/65765c65acc86bde3d305b2594229af0839b3636aabea49e7255521412baede2}{65765c$\cdots$baede2} & \num{658898}                        & F2Pool                 & 73\tsup{rd}                          & 1           & 0.001254  & 110    & 2             & \num{20998}        & 32.55   & 2020-11-27 11:06 \\
\href{https://explorer.btc.com/btc/transaction/0c2098e3b3c993f5fc1d188da3b9d0a8731961bb946c4048d7a99fa83129fbf0}{0c2098$\cdots$29fbf0} & \num{658912}                        & AntPool                & 2\tsup{nd}                           & 2           & 0.001254  & 110    & 1             & \num{30126}        & 38.01   & 2020-11-27 13:38 \\
\href{https://explorer.btc.com/btc/transaction/1515a78b711558a1508400b36f554d798a31bd97e3852de5bae598e020179af3}{1515a7$\cdots$179af3} & \num{658971}                        & Binance                & 2\tsup{nd}                           & 3           & 0.001254  & 110    & 1             & \num{25922}        & 37.89   & 2020-11-27 21:55 \\
\href{https://explorer.btc.com/btc/transaction/48a0a55252bc029286e4af6215d1673e6744216ffc86b3c7b36eeafe640ddaec}{48a0a5$\cdots$0ddaec} & \num{659335}                        & ViaBTC                 & 3\tsup{rd}                           & 1           & 0.001045  & 110    & 1             & \num{15605}        & 9.82    & 2020-11-30 10:09 \\
\href{https://explorer.btc.com/btc/transaction/9a17cfef7e7bda668415a4a4918195669086f0507786a0c971df24a1c3f3734c}{9a17cf$\cdots$f3734c} & \num{659341}                        & Huobi                  & 2\tsup{nd}                           & 2           & 0.001045  & 110    & 1             & \num{14945}        & 9.41    & 2020-11-30 10:28 \\
\href{https://explorer.btc.com/btc/transaction/831b246f748db46d4f52318e39171b0b587165282be3f07135d978ef0795d421}{831b24$\cdots$95d421} & \num{659351}                        & AntPool                & 2\tsup{nd}                           & 1           & 0.001045  & 110    & 1             & \num{10990}        & 8.66    & 2020-11-30 12:22 \\
\href{https://explorer.btc.com/btc/transaction/1f59bfc1ef2de7b2bc9d3dd3f3e35dba437c25a93d53533a76d604284047096c}{1f59bf$\cdots$47096c} & \num{659355}                        & F2Pool                 & 111\tsup{th}                         & 3           & 0.001045  & 110    & 1             & \num{17093}        & 11.40   & 2020-11-30 12:58 \\
\href{https://explorer.btc.com/btc/transaction/6942e0751586aa8f37b6cad4eb036373035d74f40ba36277a7d1ef17ca8c06c3}{6942e0$\cdots$8c06c3} & \num{659362}                        & Huobi                  & 2\tsup{nd}                           & 2           & 0.001045  & 110    & 1             & \num{30836}        & 19.06   & 2020-11-30 14:49 \\
\href{https://explorer.btc.com/btc/transaction/8e49e27c5eb6959e26dec8ab36d4dc6508105447ce8892d71c2837934eae825f}{8e49e2$\cdots$ae825f} & \num{659481}                        & ViaBTC                 & 6\tsup{th}                           & 1           & 0.001254  & 110    & 2             & \num{30935}        & 22.59   & 2020-12-01 10:40 \\ \bottomrule
\end{tabular}
}
\end{table}

\begin{figure*}[t]
	\centering
		\includegraphics[width={\textwidth}]{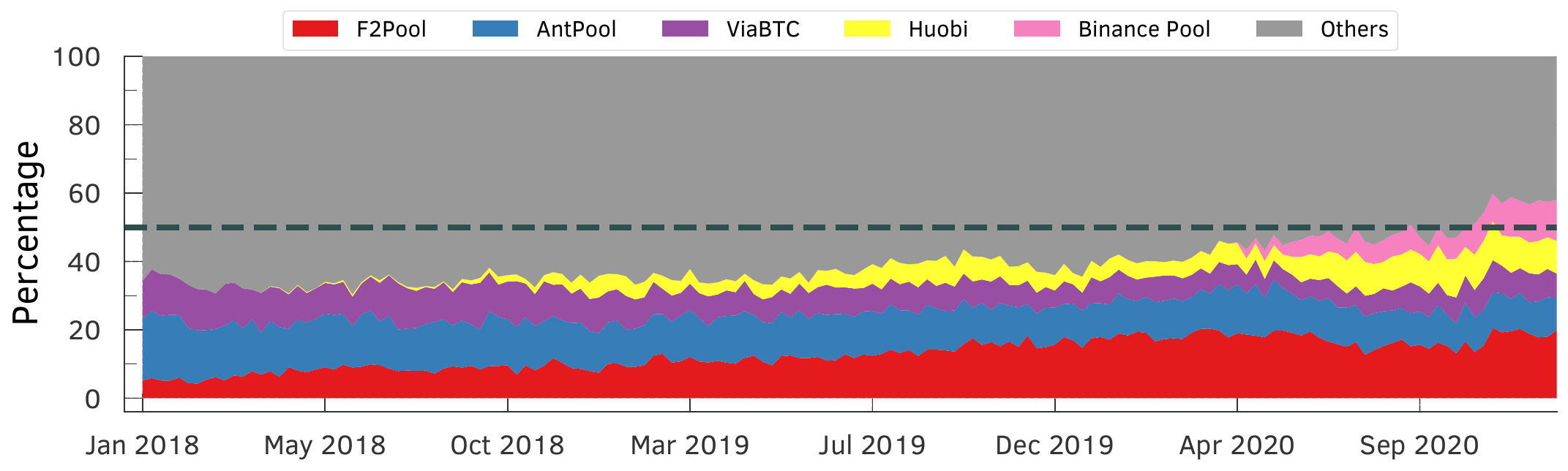}
	\caption{
  Active vs. others experiment: Bitcoin mining pools in the active experiment (i.e., mining pools that included transactions accelerated by ourselves) increased their hash rate in 2020. Together, they accounted for more than $55\%$ of the overall hash rate. The plot shows the weekly average percentage of the mining pool's hash-rate over 3 years.}
\label{fig:tx-acceleration-active-overtime-month}
\end{figure*}

\begin{figure*}[t]
	\centering
		\includegraphics[width={\textwidth}]{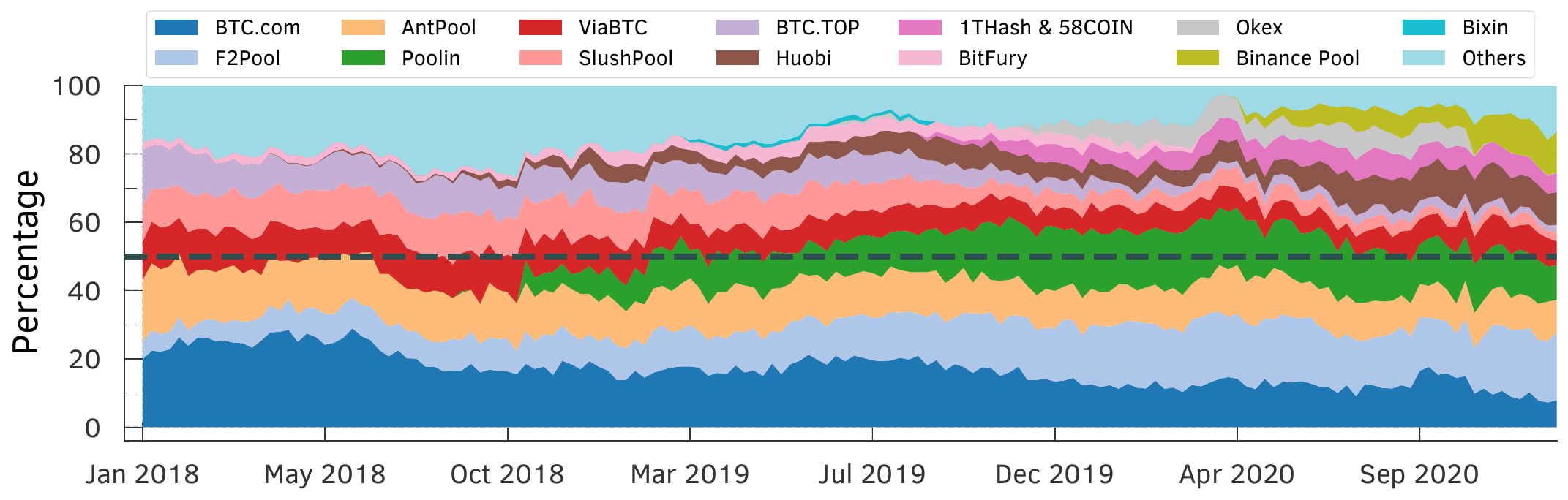}
	\caption{
  Passive + active vs. others experiment: Bitcoin mining pools in the active experiment (i.e., mining pools that included transactions accelerated by ourselves) and passive experiment (mining pools that included transactions inferred to be accelerated using the BTC.com API) increased their hash rate in 2020. The plot shows the weekly average percentage of the mining pool's hash-rate over 3 years.}
\label{fig:tx-acceleration-passive-active-overtime-month}
\end{figure*}

\begin{figure*}[t]
	\centering
		\includegraphics[width={\textwidth}]{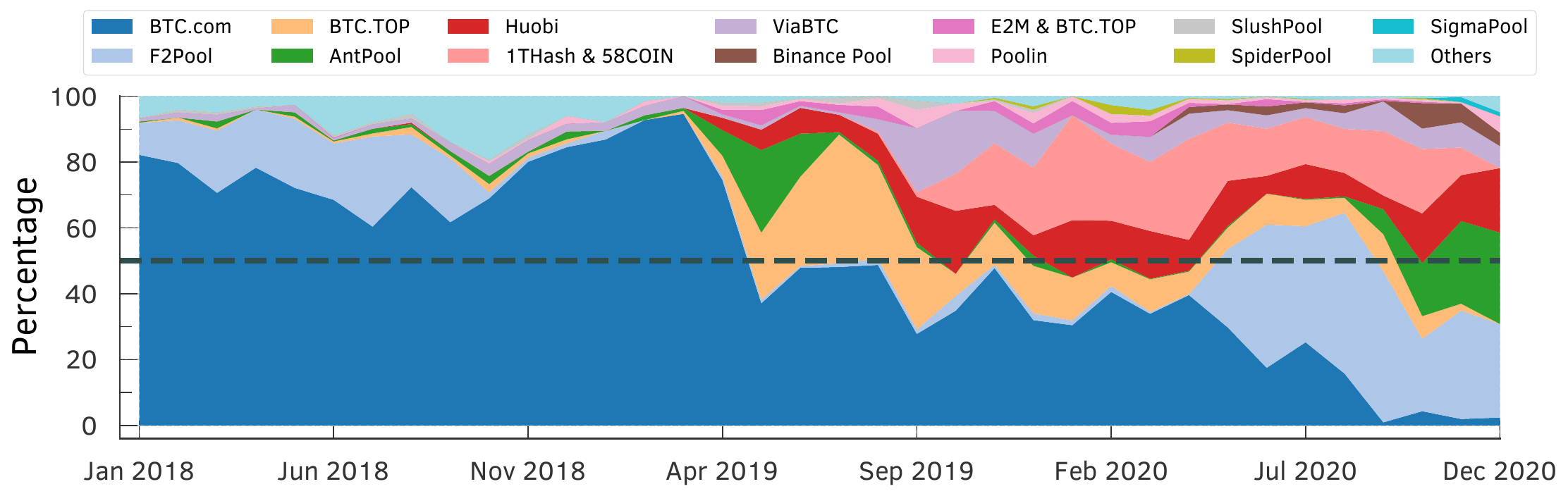}
	\caption{
  The plot shows the monthly average percentage of accelerated Bitcoin transactions inclusion by each mining pool over 3 years. Transaction acceleration services or simply Front-running as a Services (FRaaS) are becoming popular across all mining pools.}
\label{fig:tx-acceleration-overtime-month}
\end{figure*}

We ran an active Bitcoin transaction acceleration experiment where we paid $205$ EUR to ViaBTC~\cite{ViaBTC@accelerator} to accelerated $10$ transactions from $10$ different snapshots of our \mpool.
To select these transactions, we checked whether the \mpool was congested (i.e., having more transactions waiting for inclusion than the
next block would be able to include), with its size being at least $\uMB{8}$.
Then, we
considered only transactions with low fee rates---less than or equal to
2 sat-per-byte---to ensure that these transactions would be highly unlikely to be included soon in a subsequent block.
Next, we sorted the remaining transactions by size to limit the experiment cost
as the acceleration-service costs grow proportional to the transaction size. Finally, 
we select the transaction with the smallest size in bytes for our active experiment.

Most of these $10$ accelerated transactions were included nearly in the next block, demonstrating the acceleration efficiency. Also, these transactions were wrongly positioned in the block: They appeared, for instance, at the top of the block, i.e., higher than the non-accelerated transactions, showing that miners indeed prioritized them (see Table~\ref{tab:active-experiment-delay-position}).
Further, we observed that although we had only accelerated transactions via ViaBTC, other top mining pools were also involved in confirming the accelerated transactions.

Table~\ref{tab:acceleration-experiment} shows the transactions used in our experiments. At the time we conducted our experiments, if we rank the miners whose included these transactions based on their daily hash-rate power as (D) and weekly hash-rate power as (W), we would have Huobi (D: $8.1\%$; W: $9.3\%$), Binance (D: $9.6\%$; W: $10.3\%$), F2Pool (D: $19.9\%$; W: $18.7\%$), AntPool (D: $12.5\%$; W: $10.6\%$), ViaBTC (D: $5.1\%$; W: $7.1\%$). Together these mining pools corresponds to a hash-rate power of (D: $55.2\%$; W: $56\%$). Figures~\ref{fig:tx-acceleration-active-overtime-month} and \ref{fig:tx-acceleration-passive-active-overtime-month} show the hash-rate of mining pools in the active experiment and considering the passive experiment (inferred to be accelerated by BTC.com API), respectively.

Furthermore, BTC.com~\cite{BTC@accelerator}, one of the leading Bitcoin mining pools, provides transaction acceleration services and allows users to verify if transactions have been accelerated through their platform or partner services.
From our dataset, we selected those with a SPPE greater than or equal to \num{1}\% (\num{12983282} transactions in total) and checked if they were said to be accelerated by BTC.com's API.
Of these transactions, \num{14104} were found to have been accelerated.
Our findings also show that transaction acceleration services are becoming quite common among Bitcoin mining pools (as shown in Figure~\ref{fig:tx-acceleration-overtime-month}).
Between 2018 and April 2019, only BTC.com and F2Pool alone accounted for most of the accelerated transactions.
However, as of December 2020, we see that BTC.com accounts for a very small fraction of accelerated transactions, with AntPool, Huobi, and F2Pool accounting for most of the accelerated transactions. 

\clearpage

%

\chapter{Additional Analysis of Distribution of Voting Power}
\label{appendix:governance}

\section{Compound proposals categorization}\label{sec:proposal_category}

\begin{figure*}[t]
	\centering
		\includegraphics[width={1.2\onecolgrid}]{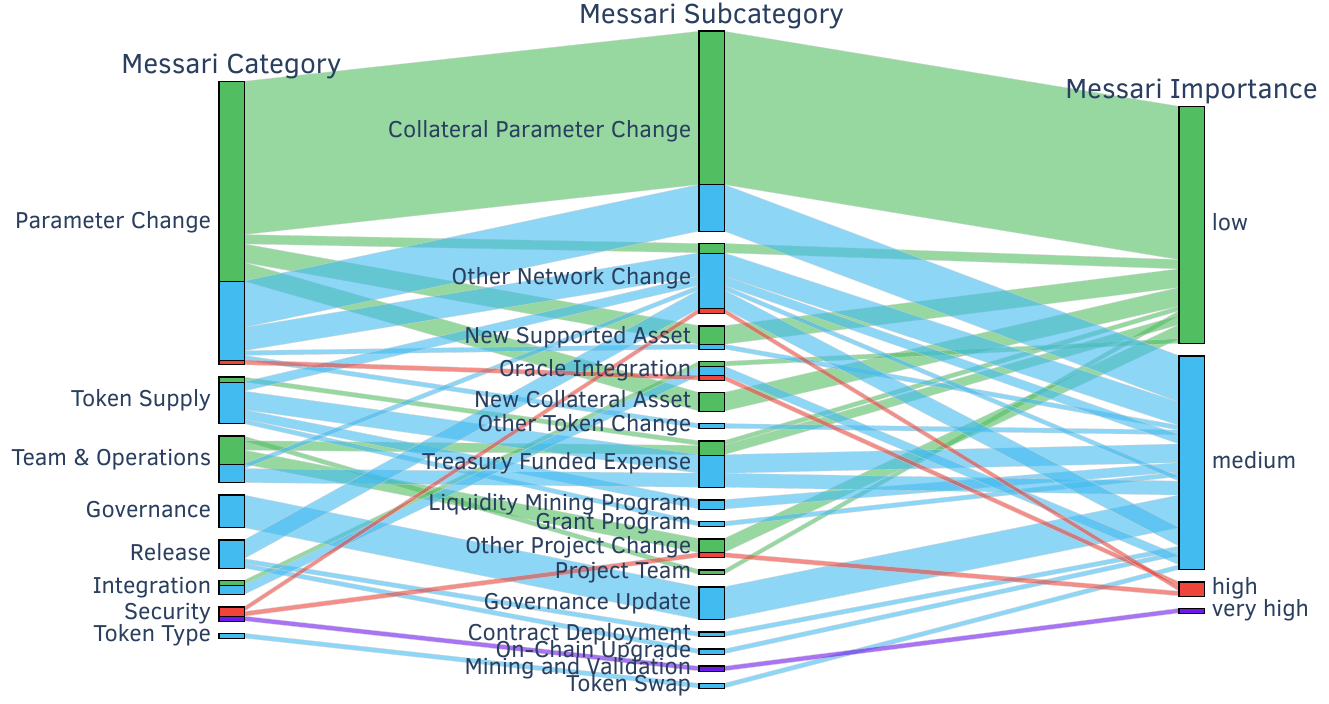}
	\caption{
  Categorization of executed proposals. Most of the proposals (60.4\%) are related to ``Parameter Change''. We also show the importance level (low in \green{green}, medium in \blue{blue}, high in \red{red}, and very high in \purple{purple} color) for each proposal according to Messari~\cite{Compound@Messari}.}
\label{fig:messari-compound-proposal-categorization}
\end{figure*}

We gathered data from Messari~\cite{Compound@Messari} to determine the categories, subcategories, and the level of importance associated with each Compound proposal.
Figure~\ref{fig:messari-compound-proposal-categorization} shows the distribution of \num{101} executed Compound proposals across different categories and subcategories.
We show the degree of importance for each proposal according to Messari divided into ``low'', ``medium'', ``high'', and ``very high''.
As a result, a few proposals categorized as ``Parameter Change'' and ``Security'' demonstrate a high level of importance.
Furthermore, proposals with the highest level of importance are found within the ``Security'' category, specifically within the ``Mining and Validation'' subcategory.
This refers to the proposal 64 that was created to fix a bug introduced by proposal \#62~\cite{Proposal-62@Compound,Proposal-64@Compound}.

The majority of the proposals (\num{61} proposals, accounting for \num{60.4}\%) are related to ``Parameter Change'' followed by ``Team and Operations'' and ``Token Supply'' accounting for \num{10} (\num{9.9}\%) each, and ``Governance'' with \num{7} (\num{6.93}\%) proposals.
According to the level of importance reported by Messari, out of the total of \num{101} executed proposals, \num{51} proposals (\num{50.5}\%) are classified as low importance, \num{46} proposals (\num{45.54}\%) as medium importance, \num{3} proposals as high importance, and \num{1} proposal as very high importance.

\section{Filtering events to construct our Compound data Set}\label{sec:filtering-events}

\begin{figure*}[t]
	\centering
		\includegraphics[width={\textwidth}]{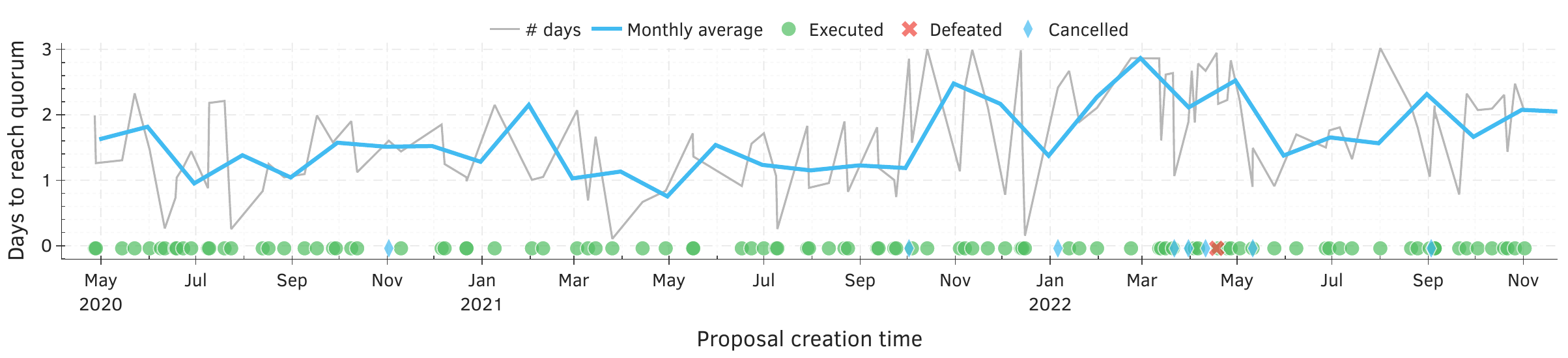}
	\caption{
  Compound proposals typically reach the quorum after 1.64 days on average.}
\label{fig:compound-time-to-quorum}
\end{figure*}

\clearpage
\begin{table*}[t]
\centering
\caption{A comparison of voting mechanisms in decentralized governance protocols such as AAVE~\cite{Governance@AAVE}, Balancer~\cite{Governance@Balancer}, Compound~\cite{leshner2019compound}, Convex Finance~\cite{Governance@ConvexFinance}, Curve~\cite{Governance@Curve}, Maker~\cite{Governance@MakerDAO}, and Uniswap~\cite{adams2021uniswap}. \stress{SC} stands for smart contract.}
\label{table:protocol-types}
\resizebox{\textwidth}{!}{%
\begin{tabular}{rrrp{5cm}rrp{4.5cm}}
\toprule
\thead{Protocol} & \thead{Type} & \thead{Voting} & \thead{Who can vote?} & \thead{Delegation} & \thead{Voting Aggregation} & \thead{How proposals are implemented} \\
\midrule
AAVE & Lending & on-chain & addresses with delegated tokens & yes & on-chain & on-chain via an SC call.\\
Balancer & DEX & off-chain & stakers with locked tokens & yes (off-chain) & off-chain  & via 6-of-11 multisig. \\
Compound & Lending & on-chain & addresses with delegated tokens & yes & on-chain & on-chain via an SC call. \\
Convex Finance & Yield Farming & off-chain & stakers with locked tokens & yes (off-chain) & off-chain & via 3-of-5 multisig.\\
Curve & DEX & on-chain & holders & yes & on-chain & on-chain through an SC call.\\
Maker Executive & Stablecoin & on-chain & holders & no & on-chain & New Governance Contract requires more MKR staked than previous. \\
Maker Polling & Stablecoin & on-chain & addresses with delegated tokens & yes & off-chain & Engineers at Maker create the governance contract based on the voting outcome. \\
Uniswap & DEX & on-chain & addresses with delegated tokens & yes &  on-chain & on-chain via an SC call.\\
\bottomrule
\end{tabular}}
\label{tab:voting-mechanisms}
\end{table*}

This section describes the details required to filter and collect transactions data that triggered events of interest from any smart contract on the Ethereum blockchain.
Before creating a filter, we need the address of our target contract and its Application Binary Interface (ABI).
The ABI is a JSON file that specifies the functions available in the contract, their signatures, and the available events. We can retrieve this information by calling the Etherscan API~\cite{Etherscan_API_Contracts}.
Once we have the contract address and ABI, we can create a filter to track the contract's activity on the Ethereum blockchain using an important Python library for interacting with Ethereum nodes called Web3.py~\cite{web3py} to facilitate the communication with our node's API.  

The Web3.py library provides a filtering function called \stress{createFilter}.
This function can be used to filter transactions that triggered events of interest from a specific contract within a range of block numbers.
We use this function to efficiently collect all transactions that triggered these events from the Compound~\cite{leshner2019compound} smart contract.

\begin{figure*}[tb]
	\centering
		\includegraphics[width={\textwidth}]{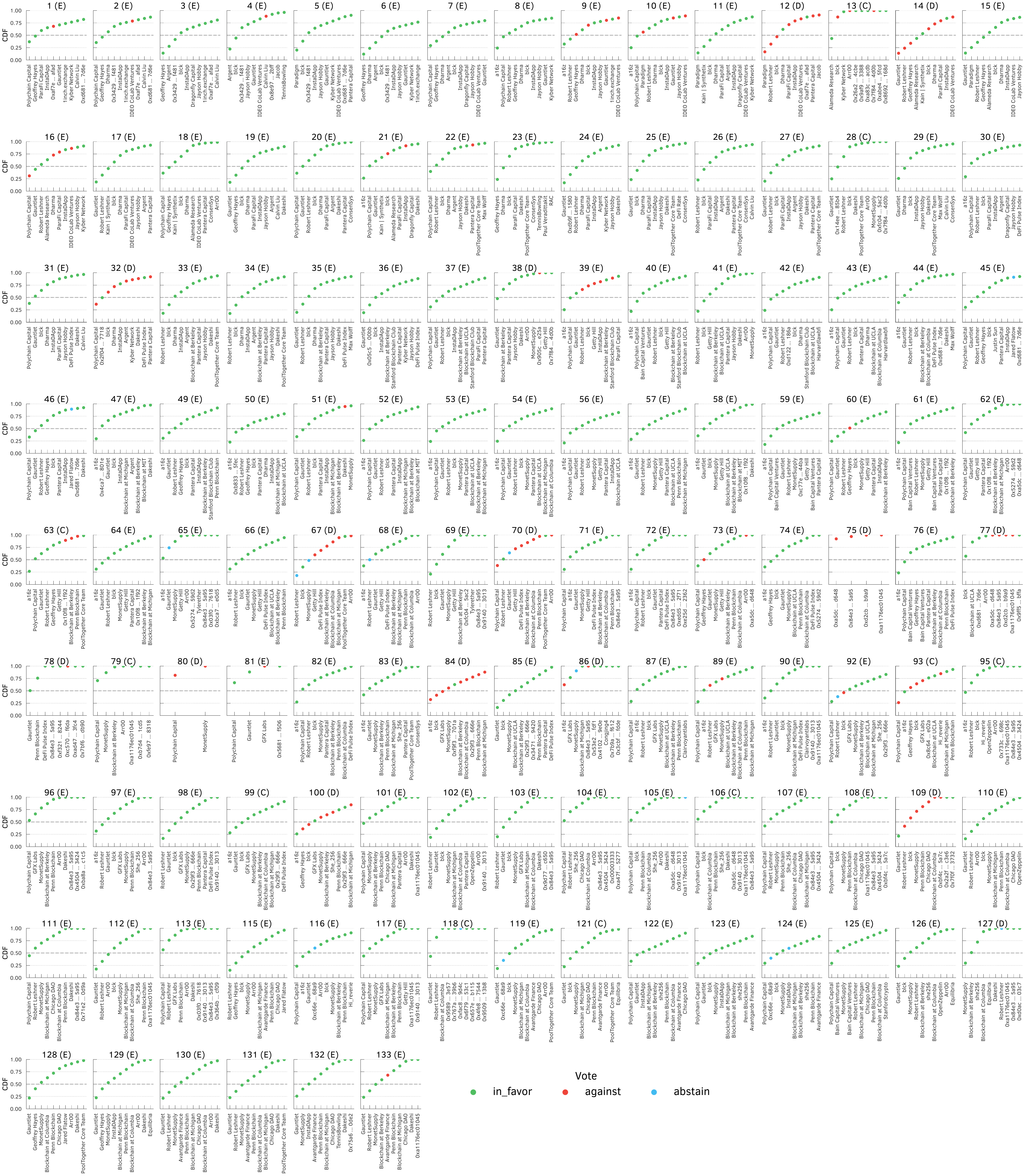}
	\caption{
  Cumulative voting power distribution of the top-10 Compound voters per proposal. On average, proposals required 2.84 voters (std. of 0.97) to reach at least 50\% of their total votes. The median was 3 voters, with a range of 1 to 5 votes. This indicates a concentrated amount of voting power. The subtitles indicate the proposal ID and outcome (``E'' for executed, ``D'' for defeated, and ``C'' for cancelled).}
\label{fig:compound-top-voters}
\end{figure*}

\begin{figure*}[tb]
	\centering
		\includegraphics[width={\textwidth}]{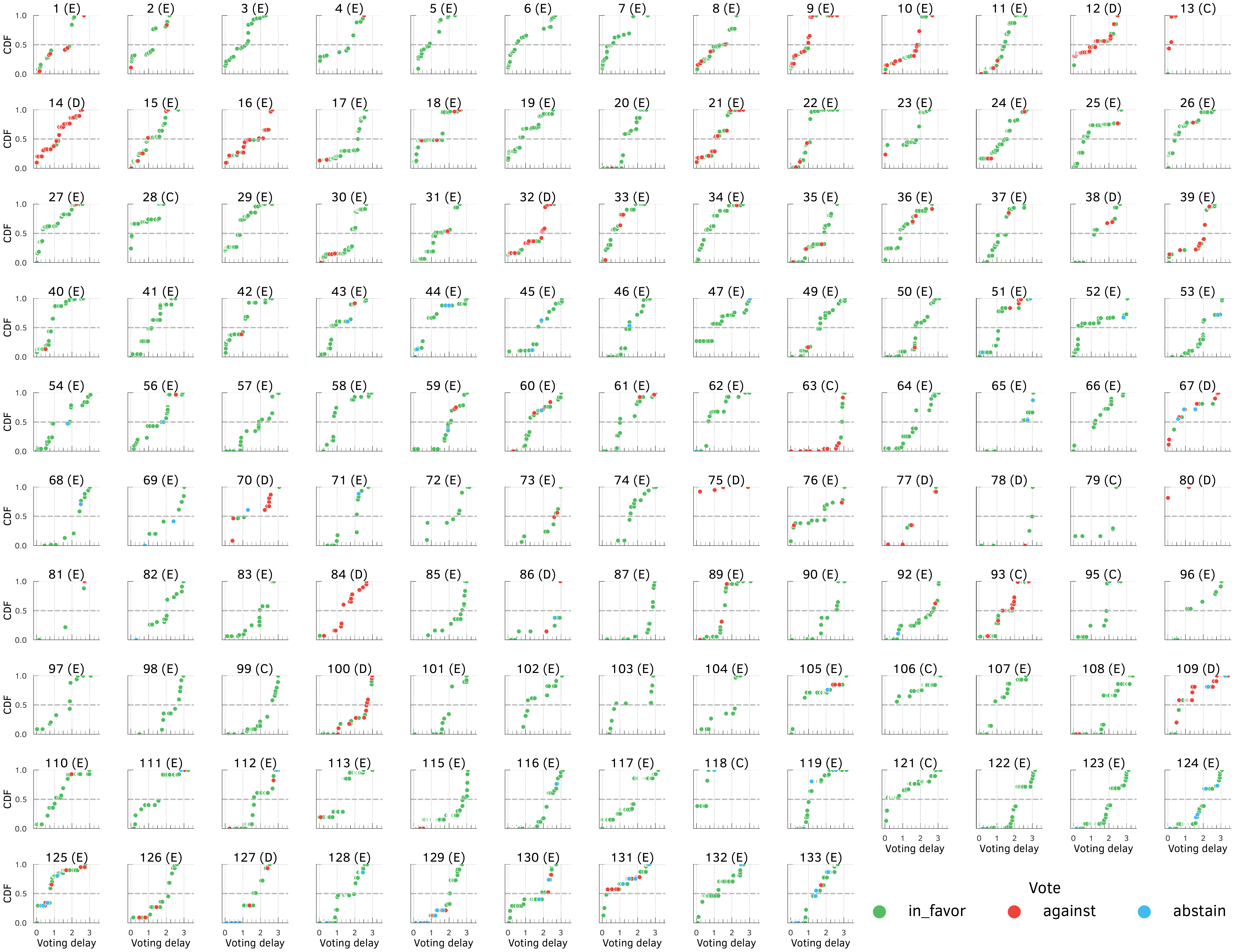}
	\caption{
  Voting delays for all votes cast per proposal in chronological order of vote. On average, voters took 1.4 days (with a standard deviation of 0.95 and a median of 1.34 days) to cast their votes after the voting period began. The subtitles indicate the proposal ID and outcome (``E'' for executed, ``D'' for defeated, and ``C'' for cancelled).}
\label{fig:compound-top-voters-delay-per-day}
\end{figure*}

\section{Inferring wallet addresses ownership}\label{sec:inferring_addresses}

We aim to identify the ownership of public wallet addresses on the Ethereum blockchain.
Due to the inherent anonymity of blockchain addresses, this proves to be a challenging task as we can only know the owners of an address if the owner chooses to disclose it.
However, popular blockchain explorers such as Etherscan~\cite{Etherscan@ETH-explorer} often provide information on the top holders of specific cryptocurrencies, which allows us to partially overcome this obstacle.

Then, we first obtained the lists of the top \num{10000} Ether holders from which there are \num{290} (\num{2.9}\%) identified addresses and the top \num{1000} COMP holders from which there are \num{82} (\num{8.2}\%) identified addresses from Etherscan.
By comparing these lists to our data set, we were able to identify most of the top COMP holder addresses in our sample.
However, this method did not work for the top delegated accounts, as most of them were not included in the list of top COMP holders on Etherscan.
This means that most of the delegated accounts does not hold many tokens. 
Further, we also used the list of top \num{100} delegated Compound addresses by voting weight available on the Compound website~\cite{Voting-weight@Compound} from which there are \num{66} identified addresses.

Furthermore, to extend the available identified addresses in our analysis, we obtained the addresses of \num{2783} identified users from the Sybil-List~\cite{Addresses@Sybil}, a project maintained by Uniswap that uses cryptographic proofs to verify wallet addresses ownership.
By combining the identified addresses from both sources, we were able to obtain the ownership of \num{3191} inferred public wallet addresses to use in our analysis.
We were able to infer \num{114} (\num{3.41}\%) out of the \num{3341} unique addresses in our data set.
Considering the top 10 most powerful voters for each proposal (refer to Figure~\ref{fig:compound-top-voters} in \S\ref{sec:top-voters}), we were able to infer \num{67} (\num{50.37}\%) of the \num{133} unique addresses.
Overall, our methodology allowed us to partially overcome the anonymity of public wallet addresses on the Ethereum blockchain and shed light on the ownership of these addresses in our data set.
Finally, as an entity can control more than one address, we grouped the addresses we identified belonging to the same entity together to conduct our analysis.

\section{Types of existing governance protocols}\label{sec:types-of-governance}

There are various smart contract applications that utilize decentralized governance protocols for decision-making, including those for lending, decentralized exchanges (DEXes), and stablecoins, among others.
An example of such protocols can be found on the Ethereum blockchain, where a number of these applications are available.
We have selected some of the most protocols that use decentralized governance for decision-making.
Table~\ref{table:protocol-types} presents 8 protocols, including Maker Executive and Maker Pooling, which are part of the MakerDAO~\cite{Governance@MakerDAO} stablecoin protocol responsible for the DAI token.
These protocols use decentralized governance mechanisms, and we characterize them based on whether their votes are cast on- or off-chain, the delegation methods they use, how they aggregate the votes, and how the proposal outcome take effect.

\section{How voters cast their votes}\label{sec:top-voters}

This section examines how each of the top-10 voters of Compound and Uniswap cast their votes.
Some proposals may not have received any votes if they were cancelled before the voting period began.
See~\S\ref{subsec:margin_victory_defeat} for details.
Figure~\ref{fig:compound-top-voters} shows how each of the top-10 voters cast their votes in each of the \num{126} (\num{94.74}\%) out of \num{133} Compound proposals.

Figure~\ref{fig:compound-top-voters-delay-per-day} shows the all votes cast in chronological order per proposal. On average, voters took 1.4 days (with a standard deviation of 0.95 and a median of 1.34 days) to cast their votes after the voting period began.

\section{Time until reaching the quorum in Compound}\label{sec:compound-quorum}

For a proposal to pass, it must receive a majority of in favor votes and at least \num{400000} (\num{4}\%) votes in favor from the total supply of Compound tokens.
This minimum number of in favor votes is referred to as the \stress{quorum} and is defined by the Compound Governor Bravo contract.

We analyzed how long it takes for these proposals to reach the required quorum.
Figure~\ref{fig:compound-time-to-quorum} shows the number of days it took each of the evaluated Compound proposals to reach the quorum.
On average, it takes \num{1.64} days with a standard deviation of \num{0.72} days for the proposals to reach the quorum.
The cumulative distribution function of our results, where \num{32}\% take more than \num{2} days to reach the quorum. The shortest and longest time it took was \num{0.11} and \num{3} days, respectively.

\clearpage

\end{appendices}


\clearpage
\phantomsection
{
\addcontentsline{toc}{chapter}{Bibliography}

\bibliographystyle{apalike} 

\bibliography{thesis}
}

\end{document}